\documentclass[number]{elsarticle}
\biboptions{sort&compress}

\makeatletter \RequirePackage[bookmarks,unicode,colorlinks=true]{hyperref}%
\def\@citecolor{blue}%
\def\@urlcolor{blue}%
\def\@linkcolor{blue}%

\def\orcidID#1{\href{http://orcid.org/#1}{\smash{\protect\raisebox{-1.25pt}{\protect\includegraphics{orcid_color.eps}}}}}
\makeatother

\usepackage{amsthm}
\usepackage[usenames,dvipsnames]{xcolor,colortbl}
\usepackage{mathtools}
\usepackage{multirow}
\usepackage{stmaryrd}
\usepackage{cancel}
\usepackage{amsmath,amssymb}
\usepackage{thmtools, thm-restate}
\usepackage[shortlabels]{enumitem}
\usepackage{wrapfig}
\usepackage{pbox}
\usepackage{marvosym}
\usepackage{listings}
\usepackage[ruled, vlined, linesnumbered]{algorithm2e}
\usepackage{placeins}
\usepackage{float}
\usepackage{tikz}
\usepackage{caption}
\usepackage{subcaption}
\usepackage{booktabs}
\usepackage{array,longtable}
\newcolumntype{C}{>{$}c<{$}}
  \newcolumntype{L}{>{$}l<{$}}
\newcolumntype{R}{>{$}r<{$}}
\usepackage{nicefrac,xfrac}
\usepackage{hyperref}
\usepackage[doipre={doi:~}]{uri}
\usetikzlibrary{shapes,calc,arrows,automata,decorations.pathmorphing,backgrounds,arrows.meta,shapes.geometric,shapes.multipart,shapes.misc}
\pgfdeclarelayer{edgelayer}
\pgfdeclarelayer{nodelayer}
\pgfsetlayers{background,edgelayer,nodelayer,main}
\tikzstyle{none}=[inner sep=0mm] \tikzstyle{moveBlock}=[fill=white, draw=black, shape=rectangle] \tikzstyle{target}=[fill=white, draw=black, shape=circle]

\tikzstyle{dotHead}=[dotted, ->] \tikzstyle{dotWithoutHead}=[dotted, -] \tikzstyle{dashHead}=[dashed,->] \tikzstyle{dashWithoutHead}=[dashed,-] \tikzstyle{arrow}=[->]

\RequirePackage{makecell}

\usepackage[capitalize,nameinlink]{cleveref}
\usepackage{IEEEtrantools}

\newenvironment{myproof}[1][\unskip]{
  \noindent{\it Proof #1.}
}{\qed
  \medskip
}
\newenvironment{myproofsketch}{
	\noindent{\it Proof Sketch.}
}{\qed
	\medskip
}

\SetKwInOut{Input}{input}
\SetKwInOut{Output}{output}

\def\namedlabel#1#2{\begingroup #2%
  \def\@currentlabel{#2}%
  \phantomsection\label{#1}
  \endgroup
}

\renewcommand{\appendix}{\par \setcounter{section}{0}
  \setcounter{subsection}{0}
  \gdef\thesection{\Alph{section}}
}

\newcommand{\disabledcomment}[1]{}
\newcommand{\oldcomment}[1]{}

\renewcommand\cellgape{\Gape[4pt]}

\theoremstyle{plain}

\newtheorem{thm}{Theorem}[section] 
\newtheorem{lem}[thm]{Lemma}

\theoremstyle{definition}

\newtheorem{rem}[thm]{Remark}

\newtheorem{exa}[thm]{Example}

\newtheorem{defi}[thm]{Definition}

\newtheorem{com}[thm]{Comment}

\newcommand{\tool}[1]{\textsf{#1}}

\newcommand{\dash}{\textthreequartersemdash\xspace}

\newcommand{\aprove}{\textsf{AProVE}}
\newcommand{\muterm}{\textsf{MU-TERM}}
\newcommand{\natt}{\textsf{NaTT}}
\newcommand{\ttttwo}{\textsf{T\kern-0.15em\raisebox{-0.55ex}T\kern-0.15emT\kern-0.15em\raisebox{-0.55ex}2}}

\newcommand{\ceta}{\textsf{CeTA}}
\newcommand{\matchbox}{\textsf{Matchbox}}
\newcommand{\multumnonmulta}{\textsf{MultumNonMulta}}
\newcommand{\nti}{\textsf{NTI}}

\renewcommand{\emph}[1]{\index{#1}\textit{#1}}

\renewcommand{\emptyset}{\varnothing}

\newcommand{\IN}{\mathbb{N}}
\newcommand{\IR}{\mathbb{R}}

\newcommand{\IE}{\ensuremath{\mathbb{E}}}

\newcommand{\F}[1]{\mathfrak{#1}}

\makeatletter \def\moverlay{\mathpalette\mov@rlay}
\def\mov@rlay#1#2{\leavevmode\vtop{%
    \baselineskip\z@skip \lineskiplimit-\maxdimen \ialign{\hfil$\m@th#1##$\hfil\cr#2\crcr}}}
\newcommand{\charfusion}[3][\mathord]{
  #1{\ifx#1\mathop\vphantom{#2}\fi \mathpalette\mov@rlay{#2\cr#3}
    }
  \ifx#1\mathop\expandafter\displaylimits\fi}
\makeatother

\renewcommand{\P}{\mathcal{P}}

\newcommand{\Proc}{\operatorname{Proc}}

\newcommand{\PosDPoss}{\pos_{\mathcal{D} \land \lnot \mathtt{NF}_{\R}}}

\newcommand{\flateq}{\stackrel{\flat}{=}}
\newcommand{\Junk}{\mathrm{Junk}}

\newcommand{\TSet}[2]{\mathcal{T}\left(#1,#2\right)}

\newcommand{\VSet}{\mathcal{V}}

\newcommand{\R}{\mathcal{R}}

\newcommand{\DPair}[1]{\mathcal{DP}(#1)}

\newcommand{\FDist}{\operatorname{FDist}}
\newcommand{\Supp}{\operatorname{Supp}}

\newcommand{\rootsym}{\operatorname{root}}
\newcommand{\capterm}{\mathit{Cap}}

\newcommand{\renterm}{\mathit{Ren}}
\newcommand{\rules}{\operatorname{Rules}}
\newcommand{\urules}{\mathcal{U}}

\newcommand{\vr}{\operatorname{vr}}
\newcommand{\capt}{\operatorname{cov}}

\newcommand{\Pol}{\operatorname{Pol}}

\newcommand{\nonprob}{\normalfont{\text{np}}}
\newcommand{\nonprobDP}{\normalfont{\text{dp}}}
\newcommand{\PP}{\mathcal{P}}

\renewcommand{\AA}{\mathcal{A}}

\newcommand{\QQ}{\mathcal{Q}}

\newcommand{\tex}{\mathsf{ex}}

\newcommand{\tplus}{\mathsf{plus}}
\newcommand{\ts}{\mathsf{s}}
\newcommand{\tz}{\mathsf{0}}

\newcommand{\tf}{\mathsf{f}}
\newcommand{\tg}{\mathsf{g}}
\renewcommand{\th}{\mathsf{h}}
\newcommand{\ta}{\mathsf{a}}
\newcommand{\tb}{\mathsf{b}}
\newcommand{\tc}{\mathsf{c}}
\newcommand{\td}{\mathsf{d}}
\newcommand{\te}{\mathsf{e}}

\newcommand{\tq}{\mathsf{q}}

\newcommand{\tffg}{\mathsf{ffg}}

\newcommand{\tic}{\mathsf{incpl}}
\newcommand{\tcons}{\mathsf{cons}}
\newcommand{\tnil}{\mathsf{nil}}

\newcommand{\tmoveelements}{\mathsf{moveElements}}
\newcommand{\tor}{\mathsf{or}}

\newcommand{\tif}{\mathsf{if}}
\newcommand{\ttrue}{\mathsf{true}}
\newcommand{\tfalse}{\mathsf{false}}

\newcommand{\tgt}{\mathsf{gt}}
\newcommand{\tp}{\mathsf{p}}

\newcommand{\tD}{\mathsf{D}}
\newcommand{\tF}{\mathsf{F}}
\newcommand{\tG}{\mathsf{G}}
\newcommand{\tH}{\mathsf{H}}
\newcommand{\tA}{\mathsf{A}}

\newcommand{\trotate}{\mathsf{rotate}}
\newcommand{\tapp}{\mathsf{app}}
\newcommand{\tqs}{\mathsf{qsrt}}

\newcommand{\tifhigh}{\mathsf{ifHigh}}
\newcommand{\thigh}{\mathsf{high}}
\newcommand{\tiflow}{\mathsf{ifLow}}
\newcommand{\tlow}{\mathsf{low}}
\newcommand{\tleq}{\mathsf{leq}}
\newcommand{\tinit}{\mathsf{init}}

\newcommand{\tloopOne}{\mathsf{loop1}}
\newcommand{\tloopTwo}{\mathsf{loop2}}
\newcommand{\tdouble}{\mathsf{double}}
\newcommand{\ttriple}{\mathsf{triple}}

\newcommand{\tLoopOne}{\mathsf{L1}}
\newcommand{\tLoopTwo}{\mathsf{L2}}
\newcommand{\tDouble}{\mathsf{D}}
\newcommand{\tTriple}{\mathsf{T}}

\newcommand{\thd}{\mathsf{hd}}
\newcommand{\ttail}{\mathsf{tl}}
\newcommand{\tisempty}{\mathsf{empty}}

\newcommand{\teven}{\mathsf{even}}
\newcommand{\tloop}{\mathsf{loop}}
\newcommand{\tevenif}{\mathsf{if}}
\newcommand{\tstop}{\mathsf{stop}}
\newcommand{\xs}{\mathit{xs}}
\newcommand{\ys}{\mathit{ys}}

\newcommand{\tbot}{\mathsf{bot}}

\newcommand{\tsum}{\mathsf{sum}}
\newcommand{\taddNum}{\mathsf{addNum}}
\newcommand{\tcreateL}{\mathsf{createL}}
\newcommand{\ttree}{\mathsf{tree}}
\newcommand{\tleaf}{\mathsf{leaf}}
\newcommand{\tconcat}{\mathsf{concat}}
\newcommand{\tcreateT}{\mathsf{createT}}
\newcommand{\tlessleaves}{\mathsf{lessleaves}}

\newcommand{\ctroot}{\F{r}}
\newcommand{\aV}{\Gamma}
\newcommand{\nV}{\overline{\Gamma}}
\newcommand{\ctleaf}{\operatorname{Leaf}}

\newcommand{\ctdepth}{\operatorname{d}}

\newcommand{\ctlevelTwo}{\mathcal{L}}
\newcommand{\ctlevelTwowithborder}{\F{L}}

\newcommand{\tval}{\mathit{\mathcal{V}al}_{\F{T}}}
\newcommand{\adtval}{\overline{\mathit{\mathcal{V}al}}_{\F{T}}}
\newcommand{\adtvaltwo}{\widehat{\mathit{\mathcal{V}al}}_{\F{T}}}

\newcommand{\proj}{\mathit{proj}}

\crefname{definition}{Def.}{Def.}
\crefname{defi}{Def.}{Def.}
\crefname{example}{Ex.}{Ex.}
\crefname{exa}{Ex.}{Ex.}
\crefname{counterexample}{Counterex.}{Counterex.}
\crefname{appendix}{App.}{App.}
\crefname{ex}{Ex.}{Ex.}
\crefname{theorem}{Thm.}{Thm.}
\crefname{thm}{Thm.}{Thm.}
\crefname{lemma}{Lemma}{Lemmas}
\crefname{remark}{Remark}{Remarks}
\crefname{rem}{Remark}{Remarks}
\crefname{section}{Sect.}{Sect.}
\crefname{subsection}{Sect.}{Sect.}
\crefname{subsubsection}{Sect.}{Sect.}
\crefname{line}{Line}{Lines}
\crefname{corollary}{Cor.}{Cor.}
\crefname{figure}{Fig.}{Fig.}
\crefname{enumi}{}{}
\crefname{algorithm}{Alg.}{Alg.}

\makeatletter \NewDocumentCommand{\dparrow}{+O{} +O{0.5cm}}{%
\begin{tikzpicture}[baseline=-0.5ex]
  {
  \node[inner sep=0](@1) at (0,0) {};
  \node[inner sep=0](@2) at (#2,0) {};
  \draw [arrows={-Triangle[open]},shorten >= 1pt,shorten <= 1pt](@1)--(@2) node[pos=.5,above,inner sep=1pt] {\ensuremath{#1}};}
\end{tikzpicture}
\xspace
}

\NewDocumentCommand{\myto}{+O{} +O{0.5cm}}{%
\begin{tikzpicture}[baseline=-0.5ex]
  {
    \node[inner sep=0](@1) at (0,0) {};
    \node[inner sep=0](@2) at (#2,0) {};
    \draw [arrows={-to}](@1)--(@2) node[pos=.5,above,inner sep=1pt] {\ensuremath{#1}};}
\end{tikzpicture}
\xspace
}

\NewDocumentCommand{\paraarrow}{+O{} +O{0.4cm}}{%
\begin{tikzpicture}[baseline=-0.5ex]
  {
    \node[inner sep=0](@1) at (0,0) {};
    \node[inner sep=0](@2) at (#2,0) {};
    \node[inner sep=0](@3) at (0.07,0) {};
    \draw [arrows={-to}](@1)--(@2) node[pos=.5,above,inner sep=1pt] {\ensuremath{#1}};
    \draw [arrows={-to}](@1)--(@3);}
\end{tikzpicture}
\xspace
}

\NewDocumentCommand{\paradparrow}{+O{} +O{0.4cm}}{%
\begin{tikzpicture}[baseline=-0.5ex]
  {
  \node[inner sep=0](@1) at (0,0) {};
  \node[inner sep=0](@2) at (#2,0) {};
  \node[inner sep=0](@3) at (0.07,0) {};
  \draw [arrows={-Triangle[open]}](@1)--(@2) node[pos=.5,above,inner sep=1pt] {\ensuremath{#1}};
  \draw [arrows={-to}](@1)--(@3);}
\end{tikzpicture}
\xspace
}

\newcommand{\oset}[2]{%
  {\mathop{#2}\limits^{\vbox to 1\ex@{\kern-\tw@\ex@ \hbox{\scriptsize #1}\vss}}}}

\newcommand{\osetthree}[2]{%
  {\mathop{#2}\limits^{\vbox to 3\ex@{\kern-\tw@\ex@ \hbox{\scriptsize #1}\vss}}}}

\newcommand{\osetfive}[2]{%
  {\mathop{#2}\limits^{\vbox to 5\ex@{\kern-\tw@\ex@ \hbox{\scriptsize #1}\vss}}}}

\newcommand{\osetminus}[2]{%
  {\mathop{#2}\limits^{\vbox to -2\ex@{\kern-\tw@\ex@ \hbox{\scriptsize #1}\vss}}}}
\makeatother

\newcommand{\epsto}{\mathrel{\smash{\stackrel{\raisebox{3.4pt}{\tiny $\varepsilon\:$}}{\smash{\rightarrow}}}}}
\newcommand{\ito}{\mathrel{\smash{\stackrel{\raisebox{3.4pt}{\tiny $\mathsf{i}\:$}}{\smash{\rightarrow}}}}}
\newcommand{\itomaybe}{\mathrel{\smash{\stackrel{\raisebox{3.4pt}{\tiny $(\mathsf{i})\:$}}{\smash{\rightarrow}}}}}
\newcommand{\iepsto}{\mathrel{\smash{\stackrel{\raisebox{3.4pt}{\tiny $\varepsilon\:\mathsf{i}\:$}}{\smash{\longrightarrow}}}}}

\newcommand{\itorstar}{\mathrel{\ito_{\R}^{*}}}

\newcommand{\itor}{\mathrel{\ito_{\R}}}

\newcommand{\fs}[1]{\mathsf{#1}}
\newcommand{\fun}[1]{\mathrm{#1}}

\renewcommand{\phi}{\varphi}
\renewcommand{\emptyset}{\varnothing}
\newcommand{\true}{\fs{true}}
\newcommand{\false}{\fs{false}}

\newcommand{\pos}{\fun{Pos}}

\newcommand{\posT}{\fun{Pos}_{\SignatureA}}

\newcommand{\posD}{\fun{Pos}_{\SignatureD}}
\newcommand{\posDT}{\fun{Pos}_{\SignatureD \cup \SignatureA}}

\newcommand{\anno}{\sharp} 

\newcommand{\annoEps}{\anno_{\varepsilon}}
\newcommand{\annoD}{\anno_{\SignatureD}}
\newcommand{\disannoPos}[1]{\flat_{#1}^{\uparrow}}

\newcommand{\TT}{\mathcal{T}}

\newcommand{\VV}{\mathcal{V}}

\newcommand{\SignatureDC}{\Sigma}
\newcommand{\SignatureADC}{\Sigma^\sharp}

\newcommand{\SignatureC}{\mathcal{C}}
\newcommand{\SignatureD}{\mathcal{D}}
\newcommand{\SignatureA}{\mathcal{D}^\sharp}

\newcommand{\NF}{\mathtt{NF}}
\newcommand{\ANF}{\mathtt{ANF}}

\newcommand{\NN}{\mathbb{N}}

\newcommand{\ruleArr}[3]{
\mathrel{
\xrightarrow{{}_{\scriptstyle #1}}
\!\!{}^{#2}_{#3}
}
}
\newcommand{\tored}[3]{
\mathrel{
\xhookrightarrow{{}_{\scriptstyle #1}}
\!\!{}^{#2}_{#3}
}
}
\newcommand{\itored}[3]{
  \mathrel{
    \smash{\stackrel{\raisebox{3.4pt}{\tiny $\mathsf{i}\:$}}{\smash{\hookrightarrow}}}^{#2}_{#3}
  }
}

\newcommand{\defemph}[1]{{\rm #1}}

\journal{Science of Computer Programming}

\begin{document}

\begin{frontmatter}

  \title{The Annotated Dependency Pair Framework\\
    for Almost-Sure Termination\\
    of Probabilistic Term Rewriting}
  \tnotetext[]{funded by the Deutsche Forschungsgemeinschaft (DFG) Research Training Group 2236 UnRAVeL}

  \author{Jan-Christoph Kassing}
  \ead{kassing@cs.rwth-aachen.de}
  
  \author{Jürgen Giesl}
  \ead{giesl@cs.rwth-aachen.de}

  \affiliation{organization={RWTH Aachen University},
      city={Aachen},
      country={Germany}}

  \begin{abstract}
    Dependency pairs are one of the most powerful techniques to analyze termination of term rewrite systems automatically.
    We adapt dependency pairs to the probabilistic setting and develop an \emph{annotated dependency pair} framework for automatically proving almost-sure termination of \emph{probabilistic term rewrite systems}, both for full and innermost rewriting.
    To evaluate its power, we implemented our framework in the tool \aprove{}.
\end{abstract}

  \begin{keyword}
    Dependency Pairs \sep Term Rewriting \sep Probabilistic Programming \sep Almost-Sure Termination
  \end{keyword}

\end{frontmatter}

\section{Introduction}
\label{Introduction}

\noindent
Term rewrite systems (TRSs) are used in many areas of computer science, like symbolic computation, automated theorem proving, and automated analysis of\linebreak
programs in different languages.
Termination is one of the most crucial pro\-perties of programs, and numerous tools have been developed to analyze termination of TRSs, e.g., \aprove{}~\cite{JAR-AProVE2017}, \matchbox{}~\cite{matchbox_sys}, \multumnonmulta{}~\cite{GHW_FSCD19}, \muterm{}~\cite{gutierrez_mu-term_2020}, \natt{}~\cite{natt_sys_2014}, \nti{}~\cite{Payet2024}, \ttttwo{}~\cite{ttt2_sys}, etc.
One of the most powerful techniques integrated in essentially all current termination tools for TRSs is the \emph{dependency pair} (DP) framework \cite{arts2000termination,gieslLPAR04dpframework,giesl2006mechanizing,hirokawa2005automating,DBLP:journals/iandc/HirokawaM07} which allows modular proofs that apply different techniques in different sub-proofs.

The DP framework also allows for certification of proofs via tools like, e.g., \ceta{}~\cite{ceta_sys},
and DPs have been formalized in several proof assistants 
(e.g., in \tool{Rocq} (formerly \tool{Coq})~\cite{Contejean07,Blanqui11},
\tool{Isabelle}~\cite{ceta_sys}, and \tool{PVS}
\cite{almeida_formalizing_2020}).

\emph{Probabilistic} programs describe randomized algorithms and probability distributions, with applications in many areas, see, e.g.,~\cite{Gordon14}.
For instance, well-known examples are variants of the probabilistic quicksort algorithm, see, e.g., \Cref{Evaluation}.
If one uses a version where every list element has the same probability of becoming the next pivot element, then the algorithm has an expected runtime of $\mathcal{O}(n \cdot \log(n))$, for every possible input list of length~$n$.
In contrast, for a non-probabilistic version of quicksort, there always exists a ``worst-case'' input of length $n$ such that the algorithm takes at least $\mathcal{O}(n^2)$ steps.
Several probabilistic programming languages have been developed.
An example is\linebreak
the probabilistic guarded command language (\textsf{pGCL})~\cite{Kozen85,McIverMorgan05}, whose basic form however cannot express arbitrary data structures.
In contrast, TRSs are especially suitable for modeling and analyzing algorithms like quicksort that operate on (user-defined) data structures like lists, and they were first extended to \emph{probabilistic term rewrite systems} (PTRSs) in~\cite{BournezRTA02,bournez2005proving}.

In the probabilistic setting, there are several notions of ``termination''.
One of the most important notions is \emph{almost-sure termination} ($\mathtt{AST}$), which means that the probability of termination is $1$.
A strictly stronger notion is \emph{positive $\mathtt{AST}$} ($\mathtt{PAST}$), which requires that the expected runtime is finite.
Moreover, there are several different rewrite strategies.
For example, \emph{innermost} rewriting only allows rewrite steps at innermost positions in a term.
This strategy is the easiest to analyze for termination, and it corresponds to a call-by-value strategy as used by many programming languages.
Thus, in this paper we will not only investigate $\mathtt{AST}$ for PTRSs, but we also consider innermost $\mathtt{AST}$ ($\mathtt{iAST}$, allowing only innermost rewrite sequences).

\begin{wrapfigure}[14]{r}{0.35\textwidth}
  \vspace*{-.6cm}
  \begin{minipage}{0.35\textwidth}
    \hspace*{-.1cm}%
    \begin{algorithm}[H]
      \DontPrintSemicolon
      \DecMargin{1cm}
      \caption{}
      \label{alg1}
      $x \gets 0$\;
      \While{$x = 0$}
      {
        $\{$\;
        $\phantom{\{} x \gets 0 \oplus_{\nicefrac{1}{2}} x \gets 1;\hspace*{-3.8cm}$ \hspace*{-3.8cm}\;
        $\phantom{\{} y \gets 2 \cdot y;$\;
        $\} \square \{$\;
        $\phantom{\{} x \gets 0 \oplus_{\nicefrac{1}{3}} x \gets 1;$\hspace*{-3.8cm}\;
        $\phantom{\{} y \gets 3 \cdot y;$\;

        $\}$
      }
      \While{$y > 0$}
      {
        $y \gets y - 1;$\;
      }
    \end{algorithm}
  \end{minipage}
\end{wrapfigure}
As a running example in the first part of this\linebreak
paper (\Cref{Introduction} - \ref{The Probabilistic ADP Framework}), we transform \Cref{alg1} on the right (written in \textsf{pGCL}) into an equivalent PTRS and show how our framework proves $\mathtt{AST}$.
Here, $\oplus_{\nicefrac{1}{2}}$ denotes probabilistic choice, and $\square$ denotes demonic non-determinism.
In every iteration of the first loop, we have a chance of at least
$\nicefrac{1}{3}$ 
to leave the first loop and enter the second one, which in turn
terminates after a finite number of steps.
Hence, the termination probability for \Cref{alg1} is at least
$\nicefrac{1}{3} +
\nicefrac{2}{3} \cdot \nicefrac{1}{3} +
(\nicefrac{2}{3})^2 \cdot \nicefrac{1}{3} + \ldots =
\nicefrac{1}{3} \cdot
\sum_{i = 0}^{\infty} (\nicefrac{2}{3})^n = 1$. 
Note that while \Cref{alg1} is $\mathtt{AST}$, its expected runtime is infinite, i.e., it is not $\mathtt{PAST}$.
This already holds for the program where only the first possibility of the first \textbf{while}-loop is considered (i.e., where $y$ is always doubled in its body).
Then for the initial value $y = 1$, the expected number of iterations of the second \textbf{while}-loop which decrements $y$ is $\tfrac{1}{2} \cdot 2 + \tfrac{1}{4} \cdot 4 + \tfrac{1}{8} \cdot 8 + \ldots = 1 + 1 + 1 + \ldots = \infty$.
In this paper, 
we use
\Cref{alg1} as a leading example to illustrate
how to handle algorithms with non-determinism. While the non-deterministic choice does not
affect the termination behavior of \Cref{alg1},  
in general it can indeed influence termination.

There exist numerous techniques to prove ($\mathtt{P}$)$\mathtt{AST}$ of imperative programs on numbers like \Cref{alg1} (e.g., \cite{kaminski2018weakest,mciver2017new,TACAS21,lexrsm,FoundationsTerminationMartingale2020,FoundationsExpectedRuntime2020,rsm,cade19,dblp:journals/pacmpl/huang0cg19,amber,ecoimp,absynth}).
In fact, there are proof rules (e.g.,~\cite{mciver2017new}) and tools (e.g.,~\cite{amber}) that can prove $\mathtt{AST}$ for both loops of \cref{alg1} individually, and hence for the whole algorithm.
Moreover, the tool \textsf{Caesar}~\cite{DBLP:journals/pacmpl/SchroerBKKM23}
can prove $\mathtt{AST}$ if one provides super-martingales for the two loops.
However, there are only few automated methods available to analyze $\mathtt{(P)AST}$ for programs with complex non-tail recursive structure \cite{beutner2021probabilistic,Dallago2017ProbSizedTyping,lago_intersection_2021}.
The approaches that are suitable for algorithms on recursive data structures \cite{wang2020autoexpcost,LeutgebCAV2022amor,KatoenPOPL23}
are mostly tailored to specific structures, making it challenging to adapt them to other (possibly user-defined) data structures, or are not yet fully automated.
To the best of our knowledge, there exist no automatic techniques to handle algorithms similar to \Cref{alg1}
that operate on arbitrary algebraic data structures, i.e., (non-deterministic) algorithms that first create a random data object $y$ in a first loop and then access or modify it in a second loop using auxiliary functions, whereas this is possible with our \emph{annotated dependency pair}
framework.\footnote{Such examples can be found in our benchmark set, see \Cref{Evaluation}
  and App.\ \ref{Examples}.}
In this paper we develop an approach for fully automatic termination analysis of (arbitrary) probabilistic TRSs that can successfully analyze probabilistic programs like the probabilistic quicksort algorithm mentioned above.

Up to now, there were only two approaches for automatic termination analysis of PTRSs.
In~\cite{avanzini2020probabilistic}, orders based on interpretations were adapted to prove $\mathtt{PAST}$, and we presented a related technique to prove $\mathtt{AST}$ in~\cite{kassinggiesl2023iAST}.
However, already for non-probabilistic TRSs such a direct application of orders is limited in power.
To obtain a powerful approach, one should combine such orders in a modular way, as in the DP framework.
Thus, we presented a first version of a probabilistic DP framework for innermost rewriting in~\cite{kassinggiesl2023iAST}.
However, in contrast to the DP framework for ordinary TRSs, the framework of~\cite{kassinggiesl2023iAST} was \emph{incomplete}, i.e., there are PTRSs which are $\mathtt{iAST}$ but where this cannot be proved with DTs.

Therefore, in this paper we introduce the new concept of \emph{annotated dependency pairs} (ADPs) to analyze termination of PTRSs.
In this way, we obtain a novel complete criterion for $\mathtt{iAST}$ via DPs while maintaining soundness for all processors that were developed in the probabilistic framework of~\cite{kassinggiesl2023iAST}.
Moreover, our improvement allows for additional more powerful ``transformational''
probabilistic DP processors which were not possible in the framework
of~\cite{kassinggiesl2023iAST}.

In~\cite{FOSSACS24} we showed that there exist classes of PTRSs where it suffices to analyze $\mathtt{iAST}$ in order to prove $\mathtt{AST}$ for full rewriting.
However, these classes are quite restrictive (they exclude PTRSs with non-probabilistic non-determinism and impose linearity restrictions on both sides of the rewrite rules), 
e.g., a PTRS corresponding to \Cref{alg1} would not be contained in this class. 
Thus, up to now there were no powerful techniques to prove full $\mathtt{AST}$ for more general forms of PTRSs.
Therefore, in the current paper we show that our ADP framework 
can also be adapted to analyze full (instead of innermost) $\mathtt{AST}$.
In particular, our novel ADP framework can also be applied to prove $\mathtt{AST}$ for overlapping PTRSs and it weakens the linearity requirements considerably.

We start with preliminaries on (probabilistic) term rewriting and recapitulate the most important notions of the DP framework for non-probabilistic TRSs in \cref{Preliminaries}.
Then, we present our novel ADPs for probabilistic TRSs in \cref{Probabilistic Annotated Dependency Pairs}.
In \cref{The Probabilistic ADP Framework}, we explain the general idea of the probabilistic ADP framework, and present its fundamental processors.
In addition, our framework allows for the definition of new processors which \emph{transform} ADPs.
To increase its power, we extend the probabilistic ADP framework by \emph{transformational processors} in \Cref{ADP Transformations}.
The implementation of our approach in the tool \aprove{} is evaluated in \cref{Evaluation}.
In App.\ \ref{Examples},
we illustrate our approach on several examples, including PTRSs with non-numerical data
structures like lists or trees. In the main part of the paper, we give proof sketches for
all theorems and we
refer to App.~\ref{Proofs} for the full proofs.

The current paper is based on our earlier conference papers~\cite{FLOPS2024}
and~\cite{JPK60}, and it extends them by:

\begin{itemize}
  \item A unified formalization and presentation of the ADP framework with all processors for both innermost and full rewriting, whereas these strategies were previously handled in different frameworks for innermost \cite{FLOPS2024} and full rewriting \cite{JPK60}.
        \vspace*{-.15cm}
  \item A new more general version of the \emph{reduction pair processor}
        (\Cref{sec:Reduction Pair Processor}) that allows us to use arbitrary algebras where addition and the barycentric operation $\IE$ are defined (whereas this processor was restricted to polynomial interpretations in \cite{FLOPS2024,JPK60}).
        \vspace*{-.15cm}
  \item A new processor based on the \emph{subterm criterion} of \cite{DBLP:journals/iandc/HirokawaM07} (\Cref{sec:Subterm Criterion}) in the ADP framework for innermost $\mathtt{AST}$.
        \vspace*{-.15cm}
  \item A new discussion and counterexamples to show why the linearity requirement is needed for the \emph{rewriting processor} in the ADP framework (\Cref{sec:Rewriting Processor}).
        \vspace*{-.15cm}
  \item A new \emph{instantiation processor} (\Cref{sec:Instantiation Processor}), which is applicable both in the ADP framework for full and innermost $\mathtt{AST}$.
        \vspace*{-.15cm}
  \item A new \emph{forward instantiation processor} (\Cref{sec:Forward Instantiation Processor}), which is also applicable in the ADP framework for full and innermost $\mathtt{AST}$.
        \vspace*{-.15cm}
  \item A new \emph{rule overlap instantiation processor} (\Cref{sec:Rule Overlap Processor}), which is applicable in the ADP framework for innermost $\mathtt{AST}$.
        \vspace*{-.15cm}
        \item Proofs for all our lemmas and theorems.        
          \vspace*{-.15cm}
        \item Numerous additional explanations, examples, and remarks. \vspace*{-.15cm}
  \item An improved implementation and evaluation with our tool \aprove{}.
\end{itemize}

\section{Preliminaries}
\label{Preliminaries}

\noindent
In this section, we recapitulate the basics of term rewriting and the DP framework for ordinary termination in \Cref{DP Framework}, and probabilistic term rewriting in \Cref{Probabilistic Rewriting}.

\subsection{Term Rewriting and the DP Framework}
\label{DP Framework}

\noindent
We assume some familiarity with term rewriting \cite{baader_nipkow_1999}, but recapitulate all important notions.
We regard finite signatures $\Sigma = \biguplus_{n \in \IN} \Sigma_n$, where $\Sigma_n$ is the set of \emph{$n$-ary functions symbols} and we require $\Sigma_0 \neq \emptyset$.\footnote{This requirement ensures that there exist ground terms.
  This is not a restriction, since we showed in \cite{FOSSACS24} that almost sure termination of PTRSs is preserved under signature extensions.}
Let $\TT = \TSet{\Sigma}{\VSet}$ denote the set of all \emph{terms} over $\Sigma$ and a set of variables $\VSet$, which is recursively defined as the smallest set such that $\VSet \subseteq \TT$, and if $f \in \Sigma_n$ and $t_1, \ldots, t_n \in \TT$, then $f(t_1, \ldots, t_n) \in \TT$.
A \emph{substitution} is a function $\sigma:\VSet \to \TT$ with $\sigma(x) = x$ for all but finitely many $x \in \VSet$.
We lift substitutions to terms by defining $\sigma(t) = f(\sigma(t_1),\dots,\sigma(t_n))$ for a term $t=f(t_1,\dots,t_n)\in \TT$, and we often write $t \sigma$ instead of $\sigma(t)$.
For a term $t \in \TT$, the set of \emph{positions} $\pos(t)$ is the smallest subset of $\IN^*$ satisfying $\varepsilon \in \pos(t)$, and if $t=f(t_1,\dots,t_n)$ then for all $1 \leq j \leq n$ and all $\pi \in \pos(t_j)$ we have $j.\pi \in \pos(t)$.
Let $<_{\IN^*}$ be the \emph{prefix ordering} on positions and let $\leq_{\IN^*}$ be its reflexive
closure.
We say that a position $\pi$ is  \emph{above} (strictly above)
another position $\tau$ if $\pi \leq_{\IN^*} \tau$ ($\pi <_{\IN^*} \tau$),
i.e., 
$\pi$ is a (strict) prefix of $\tau$.
Two positions
$\pi$ and $\tau$ are \emph{orthogonal} if both $\pi \not\leq_{\IN^*} \tau$ and
$\tau \not\leq_{\IN^*} \pi$ hold.

If $\pi \in \pos(t)$ then $t|_{\pi}$ denotes the \emph{subterm}
at position $\pi$, i.e., $t|_{\varepsilon} = t$ for the \emph{root position} $\varepsilon$ and $t|_{j.\pi} = t_j|_{\pi}$ for $t=f(t_1,\dots,t_n)$, $1 \leq j \leq n$, and $\pi \in \IN^*$.
The \emph{root symbol} (or variable) at position $\varepsilon$ is also denoted by $\rootsym(t) \in \Sigma \cup \VSet$.
We write $t \trianglerighteq s$ if $s$ is a subterm of $t$ and $t \triangleright s$ if $s$ is a \emph{proper} subterm of $t$ (i.e., if $t \trianglerighteq s$ and $t \neq s$).
Let $t[r]_{\pi}$ denote the term that results from replacing the subterm $t|_{\pi}$ at position $\pi$ with the term $r \in \TT$, i.e., $t[r]_{\varepsilon} = r$ and $t[r]_{j.\pi} = f(t_1,\dots, t_j[r]|_{\pi}, \dots,t_n)$ for $t=f(t_1,\dots,t_n)$, $1 \leq j \leq n$, and $\pi \in \IN^*$.
A \emph{context} is a term $C \in \TSet{\Sigma \uplus \{\square\}}{\VSet}$ containing the $0$-ary function symbol $\square$ exactly once.
For a context $C$ and a term $t$, let $C[t]$ denote the term resulting from $C$ by replacing the only occurrence of $\square$ by $t$.

A \emph{rewrite rule} $\ell \to r$ is a pair of terms $\ell, r \in \TT$ such that $\VSet(r) \subseteq \VSet(\ell)$ and $\ell \notin \VSet$, where $\VSet(t)$ denotes the set of all variables occurring in $t \in \TT$.
The restrictions $\VSet(r) \subseteq \VSet(\ell)$ and $\ell \notin \VSet$ are imposed to exclude trivially non-terminating systems.
Here, $\ell$ is called the \emph{left-hand side} and $r$ the \emph{right-hand side} of the rule.
A \emph{term rewrite system} (TRS) is a finite set of rewrite rules.
As an example, consider $\R_{\tex}\!=\!\{ \eqref{R-ex-1}, \eqref{R-ex-2} \}$.

\vspace*{-.3cm}
\noindent
\begin{minipage}{0.5\textwidth}
  \begin{align}
    \label{R-ex-1} \tf(\ts(x)) & \!\to\! \tc(\tf(\tg(x)))\!
  \end{align}
\end{minipage}
\hfill
\begin{minipage}{0.4\textwidth}
  \begin{align}
    \label{R-ex-2} \tg(x) & \!\to\! \ts(x)\!
  \end{align}
\end{minipage}

\vspace*{.2cm}
A TRS $\R$ induces a \emph{rewrite relation} ${\to_{\R}} \subseteq \TT \times \TT$ on
terms where $s \to_{\R} t$ holds if there is a position $\pi \in \pos(s)$, a rule $\ell
\to r \in \R$, and a substitution $\sigma$ such that $s|_{\pi}=\ell\sigma$ (i.e., $\ell$
\emph{matches} $s|_{\pi}$)  and $t = s[r\sigma]_{\pi}$.
A rewrite step $s \to_{\R} t$ is an \emph{innermost} rewrite step (denoted $s \itor t$) if all proper subterms of the used \emph{redex} $\ell\sigma$ are in \emph{normal form} w.r.t.\ $\R$, i.e., the proper subterms of $\ell\sigma$ do not contain redexes themselves and thus, they cannot be reduced with $\to_\R$.
For example, we have $\tf(\tg(x)) \ito_{\R_{\tex}} \tf(\ts(x))$.
Let $\mathtt{NF}_{\R}$ denote the set of all terms that are in normal form w.r.t.\ $\R$, and let $\mathtt{ANF}_{\R}$ denote the set of all terms that are in \emph{argument normal form} w.r.t.\ $\R$, i.e., all terms where every proper subterm is in normal form.

Two rules $\ell_1 \to r_1, \ell_2 \to r_2 \in \R$ with renamed variables such that $\VSet(\ell_1) \cap \VSet(\ell_2) = \emptyset$ are \emph{overlapping}
if there exists a non-variable position $\pi$ of $\ell_1$ such that $\ell_1|_{\pi}$ and $\ell_2$ are \emph{unifiable}, i.e., there exists a substitution $\sigma$ such that $\ell_1|_{\pi} \sigma = \ell_2 \sigma$.
If $(\ell_1 \to r_1) = (\ell_2 \to r_2)$, then we require that $\pi \neq \varepsilon$.
$\R$ is \emph{non-overlapping} (NO) if it has no overlapping rules.
As an example, the TRS $\R_{\tex}$ is non-overlapping.
A TRS is \emph{left-linear} (\emph{right-linear}) if every variable occurs at most once in the left-hand side (right-hand side) of a rule, and it is \emph{linear} if it is both left- and right-linear.
Finally, a TRS is \emph{non-duplicating}
if for every rule, every variable occurs at most as often in the right-hand side as in the left-hand side.
As an example, $\R_{\tex}$ is both left- and right-linear, and hence non-duplicating.

We call a TRS $\R$ \emph{strongly normalizing} (or \emph{terminating}) or $\mathtt{SN}$ for short if $\to_{\R}$ is well founded.
$\R$ is \emph{strongly innermost normalizing} (or \emph{innermost terminating}) or $\mathtt{iSN}$ for short if $\ito_{\R}$ is well founded.

Next, we recapitulate the DP framework with its core processors to analyze (innermost) termination (see, e.g., \cite{arts2000termination,gieslLPAR04dpframework,giesl2006mechanizing,hirokawa2005automating,DBLP:journals/iandc/HirokawaM07}
for further details).
We decompose the signature $\SignatureDC = \SignatureD \uplus \SignatureC$ such that $f \in \SignatureD$ if $f = \rootsym(\ell)$ for some $\ell \to r \in \R$.
The symbols in $\SignatureD$ are called \emph{defined symbols}
(representing executable functions) and the symbols in $\SignatureC$ are \emph{constructors} (representing data).
For $\R_{\tex}$ we have $\SignatureD = \{\tf, \tg\}$ and $\SignatureC = \{\ts, \tc\}$.
For every $f \in \SignatureD$ we introduce a fresh \emph{annotated symbol} $f^{\sharp}$ of the same 
arity.\footnote{The symbols $f^{\sharp}$ were also called \emph{tuple symbols} (or \emph{marked symbols}) in the literature.}
Let $\SignatureA$ denote the set of all annotated symbols, $\SignatureADC = \SignatureA \uplus \Sigma$, and $\TT^{\sharp} = \TSet{\SignatureADC}{\VSet}$.
For any $t = f(t_1, \ldots,t_n) \in \TT$ with $f \in \SignatureD$, let $t^{\sharp} = f^{\sharp}(t_1,\ldots,t_n)$.
The annotated symbols are used to compare the arguments of two subsequent ``function
calls''. For example,
if we want to prove termination of a TRS containing only the rule $\tplus(\ts(x),y) \to
\ts(\tplus(x,
y))$, 
then we need to show that the arguments of the defined function symbol $\tplus$ ``decrease'' 
w.r.t.\ some well-founded order in every recursive call.
To avoid handling tuples of arguments, we use the annotated symbol
$\tplus^\sharp$ to combine the arguments of $\tplus$, i.e., 
we compare $\tplus^\sharp(\ts(x),y)$ with $\tplus^\sharp(x,y)$.
More precisely, for every rule $\ell \to r$ and every (not necessarily proper) subterm $t$ of $r$ with defined root symbol, 
one obtains a \emph{dependency pair} (DP) $\ell^\sharp \to t^\sharp$ that represents two possible subsequent ``function calls'' in a rewrite sequence.
$\DPair{\R}$ denotes the set of all dependency pairs of $\R$.
As an example, we have $\DPair{\R_\tex}=\{\eqref{R-ex-3}, \eqref{R-ex-4}\}$.
To ease readability, we often use capital letters like $\tF$ instead of $\tf^\sharp$.

\vspace*{-.2cm}
\noindent
\begin{minipage}{0.5\textwidth}
  \begin{align}
    \label{R-ex-3} \tF(\ts(x)) & \!\to\! \tF(\tg(x))\!
  \end{align}
\end{minipage}
\hfill
\begin{minipage}{0.4\textwidth}
  \begin{align}
    \label{R-ex-4} \tF(\ts(x)) & \!\to\! \tG(x)\!
  \end{align}
\end{minipage}

\vspace*{.2cm}

The DP framework uses \emph{DP problems}
$(\PP, \R)$ where $\PP$ is a (finite) set of DPs and $\R$ is a TRS\@.
A (possibly infinite) sequence $t_0^{\sharp}, t_1^{\sharp}, t_2^{\sharp}, \ldots$ with $t_j^{\sharp} \epsto_{\PP} \circ \to_{\R}^* t_{j+1}^{\sharp}$ for all $j$ is a $(\PP, \R)$-\emph{chain}
which represents sequences of subsequent ``function calls'' in evaluations.
Here, for any binary relation $\to$, ``$\to^*$'' denotes its reflexive and transitive closure.
Moreover, ``$\epsto_{\PP}$'' denotes rewrite steps with DPs from $\PP$ at the root and ``$\circ$'' denotes composition of relations.
We refer to steps with $\epsto_{\PP}$ as \emph{$\mathbf{a}$-steps}
($\mathbf{a}$ for \emph{annotation}, as we rewrite at the position of an annotated symbol, since rewrite steps with $\epsto_{\PP}$ require an annotated symbol at the root), and we call steps with $\to_{\R}$ \emph{$\mathbf{n}$-steps}
($\mathbf{n}$ for \emph{no-annotation}, as we rewrite at the position of a symbol without annotation).
The latter are used to evaluate the arguments of an annotated function symbol.
So an infinite chain consists of an infinite number of $\mathbf{a}$-steps with a finite number of $\mathbf{n}$-steps between consecutive $\mathbf{a}$-steps.
A (possibly infinite) sequence $t_0^\sharp, t_1^\sharp, t_2^\sharp, \ldots$ with $t_j^\sharp \iepsto_{\PP,\R} \circ \itorstar t_{j+1}^\sharp$ for all $j$ is an \emph{innermost} $(\PP, \R)$-\emph{chain}.
Here, ``$\iepsto_{\PP,\R}$'' is the restriction of $\epsto_{\PP}$ to rewrite steps where the used redex is in $\NF_{\R}$.
For example, $\tF(\ts(x)), \tF(\ts(x)), \ldots$ is an infinite innermost $(\DPair{\R_{\tex}}, \R_{\tex})$-chain, as $\tF(\ts(x)) \iepsto_{\DPair{\R_{\tex}},\R_{\tex}} \tF(\tg(x)) \ito_{\R_{\tex}}^* \tF(\ts(x))$.

A DP problem $(\PP, \R)$ is called \emph{terminating} (or $\mathtt{SN}$) if there is no infinite $(\PP, \R)$-chain.
It is \emph{innermost terminating} (or $\mathtt{iSN}$) if there is no infinite innermost $(\PP, \R)$-chain.
The main result on DPs is the \emph{chain criterion} which states that there is no infinite (innermost) sequence $t_0 \itomaybe_{\R} t_1 \itomaybe_{\R} \ldots$ (i.e., $\R$ is $\mathtt{(i)SN}$) iff there is no infinite (innermost) $(\DPair{\R},\R)$-chain (i.e., $(\DPair{\R},\R)$ is $\mathtt{(i)SN}$).
The DP framework is a \emph{divide-and-conquer} approach, which applies \emph{DP processors} $\Proc$ of the form $\Proc(\PP, \R) = \{(\PP_1,\R_1), \ldots, (\PP_n,\R_n)\}$, where $\PP, \PP_1, \ldots, \PP_n$ are (finite) sets of DPs and $\R, \R_1, \ldots, \R_n$ are TRSs, to transform DP problems into simpler sub-problems.
A processor $\Proc$ is \emph{sound} for termination (for innermost termination) if $(\PP, \R)$ is $\mathtt{SN}$ ($\mathtt{iSN}$) whenever $(\PP_j,\R_j)$ is $\mathtt{SN}$ ($\mathtt{iSN}$) for all $1 \leq j \leq n$.
It is \emph{complete} for termination (for innermost termination) if $(\PP_j,\R_j)$ is $\mathtt{SN}$ ($\mathtt{iSN}$) for all $1 \leq j \leq n$ whenever $(\PP, \R)$ is $\mathtt{SN}$ ($\mathtt{iSN}$).

So given a TRS $\R$, one starts with the initial DP problem $(\DPair{\R}, \R)$ and applies sound (and preferably complete) DP processors repeatedly until all sub-problems are ``solved'', i.e., sound processors transform them to the empty set.
This yields a modular framework for termination and innermost termination proofs, as different techniques can be used for different sub-problems $(\PP_j,\R_j)$.
To give an intuition for the framework, the following three theorems recapitulate the three most important processors of the DP framework.
Later in the paper, we will also introduce additional processors which we adapt to the probabilistic setting.

The (innermost) \emph{$(\PP, \R)$-dependency graph} is a control flow graph that indicates which DPs can be used after each other in an (innermost) chain.
Its set of nodes is $\PP$ and there is an edge from $\ell_1^\sharp \to t_1^\sharp$ to $\ell_2^\sharp \to t_2^\sharp$ if there exist substitutions $\sigma_1, \sigma_2$ such that $t_1^\sharp \sigma_1 \to_{\R}^* \ell_2^\sharp \sigma_2$ ($t_1^\sharp \sigma_1 \ito_{\R}^* \ell_2^\sharp \sigma_2$ and $\ell_1^\sharp \sigma_1, \ell_2^\sharp \sigma_2 \in \NF_{\R}$ for the innermost dependency graph).
Any infinite (innermost) $(\PP, \R)$-chain corresponds to an infinite path in the (innermost) dependency graph, and since the graph is finite, this infinite path must end in some strongly connected component (SCC).\footnote{Here, a set $\PP'$ of DPs is an \emph{SCC} if it is a maximal cycle, i.e., it is a maximal set such that for any $\ell_1^\sharp \to t_1^\sharp$ and $\ell_2^\sharp \to t_2^\sharp$ in $\PP'$ there is a non-empty path from $\ell_1^\sharp \to t_1^\sharp$ to $\ell_2^\sharp \to t_2^\sharp$ which only traverses nodes from $\PP'$.}
Hence, it suffices to consider the SCCs of this graph independently.

\begin{restatable}[Dependency Graph Processor]{thm}{depgraph}
  \label{DGP}
  Let $\PP_1, \ldots, \PP_n$ be the SCCs of the (innermost) $(\PP, \R)$-dependency graph.
  Then $\Proc_{\mathtt{DG}}(\PP,\R) = \{(\PP_1,\R), \ldots,\linebreak
    (\PP_n,\R)\}$ is sound and complete for (innermost) termination.
\end{restatable}

\begin{exa}[Dependency Graph]
  \label{DG-example}
  Consider the TRS $\R_{\tffg} \!=\! \{ \eqref{R-fgf-1}\}$ with $\DPair{\R_{\tffg}} \!=\! \{ \eqref{R-fgf-2}, \eqref{R-fgf-3}, \eqref{R-fgf-4} \}$.
  The (innermost) $(\DPair{\R_{\tffg}}, \R_{\tffg})$-dependency graph is on the right.

    {\footnotesize \vspace*{-.2cm}
      \noindent
      \hspace*{-.4cm}
      \begin{minipage}{0.38\textwidth}
        \begin{align}
          \label{R-fgf-1} \tf(\tf(\tg(x))) & \!\to\! \tf(\tg(\tf(\tg(\tf(x)))))
        \end{align}
      \end{minipage}
      \hspace{.2cm}
      \begin{minipage}{0.4\textwidth}
        \begin{align}
          \label{R-fgf-2} \tF(\tf(\tg(x))) & \!\to\! \tF(\tg(\tf(\tg(\tf(x))))) \\
          \label{R-fgf-3} \tF(\tf(\tg(x))) & \!\to\! \tF(\tg(\tf(x))) \\
          \label{R-fgf-4} \tF(\tf(\tg(x))) & \!\to\! \tF(x)
        \end{align}
      \end{minipage}
      \hspace{.2cm}
      \begin{minipage}{0.15\textwidth}
        \scriptsize \vspace*{.5cm}
        \hspace*{.5cm}
        \begin{tikzpicture}
          \node[shape=rectangle,draw=black!100] (A) at (0,1.4) {\eqref{R-fgf-2}};
          \node[shape=rectangle,draw=black!100] (B) at (0,.7) {\eqref{R-fgf-3}};
          \node[shape=rectangle,draw=black!100] (C) at (1,1.4) {\eqref{R-fgf-4}};

          \path [->,in=290,out=250,looseness=5] (C) edge (C); \path [->] (C) edge (A); \path [->] (C) edge (B);
        \end{tikzpicture}
      \end{minipage}
          }
\end{exa}

\noindent
Here, $\Proc_{\mathtt{DG}}(\DPair{\R_{\tffg}}, \R_{\tffg})$ yields the DP problem $(\{ \eqref{R-fgf-4} \}, \R_\tffg)$.

While the exact (innermost) dependency graph is not computable in general, there exist several techniques to over-approximate it automatically, see, e.g., \cite{arts2000termination,giesl2006mechanizing,hirokawa2005automating}.
A basic approximation technique that we will also use for the novel transformational processors in \Cref{ADP Transformations} works as follows.
To find out whether there is an edge from $\ell^\sharp_1 \to t^\sharp_1$ to $\ell^\sharp_2 \to t^\sharp_2$ we compute the skeleton $\capterm_\R(t^\sharp_1)$ of $t^\sharp_1$ that remains unchanged when reducing $t^\sharp_1 \sigma_1$ to $\ell^\sharp_2 \sigma_2$ for arbitrary substitutions $\sigma_1$ and $\sigma_2$.
$\capterm_\R(t^\sharp_1)$ results from replacing all those subterms of $t^\sharp_1$ by different fresh variables whose root is a defined symbol of $\R$.
Here, multiple occurrences of the same subterm are also replaced by pairwise different variables.
So if $\tf \in \SignatureD$ and $\tc \in \SignatureC$, then $\capterm_\R(\tc(\tf(x),\tf(x))) = \tc(x_1, x_2)$.
Afterwards, we check whether the skeleton $\capterm_\R(t^\sharp_1)$ and $\ell^\sharp_2$ are unifiable by a substitution $\sigma$ where $\ell^\sharp_1 \sigma$ and $\ell^\sharp_2 \sigma$ are in $\mathtt{ANF}_{\R}$ (and thus, also in $\mathtt{NF}_{\R}$).

If they are, then we create an edge from $\ell^\sharp_1 \to t^\sharp_1$ to $\ell^\sharp_2 \to t^\sharp_2$ in the estimated innermost dependency graph.
If they cannot be unified, then there is no innermost rewrite sequence from an instantiation of $t^\sharp_1$ to an instantiation of $\ell^\sharp_2$ such that both instantiated left-hand sides $\ell^\sharp_1, \ell^\sharp_2$ are in argument normal form.
Thus, this leads to a sound over-approximation of the innermost dependency graph.
To check whether we can reduce $t^\sharp_1 \sigma_1$ to $\ell^\sharp_2 \sigma_2$ in non-innermost chains, we add another transformation $\renterm$ that replaces all variable occurrences (including multiple occurrences of the same variable) by fresh variables.
The reason for this additional transformation is that for non-innermost chains, variables in $t^\sharp_1$ may be instantiated by terms that are not in normal form.
Hence, they can be evaluated to different terms during the reduction of the arguments of $t^\sharp_1$.
Thus, we check unifiability of $\renterm(\capterm_\R(t^\sharp_1))$ and $\ell^\sharp_2$ for the (non-innermost) dependency graph.

The next processor removes rules that cannot be used for right-hand sides of dependency pairs when their variables are instantiated with normal forms.

\begin{restatable}[Usable Rules Processor]{thm}{usablerules}
  \label{URP}
  Let $\R$ be a TRS.
  For every $f\!\in\!\SignatureADC$, let $\rules_\R(f) = \{\ell \to r \in \R \mid \rootsym(\ell) = f\}$.
  Moreover, for every $t \in \TSet{\SignatureADC}{ \VSet}$, the \defemph{usable rules} $\urules_\R(t)$ of $t$ w.r.t.\ $\R$ are recursively defined as follows:

  \vspace*{-0.4cm}
  \[ \small
  \displaystyle 
  \begin{array}{r@{\;}l@{\;}ll}
    \urules_\R(t) &=& \emptyset, & \text{if } t \in \VSet \text{ or } \R = \emptyset\\
    \urules_\R(f(t_1, \ldots, t_n)) &=& \rules_{\R}(f) 
    \cup \displaystyle\bigcup_{j = 1}^n \urules_{\R'}(t_j)
    \cup \hspace*{-.3cm} \displaystyle\bigcup_{\ell \to r \in
      \rules_\R(f)} \hspace*{-.3cm} \urules_{\R'}(r),
    & \text{otherwise,}\!
  \end{array}
  \]
  \vspace*{-0.4cm}
  
  \noindent
where $\R' = \R \setminus \rules_\R(f)$.

  The \defemph{usable rules} for the DP problem $(\PP, \R)$ are $\urules(\PP,\R) = \bigcup_{\ell^\sharp \to t^\sharp \in \PP} \, \urules_\R(t^\sharp)$.
  Then $\Proc_{\mathtt{UR}}(\PP,\R) = \{(\PP,\urules(\PP,\R))\}$ is sound but not complete for innermost termination.\footnote{\label{CompletenessUsableRules}
  See \cite{gieslLPAR04dpframework} for a complete version of this processor.
  It extends DP problems by an additional set which stores the left-hand sides of all rules (including the non-usable ones) to determine whether a rewrite step is innermost.
  We omit this refinement here for readability.}
\end{restatable}

To continue \Cref{DG-example} for innermost rewriting, $\Proc_{\mathtt{UR}}\bigl(\{ \eqref{R-fgf-4} \}, \R_{\tffg}\bigr)$ yields the problem $(\{ \eqref{R-fgf-4} \}, \emptyset)$, i.e., it removes all rules, because the right-hand side of $\eqref{R-fgf-4}$ does not contain the defined symbol $\tf$.

Finally, the \emph{reduction pair processor} allows us to use term orders to remove certain DPs.
As our probabilistic adaption in \Cref{The Probabilistic ADP Framework}
needs to consider ``expected values'', we focus on reduction pairs derived from a $\Sigma^\sharp$-algebra.
An (ordered) $\Sigma^\sharp$-algebra is a pair $(\mathcal{A}, \succ)$, where $\mathcal{A}$ maps every function symbol $f \in \Sigma_n^\sharp$ of arity $n$ to a function $\mathcal{A}(f) = f_{\mathcal{A}} : A^n \to A$ for a non-empty carrier set $A$.
The mapping $\mathcal{A}$ is extended to ground terms, i.e., $\mathcal{A}(f(t_1,\ldots,t_n)) = f_{\mathcal{A}}(\mathcal{A}(t_1), \ldots, \mathcal{A}(t_n))$ for $f \in \Sigma_n^\sharp$ and ground terms $t_1,\ldots,t_n \in \mathcal{T}^\sharp$.
Furthermore, $\succ$ is a well-founded order on $A$, i.e., $\succ$ is well founded and transitive.
Now one can obtain a well-founded order on ground terms by considering $t_1$ to be greater than $t_2$ iff $\mathcal{A}(t_1) \succ\mathcal{A}(t_2)$.
We extend this to a well-founded order on terms with variables by writing $\mathcal{A}(t_1) \succ\mathcal{A}(t_2)$ iff $\mathcal{A}(t_1\sigma) \succ\mathcal{A}(t_2\sigma)$ holds for all substitutions $\sigma$ that instantiate the variables of $t_1$ and $t_2$ by ground terms.
Let $\succcurlyeq$ be the reflexive closure of $\succ$.
We say that $(\mathcal{A}, \succ)$ is \emph{monotonic}
if $\mathcal{A}(t_1) \succ \mathcal{A}(t_2)$ implies $\mathcal{A}(C[t_1]) \succ \mathcal{A}(C[t_2])$, and \emph{weakly monotonic} if $\mathcal{A}(t_1) \succcurlyeq \mathcal{A}(t_2)$ implies $\mathcal{A}(C[t_1]) \succcurlyeq \mathcal{A}(C[t_2])$ for every context $C$.
The reduction pair processor searches for a weakly monotonic $\Sigma^\sharp$-algebra $(\mathcal{A}, \succ)$ such that all rules and DPs are ordered by $\succcurlyeq$, and it removes those DPs that are ordered by~$\succ$.

\begin{restatable}[Reduction Pair Processor]{thm}{rpp}
  \label{RPP}
  Let $(\mathcal{A},\succ)$ be a weakly monotonic $\Sigma^\sharp$-algebra.
  Let $\PP = \PP_{\succcurlyeq} \uplus \PP_{\succ}$ with $\PP_{\succ} \neq \emptyset$ where:
  {\small
  \[\begin{array}{clll}
      (1) & \forall \ell^\sharp \to r^\sharp \in \PP         & : & \mathcal{A}(\ell^\sharp) \succcurlyeq \mathcal{A}(r^\sharp) \\
      (2) & \forall \ell^\sharp \to r^\sharp \in \PP_{\succ} & : & \mathcal{A}(\ell^\sharp) \succ \mathcal{A}(r^\sharp)        \\
      (3) & \forall \ell \to r \in \R                & : &
      \mathcal{A}(\ell) \succcurlyeq \mathcal{A}(r)
    \end{array}\]}

  \noindent
  Then $\Proc_{\mathtt{RP}}(\PP,\R) = \{(\PP_{\succcurlyeq}, \R)\}$ is sound and complete for both termination and innermost termination.
\end{restatable}

Examples for $\Sigma^\sharp$-algebras are \emph{polynomial} \cite{lankford1979proving}
or \emph{matrix interpretations} \cite{Endrullis08}.
A polynomial interpretation $(\mathcal{A},\succ) = (\Pol, >)$
maps every function symbol $f \in \Sigma^\sharp$ to a polynomial $f_{\Pol}~\in~\IN[\VSet]$ 
over the variables $\VSet$ with coefficients from $\IN$, see \cite{lankford1979proving}.
The order $>$ is simply the natural order on $\IN$.
The reduction pair processor allows us to use weakly monotonic polynomial interpretations that do not have to depend on all of their arguments, and SMT solvers can be used to search for an appropriate interpretation automatically.
For $(\{ \eqref{R-fgf-4} \}, \emptyset)$, one can use the reduction pair processor with the polynomial interpretation that maps $\tf(x)$ to $x+1$ and both $\tF(x)$ and $\tg(x)$ to $x$.
Then, $\Proc_{\mathtt{RP}}\bigl(\{ \eqref{R-fgf-4} \}, \emptyset\bigr) = \{\bigl(\emptyset, \emptyset\bigr)\}$.
As $\Proc_{\mathtt{DG}}(\emptyset, \emptyset) = \emptyset$ and all processors used are sound, this means that there is no infinite innermost chain for the initial DP problem $(\DPair{\R_{\tffg}}, \R_{\tffg})$ and thus, $\R_{\tffg}$ is innermost terminating.
This implies that $\R_{\tffg}$ is terminating, since $\R_{\tffg}$ is non-overlapping, 
hence, innermost termination implies termination \cite{Gramlich1995AbstractRB}.

\subsection{Probabilistic Rewriting}
\label{Probabilistic Rewriting}

\noindent
Next, we recapitulate the basics of probabilistic rewriting as introduced in \cite{avanzini2020probabilistic,BournezRTA02,bournez2005proving,kassinggiesl2023iAST}.
A \emph{probabilistic rule} has (finite) multi-distributions on the right-hand sides of its rewrite rules instead of a single term.
A finite \emph{multi-distribution}
$\mu$ on a set $A \neq \emptyset$ is a finite multiset of pairs $(p:a)$, 
where $0 < p \leq 1$ is a probability and $a \in A$, such that $\sum_{(p:a) \in \mu} \, p = 1$.
$\FDist(A)$ is the set of all finite multi-distributions on $A$.
For $\mu\in\FDist(A)$, its \emph{support} is the multiset $\Supp(\mu)\!=\!\{a \mid (p\!:\!a)\!\in\!\mu$ for some $p\}$.
A \emph{probabilistic rewrite rule} $\ell \to \mu \in \TT \times \FDist(\TT)$ is a pair 
such that $\ell \not\in \VSet$ and $\VSet(r) \subseteq \VSet(\ell)$ for every $r \in \Supp(\mu)$.
A \emph{probabilistic TRS} (PTRS) is a finite set $\R$ of probabilistic rewrite rules.

\begin{exa}
  \label{S1S2S3}

  Consider the following probabilistic rewrite rules:

  \vspace*{-.55cm}

  \begin{minipage}[t]{0.47\textwidth}
    \begin{align}
      \tg    & \to \{\nicefrac{3}{4}:\td(\tg), \, \nicefrac{1}{4}:\tz\} \label{rule-01} \\
      \td(x) & \to \{1:\tz\} \label{rule-02}\!
    \end{align}
  \end{minipage}
  \hfill
  \begin{minipage}[t]{0.47\textwidth}
    \begin{align}
      \td(x)      & \to \{1:\tc(x,x)\} \label{rule-03} \\
      \td(\td(x)) & \to \{1:\tc(x,\tg)\} \label{rule-03nd}
    \end{align}
  \end{minipage}

  \vspace*{.2cm}

  \noindent
  and the PTRSs $\R_1 = \{\eqref{rule-01}\}$, $\R_2 = \{\eqref{rule-01}, \eqref{rule-03}\}$, and $\R_3 = \{\eqref{rule-01}, \eqref{rule-02}, \eqref{rule-03nd}\}$.
\end{exa}

Similar to TRSs, a PTRS $\R$ induces a  \emph{probabilistic rewrite relation}
${\to_{\R}} \subseteq \TT \times \FDist(\TT)$ where 	$s \to_{\R} \{p_1:t_1, \ldots, p_k:t_k\}$ if there are an $\ell \to \{p_1:r_1, \ldots, p_k:r_k\} \in \R$, a substitution $\sigma$, and a $\pi \in \pos(s)$ such that $s|_{\pi}=\ell\sigma$ and $t_j = s[r_j\sigma]_{\pi}$ for all $1 \leq j \leq k$.
$\NF_{\R}$ and $\ANF_{\R}$ are defined as for TRSs.
The rewrite step is \emph{innermost} (denoted $s \itor \{p_1:r_1, \ldots, p_k:r_k\}$) if $\ell \sigma \in \ANF_{\R}$.
So the PTRS $\R_1$ can be interpreted as a biased coin flip that terminates in each step with a probability of $\nicefrac{1}{4}$.

To track all possible rewrite sequences (up to non-determinism) with their corresponding probabilities, we \emph{lift} $\to_{\R}$ to \emph{rewrite sequence trees (RSTs)}.

\begin{defi}[Rewrite Sequence Tree, Almost-Sure Termination]\label{def:RST-AST}
  Let $\R$ be a PTRS.
 We say that $\F{T}\!=\!(V,E,L)$ is an $\R$\emph{-rewrite sequence tree ($\R$-RST)} if
  \begin{enumerate}
  \item[(1)] $(V, E)$ is a (possibly infinite) directed tree with nodes 
    $V \neq \emptyset$  and directed edges $E \subseteq V \times V$, where $vE = \{ w \mid (v,w) \in E \}$ is finite for every $v \in V$.
  \item[(2)]  $L : V \rightarrow (0,1] \times \TT$ labels every node $v$ by a probability $p_v$ and a term $t_v$.
  For the root $v \in V$ of the tree, we have $p_v = 1$.
  \item[(3)] For all $v \in V$: If $vE = \{w_1, \ldots, w_k\}$, then $t_v \to_\R \{\tfrac{p_{w_1}}{p_v}:t_{w_1}, \ldots, \tfrac{p_{w_k}}{p_v}:t_{w_k}\}$.
  \end{enumerate}
  An $\R$-RST is an \emph{innermost} $\R$-RST if the edge relation represents only innermost steps.
  We say that an RST $\F{T}$ \emph{converges} (or \emph{terminates}) with probability $|\F{T}| = \sum_{v \in \ctleaf} \, p_v$, where $\ctleaf$ is the set of all its leaves.
  A PTRS $\R$ is \emph{almost-surely (innermost) terminating} (or $\mathtt{(i)AST}$ for short) if $|\F{T}| = 1$ holds for all (innermost) $\R$-RSTs $\F{T}$.
\end{defi}

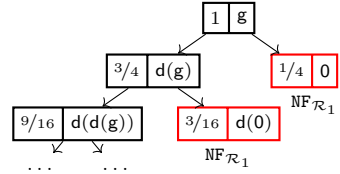
\begin{wrapfigure}[7]{r}{0.35\textwidth}
  \setlength{\abovecaptionskip}{5pt}
  \scriptsize \vspace*{-0.55cm}
  \hspace*{-.12cm}
  \begin{tikzpicture}
    \tikzstyle{adam}=[thick,draw=black!100,fill=white!100,minimum size=4mm, shape=rectangle split, rectangle split parts=2,rectangle split horizontal] \tikzstyle{adam2}=[thick,draw=red!100,fill=white!100,minimum size=4mm, shape=rectangle split, rectangle split parts=2,rectangle split horizontal] \tikzstyle{empty}=[rectangle,thick,minimum size=4mm]

    \node[adam] at (-4, 0) (a) {$1$
      \nodepart{two}$\tg$};
    \node[adam] at (-5, -0.7) (b) {$\nicefrac{3}{4}$
      \nodepart{two}$\td(\tg)$};
    \node[adam2,label=below:{$\quad \mathtt{NF}_{\R_{1}}$}] at (-3, -0.7) (c) {$\nicefrac{1}{4}$
        \nodepart{two}$\tz$};
    \node[adam] at (-6, -1.4) (d) {$\nicefrac{9}{16}$
      \nodepart{two}$\td(\td(\tg))$};
    \node[adam2,label=below:{$\mathtt{NF}_{\R_{1}}$}] at (-4, -1.4) (e) {$\nicefrac{3}{16}$
        \nodepart{two}$\td(\tz)$};
    \node[empty] at (-6.5, -2) (f) {$\ldots$};
    \node[empty] at (-5.5, -2) (g) {$\ldots$};

    \draw (a) edge[->] (b);
    \draw (a) edge[->] (c);
    \draw (b) edge[->] (d);
    \draw (b) edge[->] (e);
    \draw (d) edge[->] (f);
    \draw (d) edge[->] (g);
  \end{tikzpicture}
 \vspace*{-.5cm} \caption{Infinite $\R_1$-RST}\label{fig:R1}
\end{wrapfigure}
While $|\F{T}| = 1$ for every finite RST $\F{T}$, for infinite RSTs $\F{T}$ we may have $|\F{T}| <1$ or even $|\F{T}| = 0$ if $\F{T}$ has no leaf at all.
The notion of $\mathtt{AST}$ from \Cref{def:RST-AST}
is equivalent to the ones in
\cite{bournez2005proving,avanzini2020probabilistic,kassinggiesl2023iAST}, 
where $\mathtt{AST}$ is defined via a lifting of $\to_{\R}$ to multisets or via stochastic processes.
For example, we have $|\F{T}| = 1$ for the infinite $\R_1$-RST $\F{T}$ in \Cref{fig:R1}.
As this holds for all $\R_1$-RSTs, $\R_1$ is $\mathtt{AST}$.
In \Cref{fig:R1}, we split the nodes $v$ of the RST into the corresponding probability $p_v$ and the corresponding term $t_v$
to ease readability instead of writing $(p_v:t_v)$ in a single node.
Moreover, we marked a node $v$ in red whenever
the corresponding term $t_v$ is in normal form.

\begin{exa}
  \label{example:running-1}
  $\R_2$ from \Cref{S1S2S3} is not $\mathtt{AST}$.
  If we always apply $\eqref{rule-03}$ directly after $\eqref{rule-01}$ (in a non-innermost step), 
  then this corresponds to the rule  
  \begin{align}
    \tg \to \{\nicefrac{3}{4}:\tc(\tg, \tg), \nicefrac{1}{4}:\tz\}\label{rule-01and03}\!
  \end{align}
  which represents a random walk on the number of $\tg$'s in a term biased towards non-termination 
  (as $\nicefrac{3}{4} > \nicefrac{1}{4}$).
  $\R_3$ is not $\mathtt{AST}$ either, because if we always apply \eqref{rule-03nd}
  after two applications of \eqref{rule-01}, this corresponds to
  \begin{align*}
    \tg \to \{\nicefrac{9}{16}:\tc(\tg,\tg), \; \nicefrac{3}{16}:\tz, \; \nicefrac{1}{4}:\tz \}\!
  \end{align*}
  which is also a biased random walk towards non-termination (as $\nicefrac{9}{16} > \nicefrac{3}{16} \, + \, \nicefrac{1}{4}$).

  However, in innermost evaluations, the $\td$-rule \eqref{rule-03} can only duplicate normal forms, 
  i.e., we cannot apply the rule $\eqref{rule-03}$ directly after $\eqref{rule-01}$,
  and hence $\R_2$ is $\mathtt{iAST}$, see \cite{FOSSACS24}.
  $\R_3$ is $\mathtt{iAST}$ as well, as \eqref{rule-03nd}
  is not applicable in innermost evaluations.
  For both $\R_2$ and $\R_3$, we will see that $\mathtt{iAST}$ can be proved automatically by our implementation 
  of the ADP framework for $\mathtt{iAST}$ introduced in \Cref{The Probabilistic ADP Framework}.
\end{exa}

In \Cref{example:running-1} we can see the reason for using multi-distributions instead of
distributions.
Consider $\tg \to \{\nicefrac{9}{16}:\tc(\tg,\tg), \; \nicefrac{3}{16}:\tz, \; \nicefrac{1}{4}:\tz \}$ again.
If we had two different rules that can rewrite $\tz$, 
e.g., $\tz \to \{1:\ta\}$ and $\tz \to \{1:\tb\}$, then we would be able to rewrite both occurrences of $\tz$ differently.
Hence, we do not ``merge'' those terms in the multi-distribution to $\{\nicefrac{9}{16}:\tc(\tg,\tg), \; \nicefrac{3}{16} + \nicefrac{1}{4}:\tz\}$.

\begin{exa}
  \label{example:running-2}
  The following PTRS $\R_{\textsf{alg}}$ corresponds to \Cref{alg1}.
  Here, natural numbers are represented via the constructors $\tz$ and $\ts$, and the non-deter\-minism is modeled by the non-deterministic choice between the overlapping rules \eqref{loopOne-1} and \eqref{loopOne-2}.
  In \Cref{The Probabilistic ADP Framework}, we will prove that $\R_{\textsf{alg}}$ is $\mathtt{AST}$.

  \vspace*{-.3cm}

  {\small
    \begin{align}
      \tloopOne(y) & \to \{\nicefrac{1}{2}:\tloopOne(\tdouble(y)), \;
      \nicefrac{1}{2}:\tloopTwo(\tdouble(y))\} \label{loopOne-1} \\
      \tloopOne(y) & \to \{\nicefrac{1}{3}:\tloopOne(\ttriple(y)), \;
      \nicefrac{2}{3}:\tloopTwo(\ttriple(y))\}\! \label{loopOne-2}
    \end{align}
    \begin{minipage}[t]{0.49\linewidth}
      \vspace*{-0.8cm}
      \begin{align}
        \tloopTwo(\ts(y)) & \to \{1:\tloopTwo(y)\} \label{run2-rule-3}\\
        \tdouble(\ts(y))  & \to \{1:\ts(\ts(\tdouble(y)))\} \label{run2-rule-4}\\
        \tdouble(\tz)     & \to \{1:\tz\} \label{run2-rule-5}\!
      \end{align}
    \end{minipage}
    \hfill
    \begin{minipage}[t]{0.49\linewidth}
      \vspace*{-0.6cm}
      \begin{align}
        \ttriple(\ts(y)) & \to \{1:\ts(\ts(\ts(\ttriple(y))))\} \label{run2-rule-6}\\
        \ttriple(\tz)    & \to \{1:\tz\} \label{run2-rule-7}\!
      \end{align}
    \end{minipage}
  }
\end{exa}

\section{Probabilistic Annotated Dependency Pairs}
\label{Probabilistic Annotated Dependency Pairs}

\noindent
In this section, we extend DPs to the probabilistic setting, which results in the notion of \emph{annotated dependency pairs}.
Compared to our first adaption of DPs to the probabilistic setting via dependency tuples in \cite{kassinggiesl2023iAST}, ADPs are easier, more elegant, and lead to a sound and \emph{complete} chain criterion.
Note that ordinary DPs do not suffice in the probabilistic setting.
Similar to the dependency tuples for complexity analysis \cite{noschinski2013analyzing}, we cannot consider each defined symbol in a right-hand side individually, but we have to consider all of them simultaneously.

\begin{exa}
  A natural idea to define dependency pairs for a probabilistic rule $\ell \to \{p_1:r_1,\dots, p_k:r_k\} \in \R$ would be 
  \eqref{dp A} or \eqref{dp B}:

  \vspace*{-.4cm}

  {\small \begin{align}
      &\{  \ell^{\sharp} \to \{p_1:r_1, \ldots, p_i:t_j^\sharp, \dots, p_k:r_k\} \, \mid
      \, \exists j \in \{1, \ldots, k\}: t_j \trianglelefteq 
      r_j, \rootsym(t_j) \in \SignatureD  \} \label{dp A}\\
       &\{  \ell^{\sharp} \to \{p_1:t_1^\sharp, \ldots, p_k:t_k^\sharp\} \, \mid \,
      \forall
      j \in \{1, \ldots, k\}: t_j \trianglelefteq
      r_j, \rootsym(t_j) \in \SignatureD  \} \label{dp B}\!
  \end{align}
  }

  \vspace*{-.1cm}
  
  For \eqref{dp B}, if some $r_j$ does not contain any defined symbol, 
  then instead of $t_j^\sharp$
  we use a fresh constructor $\tbot$ that does not occur in $\R$ instead.
  So in both \eqref{dp A} and \eqref{dp B}, we replace $r_j$ by
  a single term $t_j^\sharp$ in the right-hand side, as in the non-probabilistic DP framework.
  A counterexample to such an approach is the PTRS $\R_{rw}$ containing only the rule
  $\tg \to \{\nicefrac{3}{4}:\tc(\tg, \tg), \nicefrac{1}{4}:\tz\}$ \eqref{rule-01and03} of \Cref{example:running-1}, which is a biased random walk towards non-termination.
  When using \eqref{dp A} or \eqref{dp B}, 
  we would just get the dependency pair $\tG \to \{\nicefrac{3}{4}:\tG, \nicefrac{1}{4}:t\}$,
  where $t$ is either $\tz$ or $\tbot$.
  However, with this dependency pair we lose the information about the second call of the defined symbol $\tg$, 
  which is the reason for non-termination of the random walk.
  In fact, the PTRS $\R_1 = \{ \eqref{rule-01} \}$ from \Cref{S1S2S3}
  would yield the same DP $\tG \to \{\nicefrac{3}{4}:\tG, \nicefrac{1}{4}:t\}$, 
  although $\R_1$ is $\mathtt{AST}$, whereas $\R_{rw}$ is not.
  Thus, using such a direct adaption of dependency pairs to the probabilistic setting is unsound.
\end{exa}

The counterexample $\R_{rw}$ shows that having only one annotation in the right-hand side of each dependency pair is not expressive enough, but instead we need a rule where all defined symbols in the right-hand side are annotated, called an \emph{annotated dependency pair}.
As we showed in \cite{IJCAR2024}, ADPs are not only useful to analyze probabilistic rewriting, 
but a suitable variant of ADPs can also be used to extend the DP framework to (non-probabilistic) \emph{relative} rewriting.
In the following, we first define annotated dependency pairs (\Cref{subsec:ADP}) formally.
Then we explain how to perform innermost rewriting with ADPs (\Cref{subsec:Innermost})
and extend this to an arbitrary evaluation strategy in
\Cref{subsec:Full}.
Finally, we adapt the notion of chains to the probabilistic setting (\Cref{subsec:Chains}).

\subsection{Annotated Dependency Pairs}\label{subsec:ADP}
\noindent
We start with some basic definitions regarding the positions in a term, in order to obtain all positions of defined symbols and to replace certain symbols with their annotated or non-annotated version.
For $t \in \TT^{\sharp}$ and $\mathcal{X} \subseteq \SignatureADC \cup \VSet$, let $\pos_{\mathcal{X}}(t)$ be all positions of $t$ with symbols or variables from $\mathcal{X}$.
For a set of positions $\Phi \subseteq \posDT(t)$, let $\anno_\Phi(t)$ be the variant of $t$ where the symbols at positions from $\Phi$ in $t$ are annotated, and all other annotations are removed.
Thus, $\posT(\anno_\Phi(t)) = \Phi$, and $\anno_\emptyset(t)$ removes all annotations from
$t$, where we often write $\flat(t)$ instead of $\anno_\emptyset(t)$.\footnote{Instead of using
$\natural$ as an inverse operator to  $\sharp$ (like in musical notation), we use $\flat$
as in the musical isomorphism from differential geometry.
}
Moreover, let $\disannoPos{\pi}(t)$ result from removing all annotations from $t$ that are strictly above the position $\pi$.
So for $\R_2$, we have $\anno_{\{1\}}(\td(\tg)) = \anno_{\{1\}}(\tD(\tG)) = \td(\tG)$, $\flat(\tD(\tG)) = \td(\tg)$, and $\disannoPos{1}(\tD(\tG)) = \td(\tG)$.
To transform the rules of a PTRS into ADPs, initially we annotate all defined symbols $f \in \SignatureD$ occurring in right-hand sides, 
as we need to consider all these defined symbols simultaneously.
The left-hand side of the resulting ADP is just the left-hand side of the original rule.

The original DP framework for TRSs initially starts with the DP problem $(\PP, \R)$, where $\PP$ is the set of all DPs and $\R$ contains the original rewrite rules.
Instead of additionally considering the original rewrite rules in our ADP framework, every ADP has a flag $m \in \{\ttrue, \tfalse\}$ to indicate whether this ADP may also be used as an ordinary probabilistic rewrite rule without annotations to rewrite at a position below an annotated symbol.
This flag will be modified and used by the processors in \Cref{The Probabilistic ADP
  Framework}, similar to the ordinary rewrite rules $\R$ in the DP problem $(\PP, \R)$.
So an ADP can  be used as a collection of all DPs of a rule, but it can also be used as
the original rule itself. 
We call this  the \emph{duality} of ADPs.\footnote{Instead of the flag $m$, one could of
course also regard pairs $(\PP, \R)$ where $\PP$ is 
a set of  ADPs and $\R$  contains the same ADPs like $\PP$ but without any
annotations. We did not choose this representation since it ``duplicates'' information
(i.e., every non-annotated original rule can be directly obtained from the corresponding ADP in
$\PP$ by removing all annotations and thus, it is superfluous to store both the ADP and
its corresponding rule in two separate components). 
This is in contrast to ordinary DPs where the original rule cannot be reconstructed from
the DPs.

Moreover, a representation with pairs $(\PP, \R)$
has the disadvantage
that a processor like the usable rules processor would have to
delete rules from $\R$ which affects the innermost evaluation strategy and would result in
a processor that is not complete. Therefore, instead of pairs  $(\PP, \R)$ we only regard
a set of ADPs and use a flag for every ADP to indicate whether this ADP  can also be used
as an ordinary rule.}

\begin{defi}[ADPs, ADP Problem]
  \label{def:canonical-ADPs}
  An \emph{annotated dependency pair (ADP)} has the form $\ell \ruleArr{}{}{} \{ p_1:r_{1}, \ldots, p_k: r_k\}^{m}$, where $\ell \in \TT$ with $\ell \notin \VSet$, $m \in \{\ttrue, \tfalse\}$, and for all $1 \leq j \leq k$ we have $r_{j} \!\in\! \TT^{\sharp}$ with $\VSet(r_j) \subseteq \VSet(\ell)$.
  
  For a rule $\ell \to \mu = \{ p_1 : r_1, \ldots, p_k : r_k \}$, its \emph{canonical annotated dependency pair} is $\DPair{\ell \to \mu} = \ell \to \{ p_1 : \anno_{\pos_{\SignatureD}(r_1)}(r_1), \ldots, p_k : \anno_{\pos_{\SignatureD}(r_k)}(r_k)\}^{\ttrue}$.
  The canonical ADPs of a PTRS $\R$ are $\DPair{\R} = \{\DPair{\ell \to \mu} \mid \ell \to \mu \in \R\}$.
  An \emph{ADP problem} is a (finite) set of ADPs $\PP$.
\end{defi}

In the following, we fix an ADP problem $\PP$.
Note that in contrast to ordinary dependency pairs as in \Cref{DP Framework}, due to the
duality
of ADPs,
the
left-hand sides of ADPs are not annotated
and the matching used to perform rewriting with ADPs in \Cref{def:ADPs-and-Rewriting-Innermost-main,def:ADPs-and-Rewriting-full-main} 
will ignore the annotations as well.

\begin{exa}
  \label{example:running-1-ADPs}
  For $\R_2 = \{\eqref{rule-01}, \eqref{rule-03}\}$ and $\R_3 = \{\eqref{rule-01},
  \eqref{rule-02}, \eqref{rule-03nd}\}$ from \Cref{S1S2S3}, we obtain $\DPair{\R_2} =
  \{\eqref{run1-ADP-1}, \eqref{run1-ADP-2} \}$ and $\DPair{\R_3} = \{\eqref{run1-ADP-1},
  \eqref{run1-ADP-3}, \eqref{run1-ADP-2nd} \}$ with

  \vspace*{-.5cm}

  \hspace*{-.7cm}{\small
    \begin{minipage}[t]{0.47\textwidth}
      \begin{align}
        \tg    & \to \{\nicefrac{3}{4}:\tD(\tG), \nicefrac{1}{4}:\tz\}^{\ttrue} \tag{\ensuremath{\sharp}\ref{rule-01}} \label{run1-ADP-1} \\
        \td(x) & \to \{1:\tz\}^{\ttrue} \tag{\ensuremath{\sharp}\ref{rule-02}} \label{run1-ADP-3}
      \end{align}
    \end{minipage}
    \hfill
    \begin{minipage}[t]{0.47\textwidth}
      \begin{align}
        \td(x)      & \to \{1:\tc(x,x)\}^{\ttrue} \tag{\ensuremath{\sharp}\ref{rule-03}} \label{run1-ADP-2} \\
        \td(\td(x)) & \to \{1:\tc(x,\tG)\}^{\ttrue} \tag{\ensuremath{\sharp}\ref{rule-03nd}} \label{run1-ADP-2nd}
      \end{align}
    \end{minipage}
  }
\end{exa}

Similar to the DPs in the non-probabilistic setting, the annotated subterms of a right-hand side $r_j$ correspond to those subsequent ``function calls'' 
where we need to prove that the arguments are decreasing to conclude $\mathtt{AST}$.
For example, for the ADP $ \tg \to \{\nicefrac{3}{4}:\tD(\tG), \nicefrac{1}{4}:\tz\}^{\ttrue}$ \eqref{run1-ADP-1},
we have to compare the annotated left-hand side $\tG$ to both $\tD(\tg)$ and $\tG$ from
the first term in the right-hand side.
When adapting the
reduction pair processor to the probabilistic setting in \Cref{sec:Reduction Pair
  Processor}, we will show how to compare a single term with several annotated subterms
on the right-hand side. 

\begin{exa}
  \label{example:running-2-ADPs}
  For $\R_{\textsf{alg}}$ from \Cref{example:running-2}, the canonical ADPs
  are\footnote{For reasons of space, we write $\tLoopOne$, $\tLoopTwo$, $\tTriple$, and $\tDouble$, instead of $\mathsf{Loop1}$, $\mathsf{Loop2}$, $\mathsf{Triple}$, and $\mathsf{Double}$, respectively.}
    {\small
      \begin{align}
        \tloopOne(y) & \to \{\nicefrac{1}{2}:\tLoopOne(\tDouble(y)), \; \nicefrac{1}{2}:\tLoopTwo(\tDouble(y))\}^{\ttrue} \tag{\ensuremath{\sharp}\ref{loopOne-1}} \label{run2-ADP-1} \\
        \tloopOne(y) & \to \{\nicefrac{1}{3}:\tLoopOne(\tTriple(y)), \; \nicefrac{2}{3}:\tLoopTwo(\tTriple(y))\}^{\ttrue} \tag{\ensuremath{\sharp}\ref{loopOne-2}} \label{run2-ADP-2}\!
      \end{align}

      \smallskip

      {\footnotesize \hspace*{-.8cm}
        \begin{minipage}[t]{0.49\textwidth}
          \vspace*{-0.8cm}
          \begin{align}
            \tloopTwo(\ts(y)) & \to \{1:\tLoopTwo(y)\}^{\ttrue} \tag{\ensuremath{\sharp}\ref{run2-rule-3}} \label{run2-ADP-3} \\
            \tdouble(\ts(y))  & \to \{1:\ts(\ts(\tDouble(y)))\}^{\ttrue} \tag{\ensuremath{\sharp}\ref{run2-rule-4}} \label{run2-ADP-4} \\
            \tdouble(\tz)     & \to \{1:\tz\}^{\ttrue} \tag{\ensuremath{\sharp}\ref{run2-rule-5}} \label{run2-ADP-5}\!
          \end{align}
        \end{minipage}
        \hfill
        \begin{minipage}[t]{0.49\textwidth}
          \vspace*{-0.6cm}
          \begin{align}
            \ttriple(\ts(y)) & \to \{1:\ts(\ts(\ts(\tTriple(y))))\}^{\ttrue} \tag{\ensuremath{\sharp}\ref{run2-rule-6}} \label{run2-ADP-6} \\
            \ttriple(\tz)    & \to \{1:\tz\}^{\ttrue} \tag{\ensuremath{\sharp}\ref{run2-rule-7}} \label{run2-ADP-7}\!
          \end{align}
        \end{minipage}
      }}
  \vspace*{.1cm}

  \noindent
\end{exa}

\noindent
When defining the rewrite relation for ADPs, the essential part is to determine how to handle annotations 
if we rewrite above or below them.
We start with defining innermost rewriting.

\subsection{Innermost Rewriting with ADPs}\label{subsec:Innermost}
\noindent
As in the classical non-probabilistic DP framework, 
our goal is to track reduction sequences with \textbf{a}-steps where the root symbol of the redex is annotated, 
and between two \textbf{a}-steps there can be several \textbf{n}-steps.
Our novel rewrite relation for ADPs refines \textbf{a}-steps into (\textbf{at})- and
(\textbf{af})-steps
in order to distinguish between
rewriting at an \underline{\textbf{a}}nnotation by an ADP with flag
\underline{\textbf{t}}\textsf{rue} or \underline{\textbf{f}}\textsf{alse}. Similarly, we
also refine
\textbf{n}-steps into (\textbf{nt})- and
(\textbf{nf})-steps depending on the flag of the ADP that is used to rewrite at a
\textbf{n}on-annotated position.
Still, the (\textbf{at})- and (\textbf{af})-steps are the crucial ones, and when extending
chains to the probabilistic setting, we prohibit
infinite rewrite sequences only consisting of \textbf{n}-steps.
Similar to the 
 non-probabilistic setting (\Cref{DP Framework}),
this is needed for the reduction pair processor (\Cref{sec:Reduction Pair Processor}) 
to guarantee termination if we orient the \textbf{a}-steps strictly with
a well-founded order,
while only orienting the \textbf{n}-steps weakly.

Rewriting with an ADP problem
$\PP$ is like ordinary probabilistic term rewriting while considering and modifying
annotations
that indicate where a non-$\mathtt{AST}$ evaluation may arise.
If we rewrite a term $s$ at an innermost position $\pi$ with an ADP $\ell \ruleArr{}{}{} \{ p_1:r_{1}, \ldots, p_k: r_k\}^{m}$, 
then the following properties should hold:
\begin{itemize}
\item[(a)] Due to the duality of ADPs,
the matching of left-hand
sides should ignore the annotations. So for example, 
  we have to be able to rewrite both $\td(\td(x))$ and
    $\tD(\tD(x))$  
with an ADP $\td(\td(x)) \to \{1:\tc(x,\tG)\}^{\ttrue}$.
\item[(b)] Annotations below the position $\pi$ should be removed, as terms below the redex have
    to be normal forms due to the innermost evaluation strategy.
  \item[(c)] Annotations at positions orthogonal to $\pi$ should be kept.
  \item[(d)] Annotations at positions strictly above $\pi$ should only be kept if the ADP
    has
    the flag $m = \ttrue$.
  \item[(e)] Annotations in the right-hand sides $r_1, \ldots, r_k$ should only be kept
    if the redex was annotated at the root, i.e., if  $\pi \in \posT(s)$.  
\end{itemize}
This leads to the following probabilistic rewrite relation.

\begin{defi}[Innermost Rewriting with ADPs, $\itored{}{}{\PP}$]\label{def:ADPs-and-Rewriting-Innermost-main}
  A term $s \in \TT^{\sharp}$ \emph{rewrites innermost} with $\PP$ to $\mu =
  \{p_1:t_1,\ldots,p_k:t_k\}$ (denoted $s \itored{}{}{\PP} \mu$) if there are an $\ell
  \ruleArr{}{}{} \{ p_1:r_{1}, \ldots, p_k: r_k\}^{m} \in \PP$, a substitution $\sigma$,
  and a $\pi \in \pos_{\SignatureD \cup \SignatureA}(s)$ such that
  $\flat(s|_\pi)=\ell\sigma \in \ANF_{\PP}$, and for all $1 \leq j \leq k$ the term $t_j$
  is defined as follows, depending on the flag $m$ and on whether $\pi \in \posT(s)$ holds:
  \begin{equation*}
    \begin{array}{c | ll | ll |}
     & \pi \in \posT(s) & & \pi \not\in \posT(s) &\\
      \hline
      m = \ttrue & t_j = \quad\; s[r_j\sigma]_{\pi} & (\mathbf{at}) & t_j = \quad\; s[\flat(r_j)\sigma]_{\pi} & (\mathbf{nt}) \\
      \hline
      m = \tfalse & t_j = \disannoPos{\pi}( s[r_j\sigma]_{\pi}) & (\mathbf{af}) & t_j = \disannoPos{\pi}( s[\flat(r_j)\sigma]_{\pi}) & (\mathbf{nf}) \\
      \hline 
    \end{array}
  \end{equation*}
  We sometimes indicate explicitly which case of the above definition is used, so,
  e.g., we write $s \itored{}{(\mathbf{at})}{\PP} \mu$.
\end{defi}

The above definition indeed corresponds to the properties (a)-(e) mentioned above:
\begin{itemize}
  \item[(a)] When computing the matching substitution $\sigma$, we only consider the terms
    $\ell$ and 
    $\flat(s|_\pi)$ without annotations.
    \item[(b)] Since all annotations of $s|_\pi$ below the root correspond to variables of
      $\ell$, the substitution $\sigma$ removes these annotations.
      \item[(c)] All annotations in $s$ on positions orthogonal to $\pi$ are not affected
        when computing $t_j$.
        \item[(d)] If $m = \tfalse$, then $\disannoPos{\pi}$ removes all annotations at
          positions strictly above $\pi$.
          \item[(e)] If $\pi \notin \posT(s)$, then we use $\flat(r_j)$ to remove all
            annotations from the right-hand side $r_j$.
\end{itemize}
Let us consider an example for each individual case of $\itored{}{}{\PP}$.

A step of the form $(\mathbf{at})$ (\underline{\textbf{a}}nnotation and \underline{\textbf{t}}\textsf{rue}) is performed at the position of an annotation, i.e., this can potentially lead to a non-$\mathtt{AST}$ evaluation.
Hence, all annotations from the right-hand side $r_j$ of the used ADP are kept during the rewrite step.
First, consider 
$s = \tD(\tG) \itored{}{(\mathbf{at})}{\DPair{\R_3}} \{\nicefrac{3}{4}:\tD(\tD(\tG)), \nicefrac{1}{4}:\tD(\tz)\}$.
Here, we use the ADP $\tg \to \{\nicefrac{3}{4}:\tD(\tG), \nicefrac{1}{4}:\tz\}^{\ttrue}$ \eqref{run1-ADP-1} at position $\pi=1$.
We get $\flat(s|_\pi) = \flat(\tD(\tG)|_{1}) = \tg = \ell$, $t_1 = s[r_1]_{\pi} =
\tD(\tD(\tG))$ and $t_2 = s[r_2]_{\pi} = \tD(\tz)$. 
Note that all annotations below the redex are removed. So if we had the ADPs $\tf(\ta) \to
\{ 1: \tb \}^{\true}$ and $\tf(\tf(x)) \to \{ 1: x \}^{\true}$, then we would have
the rewrite step $\tF(\tF(\tb)) \itored{}{(\mathbf{at})}{} \{ 1: \tf(\tb) \}$,
where we remove the annotation of the normal form $\tF(\tb)$.

A step of the form $(\mathbf{af})$ (\underline{\textbf{a}}nnotation and \underline{\textbf{f}}\textsf{alse}) 
is similar but due to the flag $m = \tfalse$, this ADP cannot be used below an annotation in a non-$\mathtt{AST}$ evaluation.
Hence, we remove all annotations above the used redex.
Using an ADP of the form $\tg \to \{\nicefrac{3}{4}:\tD(\tG), \nicefrac{1}{4}:\tz\}^{\tfalse}$ on the term $\tD(\tG)$ yields $\tD(\tG) \itored{}{(\mathbf{af})}{}
\{\nicefrac{3}{4}:\td(\tD(\tG)), \nicefrac{1}{4}:\td(\tz)\}$, i.e., we remove the annotation of $\tD$ at the root.

A step of the form $(\mathbf{nt})$ (\underline{\textbf{n}}o annotation and \underline{\textbf{t}}\textsf{rue}) 
is performed at the position of a subterm without annotation.
Hence, the subterm cannot lead to a non-$\mathtt{AST}$ evaluation, but this rewrite step may be needed for an annotation at a position above.
As an example, one could rewrite the non-annotated subterm $\tg$ in $\tD(\tg) \itored{}{(\mathbf{nt})}{\DPair{\R_3}}
\{\nicefrac{3}{4}:\tD(\td(\tg)), \nicefrac{1}{4}:\tD(\tz)\}$ using the ADP
$\eqref{run1-ADP-1}$.

Finally, a step of the form $(\mathbf{nf})$ (\underline{\textbf{n}}o annotation and \underline{\textbf{f}}\textsf{alse}) 
is irrelevant for non-$\mathtt{AST}$ evaluations, because the redex is not annotated and due to $m = \tfalse$, 
afterwards one cannot rewrite an annotated term at a position above.
For example, if one had the ADP $\tg \to \{\nicefrac{3}{4}:\tD(\tG), \nicefrac{1}{4}:\tz\}^{\false}$, 
then we would obtain $\tD(\tg) \itored{}{(\mathbf{nf})}{} \{\nicefrac{3}{4}:\td(\td(\tg)), \nicefrac{1}{4}:\td(\tz)\}$.
The case $(\mathbf{nf})$ is  needed for innermost rewriting to ensure that normal forms always remain the same, 
even if we remove or add annotations in ADPs or if we modify their flag $m$.
The flag $\false$ and steps of the form
 $(\mathbf{nf})$ will allow us to obtain a usable rules processor which is \emph{complete} (see \Cref{thm:prob-iURP}), 
in contrast to the incomplete usable rules processor from \Cref{URP} in the non-probabilistic setting.

\subsection{Full Rewriting with ADPs}\label{subsec:Full}
\noindent
Next, we consider rewriting with ADPs w.r.t.\ an arbitrary evaluation strategy.
As shown in \Cref{example:running-1}, the PTRS $\R_3$ is $\mathtt{iAST}$, but not $\mathtt{AST}$.
To simulate the corresponding non-$\mathtt{AST}$ reduction from \Cref{example:running-1} with ADPs, it is crucial that if the variables of an ADP are instantiated by annotated terms, then these annotations are not all removed when rewriting with this ADP.
For example, if one uses the ADP \eqref{run1-ADP-2nd} to rewrite the redex $\tD(\tD(\tG))$ (whose subterm $\tG$ starts a non-$\mathtt{AST}$ evaluation), then it should still be possible to continue the evaluation of the subterm $\tG$ afterwards which was ``completely inside'' the substitution of the applied rewrite step.
(We will illustrate this in detail in \Cref{example:ADPs-full-rewriting-fail}.)

So for full rewriting with ADPs, we have to keep certain annotations below the used redex.
To this end, we use \emph{variable reposition functions (VRFs)} to relate positions of variables in the left-hand side of an ADP to those positions of the same variables in the right-hand sides where we want to keep the annotations of the instantiated variables.
So for an ADP $\ell \to \mu^m$ with $\ell|_\pi = x$, we indicate which occurrence of $x$ in $r \in \Supp(\mu)$ should keep the annotations if one rewrites an instance of $\ell$ where the subterm at position $\pi$ contains annotations.\footnote{VRFs were introduced in \cite{IJCAR2024} when adapting ADPs to full \emph{relative} rewriting.
However, due to the probabilistic setting, our definition here is slightly different.}

\begin{defi}[Variable Reposition Functions]
  \label{def:Var-Repos-Func}
  Let $\ell\to\{p_1\!:\!r_1, \ldots, p_k\!:\!r_k\}^m$\linebreak
  be an ADP.
  A family of functions $(\varphi_j)_{1 \leq j \leq k}$ with $\varphi_j: \pos_{\VSet}(\ell) \to \pos_{\VSet}(r_j) \uplus \{\bot\}$  is called a family of \emph{variable reposition functions} (VRF) for the ADP iff for all $1 \leq j \leq k$ we have $\ell|_\pi = r_j|_{\varphi_j(\pi)}$ whenever $\varphi_j(\pi) \neq \bot$.
\end{defi}

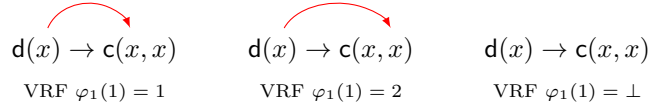
\begin{figure}
    \centering
    \begin{tikzpicture}
        \node (formula) [label=below:{\scriptsize VRF $\varphi_1(1) = 1$}] {$\td(x) \to \tc(x,x)$};
        \draw[-latex,red] ($(formula.north west)+(.6,0)$) arc
        [
            start angle=160,
            end angle=20,
            x radius=0.58cm,
            y radius =0.5cm
        ] ;
    
    \end{tikzpicture}\qquad
    \begin{tikzpicture}
        \node (formula) [label=below:{\scriptsize VRF $\varphi_1(1) = 2$}] {$\td(x) \to \tc(x,x)$};
        \draw[-latex,red] ($(formula.north west)+(.6,0)$) arc
        [
            start angle=160,
            end angle=20,
            x radius=0.75cm,
            y radius =0.5cm
        ] ;
    
    \end{tikzpicture}\qquad
      \begin{tikzpicture}
        \node (formula) at (0,0) [label=below:{\scriptsize VRF $\varphi_1(1) =  \bot$}] {$\td(x) \to \tc(x,x)$};
    \end{tikzpicture}
  \caption{The three possible VRFs for the ADP $\td(x) \to \{1:\tc(x,x)\}^{\ttrue}$ indicated by the red arrows.}\label{fig:var-repos}
\end{figure}

\begin{exa}
  Let us first consider the left-linear ADP $\td(x) \to \{1:\tc(x,x)\}^{\ttrue}$ \eqref{run1-ADP-2}. 
  Here, there are three possible VRFs, viz.,  $\varphi_1(1) =  1$,
  $\varphi_1(1) =  2$, or
  $\varphi_1(1) =  \bot$  
  to indicate whether \eqref{run1-ADP-2} should rewrite a term like $\tD(\tG)$ to $\tc(\tG,\tg)$, $\tc(\tg,\tG)$, or $\tc(\tg,\tg)$.
  See \Cref{fig:var-repos} for a graphical representation of these VRFs.
\end{exa}

A VRF indicates at which positions the annotations should be kept.
However, since ADPs do not have to be left-linear, this movement of annotations from the left- to
the right-hand side is a bit more involved.
Note that our VRFs do not have to be injective, i.e., for the ADP $\tc(x,x) \to \{1:\td(x)\}^{\ttrue}$ 
a valid VRF would be $\varphi_1$ with $\varphi_1(1) = 1$ and $\varphi_1(2) = 1$.
Now consider the term $\tc(\tg(\tF), \tG(\tf))$.
Which annotations should be kept when
rewriting at the root position with the ADP $\tc(x,x) \to \{1:\td(x)\}^{\ttrue}$?
Since both the subterm $\tG(\tf)$ and the subterm
$\tF$ may lead to a non-$\mathtt{AST}$ evaluation, this should still be possible
after the rewrite step, 
since both subterms are completely ``inside'' the substitution used for matching.
Hence, our rewrite relation allows for the step $\tc(\tg(\tF), \tG(\tf)) \tored{}{}{\PP} \td(\tG(\tF))$, 
i.e., we consider all terms corresponding to the same variable when
applying an ADP and \emph{merge} their annotations.
In this way, we obtain a term that still contains all annotations that may lead to
non-$\mathtt{AST}$ evaluations.

So whenever there is a position $\rho \in \pos_{\VSet}(\ell)$ of a variable in the
left-hand side $\ell$ of an ADP and
$\rho$ corresponds to a position 
$\varphi_j(\rho)\neq\bot$ of the same variable in the right-hand side $r_j$,
then all annotations
below the position $\rho$ in the redex $s|_{\pi}$ must be kept
below the position $\varphi_j(\rho)$ in the resulting term $r_j \sigma$.
In other words, whenever there is a position $\tau$ with an annotated symbol at position
$\rho.\tau$ in the redex $s|_{\pi}$, then the symbol at position 
$\varphi_j(\rho).\tau$ in the resulting term $r_j \sigma$ should also be annotated.
We collect all these positions in a set $\Psi_j$.

In our example
we have $\ell = \tc(x,x)$ and $r_1 = \td(x)$. The
redex $s|_{\pi}$ is $\tc(\tg(\tF), \tG(\tf))$ and thus, for the matching substitution
$\sigma$ with
$\ell\sigma =
\flat(s|_{\pi})$,
we obtain
$r_1 \sigma =
\td(\tg(\tf))$. To find out which symbols should be annotated in $r_1 \sigma$,
we consider the VRF
 $\varphi_1(1) = 1$ and $\varphi_1(2) = 1$ which maps variable positions $\rho$ of $\ell$ to
variable positions of $r_1$.
For  $\rho = 1$, the redex
$\tc(\tg(\tF), \tG(\tf))$ 
has  an annotation at the position $\rho.\tau$ with $\tau =
1$. Hence, the position 
 $\varphi_1(\rho).\tau = 1.1$  in $r_1 \sigma$ should be annotated. Moreover, for 
$\rho' = 2$,  the redex
$\tc(\tg(\tF), \tG(\tf))$ has
an annotation at the position $\rho'.\tau'$ with $\tau' =
\epsilon$. Hence, the position 
$\varphi_1(\rho').\tau' = 1$  in $r_1 \sigma$ should be annotated as well.
Thus, for $\Psi_1 = \{1.1, 1 \}$ we obtain $\sharp_{\Psi_1}(r_1 \sigma) = \sharp_{\{1.1, 1
  \}}(\td(\tg(\tf))) = \td(\tG(\tF))$.

Now we can define (full) rewriting with ADPs. As in the definition of innermost rewriting
(\Cref{def:ADPs-and-Rewriting-Innermost-main}),
all annotations of $r_j$ should be deleted whenever $\pi \notin \posT(s)$. Thus, we use
$\sharp_{\Psi_j}(r_j \sigma)$ which removes
all annotations from $r_j$ (and only adds certain annotations on the terms introduced by
$\sigma$). In contrast, if $\pi \in \posT(s)$, then we use
$\sharp_{\posT(r_j) \cup
  \Psi_j}(r_j \sigma)$ in order to keep the annotations of $r_j$.

\begin{defi}[Rewriting with ADPs, $\tored{}{}{\PP}$]
  \label{def:ADPs-and-Rewriting-full-main}
   A term $s \in \TT^{\sharp}$ \emph{rewrites} with $\PP$ to $\mu = \{p_1:t_1,\ldots,p_k:t_k\}$ (denoted $s \tored{}{}{\PP} \mu$) 
  if there are an $\ell \ruleArr{}{}{} \{ p_1:r_{1}, \ldots, p_k: r_k\}^{m} \in \PP$, a VRF $(\varphi_j)_{1 \leq j \leq k}$ for this ADP, 
  a substitution $\sigma$, and a $\pi \in \pos_{\SignatureD \cup \SignatureA}(s)$ 
  such that $\flat(s|_\pi)=\ell\sigma$, 
  and for all $1 \leq j \leq k$ the term $t_j$ is defined
as follows,  depending on the flag $m$ and on whether $\pi \in \posT(s)$ holds:
  \begin{equation*}
    \begin{array}{c | ll | ll |}
    & \pi \in \posT(s) & & \pi \not\in \posT(s) &\\
      \hline
      m = \ttrue & t_j = \quad\; s[\sharp_{\posT(r_j) \cup   \Psi_j}(r_j\sigma)]_{\pi}
      & (\mathbf{at}) & t_j = \quad\; s[\sharp_{\Psi_j}(r_j\sigma)]_{\pi}& (\mathbf{nt}) \\
      \hline
      m = \tfalse & t_j = \disannoPos{\pi}(s[\sharp_{\posT(r_j) \cup   \Psi_j}(r_j\sigma)]_{\pi})
      & (\mathbf{af}) & t_j = \disannoPos{\pi}(s[\sharp_{\Psi_j}(r_j\sigma)]_{\pi})
& (\mathbf{nf}) \\
      \hline 
    \end{array}
  \end{equation*}
  Here, $\Psi_j\!=\!\{\varphi_j(\rho).\tau \mid \rho \!\in\! \pos_{\VSet}(\ell), \, \varphi_j(\rho)\!\neq\!\bot, \, \rho.\tau \!\in\! \posT(s|_{\pi}) \}$.
\end{defi}

Thus, innermost rewriting is the special case of \Cref{def:ADPs-and-Rewriting-full-main} where we require
that  $\flat(s|_\pi)=\ell\sigma$ is in argument normal form and where we use the VRF which
maps all variable positions to $\bot$ (i.e.,  all annotations below the redex are
removed).

\subsection{Chains}\label{subsec:Chains}
\noindent
For \emph{chains} in the probabilistic setting we now consider specific RSTs, called \emph{chain trees}.
Chain trees are defined analogously to RSTs, 
but the crucial requirement is that every infinite path of the tree must contain infinitely many steps 
of the forms $(\mathbf{at})$ or $(\mathbf{af})$.

\begin{defi}[Chain Tree, $|\F{T}|$]
  \label{def:chaintree}
  We say that $\F{T}=(V,E,L)$ is a $\PP$\emph{-chain tree} (CT) if
  \begin{enumerate}
    \item[(1)] $(V, E)$ is a (possibly infinite) directed tree with nodes $V \neq \emptyset$ and directed edges $E \subseteq V \times V$ where $vE = \{ w \mid (v,w) \in E \}$ is finite for every $v \in V$.
          \vspace*{-.15cm}
    \item[(2)] $L:V\rightarrow(0,1]\times\TT^{\sharp}$ labels every node $v$ by a probability $p_v$ and a term $t_v$.
          For the root $v \in V$ of the tree, we have $p_v = 1$.
          \vspace*{-.15cm}
    \item[(3)] If $vE = \{w_1, \ldots, w_k\}$, then $t_v \tored{}{}{\PP}
            \{\tfrac{p_{w_1}}{p_v}:t_{w_1}, \ldots, \tfrac{p_{w_k}}{p_v}:t_{w_k}\}$. \vspace*{-.15cm}
    \item[(4)] Every infinite path in $\F{T}$ contains infinitely many rewrite steps with Case $(\mathbf{at})$ or $(\mathbf{af})$. 
  \end{enumerate}
  $\F{T}$ is an \emph{innermost} chain tree if the edges represent innermost rewrite steps.
  As for RSTs, we say that a CT \emph{converges} (or \emph{terminates}) with probability $|\F{T}| = \sum_{v \in \ctleaf}
    \, p_v$, where $\ctleaf$ is the set of its leaves.
\end{defi}

With this notion of chain trees, we can now define almost-sure termination for ADP problems.

\begin{defi}[$\mathtt{AST}$ and $\mathtt{iAST}$ for ADP Problems]
  An ADP problem $\PP$ is called \emph{almost-surely terminating} ($\mathtt{AST}$) if we have $|\F{T}| = 1$ for every $\PP$-CT $\F{T}$.
  It is called \emph{almost-surely innermost terminating} ($\mathtt{iAST}$) if we have $|\F{T}| = 1$ for every innermost $\PP$-CT $\F{T}$.
\end{defi}

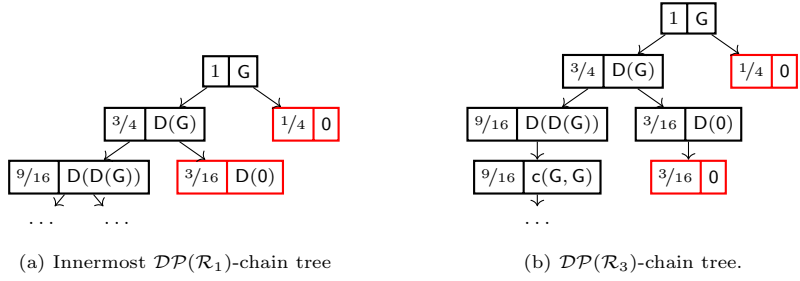
\begin{figure}
  \begin{subfigure}{0.5\textwidth}
    \scriptsize \center \vspace*{-0.1cm}
    \hspace*{-.12cm}
    \begin{tikzpicture}
      \tikzstyle{adam}=[thick,draw=black!100,fill=white!100,minimum size=4mm, shape=rectangle split, rectangle split parts=2,rectangle split horizontal] \tikzstyle{adam2}=[thick,draw=red!100,fill=white!100,minimum size=4mm, shape=rectangle split, rectangle split parts=2,rectangle split horizontal] \tikzstyle{empty}=[rectangle,thick,minimum size=4mm]

      \node[adam] at (-4, 0) (a) {$1$
        \nodepart{two}$\tG$};
      \node[adam] at (-5, -0.7) (b) {$\nicefrac{3}{4}$
        \nodepart{two}$\tD(\tG)$};
      \node[adam2] at (-3, -0.7) (c) {$\nicefrac{1}{4}$
        \nodepart{two}$\tz$};
      \node[adam] at (-6, -1.4) (d) {$\nicefrac{9}{16}$
        \nodepart{two}$\tD(\tD(\tG))$};
      \node[adam2] at (-4, -1.4) (e) {$\nicefrac{3}{16}$
        \nodepart{two}$\tD(\tz)$};
      \node[empty] at (-6.5, -2) (f) {$\ldots$};
      \node[empty] at (-5.5, -2) (g) {$\ldots$};

      \draw (a) edge[->] (b);
      \draw (a) edge[->] (c);
      \draw (b) edge[->] (d);
      \draw (b) edge[->] (e);
      \draw (d) edge[->] (f);
      \draw (d) edge[->] (g);
    \end{tikzpicture}
    \caption{Innermost $\DPair{\R_1}$-chain tree}\label{fig:CT-1}
  \end{subfigure}%
  \begin{subfigure}{0.5\textwidth}
    \scriptsize \center \vspace*{-0.1cm}
    \hspace*{-.12cm}
    \begin{tikzpicture}
      \tikzstyle{adam}=[thick,draw=black!100,fill=white!100,minimum size=4mm, shape=rectangle split, rectangle split parts=2,rectangle split horizontal] \tikzstyle{adam2}=[thick,draw=red!100,fill=white!100,minimum size=4mm, shape=rectangle split, rectangle split parts=2,rectangle split horizontal] \tikzstyle{empty}=[rectangle,thick,minimum size=4mm]

      \node[adam] at (-4, 0) (a) {$1$
        \nodepart{two}$\tG$};
      \node[adam] at (-5, -0.7) (b) {$\nicefrac{3}{4}$
        \nodepart{two}$\tD(\tG)$};
      \node[adam2] at (-3, -0.7) (c) {$\nicefrac{1}{4}$
        \nodepart{two}$\tz$};
      \node[adam] at (-6, -1.4) (d) {$\nicefrac{9}{16}$
        \nodepart{two}$\tD(\tD(\tG))$};
      \node[adam] at (-4, -1.4) (e) {$\nicefrac{3}{16}$
        \nodepart{two}$\tD(\tz)$};
      \node[adam2] at (-4, -2.1) (e2) {$\nicefrac{3}{16}$
        \nodepart{two}$\tz$};
      \node[adam] at (-6, -2.1) (f) {$\nicefrac{9}{16}$
        \nodepart{two}$\tc(\tG,\tG)$};
      \node[empty] at (-6, -2.7) (g) {$\ldots$};

      \draw (a) edge[->] (b);
      \draw (a) edge[->] (c);
      \draw (b) edge[->] (d);
      \draw (b) edge[->] (e);
      \draw (d) edge[->] (f);
      \draw (e) edge[->] (e2);
      \draw (f) edge[->] (g);
    \end{tikzpicture}
    \caption{$\DPair{\R_3}$-chain tree.}\label{fig:CT-2}
  \end{subfigure}
  \caption{Two chain trees. Nodes labeled by normal forms are marked in red.}
\end{figure}

The corresponding $\DPair{\R_1}$-chain tree $\F{T}$ for the $\R_1$-RST from \Cref{fig:R1}
is shown in \Cref{fig:CT-1}.
Here, we again have $|\F{T}| = 1$.

\begin{exa}
  \label{example:ADPs-full-rewriting-fail}
  To see why VRFs are needed to obtain a sound chain criterion for full rewriting, consider $\R_3$ with $\DPair{\R_3} = \{\eqref{run1-ADP-1}, \eqref{run1-ADP-3}, \eqref{run1-ADP-2nd} \}$ from \Cref{example:running-1-ADPs}:
  \vspace*{-.9cm}

  \hspace*{-.7cm}{\small
    \begin{minipage}[t]{0.47\textwidth}
      \begin{align*}
        \tg    & \to \{\nicefrac{3}{4}:\tD(\tG), \nicefrac{1}{4}:\tz\}^{\ttrue} \quad \eqref{run1-ADP-1} \\
        \td(x) & \to \{1:\tz\}^{\ttrue} \quad \eqref{run1-ADP-3}\!
      \end{align*}
    \end{minipage}
    \hfill
    \begin{minipage}[t]{0.47\textwidth}
           \begin{align*}
        \td(\td(x)) & \to \{1:\tc(x,\tG)\}^{\ttrue} \quad \eqref{run1-ADP-2nd}
      \end{align*}
    \end{minipage}
  }\vspace*{.2cm}

  As noted in \Cref{example:running-1}, $\R_3$ is $\mathtt{iAST}$, but not $\mathtt{AST}$.
  For $\DPair{\R_3}$, applying two rewrite steps with the ADP \eqref{run1-ADP-1} to $\tG$ would result in a chain tree with the leaves $\nicefrac{9}{16}:\tD(\tD(\tG))$, $\nicefrac{3}{16}:\tD(\tz)$ (which can be extended by the child $\nicefrac{3}{16}:\tz$), and $\nicefrac{1}{4}:\tz$.
  Now it is important that the next application of the ADP \eqref{run1-ADP-2nd}
  to $\nicefrac{9}{16}:\tD(\tD(\tG))$ allows us to keep the annotation of its argument $\tG$, 
  i.e., it can yield $\nicefrac{9}{16}:\tc(\tG,\tG)$.
  The corresponding chain tree is depicted in \Cref{fig:CT-2}.
  Otherwise (i.e., if it could only yield $\nicefrac{9}{16}:\tc(\tg,\tG)$), the number of $\tG$-symbols would never be increased.
  Since non-annotated symbols like $\tg$ do not result in any $(\mathbf{at})$- or $(\mathbf{af})$-steps, 
  for all such chain trees $\F{T}$ we would have $|\F{T}| = 1$, 
  i.e., then the chain criterion would falsely conclude that $\R_3$ is $\mathtt{AST}$.
\end{exa}

For innermost rewriting, we now indeed obtain a sound and complete chain criterion as in the non-probabilistic setting, i.e., to analyze $\mathtt{iAST}$ for a PTRS $\R$ one can analyze $\mathtt{iAST}$ of its canonical ADP problem $\DPair{\R}$ instead.

\begin{restatable}[Chain Criterion for $\mathtt{iAST}$]{thm}{ProbChainCriterionInnermost}
  \label{thm:prob-chain-criterion-innermost}
  A PTRS $\R$ is $\mathtt{iAST}$ iff $\DPair{\R}$ is $\mathtt{iAST}$.
\end{restatable}

\begin{myproofsketch}
  \textit{Soundness}: Each innermost $\R$-RST $\F{T}$ can be transformed into an innermost $\DPair{\R}$-CT $\F{T}'$ with $|\F{T}| = |\F{T}'|$ by annotating at least all defined symbols at the root of subterms that are not in normal form. 
  This is possible
  since the relation $\itored{}{}{\DPair{\R}}$ only removes annotations of normal forms if
  all ADPs have the flag $\true$.

  \textit{Completeness}: Each innermost $\DPair{\R}$-CT $\F{T}$ can be transformed into an
  innermost $\R$-RST
  $\F{T}'$ with
   $|\F{T}| = |\F{T}'|$ by removing all annotations. 
\end{myproofsketch}

For the chain criterion for full $\mathtt{AST}$, we have to impose a restriction on the initial PTRS.
Note that our VRFs in \Cref{def:Var-Repos-Func}
map a position of the left-hand side $\ell$ to at most one position in each right-hand side $r_j$ of an ADP, even if the ADP is duplicating.
A probabilistic rule or ADP $\ell \to \mu$ is \emph{non-duplicating} if all rules in $\{\ell \to r \mid r \in \Supp(\mu)\}$ are, and a PTRS or ADP problem is non-duplicating if all of its rules are (disregarding the flag for ADPs).
For example, as mentioned, for the duplicating ADP $\td(x) \to \{1:\tc(x,x)\}^{\ttrue}\,$ \eqref{run1-ADP-2}, we have three different VRFs which map position $1$ to either $\bot$, $1$, or $2$, but we cannot map it to both positions $1$ and $2$.

Therefore, our VRFs (and the corresponding rewrite relation of ADPs) cannot handle duplicating rules correctly.
With VRFs as in \Cref{def:Var-Repos-Func}, $\DPair{\R_2}$ would be considered to be $\mathtt{AST}$, as $\tD(\tG)$ only rewrites to $\{ 1: \tc(\tG,\tg) \}$ or $\{ 1: \tc(\tg,\tG) \}$, but the annotation cannot be duplicated.
Hence, the chain criterion would be unsound for duplicating PTRSs like $\R_2$.

To handle duplicating rules, one can adapt the direct application of orders to prove $\mathtt{AST}$ from \cite{kassinggiesl2023iAST}
and try to remove the duplicating rules of the PTRS before constructing the canonical ADPs.

Alternatively, one could modify the definition of the rewrite relation $\tored{}{}{\PP}$ and use \emph{generalized} VRFs (GVRFs) instead of VRFs which can duplicate annotations.
This would yield a sound and complete chain criterion for $\mathtt{AST}$ of possibly duplicating PTRSs, but then one would also have to consider this modified definition of $\tored{}{}{\PP}$ for the processors of the ADP framework in \Cref{The Probabilistic ADP Framework}.
Unfortunately, the most important processors would become unsound when defining the rewrite relation $\tored{}{}{\PP}$ via GVRFs (see Ex.\ \ref{example:dup-problem} and \ref{example:running-1-RPP-iAST}).
Therefore, we use VRFs instead and restrict ourselves to non-duplicating PTRSs for the soundness of the chain criterion for full probabilistic rewriting.\footnote{A related restriction is needed when analyzing (non-probabilistic) relative termination due to the VRFs \cite{IJCAR2024}.}

\begin{restatable}[Chain Criterion for $\mathtt{AST}$]{thm}{ProbChainCriterionFull}
  \label{thm:prob-chain-criterion-full}
  A non-duplicating PTRS $\R$ is $\mathtt{AST}$ iff $\DPair{\R}$ is $\mathtt{AST}$.
\end{restatable}

\begin{myproofsketch}
  \textit{Soundness}: Each $\R$-RST $\F{T}$ can be transformed into a $\DPair{\R}$-CT
  $\F{T}'$ with
   $|\F{T}| = |\F{T}'|$ by annotating all defined symbols.
  Since $\R$ is non-duplicating and all ADPs in
$\DPair{\R}$ have the flag $\true$,
   we do not remove any annotations when rewriting with $\tored{}{}{\DPair{\R}}$.

  \textit{Completeness}: Each $\DPair{\R}$-CT $\F{T}$ can be transformed into a $\R$-RST
  $\F{T}'$ with $|\F{T}| = |\F{T}'|$ by removing all annotations. 
\end{myproofsketch}

\begin{exa}
  As an example, the PTRS $\R_{\textsf{alg}}$ from \Cref{example:running-2} is non-duplicating.
  Hence, we can not only analyze $\mathtt{iAST}$ but also $\mathtt{AST}$ via the canonical ADP problem $\DPair{\R_{\textsf{alg}}}$ from \Cref{example:running-2-ADPs}.
\end{exa}

\begin{rem}
  \label{remark:minimality}
  In the chain criterion for non-probabilistic DPs, it suffices to regard only instantiations where all terms below an annotated symbol are terminating.
  The reason is the \emph{minimality property}
  of non-probabilistic term rewriting, i.e., whenever a term starts an infinite rewrite sequence, then it also starts an infinite sequence where all proper subterms of every used redex are terminating.
  However, in the probabilistic setting such a minimality property does not hold \cite{FOSSACS24}.
  For $\R_3$ from \Cref{S1S2S3}, $\tg$ starts a non-$\mathtt{AST}$ RST, but in this RST, one has to apply Rule \eqref{rule-03nd}
  to the redex $\td(\td(\tg))$, although it contains the proper subterm $\tg$ that starts a non-$\mathtt{AST}$ RST.
\end{rem}

\section{The Probabilistic ADP Framework}
\label{The Probabilistic ADP Framework}

\noindent
The idea of the DP framework for non-probabilistic TRSs is to apply \emph{DP processors} repeatedly which transform a DP problem into simpler sub-problems as we have seen in \Cref{DP Framework}.
The same idea is used in the ADP framework, which yields the first \emph{modular} approach for proving $\mathtt{AST}$ of probabilistic TRSs automatically.
An \emph{ADP processor} $\Proc$ has the form $\Proc(\PP) = \{\PP_1, \ldots,\PP_n\}$ for ADP problems $\PP, \PP_1, \ldots, \PP_n$.
Let $\mathcal{Z} \in \{\mathtt{AST}, \mathtt{iAST}\}$.
$\Proc$ is \emph{sound} for $\mathcal{Z}$ if $\PP$ is $\mathcal{Z}$ whenever $\PP_j$ is $\mathcal{Z}$ for all $1 \leq j \leq n$.
It is \emph{complete} for $\mathcal{Z}$ if $\PP_j$ is $\mathcal{Z}$ for all $1 \leq j \leq n$ whenever $\PP$ is $\mathcal{Z}$.
Thus, one starts with the canonical ADP problem and applies sound (and preferably complete) ADP processors repeatedly until there are no more remaining ADP problems.
This implies that the canonical ADP problem is $\mathcal{Z}$ and by the chain criterion, the original PTRS is $\mathcal{Z}$ as well.

An ADP problem $\PP$ without annotations is always $\mathtt{AST}$ and $\mathtt{iAST}$, because then no rewrite step with $\tored{}{}{\PP}$ increases the number of annotations (recall that VRFs cannot duplicate annotations).
Hence, then any term with $n$ annotations only starts rewrite sequences with at most $n$ steps of the form $(\mathbf{at})$ or $(\mathbf{af})$, i.e., all $\PP$-CTs are finite.

In \Cref{sec:Dependency Graph Processor} - \ref{sec:Reduction Pair Processor}, we adapt the main processors from \Cref{DP Framework}
to our new ADP framework for $\mathtt{AST}$ and $\mathtt{iAST}$.
Whenever the respective processor can be used for proving $\mathtt{AST}$, we explain the variant of the processor for full rewriting first and afterwards, we show how one can improve the processor for $\mathtt{iAST}$.
Moreover, we introduce an additional processor to remove all occurring probabilities (\Cref{sec:Probability Removal Processor}), a novel processor for PTRSs based on the \emph{subterm criterion} of \cite{DBLP:journals/iandc/HirokawaM07}
(\Cref{sec:Subterm Criterion}), and in \Cref{sec:Switching From Full to Innermost AST} we discuss the possibility of introducing a processor to switch from full $\mathtt{AST}$ to innermost $\mathtt{AST}$.

All processors in this section are complete, because they only
remove annotations and only change flags of ADPs from $\ttrue$ to $\tfalse$.

\begin{restatable}[Completeness of Processors]{thm}{CompletenessProc}
  \label{thm:proc-complete}
  Let $\Proc$ be an ADP processor which satisfies the following for all ADP problems $\PP$:
  Whenever there is an ADP $\ell \to \{p_1:r_1', \ldots, p_k:r_k'\}^{m'} \in \PP'$ for some $\PP' \in \Proc(\PP)$,
  then there is an ADP $\ell \to \{p_1:r_1, \ldots, p_k:r_k\}^{m} \in \PP$ such that
  $\flat(r_j') = \flat(r_j)$ and $\pos_{\SignatureA}(r_j') \subseteq
  \pos_{\SignatureA}(r_j)$ for all $1 \leq j \leq k$, 
  and such that  $m' = \ttrue$ implies  $m = \ttrue$.
  Then $\Proc$ is complete for both $\mathtt{AST}$ and $\mathtt{iAST}$.
\end{restatable}
\begin{myproof}
  Each (innermost) $\PP'$-CT $\F{T}'$ can be transformed into an (innermost) $\PP$-CT $\F{T}$ 
  with $|\F{T}'| = |\F{T}|$ by performing the respective rewrite steps with
  $\ell \to \{p_1:r_1, \ldots, p_k:r_k\}^{m}  \in \PP$ instead of
  $\ell \to \{p_1:r_1', \ldots, p_k:r_k'\}^{m'} \in \PP'$.
  Since we have $\flat(r_j') = \flat(r_j)$ and $\pos_{\SignatureA}(r_j') \subseteq \pos_{\SignatureA}(r_j)$ 
  for all $1 \leq j \leq k$, and $m = \ttrue$ if $m' = \ttrue$,
  $\F{T}$ has the same terms  as $\F{T}'$ but possibly with more annotations.
  Since every infinite path in $\F{T}'$ contains infinitely 
  many rewrite steps with Case $(\mathbf{at})$ or $(\mathbf{af})$, this ensures that 
  $\F{T}$ satisfies this property as well. 
\end{myproof}

This proof structure will often be used for the following soundness proofs as well.
Given some $\PP$-CT $\F{T}$, we use a $\PP' \in \Proc(\PP)$ and construct
a $\PP'$-CT $\F{T}'$
such that $|\F{T}'| \leq |\F{T}|$.
If the processor $\Proc$ removes annotations,
then one has to ensure that
every infinite path in $\F{T}'$ still contains infinitely 
  many rewrite steps with Case $(\mathbf{at})$ or $(\mathbf{af})$.

  Moreover,
    the following two lemmas are important for the soundness proofs.
The first lemma states that
if we can partition the set of ADPs $\PP = \PP_1 \uplus \PP_2$ such that $\PP_1 \cup \flat(\PP_2)$ is $\mathtt{AST}$, 
then it suffices to consider $\PP$-CTs where every infinite path has an infinite number of annotated rewrite steps with nodes from $\PP_2$ 
(i.e., infinitely many rewrite steps where ADPs from $\PP_2$ are used with Case $(\mathbf{at})$ or $(\mathbf{af})$).

\begin{lem}[$\PP$-Partition Lemma]
  \label{lemma:p-partition-main}
  Let $\PP$ be an ADP problem and let $\F{T}$ be a $\PP$-CT that converges with probability $<1$.
  Assume that $\PP_1 \cup \flat(\PP_2)$ is $\mathtt{AST}$.
  Then there exists a $\PP$-CT $\F{T}'$ that converges with probability $<1$ 
  such that every infinite path uses an infinite number of annotated rewrite steps with $\PP_2$.
\end{lem}
    
The second crucial lemma is the \emph{Starting Lemma}.
It shows that w.l.o.g., we can assume that the root of a CT is labeled with $(1:t)$ for 
a term $t \in \TT^\sharp$ with $\posT(t) = \{\varepsilon\}$ and
$\flat(t) = s \theta$ for a substitution $\theta$ and an ADP $s \to \mu^m \in \PP$.
So $t$ is an instantiated left-hand side where only the root is annotated.
The proof for this lemma requires the above $\PP$-Partition Lemma.

\begin{restatable}[Starting Lemma]{lem}{StartingLemma}\label{lemma:starting}
    If an ADP problem $\PP$ is not $\mathtt{AST}$, then there exists a $\PP$-CT $\F{T}$ with $|\F{T}| <1$ that starts with $(1:t)$ for a term $t \in \TT^\sharp$ where $\posT(t) = \{\varepsilon\}$ and $\flat(t) = s \theta$ for a substitution $\theta$ and an ADP $s \to \mu^m \in \PP$.
    If $\PP$ is not $\mathtt{iAST}$, then we can further assume that $t
    \in \ANF_{\PP}$.
\end{restatable}

\subsection{Dependency Graph Processor}
\label{sec:Dependency Graph Processor}

\noindent
As in the non-probabilistic setting, the $\PP$-\emph{dependency graph} is a control flow graph whose nodes are now the ADPs from $\PP$.
Since a term can now contain multiple annotations, the definition is a bit more involved than in the non-probabilistic setting.
An edge of the dependency graph now indicates whether an ADP $\alpha$ may lead to an application of another ADP $\alpha'$ on an annotated subterm whose annotation was introduced by $\alpha$.
This possibility is not related to the probabilities.
Hence, here we use the \emph{non-probabilistic variant}
$\nonprob(\PP) = \{\ell \to \flat(r_j) \mid \ell \to \{p_1:r_1, \ldots, p_k:r_k\}^{\ttrue} \in \PP, 1 \leq j \leq k\}$, which is an ordinary TRS over the original signature $\Sigma$.
For $\nonprob(\PP)$ we only consider rules with the flag $\ttrue$, as only they can be used for rewriting below annotations without removing them.
We define $t \trianglelefteq_{\sharp} s$ if there is a $\pi \in \posT(s)$ and $t = \flat(s|_\pi)$, i.e., $t$ results from a subterm of $s$ with annotated root symbol by removing its annotations.

\subsubsection{Dependency Graph Processor for $\mathtt{AST}$}
\label{sect:DepGraphAST}

\noindent
We start with the dependency graph processor for full $\mathtt{AST}$.

\begin{defi}[Dependency Graph]
  \label{def:prob-dep-graph}
  The \defemph{$\PP$-dependency graph} has the nodes $\PP$ and there is an edge from $\ell_1 \ruleArr{}{}{} \{ p_1:r_{1}, \ldots, p_k: r_k\}^{m}$ to $\ell_2 \to \ldots$ if there are substitutions $\sigma_1, \sigma_2$ and a $t \trianglelefteq_{\sharp} r_{j}$ for some $1 \leq j \leq k$ with $t^\sharp \sigma_1 \to_{\nonprob(\PP)}^* \ell_2^\sharp \sigma_2$.
\end{defi}

So there is an edge from an ADP $\alpha$ to an ADP $\alpha'$ if after a $\tored{}{}{\PP}$-step of the form $(\mathbf{at})$ or $(\mathbf{af})$ with $\alpha$ at position $\pi$ there may eventually come another $\tored{}{}{\PP}$-step of the form $(\mathbf{at})$ or $(\mathbf{af})$ with $\alpha'$ at a position $\tau$ on or below $\pi$, where the root symbol of the subterm at position $\tau$ was introduced by $\alpha$.
Since every infinite path in a $\PP$-CT contains infinitely many rewrite steps with Case $(\mathbf{at})$ or $(\mathbf{af})$, every such path traverses a cycle of the dependency graph infinitely often.
Thus, as in the non-probabilistic setting, it suffices to consider its strongly connected components separately.
In the ADP framework, this means that we remove the annotations from all ADPs except the ones in the SCC that we want to analyze.

To automate the following processor, the same over-approximation techniques as for the non-probabilistic dependency graph can be used, e.g., to estimate whether $t^\sharp \sigma_1 \to_{\nonprob(\PP)}^* \ell_2^\sharp \sigma_2$ holds for some $\sigma_1, \sigma_2$, one checks whether $\renterm(\capterm_\PP(t^\sharp))$ and $\ell^\sharp_2$ are unifiable.
As in \Cref{DP Framework}, $\renterm$ replaces all variable occurrences by fresh variables and $\capterm_\PP(t^\sharp)$ replaces all those subterms $f(\ldots)$ of $t^\sharp$ by different fresh variables where $f$ occurs on the root position of some left-hand side of $\PP$.
One could also improve $\capterm_\PP$ by only regarding the root symbols $f$ of left-hand sides of ADPs with the flag $m = \true$ as ``defined''.
In the following, we extend $\flat$ to multi-distributions, ADPs, and ADP problems by removing the annotations of all occurring terms.
So $\flat(\{ p_1: r_1, \ldots, p_k:r_k \}) = \{ p_1: \flat(r_1), \ldots, p_k: \flat(r_k) \}$, $\flat(\ell \to \mu) = \ell \to \flat(\mu)$, and $\flat(\PP) = \{\flat(\ell \to \mu) \mid \ell \to \mu \in \PP\}$.

\begin{restatable}[Dependency Graph Processor for $\mathtt{AST}$]{thm}{ProbDepGraphProc}
  \label{thm:prob-DGP}
  Let $\PP_1, \ldots, \PP_n$ be the SCCs of the $\PP$-dependency graph.
  Then $\Proc_{\mathtt{DG}}(\PP)=\{\PP_1 \cup \flat(\PP \setminus \PP_1), \ldots,\PP_n \cup \flat(\PP \setminus \PP_n)\}$ is sound and complete for $\mathtt{AST}$.
\end{restatable}

\begin{myproofsketch}
  \textit{Soundness}: 
 For any subset $X \subseteq \PP$, let $\overline{X} = X \cup \flat(\PP \setminus X)$.
  Suppose that every $\overline{\PP_i}$-CT converges with probability $1$ for all $1 \leq i \leq n$.
  We prove that then also every $\PP$-CT converges with probability 1.
  Let  $\F{G}$ be the $\PP$-dependency graph
and let
  $\F{W} = \{\PP_1, \ldots, \PP_n\} \cup \{\{v\} \subseteq \PP \mid v$ is not in an SCC of $\F{G}\}$ be the set of all SCCs 
  and all singleton sets of nodes that do not belong to any SCC\@.
  For two $X_1,X_2 \in \F{W}$ we say that $X_2$ is a \emph{direct successor} of $X_1$ (denoted $X_1 >_{\F{G}} X_2$) 
  if there exist nodes $v \in X_1$ and $w \in X_2$ such that there is an edge from $v$ to $w$ in $\F{G}$.
  The core steps of the proof are the following (where the Starting Lemma (\Cref{lemma:starting}) is
  used in Step 1 and 2, and the $\PP$-Partition Lemma (\Cref{lemma:p-partition-main}) is used in Step 2):
  \begin{enumerate}
    \item[1.] We first show that every ADP problem $\overline{X}$ with $X \in \F{W}$ is $\mathtt{AST}$\@. 
      This also holds for the singleton sets. The reason is that
      if there is only a single ADP with annotations and this ADP
      is not contained in any cycle of the dependency graph, then there
cannot be infinitely many steps at annotated subterms, i.e., all paths are finite.
    \item[2.] Then we show that composing SCCs maintains the $\mathtt{AST}$ property.
    \item[3.] Finally, by induction on the direct successor relation we show that for every $X \in \F{W}$, 
    the ADP problem $\overline{\bigcup_{X >_{\F{G}}^* Y}Y}$ is $\mathtt{AST}$ (where Step
    1 is needed for the induction base and Step 2 is needed for the induction step). 
   This implies that $\PP$ must be $\mathtt{AST}$ as well.
  \end{enumerate}

  \textit{Completeness}: By \Cref{thm:proc-complete}.
\end{myproofsketch}

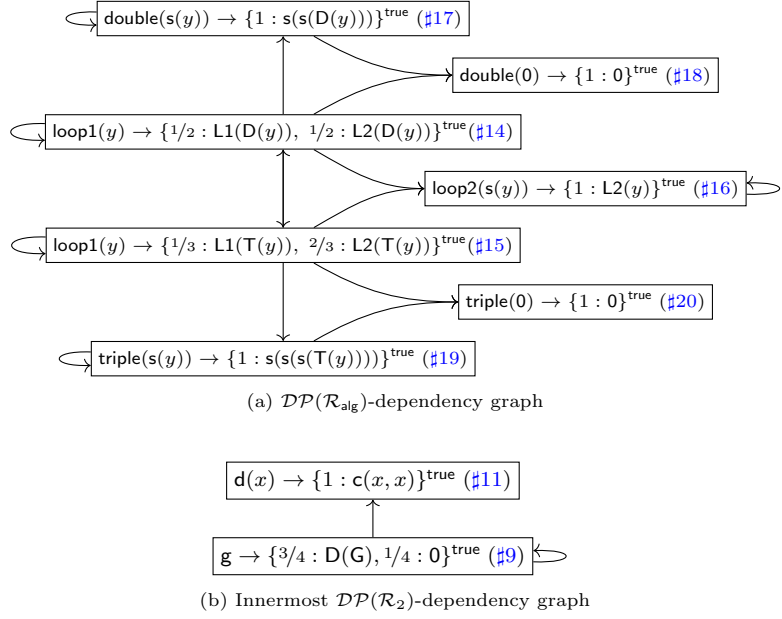
\begin{figure}
  \centering
  \begin{subfigure}{1\textwidth}
    \label{fig:dep-graph-running-2}
    \scriptsize \center
    \begin{tikzpicture}
      \node[shape=rectangle,draw=black!100, minimum size=3mm] (A) at (-2,0.75) {$\tloopOne(y) \to \{\nicefrac{1}{2}:\tLoopOne(\tDouble(y)), \; \nicefrac{1}{2}:\tLoopTwo(\tDouble(y))\}^{\ttrue} \eqref{run2-ADP-1}$};
      \node[shape=rectangle,draw=black!100, minimum size=3mm] (B) at (-2,-0.75) {$\tloopOne(y) \to \{\nicefrac{1}{3}:\tLoopOne(\tTriple(y)), \; \nicefrac{2}{3}:\tLoopTwo(\tTriple(y))\}^{\ttrue} \eqref{run2-ADP-2}$};
      \node[shape=rectangle,draw=black!100, minimum size=3mm] (C) at (-2,2.25) {$\tdouble(\ts(y)) \to \{1:\ts(\ts(\tDouble(y)))\}^{\ttrue} \; \eqref{run2-ADP-4}$};
      \node[shape=rectangle,draw=black!100, minimum size=3mm] (D) at (-2,-2.25) {$\ttriple(\ts(y)) \to \{1:\ts(\ts(\ts(\tTriple(y))))\}^{\ttrue} \; \eqref{run2-ADP-6}$};
      \node[shape=rectangle,draw=black!100, minimum size=3mm] (E) at (2,1.5) {$\tdouble(\tz) \to \{1:\tz\}^{\ttrue} \; \eqref{run2-ADP-5}$};
      \node[shape=rectangle,draw=black!100, minimum size=3mm] (F) at (2,0) {$\tloopTwo(\ts(y)) \to \{1:\tLoopTwo(y)\}^{\ttrue}  \; \eqref{run2-ADP-3}$};
      \node[shape=rectangle,draw=black!100, minimum size=3mm] (G) at (2,-1.5) {$\ttriple(\tz) \to \{1:\tz\}^{\ttrue}  \; \eqref{run2-ADP-7}$};

      \path [->] (A) edge (B); 
      \path [->] (A) edge (C); 
      \path [->, out=330, in=180] (A) edge (F); 
      \path [->, out=30, in=180] (A) edge (E); 
      \path [->, out=178, in=182, looseness=7] (A) edge (A); 
      \path [->] (B) edge (A); 
      \path [->] (B) edge (D); 
      \path [->, out=30, in=180] (B) edge (F); 
      \path [->, out=330, in=180] (B) edge (G); 
      \path [->, out=178, in=182, looseness=7] (B) edge (B); 
      \path [->, out=178, in=182, looseness=8] (C) edge (C); 
      \path [->, out=330, in=180] (C) edge (E); 
      \path [->, out=178, in=182, looseness=8] (D) edge (D); 
      \path [->, out=30, in=180] (D) edge (G); 
      \path [->, out=358, in=2, looseness=10] (F) edge (F);
    \end{tikzpicture}
    \subcaption{$\DPair{\R_{\mathsf{alg}}}$-dependency graph}
    \label{fig:dep-graph-running-2}
  \end{subfigure}%
  \vspace{0.6cm}
  \begin{subfigure}{1\textwidth}
    \label{fig:dep-graph-running-1}
    \footnotesize \center
    \begin{tikzpicture}
      \node[shape=rectangle,draw=black!100, minimum size=3mm] (A) at (0,0) {$\tg \to \{\nicefrac{3}{4}:\tD(\tG), \nicefrac{1}{4}:\tz\}^{\ttrue} \; \eqref{run1-ADP-1}$};
      \node[shape=rectangle,draw=black!100, minimum size=3mm] (B) at (0,1) {$\td(x) \to \{1:\tc(x,x)\}^{\ttrue} \; \eqref{run1-ADP-2}$};

      \path [->] (A) edge (B); \path [->, out=358, in=2, looseness=10] (A) edge (A);
    \end{tikzpicture}
    \subcaption{Innermost $\DPair{\R_{2}}$-dependency graph}
    \label{fig:dep-graph-running-1}
  \end{subfigure}
  \caption{Two examples of dependency graphs}
  \label{fig:dep-graphs}
\end{figure}

\begin{exa}
  \label{example:running-2-DPG-AST}
  Consider $\R_{\mathsf{alg}}$ and its canonical ADPs from \Cref{example:running-2-ADPs}.
  The $\DPair{\R_{\mathsf{alg}}}$-dependency graph is shown in \Cref{fig:dep-graph-running-2}.
  Its SCCs are $\{\eqref{run2-ADP-1}, \eqref{run2-ADP-2}\}$, $\{\eqref{run2-ADP-3}\}$, $\{\eqref{run2-ADP-4}\}$, and $\{\eqref{run2-ADP-6}\}$.
  For each SCC we create a separate ADP problem, where all annota\-tions outside the SCC are removed.
  This leads to the ADP problems $\{\eqref{run2-ADP-1}, \eqref{run2-ADP-2}, \flat(\ref{run2-ADP-3})$ - $\flat(\ref{run2-ADP-7})\}$, $\{\eqref{run2-ADP-3}, \flat(\ref{run2-ADP-1}), \flat(\ref{run2-ADP-2}), \flat(\ref{run2-ADP-4})$ - $\flat(\ref{run2-ADP-7})\}$, $\{\eqref{run2-ADP-4}, \flat(\ref{run2-ADP-1})$ - $\flat(\ref{run2-ADP-3}), \flat(\ref{run2-ADP-5})$ - $\flat(\ref{run2-ADP-7})\}$, and $\{\eqref{run2-ADP-6}, \flat(\ref{run2-ADP-1})$ - $\flat(\ref{run2-ADP-5}), \flat(\ref{run2-ADP-7})\}$.
\end{exa}

\begin{exa}
  \label{example:dup-problem}
  If we used GVRFs that can duplicate annotations, then the dependency graph processor would not be sound.
  The reason is that $\Proc_{\mathtt{DG}}$ maps ADP problems without annotations to the empty set.
  However, this would be unsound if we had GVRFs, because then the ADP problem consisting of $\ta \to \{1:\tb\}^{\ttrue}$ and $\td(x) \to \{1:\tc(x,\td(x))\}^{\ttrue}$ would not be $\mathtt{AST}$.
  Here, the use of GVRFs would lead to the following CT with an infinite number of $(\mathbf{at})$ steps that rewrite $\tA$ to $\tb$.

  \vspace*{-0.05cm}
  \begin{center}
    \scriptsize
    \begin{tikzpicture}
      \tikzstyle{adam}=[rectangle,thick,draw=black!100,fill=white!100,minimum size=4mm,shape=rectangle split, rectangle split parts=2,rectangle split horizontal] \tikzstyle{empty}=[rectangle,thick,minimum size=4mm]

      \node[adam] at (0, 0) (a) {$1$
        \nodepart{two}$\td(\tA)$};
      \node[adam] at (2.1, 0) (b) {$1$
        \nodepart{two}$\tc(\tA,\td(\tA))$};
      \node[adam] at (4.6, 0) (c) {$1$
        \nodepart{two}$\tc(\tb,\td(\tA))$};
      \node[empty] at (6.5, 0) (d) {$\ldots$};

      \draw (a) edge[->] (b);
      \draw (b) edge[->] (c);
      \draw (c) edge[->] (d);
    \end{tikzpicture}
  \end{center}
\end{exa}

\subsubsection{Dependency Graph Processor for $\mathtt{iAST}$}

\noindent
As in \Cref{DP Framework}, we can integrate the restriction to innermost rewriting into the dependency graph.
So instead of arbitrary rewrite steps, we now only allow innermost steps.
Moreover, the substitutions $\sigma_1, \sigma_2$ need to be chosen in such a way that both ADPs can be applied in innermost reductions, i.e., we must have $\ell_1 \sigma_1, \ell_2 \sigma_2 \in \ANF_{\PP}$.

\begin{defi}[Innermost Dep.\ Graph]
  \label{def:iDG}
  The \emph{innermost $\PP$-dependency graph} has the nodes $\PP$, and there is an edge from $\ell_1 \ruleArr{}{}{} \{ p_1:r_{1}, \ldots, p_k: r_k\}^{m}$ to $\ell_2 \to \ldots$ if there are substitutions $\sigma_1, \sigma_2$ and a $t \trianglelefteq_{\sharp}
    r_{j}$ for some $1 \leq j \leq k$ such that $t^\sharp \sigma_1 \ito_{\nonprob(\PP)}^* \ell_2^\sharp \sigma_2$ and both $\ell_1 \sigma_1$ and $\ell_2 \sigma_2$ are in $\ANF_{\PP}$.
\end{defi}

Again, the same over-approximation techniques as in \Cref{DP Framework} can also be used to estimate the innermost dependency graph in the probabilistic setting.

\begin{restatable}[Dependency Graph Processor for $\mathtt{iAST}$]{thm}{ProbDepGraphProcInnermost}
  \label{thm:prob-iDGP}
  Let $\PP_1, \ldots, \PP_n$ be the SCCs of the innermost $\PP$-dependency graph.
  Then $\Proc_{\mathtt{DG}}^{\mathbf{i}}(\PP)=\{\PP_1 \cup \flat(\PP \setminus \PP_1), \ldots, \PP_n \cup \flat(\PP \setminus \PP_n)\}$ is sound and complete for $\mathtt{iAST}$.
\end{restatable}

\begin{myproofsketch}
  \textit{Soundness}: Similar to the proof of \Cref{thm:prob-DGP}, just considering innermost rewrite steps.

  \textit{Completeness}: By \Cref{thm:proc-complete}.
\end{myproofsketch}

\begin{exa}
  \label{example:running-1-DPG-iAST}
  For $\R_{\mathsf{alg}}$, the innermost and the full $\DPair{\R_{\mathsf{alg}}}$-dependency graph coincide.
  Thus, for $\mathtt{iAST}$ we result in the same ADP problems as in \Cref{example:running-2-DPG-AST}.

  For the duplicating PTRS $\R_{2}$ and its canonical ADPs from \Cref{example:running-1-ADPs} we can only analyze $\mathtt{iAST}$.
  The innermost $\DPair{\R_2}$-dependency graph can be seen in \Cref{fig:dep-graph-running-1}.
  As the only SCC $\{\eqref{run1-ADP-1}\}$ does not contain \eqref{run1-ADP-2}, we can remove all annotations from \eqref{run1-ADP-2}.
  However, since \eqref{run1-ADP-2} already has no annotations, $\Proc_{\mathtt{DG}}^{\mathbf{i}}$ does not change $\DPair{\R_2}$.
\end{exa}

\subsection{Usable Terms Processor}
\label{sec:Usable Terms Processor}

\noindent
The dependency graph processor either removes all annotations from an ADP or none.
But an ADP can still contain terms $t^\sharp$ where no instance $t^\sharp \sigma_1$ rewrites to an instance $\ell^\sharp \sigma_2$ of a left-hand side $\ell$ of an ADP with annotations, i.e., it can only lead to ADPs outside the SCC we consider.
The \emph{usable terms processor} removes the annotation from the root of such
\emph{non-usable} terms like $\tD(\ldots)$ in $\DPair{\R_2} = \{
 \tg \to \{\nicefrac{3}{4}:\tD(\tG), \nicefrac{1}{4}:\tz\}^{\ttrue} \; \eqref{run1-ADP-1},
 \;
 \td(x) \to \{1:\tc(x,x)\}^{\ttrue} \;  \eqref{run1-ADP-2} \}$.
 So instead of whole ADPs, here we consider the subterms in the right-hand sides of an ADP individually.
Such a processor was not needed in the non-probabilistic setting, as ordinary dependency pairs only contain a single annotation.

\subsubsection{Usable Terms Processor for $\mathtt{AST}$}

\smallskip

\begin{restatable}[Usable Terms Processor for $\mathtt{AST}$]{thm}{UsableTermsProc}
  \label{thm:prob-UPP}
  We call $t \in \TT^{\sharp}$ with $\rootsym(t) \in \SignatureD^\sharp$ \defemph{usable}
  w.r.t.\ an ADP problem $\PP$ if there are substitutions $\sigma_1, \sigma_2$ and an $\ell_2 \ruleArr{}{}{} \mu_2 \in \PP$ where $\mu_2$ contains an annotated symbol, such that $\anno_{\{\varepsilon\}}(t) \sigma_1 \to_{\nonprob(\PP)}^* \ell_2^\sharp \sigma_2$.
  Let $\Delta_{\PP}(s) = \{ \pi \in \posT(s) \mid s|_\pi$ is usable w.r.t.\ $\PP\,\}$ and let $\mathcal{T}_\mathtt{UT}(\PP) \!=\! \{ \ell \!\to\! \{ p_1: \sharp_{\Delta_{\PP}(r_1)}(r_1), \ldots, p_k:\sharp_{\Delta_{\PP}(r_k)}(r_k)\}^{m}
    \mid \ell \!\to\!
    \{ p_1: r_1, \ldots, p_k:r_k\}^{m} \!\in\!
    \PP \}$ be the transformation that removes the annotations from the roots of all non-usable terms in the right-hand sides. Then $\Proc_{\mathtt{UT}}(\PP) = \{\mathcal{T}_\mathtt{UT}(\PP)\}$ is sound and complete for $\mathtt{AST}$.
\end{restatable}

\begin{myproofsketch}
  \textit{Soundness}: For every $\PP$-CT $\F{T}$ (satisfying the restrictions of the
  Starting Lemma (\Cref{lemma:starting})) we generate a $\mathcal{T}_\mathtt{UT}(\PP)$-CT $\F{T}'$
  with $|\F{T}'| = |\F{T}|$. The CT $\F{T}'$ has the same nodes and edges as $\F{T}$.
If a node $v$ is labeled by $(p_v : t_v)$ in $\F{T}$, then it is labeled by 
$(p_v : t_v')$ in  $\F{T}'$ where we construct the labeling $t_v'$ recursively and ensure
that for all nodes $v$ we have
  \begin{equation}\label{usable-terms-soundness-induction-hypothesis-in-paper}
    \parbox{.9\textwidth}{$\flat(t_v) = \flat(t_v')$ and $\posT(t_v) \setminus \Junk(t_v) \subseteq \posT(t_v')$.}
  \end{equation}
  Here, for any term $t_v$, let $\Junk(t_v)$ be the set of positions that can never be
  used for a rewrite step with an ADP that contains annotations.
  Due to \eqref{usable-terms-soundness-induction-hypothesis-in-paper},
 every infinite path of $\F{T}'$ still contains infinitely many rewrite steps at annotations.

  \textit{Completeness}: By \Cref{thm:proc-complete}.
\end{myproofsketch}

\begin{exa}
  \label{example:running-2-UTP-AST}
  For $\mathtt{AST}$, $\Proc_{\mathtt{UT}}$ transforms $\{\eqref{run2-ADP-1}, \eqref{run2-ADP-2}, \flat(\ref{run2-ADP-3})$ - $\flat(\ref{run2-ADP-7})\}$ from \Cref{example:running-2-DPG-AST} into $\{(\ref{run2-ADP-1}'), (\ref{run2-ADP-2}'), \flat(\ref{run2-ADP-3})$ - $\flat(\ref{run2-ADP-7})\}$ with
  \begin{align*}
    \tloopOne(y) & \to \{\nicefrac{1}{2}:\tLoopOne(\tdouble(y)), \;
    \nicefrac{1}{2}:\tloopTwo(\tdouble(y))\}^{\ttrue}
    \tag{$\ref{run2-ADP-1}'$} \\
    \tloopOne(y) & \to \{\nicefrac{1}{3}:\tLoopOne(\ttriple(y)), \;
    \nicefrac{2}{3}:\tloopTwo(\ttriple(y))\}^{\ttrue} \tag{$\ref{run2-ADP-2}'$}
  \end{align*}
  The reason is that the left-hand sides of the only ADPs with annotations in the ADP problem have the root $\tloopOne$.
  Thus, $\tLoopTwo$-, $\tD$-, or $\tTriple$-terms are not usable.
\end{exa}

\subsubsection{Usable Terms Processor for $\mathtt{iAST}$}

\noindent
As for the dependency graph processor, for $\mathtt{iAST}$ we only consider innermost rewriting in the definitions.

\begin{restatable}[Usable Terms Processor for $\mathtt{iAST}$]{thm}{UsableTermsProcInnermost}
  \label{thm:prob-iUPP}
  Let $\ell_1 \in \TT$ and $\PP$ be an ADP problem.
  We call $t \in \TT^{\sharp}$ with $\rootsym(t) \in \SignatureD^\sharp$ \defemph{innermost usable} w.r.t.\ $\ell_1$ and $\PP$ if there are substitutions $\sigma_1, \sigma_2$ and an $\ell_2 \ruleArr{}{}{} \mu_2 \in \PP$ where $\mu_2$ contains an annotated symbol, such that $\anno_{\{\varepsilon\}}(t) \sigma_1 \ito_{\nonprob(\PP)}^* \ell_2^\sharp \sigma_2$ and both $\ell_1 \sigma_1$ and $\ell_2 \sigma_2$ are in $\ANF_{\PP}$.
  Let $\Delta^{\mathbf{i}}_{\ell,\PP}(s) = \{ \pi \in \posT(s) \mid s|_\pi$ is innermost usable w.r.t.\ $\ell$ and $\PP\,\}$ and $\mathcal{T}_\mathtt{UT}^{\mathbf{i}}(\PP) \!=\! \{ \ell \!\to\! \{ p_1\!:\!\sharp_{\Delta^{\mathbf{i}}_{\ell,\PP}(r_1)}(r_1), \ldots, p_k\!:\!\sharp_{\Delta^{\mathbf{i}}_{\ell,\PP}(r_k)}(r_k)\}^{m}
    \mid \ell \!\to\!
    \{ p_1\!:\!r_1, \ldots, p_k\!:\!r_k\}^{m} \!\in\!
    \PP \}$. Then $\Proc_{\mathtt{UT}}^{\mathbf{i}}(\PP) = \{\mathcal{T}_\mathtt{UT}^{\mathbf{i}}(\PP)\}$ is sound and complete for $\mathtt{iAST}$.
\end{restatable}

\begin{myproofsketch}
  \textit{Soundness}: Similar to the proof of \Cref{thm:prob-UPP}, just considering innermost rewrite steps.

  \textit{Completeness}: By \Cref{thm:proc-complete}.
\end{myproofsketch}

\begin{exa}
  \label{example:running-1-UTP-iAST}
  For $\R_{\mathsf{alg}}$, up to now the proof of $\mathtt{iAST}$ is identical to the proof of $\mathtt{AST}$, i.e., here the results of $\Proc_{\mathtt{UT}}^{\mathbf{i}}$ on the ADP problems resulting from the (innermost) dependency graph processor are the same as the results of $\Proc_{\mathtt{UT}}$.

  When considering $\R_2$ and its ADPs $\DPair{\R_2}$ from \Cref{example:running-1-ADPs}
  again, $\Proc_{\mathtt{UT}}^{\mathbf{i}}$ replaces $\eqref{run1-ADP-1}$ by $\;\tg \to \{\nicefrac{3}{4}:\td(\tG), \nicefrac{1}{4}:\tz\}^{\ttrue} \quad (\ref{run1-ADP-1}')$.
\end{exa}

\subsection{Usable Rules Processor}
\label{sec:Usable Rules Processor}

\noindent
Next, we consider the \emph{usable rules processor}.
Here, we will see a major difference between the ADP frameworks for $\mathtt{AST}$ and $\mathtt{iAST}$, since it will turn out that this processor is only sound for $\mathtt{iAST}$, but not for $\mathtt{AST}$.
Therefore, this time we consider innermost rewriting first.

\subsubsection{Usable Rules Processor for $\mathtt{iAST}$}

\noindent
In an innermost rewrite step, all variables of the used rule are instantiated with normal forms.
The \emph{usable rules processor} detects rules that cannot be used below annotations in right-hand sides of ADPs when their variables are instantiated with normal forms.
For these rules we can set their flag to $\tfalse$, indicating that the annotated subterms on their right-hand sides may still lead to a non-$\mathtt{iAST}$ sequence, but the context of these annotations is irrelevant.

\begin{restatable}[Usable Rules Processor for $\mathtt{iAST}$]{thm}{UsableRulesProcInnermost}
  \label{thm:prob-iURP}
  For every $f\!\in\!\SignatureADC$ and ADP problem $\PP$, let $\rules_{\PP}(f) = \{\ell\!\to\!\mu^{\true}\!\in\!\PP \mid \rootsym(\ell)\!=\!f\}$.
  Moreover, for every $t \in \TT^{\sharp}$, the \defemph{usable rules} $\urules_{\PP}(t)$
  of $t$ w.r.t.\ $\PP$ are recursively defined as follows:

  \vspace*{-0.4cm}
\[ \small
  \displaystyle 
    \begin{array}{r@{\;}l@{\;}ll}
    \urules_\PP(t) &=& \emptyset, & \text{if } t \in \VSet \text{ or } \PP = \emptyset\\
     \urules_\PP(f(t_1, \ldots, t_n)) &=& \rules_{\PP}(f) 
    \cup \displaystyle\bigcup_{j = 1}^n\!\urules_{\PP'}(t_j) 
    \cup  \hspace*{-.5cm} \displaystyle\bigcup_{\substack{\ell \to \mu^{\true} \in \rules_{\PP}(f),\\ r \in
        \Supp(\mu)}}  \hspace*{-.7cm} \urules_{\PP'}(\flat(r)), 
    & \text{otherwise,}\!
  \end{array}
  \]
  \vspace*{-0.4cm}
  
  \noindent
 where $\PP' = \PP \setminus \rules_\PP(f)$.

  The \defemph{usable rules} of $\PP$ are $\urules(\PP) = \bigcup_{\ell \to \mu^{m} \in \PP, r \in \Supp(\mu), t \trianglelefteq_{\sharp} r}
  \urules_{\PP}(t^\sharp)$. Then, the processor $\Proc_{\mathtt{UR}}^{\mathbf{i}}(\PP) = \{
  \urules(\PP) \cup \{\ell \to \mu^{\tfalse} \mid \ell \to \mu^{m} \in \PP \setminus \urules(\PP)\} \}$ is sound and complete for $\mathtt{iAST}$, i.e., we turn the flag $m$ of all non-usable rules to $\tfalse$.
\end{restatable}

\begin{myproofsketch}
  \textit{Soundness}: 
  Due to the Starting Lemma (\Cref{lemma:starting}) we can restrict ourselves to 
  innermost
  $\PP$-CTs that start with a term $t \in \TT^\sharp$ such that $\flat(t) = s \theta \in \ANF_{\PP}$.
  By the definition of usable rules,
  rules $\ell \to \mu^m \in \PP$ that are not usable 
  are never applied below an annotated symbol in such
  $\PP$-CTs.
  Hence, we can also regard $\F{T}$ as a $\overline{\PP}$-CT for
  $\overline{\PP} = \urules(\PP) \cup \{\ell \to \mu^{\tfalse} \mid \ell \to \mu^{m}
  \in \PP \setminus \urules(\PP)\}$.
  \textit{Completeness}: By \Cref{thm:proc-complete}.
\end{myproofsketch}

Note that the innermost evaluation strategy is not affected by our usable rules processor, since it does not remove rules but only changes their flag to $\false$.
This is different from the non-probabilistic DP framework, where the usable rules processor from \cref{URP}
reduces the number of rules.
This may result in new redexes that are allowed for innermost rewriting.
So in \cite{gieslLPAR04dpframework} one has to extend DP problems by an additional component 
to achieve completeness of this processor (see \Cref{CompletenessUsableRules}), whereas the usable rules processor in our new ADP framework is complete.

\begin{exa}
  \label{example:running-1-URP-iAST}
  The ADP problem $\{(\ref{run1-ADP-1}'), \eqref{run1-ADP-2}\}$ which we obtained for the $\mathtt{iAST}$ proof of $\R_2$ in \Cref{example:running-1-UTP-iAST}
  has no subterms below annotations.
  So both rules are not usable and we set their flags to $\tfalse$ which leads to

  \vspace*{-.5cm}
  \begin{minipage}[t]{0.47\textwidth}
    \small
    \begin{align*}
      \tG & \to \{\nicefrac{3}{4}:\td(\tG), \; \nicefrac{1}{4}:\tz\}^{\tfalse}
      \tag{$\ref{run1-ADP-1}''$}
    \end{align*}
  \end{minipage}
  \hfill
  \begin{minipage}[t]{0.47\textwidth}
    \small
    \begin{align*}
      \tD(x) & \to \{1:\tc(x,x)\}^{\tfalse}
      \tag{$\ref{run1-ADP-2}'$}
    \end{align*}
  \end{minipage}
\end{exa}

\begin{exa}
  \label{example:running-2-URP-iAST}
  Consider the ADP problem $\{(\ref{run2-ADP-1}'), (\ref{run2-ADP-2}'), \flat(\ref{run2-ADP-3})$ - $\flat(\ref{run2-ADP-7})\}$ from \Cref{example:running-2-UTP-AST} that resulted from applying the innermost dependency graph processor and the innermost usable terms processor to $\DPair{\R_{\mathsf{alg}}}$.
  When proving $\mathtt{iAST}$ of $\R_{\mathsf{alg}}$, only the $\tdouble$- and $\ttriple$-ADPs $\flat(\ref{run2-ADP-4})$ - $\flat(\ref{run2-ADP-7})$ are usable, since the defined symbols $\tdouble$ and $\ttriple$ occur below an annotated symbol $\tLoopOne$.
  So we can set the flag of all other ADPs in this problem to $\tfalse$.
  All other ADP problems resulting from the innermost dependency graph and the innermost usable terms processor in this example have no usable rules, hence we can turn the flag of all ADPs in these ADP problems to $\tfalse$.
\end{exa}

\subsubsection{Usable Rules Processor for $\mathtt{AST}$}

\noindent
For full rewriting, the usable rules processor is unsound.
This is already the case for non-probabilistic rewriting, but in the classical DP framework there nevertheless exist processors for full rewriting based on usable rules which rely on taking the $C_{\mathcal{E}}$-rules $\th(x,y) \to x$ and $\th(x,y) \to y$ for a fresh function symbol $\th$ into account, see, e.g., \cite{giesl2006mechanizing,gieslLPAR04dpframework,DBLP:journals/iandc/HirokawaM07,DBLP:journals/jar/Urbain04}.
However, the following example shows that this is not possible for $\mathtt{AST}$.

\begin{exa}
  \label{example:usable-rules-ce-rules-full2}
  Consider the following ADP problem.
  \[ \begin{array}{rcl}
      \ta                       & \to & \{ \nicefrac{5}{8}:\tc(\ta,\ta), \; \nicefrac{3}{8}:\tz\}^\true \\
      \tf(\tc(x_1,x_2)) & \to & \{ 1:\tc(\tF(x_1),\tF(x_2))\}^\true
    \end{array}\]
  Although the first ADP has no annotations, the ADP problem is not $\mathtt{AST}$:

  \vspace*{-0.05cm}
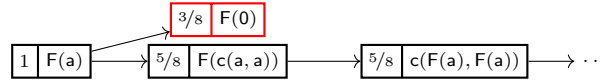
  \begin{center}
    \scriptsize \hspace*{.05cm}
    \begin{tikzpicture}
      \tikzstyle{adam}=[rectangle,thick,draw=black!100,fill=white!100,minimum size=4mm,shape=rectangle split, rectangle split parts=2,rectangle split horizontal] \tikzstyle{adam2}=[rectangle,thick,draw=red!100,fill=white!100,minimum size=4mm,shape=rectangle split, rectangle split parts=2,rectangle split horizontal] \tikzstyle{empty}=[rectangle,thick,minimum size=4mm]

      \node[adam] at (-0.2, 0) (a) {$1$
        \nodepart{two}$\tF(\ta)$};
      \node[adam2] at (2, 0.55) (b2) {$\nicefrac{3}{8}$
        \nodepart{two}$\tF(\tz)$};
      \node[adam] at (2, 0) (b) {$\nicefrac{5}{8}$
        \nodepart{two}$\tF(\tc(\ta,\ta))$};
      \node[adam] at (5, 0) (c) {$\nicefrac{5}{8}$
        \nodepart{two}$\tc(\tF(\ta),\tF(\ta))$};
      \node[empty] at (7, 0) (d) {$\ldots$};

      \draw (a) edge[->] (b);
      \draw (a) edge[->] (b2);
      \draw (b) edge[->] (c);
      \draw (c) edge[->] (d);
    \end{tikzpicture}
  \end{center}

  \vspace*{-0.05cm}
  \noindent

  This represents a random walk biased towards non-termination, where the number of $\tF(\ta)$ subterms increases by one with probability $\nicefrac{5}{8}$ or decreases by one with probability $\nicefrac{3}{8}$.

  The ADP problem has no usable rules and thus, the usable rule processor would turn the flag of all ADPs to $\tfalse$.
  However, if we now start with $\tF(\ta)$ (or $\tF(\tA)$), 
  rewriting $\ta$ (or $\tA$) would remove the annotation of $\tF$ above, 
  since the flag indicates whether the ADP can be used to rewrite arguments of an annotated symbol,
  i.e., $\tF(\ta) \tored{}{}{} \{\nicefrac{5}{8}: \tf(\tc(\ta,\ta)), \nicefrac{3}{8}:\tf(\tz)\}$
  Hence, then all CTs are finite.
  This also holds when adding the $C_{\mathcal{E}}$-ADPs $\th(x,y) \to \{1:x\}^\ttrue$ and $\th(x,y) \to \{1:y\}^\ttrue$.
\end{exa}

Thus, even integrating the $C_{\mathcal{E}}$-rules to represent non-determinism would not allow a usable rule processor for $\mathtt{AST}$.
Moreover, the corresponding proofs in the non-probabilistic setting rely on the minimality property, which does not hold in the probabilistic setting, see \Cref{remark:minimality}.

\subsection{Reduction Pair Processor}
\label{sec:Reduction Pair Processor}

\noindent
Next we consider the reduction pair processor, whose adaption is the same for $\mathtt{AST}$ and $\mathtt{iAST}$.
In the probabilistic setting, we cannot use arbitrary term orders, but as in \cite{avanzini2020probabilistic}, we have to consider algebras that allow the definition of an \emph{expected value}, called \emph{barycentric $\Sigma^\sharp$-algebras}.

\begin{defi}[Barycentric $\Sigma^\sharp$-Algebra, $\NN$-collapsible]
  \label{def:barycentric-algebra}
  A non-empty set $A$ is called a \emph{barycentric domain}
  if the addition $+ : A^2 \to A$ and the scalar multiplication with probabilities $\cdot : \IR_{\geq 0} \times A \to A$ are defined.
  The barycentric operation $\mathbb{E}_A:\FDist(A) \to A$ is then defined as $\mathbb{E}_A(\{p_1:a_1, \ldots, p_k:a_k\}) = p_1\cdot a_1 + \ldots + p_k\cdot a_k$.
  We just write $\IE$ instead of $\mathbb{E}_A$ if $A$ is clear from the context.
  A function $f: A^n \to B$ on two barycentric domains $A$ and $B$ is called \emph{concave} w.r.t.\ a relation $\succcurlyeq$ on $B$ if we have $f(\ldots, \mathbb{E}_A(\mu), \ldots) \succcurlyeq \mathbb{E}_B(f(\ldots, \mu, \ldots))$, where we define $f(\ldots, \{p_1:a_1,\ldots, p_k:a_k\}, \ldots)) = \{p_1:f(\ldots,a_1,\ldots), \ldots, p_k:f(\ldots,a_k,\ldots)\}$ for every $\mu \in \FDist(A)$.

  A \emph{barycentric $\Sigma^\sharp$-algebra} is a $\Sigma^\sharp$-algebra $(\mathcal{A}, \succ)$ on a barycentric domain $A$ such that for every $f \in \Sigma^\sharp$, $f_{\mathcal{A}}$ is concave with respect to $\succcurlyeq$.
  We say that $(\mathcal{A}, \succ)$ is \emph{$\NN$-collapsible}
  if there exists a concave function $g : A \to \IN$ that embeds $\succ$ into $>$ on the natural numbers, i.e., for all ground terms $t_1, t_2 \in \TT^\sharp$, $\mathcal{A}(t_1) \succ \mathcal{A}(t_2)$ implies $g(\mathcal{A}(t_1)) > g(\mathcal{A}(t_2))$.
\end{defi}

For example, both polynomial interpretations and matrix interpretations \cite{Endrullis08} are $\NN$-collapsible barycentric $\Sigma^\sharp$-algebras, provided that $\Pol$ maps each function symbol to a \emph{multilinear} polynomial, i.e., a polynomial whose monomials have the form $c \cdot x_1^{e_1} \cdot \ldots \cdot x_n^{e_n}$ with $c \in \NN$ and $e_1,\ldots,e_n \in \{0,1\}$, to ensure concavity.
(We will discuss the reasons for requiring concavity of $f_{\mathcal{A}}$ and for
$\NN$-collapsibility below.)

In the following, let $(\mathcal{A}, \succ)$ be an $\NN$-collapsible, barycentric $\Sigma^\sharp$-algebra.
The constraints of the reduction pair processor of \Cref{thm:prob-RPP} are based on the conditions of a ranking function for $\mathtt{AST}$ as in \cite{mciver2017new}.
If one proves $\mathtt{AST}$ by considering the rules $\ell \to \mu^m$ of a PTRS directly,
then one needs a \emph{monotonic} algebra and requires a weak decrease when comparing
$\mathcal{A}(\ell)$ to the expected value $\IE(\mathcal{A}(\mu))$ of the right-hand side,
where we extend $\mathcal{A}$ to multi-distributions by defining
$\mathcal{A}(\{p_1:t_1, \ldots, p_k:t_k\}) = \{p_1:\mathcal{A}(t_1), \ldots, p_k:\mathcal{A}(t_k)\}$.
Moreover, at least one $\mathcal{A}(r)$ with $r \in \Supp(\mu)$ must be strictly smaller than $\mathcal{A}(\ell)$, see \cite{kassinggiesl2023iAST}.

For ADPs, we do not use the interpretation $\mathcal{A}(r)$ of a term $r$ directly, but we take the occurring annotations into account.
Since only the annotated subterms may lead to non-$\mathtt{AST}$ sequences, we
only consider the interpretations of these subterms.
Hence, we compare the $\sharp$-\emph{sums}\footnote{This is the reason why we require the existence of a specific addition function $+$ on $A$ 
instead of simply requiring the existence of a barycentric operation $\IE$ as in \cite{avanzini2020probabilistic}.}
of the terms $r$ in ADPs, 
i.e., $\mathcal{A}_{sum}^{\sharp}(r) = \sum_{t \trianglelefteq_{\sharp} r}
\mathcal{A}(t^\sharp)$ sums
up the values of $r$'s annotated subterms.
So for example, we have $\mathcal{A}_{sum}^{\sharp}(\tD(\tG)) =
\mathcal{A}(\tD(\tg)) + \mathcal{A}(\tG)$.

If in each rewrite step $s \tored{}{}{\PP} \mu$ of a chain tree $\F{T}$,
the $\sharp$-\emph{sum} is weakly decreasing in expectation
($\mathcal{A}_{sum}^{\sharp}(s) \succcurlyeq \IE(\mathcal{A}_{sum}^{\sharp}(\mu))$),
and additionally,  at least one $\mathcal{A}_{sum}^{\sharp}(r)$ with $r \in \Supp(\mu)$
is strictly smaller than $\mathcal{A}_{sum}^{\sharp}(s)$, 
then we can conclude that $\F{T}$ converges with probability 1. Again,
$\mathcal{A}_{sum}^{\sharp}$ is extended to multi-distributions by defining
$\mathcal{A}_{sum}^{\sharp}(\{p_1:t_1, \ldots, p_k:t_k\}) =
\{p_1:\mathcal{A}_{sum}^{\sharp}(t_1), \ldots, p_k:\mathcal{A}_{sum}^{\sharp}(t_k)\}$.
So for a rewrite step of the form 
$\tG \tored{}{}{\PP} \{ \nicefrac{3}{4}: \tD(\tG), \nicefrac{1}{4}:\tz \}$
one would for example obtain the constraints
$\mathcal{A}_{sum}^{\sharp}(\tG) = \mathcal{A}(\tG)
\succcurlyeq \IE(\{\nicefrac{3}{4}: \mathcal{A}_{sum}^{\sharp}(\tD(\tG)),
\nicefrac{1}{4}: \mathcal{A}_{sum}^{\sharp}(\tz) \}) =
\nicefrac{3}{4} \cdot  \mathcal{A}_{sum}^{\sharp}(\tD(\tG)) + 
\nicefrac{1}{4} \cdot \mathcal{A}_{sum}^{\sharp}(\tz) =
\nicefrac{3}{4} \cdot (\mathcal{A}(\tD(\tg))
+ \mathcal{A}(\tG)) + \nicefrac{1}{4} \cdot 0$
 and
$\mathcal{A}_{sum}^{\sharp}(\tG) = \mathcal{A}(\tG) \succ \mathcal{A}_{sum}^{\sharp}(\tz)
= 0$.

\begin{lem}[Proving $\mathtt{AST}$ of Chain Trees with Algebras]
    \label{lemma:prob-RPP}
    Let
    $\F{T}$ be a $\PP$-CT and let $(\mathcal{A},\succ)$ be an $\NN$-collapsible, barycentric $\Sigma^\sharp$-algebra 
  such that for every rewrite step $s \tored{}{}{\PP} \mu$ in $\F{T}$ we have: 
  $(A):\;\mathcal{A}_{sum}^{\sharp}(s) \succcurlyeq \IE(\mathcal{A}_{sum}^{\sharp}(\mu))$ 
  and $(B):\;\exists r \in \Supp(\mu) : \mathcal{A}_{sum}^{\sharp}(s) \succ
  \mathcal{A}_{sum}^{\sharp}(r)$. 
  Then we have $|\F{T}| = 1$, i.e., $\F{T}$ converges with probability 1.
\end{lem}

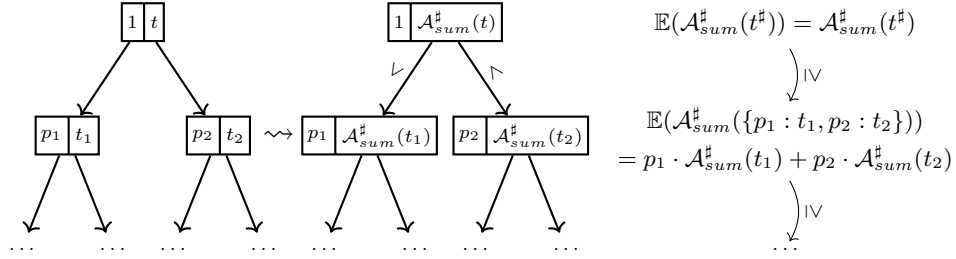
\begin{figure}[H]
    \begin{center}
        \begin{scriptsize}
            \begin{tikzpicture}
                \tikzstyle{nodeStyle}=[thick, draw=black, fill=white, minimum width=12mm, minimum height=5mm, shape=rectangle split, rectangle split parts=2, rectangle split horizontal, inner sep=2pt]
                \tikzstyle{edgeStyle}=[->, thick]
                \tikzstyle{blueLabel}=[above, blue, font=\tiny]
                \tikzstyle{empty}=[rectangle,thick,minimum size=4mm]
                \tikzstyle{evalue}=[rectangle,thick,minimum size=4mm, font=\small]

                \node[nodeStyle] (root) at (1, 3) {1 \nodepart{two} $t$};
                               \node[nodeStyle] (n1) at (0, 1.5) {$p_1$ \nodepart{two} $t_1$};
                               \node[nodeStyle] (n2) at (2, 1.5) {$p_2$ \nodepart{two} $t_2$};
                               
                \node[empty] (n3) at (-0.6, 0) {$\ldots$};
                \node[empty] (n4) at (0.6, 0) {$\ldots$};
                \node[empty] (n5) at (1.4, 0) {$\ldots$};
                \node[empty] (n6) at (2.6, 0) {$\ldots$};
                
                \draw[edgeStyle] (root) -- (n1);
                \draw[edgeStyle] (root) -- (n2);
                \draw[edgeStyle] (n1) -- (n3);
                \draw[edgeStyle] (n1) -- (n4);
                \draw[edgeStyle] (n2) -- (n5);
                \draw[edgeStyle] (n2) -- (n6);
                
                \node[nodeStyle] (rootR) at (5, 3) {1 \nodepart{two} $\mathcal{A}_{sum}^{\sharp}(t)$};
                
                \node[nodeStyle] (n1R) at (4, 1.5) {$p_1$ \nodepart{two} $\mathcal{A}_{sum}^{\sharp}(t_1)$};
                \node[nodeStyle] (n2R) at (6, 1.5) {$p_2$ \nodepart{two} $\mathcal{A}_{sum}^{\sharp}(t_2)$};
                
                \node[empty] (n3R) at (3.4, 0) {$\ldots$};
                \node[empty] (n4R) at (4.6, 0) {$\ldots$};
                \node[empty] (n5R) at (5.4, 0) {$\ldots$};
                \node[empty] (n6R) at (6.6, 0) {$\ldots$};

                \draw[edgeStyle] (rootR) -- node[sloped, above] {$<$} (n1R);
                \draw[edgeStyle] (rootR) -- node[sloped, above] {$<$} (n2R);
                \draw[edgeStyle] (n1R) -- node[sloped, above] {} (n3R);
                \draw[edgeStyle] (n1R) -- node[sloped, above] {} (n4R);
                \draw[edgeStyle] (n2R) -- node[sloped, above] {} (n5R);
                \draw[edgeStyle] (n2R) -- node[sloped, above] {} (n6R);
                
                \node (A) at (9.5, 2.75) {};
                \node (B) at (9.5, 1.85) {};
                
                \node (C) at (9.5, 1) {};
                \node (D) at (9.5, 0) {};
                
                \draw[->, bend left] (A) to node[sloped, above] {$\geq$} (B);
                \draw[->, bend left] (C) to node[sloped, above] {$\geq$} (D);
                
                \node[evalue] (ev1) at (9.5, 3) {$\IE(\mathcal{A}_{sum}^{\sharp}(t^\sharp)) = \mathcal{A}_{sum}^{\sharp}(t^\sharp)$};
                \node[evalue] (ev2) at (9.5, 1.7) {$\IE(\mathcal{A}_{sum}^{\sharp} (\{ p_1 : t_1, p_2 : t_2\}))$};
                \node[evalue] (ev21) at (9.5, 1.2) {$= p_1 \cdot \mathcal{A}_{sum}^{\sharp}(t_1) + p_2 \cdot \mathcal{A}_{sum}^{\sharp}(t_2)$};
                \node[empty] (ev3) at (9.5, 0) {$\ldots$};
                
                \node at (2.8, 1.5) {\large$\rightsquigarrow$};
                
            \end{tikzpicture}
        \end{scriptsize}
    \end{center}
    \vspace*{-2mm}
    \caption{Each term $r$ in the chain tree (left) is mapped
      to the corresponding $\sharp$-sum $\mathcal{A}_{sum}^{\sharp}(r)$ (middle).
    The expected $\sharp$-sum (right) is always weakly decreasing.}\label{fig:rpp-proof}
\end{figure}

\Cref{fig:rpp-proof} visualizes the interpretation of terms in an
example chain tree.
As required by $(A)$, 
the expected $\sharp$-sum is weakly decreasing in each step from a node to its children.
Moreover, the
 step at the root of the example CT
satisfies $(B)$, since the left successor $t_1$
has a strictly smaller $\sharp$-sum than $t$.
The $\sharp$-sum of the right successor $t_2$ may even be larger than $t$,
but the expected $\sharp$-sum needs to be weakly decreasing.

Instead of imposing
(usually infinitely many) constraints for
all  possible rewrite steps $s \tored{}{}{\PP} \mu$ as in \Cref{lemma:prob-RPP},
we only want to impose finitely many constraints which correspond
to the finitely many ADPs
in the ADP problem $\PP$. If these inequality constraints hold for the ADPs, then they
also hold for arbitrary rewrite steps if 
$(\mathcal{A},\succ)$ is \emph{weakly monotonic}. 
So as in the non-probabilistic reduction pair processor of \Cref{RPP}, 
due to the usage of $\sharp$-sums, weak instead of strict
monotonicity is sufficient.
However, 
as in \cite{avanzini2020probabilistic}, the requirement of concavity of all functions $f_{\mathcal{A}}$
is needed to ensure that we also have weak monotonicity w.r.t.\ expected values.
So $\mathcal{A}(t) \succcurlyeq \IE(\mathcal{A}(\mu))$ implies
\[ \begin{array}{rcll}
    \mathcal{A}(f(\ldots,t,\ldots)) & =            & f_{\mathcal{A}}(\ldots, \mathcal{A}(t), \ldots)        &          \\
                                    & \succcurlyeq & f_{\mathcal{A}}(\ldots, \IE(\mathcal{A}(\mu)), \ldots) & \text{by
    weak monotonicity}                                                                                                 \\
                                    & \succcurlyeq & \IE(f_{\mathcal{A}}(\ldots, \mathcal{A}(\mu), \ldots)) & \text{by
      concavity}
  \end{array}\]

Similar to \Cref{RPP}, we do not need to find an algebra that orients all ADPs strictly
(i.e., where all rewrite steps satisfy property $(B)$ of \Cref{lemma:prob-RPP}). This still allows us to simplify the
problem, because then
it suffices to consider only those 
CTs where the 
non-strictly oriented ADPs are applied infinitely often on annotated positions.

So to ensure the properties of \Cref{lemma:prob-RPP} for each reduction step, we impose
the following constraints for all ADPs (which adapt the constraints (1) - (3) from
\Cref{RPP} to the probabilistic setting):
First, (1) we require a weak decrease when comparing the annotated left-hand side with the expected value of the $\sharp$-\emph{sums} in the right-hand sides 
(this guarantees that $(A)$ holds for every $\mathbf{a}$-rewrite step performed with this ADP).
The processor then removes the annotations from those ADPs where (2a) there is at least one right-hand side $r$ whose $\sharp$-\emph{sum} is strictly decreasing.
If $m = \true$, then in addition, (2b) the corresponding non-annotated right-hand side
$\flat(r)$ must be at least weakly decreasing to
ensure that nested annotations behave monotonically.\footnote{For example, it ensures that $\mathcal{A}(G) \succ \mathcal{A}(H)$ also implies that the $\sharp$-\emph{sum} of $F(G)$ is greater than the $\sharp$-sum of $F(H)$, i.e., $\mathcal{A}(G) \succ \mathcal{A}(H)$ must imply that $\mathcal{A}_{sum}^{\sharp}(F(G)) = \mathcal{A}(F(g)) + \mathcal{A}(G) \succ \mathcal{A}(F(h)) + \mathcal{A}(H) = \mathcal{A}_{sum}^{\sharp}(F(H))$, which is ensured by $\mathcal{A}(g) \succcurlyeq \mathcal{A}(h)$.}
So (2) guarantees that  $(B)$ holds for every $\mathbf{a}$-rewrite step performed with this ADP.
Finally, (3) for every rule with the flag $\ttrue$ (which can therefore be used for steps below annotations), 
the expected value must be weakly decreasing when removing all annotations
(this guarantees that $(A)$ holds for every $\mathbf{n}$-rewrite step performed with this ADP).

\begin{restatable}[{\small Reduction Pair Processor for $\mathtt{AST}$ \& $\mathtt{iAST}$}]{thm}{ProbRedPairProc}
  \label{thm:prob-RPP}
  Let $(\mathcal{A},\succ)$ be a weakly monotonic, $\NN$-collapsible, barycentric $\Sigma^\sharp$-algebra.
  Let $\PP = \PP_{\succcurlyeq} \uplus \PP_{\succ}$ with $\PP_{\succ} \neq \emptyset$ where:
  {\small
  \[\begin{array}{cclll}
      (1) &     & \forall \ell \to \mu^{m} \in \PP                                    & : & \mathcal{A}_{sum}^{\sharp}(\ell^\sharp) \succcurlyeq \IE(\mathcal{A}_{sum}^{\sharp}(\mu))     \\
      (2) & (a) & \forall \ell \to \mu^{m} \in \PP_{\succ} : \exists r \in \Supp(\mu) & : & \mathcal{A}_{sum}^{\sharp}(\ell^\sharp) \succ \mathcal{A}_{sum}^{\sharp}(r)                   \\
          & (b) & \text{If } m = \ttrue,\text{ then we additionally have }            & : & \mathcal{A}(\ell) \succcurlyeq \mathcal{A}(\flat(r)) \\
      (3) &     & \forall \ell \to \mu^{\ttrue} \in \PP                               & : &
      \mathcal{A}(\ell) \succcurlyeq \IE(\mathcal{A}(\flat(\mu)))\!
    \end{array}\]}

  \noindent
  Then $\Proc_{\mathtt{RP}}(\PP) = \Proc_{\mathtt{RP}}^{\mathbf{i}}(\PP) = \{\PP_{\succcurlyeq} \cup \flat(\PP_{\succ})\}$ is sound and complete for $\mathtt{(i)AST}$.
\end{restatable}

\begin{myproofsketch}
  \textit{Soundness}: 
  One first lifts the properties (1) - (3)
from ADPs to the properties $(A)$ and $(B)$ of \Cref{lemma:prob-RPP}
  for rewrite steps with $\tored{}{}{\PP}$.
  Then, by adapting proof ideas of
\cite{mciver2017new} to the setting of term rewriting, one can show that
  $|\F{T}| = 1$ holds for any $\PP$-CT $\F{T}$
where every infinite path uses infinitely many steps with
ADPs from $\PP_{\succ}$ at annotated positions.
  Thus,  by the $\PP$-Partition Lemma (\Cref{lemma:p-partition-main})
  it suffices to show that $\PP_{\succcurlyeq} \cup \flat(\PP_{\succ})$ is $\mathtt{AST}$.

  \textit{Completeness}: By \Cref{thm:proc-complete}.
\end{myproofsketch}

Note that
Requirement (2) only ensures a strict decrease (i.e., Property $(B)$) for
rewrite steps of type $(\mathbf{at})$ or $(\mathbf{af})$.
This is the reason why we require an infinite number of such steps within each infinite
path of a chain tree, see \Cref{def:chaintree}. 
Moreover,
$\NN$-collapsibility implies that in addition to well-foundedness of $\succ$, for every $a_0 \in A$, the length of every sequence $a_0 \succ a_1 \succ \ldots$ is bounded.
This is needed in order to show that the properties
$(A)$ and $(B)$ are indeed sufficient for $\mathtt{AST}$.
Note that we only need a strict decrease on terms but not on their expected values.
Therefore, in contrast to \cite{avanzini2020probabilistic}, we can restrict ourselves to functions $g$ which map to the natural instead of the real numbers.

\begin{exa}
  \label{example:running-1-RPP-iAST}
  To conclude $\mathtt{iAST}$ for $\R_2$ we have to remove all remaining annotations in the ADP problem $\{(\ref{run1-ADP-1}''), (\ref{run1-ADP-2}')\}$ with

  \vspace*{-.6cm}
  \begin{minipage}[t]{0.47\textwidth}
    \small
    \begin{align*}
      \tG & \to \{\nicefrac{3}{4}:\td(\tG), \; \nicefrac{1}{4}:\tz\}^{\tfalse}
      \tag{$\ref{run1-ADP-1}''$}
    \end{align*}
  \end{minipage}
  \hfill
  \begin{minipage}[t]{0.47\textwidth}
    \small
    \begin{align*}
      \tD(x) & \to \{1:\tc(x,x)\}^{\tfalse}
      \tag{$\ref{run1-ADP-2}'$}
    \end{align*}
  \end{minipage}\vspace*{.1cm}

  \noindent from \Cref{example:running-1-URP-iAST}. Then another application of the dependency graph processor yields the empty set of ADP problems.
  Here, we can use the reduction pair processor with the polynomial interpretation $(\mathcal{A},\succ) = (\Pol, >)$ that maps $\tG$ to $1$, and all other symbols to $0$.
  Then $(\ref{run1-ADP-2}')$ is weakly decreasing, and $(\ref{run1-ADP-1}'')$ is strictly
  decreasing, since (1) $\Pol_{sum}^{\sharp}(\tG) = 
  \Pol(\tG) = 1 \geq \nicefrac{3}{4} \cdot
  \Pol_{sum}^{\sharp}(\td(\tG)) + \nicefrac{1}{4} \cdot \Pol_{sum}^{\sharp}(\tz) =
  \nicefrac{3}{4} \cdot \Pol(\tG) = \nicefrac{3}{4}$ and (2) $\Pol_{sum}^{\sharp}(\tG) =
  \Pol(\tG) = 1 > \Pol_{sum}^{\sharp}(\tz)= 0$.
  Thus, the annotation of $\tG$ in $(\ref{run1-ADP-1}'')$ is deleted.

  Note that this polynomial interpretation would also satisfy the constraints for $\DPair{\R_2} = \{ \eqref{run1-ADP-1}, \eqref{run1-ADP-2} \}$ from \Cref{example:running-1-ADPs}, i.e., it would allow us to remove the annotations from the canonical ADP directly.
  Hence, if we extended our approach for $\mathtt{AST}$ to GVRFs that can duplicate annotations, then the reduction pair processor would be unsound, as it would allow us to falsely ``prove'' $\mathtt{AST}$ of $\DPair{\R_2}$.
  The problem is that we compare terms with annotations via their $\sharp$-sum, but for
  duplicating ADPs like \eqref{run1-ADP-2}, $\Pol(\td(x)) \geq \Pol(\tc(x,x))$ does not
  imply $\Pol_{sum}^{\sharp}(\td(\tG))
  \geq \Pol_{sum}^{\sharp}(\tc(\tG,\tG))$ since $\Pol_{sum}^{\sharp}(\td(\tG)) = \Pol(\tG)$ and $\Pol_{sum}^{\sharp}(\tc(\tG,\tG)) = \Pol(\tG) + \Pol(\tG)$.
\end{exa}

\begin{exa}
  \label{example:running-2-RPP-AST}
  To prove $\mathtt{AST}$ for $\R_{\mathsf{alg}}$, we also have to remove all annotations from all remaining sub-problems.
  For instance, for the sub-problem $\PP_{(\ref{run2-ADP-1}'), (\ref{run2-ADP-2}')} = \{(\ref{run2-ADP-1}'), (\ref{run2-ADP-2}'), \flat(\ref{run2-ADP-3})$ - $\flat(\ref{run2-ADP-7})\}$ from \Cref{example:running-2-UTP-AST}, we can use the reduction pair processor with the polynomial interpretation 
  $(\mathcal{A},\succ) = (\Pol, >)$ that maps $\ts(x)$ to $x+1$, $\tdouble(x)$ to $2x$, $\ttriple(x)$ to $3x$, $\tLoopOne(x)$ to $1$, and all other symbols to $0$.
  Then $(\ref{run2-ADP-2}')$ is strictly decreasing, since (1)
  $\Pol_{sum}^{\sharp}(\tLoopOne(y)) =
  \Pol(\tLoopOne(y)) = 1 \geq \nicefrac{1}{3} \cdot
  \Pol_{sum}^{\sharp}(\tLoopOne(\ttriple(y))) + \nicefrac{2}{3} \cdot
  \Pol_{sum}^{\sharp}(\tloopTwo(\ttriple(y))) = \nicefrac{1}{3}$ and (2)
  $\Pol_{sum}^{\sharp}(\tLoopOne(y)) =
  \Pol(\tLoopOne(y)) = 1 > \Pol_{sum}^{\sharp}(\tloopTwo(\ttriple(y))) = 0$.
  Similarly, $(\ref{run2-ADP-1}')$ is also strictly decreasing, and we can remove all annotations from this ADP problem.
  One can find similar interpretations to delete the remaining annotations also from the other remaining sub-problems.
  This proves $\mathtt{AST}$ for $\DPair{\R_{\mathsf{alg}}}$, and hence for $\R_{\mathsf{alg}}$.

  If we only want to prove $\mathtt{iAST}$ for $\R_{\mathsf{alg}}$, then the application of the reduction pair processor becomes significantly easier.
  Except for $\flat(\ref{run2-ADP-4})$ - $\flat(\ref{run2-ADP-7})$ in the ADP problem $\PP_{(\ref{run2-ADP-1}'), (\ref{run2-ADP-2}')}$ above, all other ADPs have the flag $\tfalse$ after the usable rules processor, see \Cref{example:running-2-URP-iAST}.
  Therefore, (3) and (2b) from \Cref{thm:prob-RPP} do not apply, and thus, we obtain significantly fewer constraints which makes the search for interpretations easier for the SMT solver.
  For example, when proving $\mathtt{AST}$ instead of $\mathtt{iAST}$, all ADPs in the ADP problem $\{\eqref{run2-ADP-6}, \flat(\ref{run2-ADP-1})$ - $\flat(\ref{run2-ADP-5}), \flat(\ref{run2-ADP-7})\}$ have the flag $\ttrue$ and thus, here we have to find a polynomial interpretation which also makes the ADPs $\flat(\ref{run2-ADP-1})$ - $\flat(\ref{run2-ADP-3})$ weakly decreasing, which we do not require when proving $\mathtt{iAST}$.
\end{exa}

\subsection{Probability Removal Processor}
\label{sec:Probability Removal Processor}

\noindent
In proofs with the ADP framework, one may obtain ADP problems $\PP$ with a non-probabilistic structure, i.e., where every ADP has the form $\ell \to \{1:r\}^{m}$.
Then the \emph{probability removal processor} allows us to switch to ordinary (non-probabilistic) DPs.

\begin{restatable}[Probability Removal Processor]{thm}{ProbRemProc}
  \label{thm:prob-NPP}
  Let $\PP$ be an ADP problem where every ADP in $\PP$ has the form $\ell \to \{1:r\}^{m}$.
  Let $\nonprobDP(\PP) = \{\ell^\sharp \to t^\sharp \mid \ell \to \{1\!:r\}^{m} \in \PP, t \trianglelefteq_{\sharp} r\}$.
  Then $\PP$ is $\mathtt{(i)AST}$ iff the non-probabilistic DP problem $(\nonprobDP(\PP),\nonprob(\PP))$ is (innermost) terminating.
  So the processor $\Proc_{\mathtt{PR}}(\PP) =\Proc_{\mathtt{PR}}^{\mathbf{i}}(\PP) = \emptyset$ is sound and complete for $\mathtt{(i)AST}$ iff $(\nonprobDP(\PP), \nonprob(\PP))$ is $\mathtt{(i)SN}$.
\end{restatable}

\begin{myproofsketch}

  \textit{If}: Due to the trivial probabilities in each ADP $\ell \to \{1:r\}^{m}$, 
  a $\PP$-CT $\F{T}$ with $|\F{T}| < 1$ must be an infinite path, 
  where we use infinitely many rewrite steps at annotated positions.
  This path gives rise to a $(\nonprobDP(\PP), \nonprob(\PP))$-chain 
  following the construction of the non-probabilistic chain criterion.

  \textit{Only if}: Every infinite $(\nonprobDP(\PP), \nonprob(\PP))$-chain gives rise to an infinite $\PP$-CT $\F{T}$,
  and because of the trivial probabilities in $\PP$, we must have $|\F{T}| = 0$.
\end{myproofsketch}

\subsection{Subterm Criterion}
\label{sec:Subterm Criterion}

\noindent
Next, we consider the \emph{subterm criterion} of \cite{DBLP:journals/iandc/HirokawaM07}
which is a purely syntactical way of removing loops from the dependency graph.
This processor has not yet been adapted to the probabilistic setting in our previous conference papers, and we explain the problems that one has to solve for such an adaption.
In the non-probabilistic setting, the subterm criterion searches for a \emph{simple projection}
for every annotated symbol $f^\sharp$, i.e., a projection $\proj$ which maps $f^\sharp$ to one of its argument positions $i$ and replaces every term $f^\sharp(\ldots)$ by its $i$-th argument.
If the projection of the right-hand side $\proj(r^\sharp)$ is a subterm of the projected left-hand side $\proj(\ell^\sharp)$ for every dependency pair $\ell^\sharp \to r^\sharp \in \mathcal{P}$, then we can remove those dependency pairs from $\PP$ where $\proj(r^\sharp)$ is a proper subterm of $\proj(\ell^\sharp)$.

\begin{defi}[Simple Projection]
  \label{def:simple-projection}
  A \emph{simple projection} is a mapping\footnote{Projections are often denoted by $\pi$ in the literature, but we use $\pi$ for positions.} $\proj$ that assigns to every $n$-ary annotated symbol $f^\sharp \in \SignatureA$ with $n \geq 1$ an argument position $i \in \{1,\ldots,n\}$.
  We extend $\proj$ to terms with annotated root symbol as follows: For $t_1,\ldots,t_n \in \TT$ with $n\geq 1$ and $\proj(f^\sharp) = i$, we define $\proj(f^\sharp(t_1,\ldots,t_n)) = t_i$.
  If $f^\sharp$ has arity 0, then we define $\proj(f^\sharp) = f^\sharp$.
\end{defi}

\subsubsection{Subterm Criterion Processor for $\mathtt{AST}$}

\noindent
In the non-probabilistic setting, if $\proj(\ell^\sharp) \trianglerighteq \proj(r^\sharp)$ holds for all $\ell^\sharp \to r^\sharp$ in a set of DPs $\PP$, then one can remove all DPs from $\PP$ where we have $\proj(\ell^\sharp) \, \triangleright \, \proj(r^\sharp)$.
So the DP processor which transforms $(\PP,\R)$ into $(\mathcal{P} \setminus \mathcal{P}_{\triangleright}, \R)$ is sound and complete for termination, where $\mathcal{P}_{\triangleright} = \{\ell^\sharp \to r^\sharp \in \mathcal{P} \mid \proj(\ell^\sharp) \, \triangleright \, \proj(r^\sharp)\}$.
The advantage is that in contrast to the reduction pair processor, the criterion of this processor is purely syntactical (i.e., one does not have to search for interpretations or well-founded relations), and one can remove DPs without considering the rewrite rules $\R$ at all.
However, the soundness of this processor for full rewriting relies on the minimality property, which does not hold in the probabilistic setting, see \Cref{remark:minimality}.
Thus, such a processor would be unsound for $\mathtt{AST}$ as shown by the following example.

\begin{exa}
  \label{ex:subterm-crit-unsound-AST}
  The ADP problem with $\tf(\ts(x)) \to \{1:\tF(x)\}^{\ttrue}$ and $\tg \to \{1:\ts(\tg)\}^{\ttrue}$ is not $\mathtt{AST}$ as we can rewrite $\tF(\ts(\tg))$ to itself by first applying the $\tf$-ADP in an $(\mathbf{at})$-step, followed by an $(\mathbf{nt})$-step with the $\tg$-ADP.
  However, consider the simple projection $\proj$ which maps each annotated symbol to its only argument.
  Then, the projected right-hand side $\proj(\tF(x)) = x$ of the $\tf$-ADP is a proper subterm of the projected annotated left-hand side $\proj(\tF(\ts(x))) = \ts(x)$, while the $\tg$-ADP does not contain any annotations, so we can ignore it.
  Hence, we would falsely assume that we can remove the annotation of the $\tf$-ADP, leading to an ADP problem which is $\mathtt{AST}$ as it does not contain any annotations anymore.
  The problem here is the non-terminating $\tg$-rule that can generate infinitely many~$\ts$-symbols.
\end{exa}

\subsubsection{Subterm Criterion Processor for $\mathtt{iAST}$}

\noindent
For non-probabilistic innermost rewriting, the soundness of the subterm criterion does not rely on the minimality criterion.
Thus, we now show how to adapt the subterm criterion to the probabilistic setting in order to prove $\mathtt{iAST}$.
For innermost rewriting, \Cref{ex:subterm-crit-unsound-AST} is no longer a counterexample, because the ADP problem from \Cref{ex:subterm-crit-unsound-AST} is $\mathtt{iAST}$.
The reason is that terms without the symbol $\tg$ only have finite reductions, and if the start term contains the symbol $\tg$, then we have to rewrite it due to innermost strategy and we cannot rewrite any $\tF$-symbols on positions above $\tg$.

However, there is another problem when adapting the subterm criterion to the probabilistic setting.
Since we consider ADPs instead of ordinary DPs, we may have terms with several annotations in the right-hand side.
For the reduction pair processor we solved this problem by considering the $\sharp$-sum.
However, for the subterm criterion it is not clear how to combine several annotated subterms.
Therefore, we restrict ourselves to ADP problems where all terms in the right-hand sides contain at most one annotation.
This leads to the following processor.

\begin{restatable}[Subterm Criterion for $\mathtt{iAST}$]{thm}{SubtermCriterionInnermost}
  \label{thm:prob-SCP-iAST}
  Let $\PP$ be an ADP problem where all occurring terms contain at most one annotation.
  Let $\proj$ be a simple projection such that for all $\ell \to \mu^m \in \PP$ and all $r \in \Supp(\mu)$ where there exists a $t \trianglelefteq_{\sharp} r$, we have $\proj(\ell^\sharp) \trianglerighteq \proj(t^\sharp)$.
  Moreover, let $\PP_{\triangleright} \subseteq \PP$ be the set of ADPs where for all $\ell \to \mu^m \in \PP_{\triangleright}$, there exist an $r \in \Supp(\mu)$ and a term $t \trianglelefteq_{\sharp} r$ such that $\proj(\ell^\sharp) \triangleright \proj(t^\sharp)$.
  Then the processor $\Proc_{\mathtt{SC}}^{\mathbf{i}}(\PP) = \{ (\PP \setminus
  \PP_{\triangleright}) \cup \flat(\PP_{\triangleright})\}$ is sound and complete for
  $\mathtt{iAST}$. 
\end{restatable}

\begin{myproofsketch}
  \textit{Soundness}:
  When choosing $\PP_\succ$ to be $\PP_\triangleright$,
  then the conditions (1) and (2a) of the reduction pair processor that consider the $\sharp$-sums
   are satisfied with
the 
polynomial interpretation that maps each $f(x_1, \ldots, x_n)$ with $f \in \Sigma$ to $x_1
+ \ldots + x_n + 1$, and each $f^\sharp(x_1, \ldots, x_n)$ with $f^\sharp \in
\Sigma^\sharp$ to $x_{\proj(f^\sharp)}$, i.e., we map each annotated symbol to its
projected argument by the simple projection $\proj$, and all other symbols are used to
calculate the number of symbols in a term.
Due to the restriction to innermost rewriting and the restriction that ADPs only contain
at most a single annotated symbol in each term on the right-hand side,
one can show that  the conditions  (2b) and
(3) are irrelevant in this setting, since 
 we cannot rewrite below the projected position before the next $(\mathbf{at})$- or
 $(\mathbf{af})$-step at the annotated position.

  \textit{Completeness}: By \Cref{thm:proc-complete}.
\end{myproofsketch}

The idea of this processor is similar to the reduction pair processor.
Since in all ADPs, the ``value'' (which is the term structure in this case) cannot increase, we can remove those ADPs where there is a chance to decrease the ``value''.
But in contrast to the reduction pair processor, we do not have to impose any constraints which result from disregarding the annotations of ADPs, i.e., we do not obtain constraints of the form (2b) or (3) in \Cref{thm:prob-RPP}.

As noted in \cite{thiemanndiss2007}, for non-probabilistic innermost rewriting, the subterm criterion is subsumed by the reduction pair processor when integrating projections (or ``argument filterings'') and an appropriate improved notion of usable rules into the reduction pair processor \cite{giesl2006mechanizing}.
Such an improved reduction pair processor which subsumes the subterm criterion would most likely also be possible in the probabilistic setting, but the reduction pair processor of \Cref{thm:prob-RPP} does not subsume the subterm criterion, as demonstrated by the following example.

\begin{exa}
  \label{example:running-2-SCP-iAST}
  Consider the ADP problem $\{\eqref{run2-ADP-3}, \flat(\ref{run2-ADP-1}), \flat(\ref{run2-ADP-2}), \flat(\ref{run2-ADP-4})\!$ - $\!\flat(\ref{run2-ADP-7})\}$ that results from the innermost dependency graph processor when proving $\mathtt{iAST}$ of $\R_{\mathsf{alg}}$, see \Cref{example:running-2-DPG-AST}.
  Here, we can use the simple projection $\proj$ that maps $\tLoopTwo$ to its only argument.
  Then for the only ADP with annotations $\eqref{run2-ADP-3}$, the projection of the annotated subterm of the right-hand side $\proj(\tLoopTwo(y)) = y$ is a proper subterm of the projected annotated left-hand side $\proj(\tLoopTwo(\ts(y))) = \ts(y)$.
  Hence, by applying \Cref{thm:prob-SCP-iAST}
  we result in $\{\flat(\ref{run2-ADP-1})$ - $\flat(\ref{run2-ADP-7})\}$, which has no annotations anymore.
  Thus, we have directly shown that this ADP problem is $\mathtt{iAST}$.
  So in contrast to the reduction pair processor of \Cref{thm:prob-RPP}, here we can ignore the other ADPs $\flat(\ref{run2-ADP-1}), \flat(\ref{run2-ADP-2}), \flat(\ref{run2-ADP-4})$ - $\flat(\ref{run2-ADP-7})$ completely even though their flags are still $\ttrue$ at this point.
\end{exa}

\subsection{Switching From Full to Innermost AST}
\label{sec:Switching From Full to Innermost AST}

\noindent
In the non-probabilistic DP framework, there is a processor to switch from full to innermost rewriting if the DP problem satisfies certain conditions \cite[Thm.\ 32]{gieslLPAR04dpframework}, e.g., if the rules and DPs are non-overlapping.
This is useful as the DP framework for innermost termination is more powerful than the one for full termination and in this way, one can switch to the innermost case for certain sub-problems, even if the whole TRS does not belong to any class where innermost termination implies termination \cite{Gramlich1995AbstractRB}.
As we showed in \cite{FOSSACS24}, to ensure that $\mathtt{iAST}$ implies $\mathtt{AST}$ for PTRSs, in addition to non-overlappingness, one has to require linearity of the rules.
However, in the non-probabilistic setting, the soundness of this processor relies again on the minimality property which does not hold in the probabilistic setting (\Cref{remark:minimality}).
Indeed, the following example which corresponds to \cite[Ex.\ 3.15]{thiemanndiss2007}
shows that a similar processor in the ADP framework would be unsound, even if we require that all ADPs are non-overlapping and linear.
Non-overlappingness of ADPs is defined as for TRSs, as both contain only a single left-hand side.

\begin{exa}
  \label{example:from-f-to-i-unsound}
  The ADP problem containing $\tf(x) \!\to\! \{1\!:\!\tF(\ta)\}^{\ttrue}$ and $\ta \to \{1:\ta\}^{\ttrue}$ is not $\mathtt{AST}$ as we can rewrite $\tF(\ta)$ to itself with the $\tf$-ADP.
  However, it is $\mathtt{iAST}$ since in innermost evaluations, we have to rewrite the inner $\ta$, which does not terminate but does not use any annotations, i.e., any $(\mathbf{at})$ or $(\mathbf{af})$ steps.
  Note that the ADPs are non-overlapping and linear.
  Thus, a processor to switch from full to innermost $\mathtt{AST}$ cannot be applied on the level of ADP problems.
\end{exa}

Hence, to prove $\mathtt{AST}$ of PTRSs that belong to the classes of \cite{FOSSACS24} where $\mathtt{iAST}$ implies $\mathtt{AST}$, one should apply the ADP framework for $\mathtt{iAST}$, because its processors are more powerful.
But otherwise, one has to use the ADP framework for $\mathtt{AST}$.

\section{Transforming ADPs}
\label{ADP Transformations}

\noindent
Compared to our first adaption of DPs to PTRSs via dependency tuples (DTs) in \cite{kassinggiesl2023iAST}, the ADP framework is not only easier, more elegant, and yields a complete chain criterion, but it also has important practical advantages, because every processor that performs a rewrite step benefits from our novel definition of rewriting with ADPs (whereas the rewrite relation with DTs in \cite{kassinggiesl2023iAST}
was an ``incomplete over-approximation'' of the rewrite relation of the original TRS).
To illustrate this, we use the following example.

\begin{exa}
  \label{ex:R-incomplete}
  Consider the PTRS $\R_{\tic}$ with the rules

  \vspace*{-0.5cm}
  \begin{minipage}[t]{0.47\linewidth}
    \begin{align}
      \ta & \to \{1 : \tf(\th(\tg),\tg)\} \label{Rtic-rule-1} \\
      \tg & \to\{\nicefrac{1}{2} : \tb_1, \nicefrac{1}{2} : \tb_2\} \label{Rtic-rule-2}\!
    \end{align}
  \end{minipage}
  \hfill
  \begin{minipage}[t]{0.47\linewidth}
    \begin{align}
      \th(\tb_1)   & \to \{1 : \ta\} \label{Rtic-rule-3} \\
      \tf(x,\tb_2) & \to \{1 : \ta\} \label{Rtic-rule-4}\!
    \end{align}
  \end{minipage}
  \vspace*{0.1cm}

  \noindent
  and the $\R_{\tic}$-RST in \Cref{fig:running-3-RST}.
  So $\ta$ can be rewritten to the normal form $\tf(\th(\tb_2),\tb_1)$ with probability $\nicefrac{1}{4}$ and to the terms $\tf(\ta,\tb_1)$ and $\ta$ that contain the redex $\ta$ with a probability of $\nicefrac{1}{4} + \nicefrac{1}{4} = \nicefrac{1}{2}$.
  In the term $\tf(\ta,\tb_2)$, one can rewrite the subterm $\ta$, and if that ends in a normal form, one can still rewrite the outer $\tf$ which will yield $\ta$ again.
  So to over-approximate the probability of non-termination, one could consider the term $\tf(\ta,\tb_2)$ as if one had two occurrences of $\ta$.
  Then this would correspond to a random walk where the number of $\ta$ symbols is decreased by 1 with probability $\nicefrac{1}{4}$, increased by 1 with probability $\nicefrac{1}{4}$, and kept the same with probability $\nicefrac{1}{2}$.
  Such a random walk is $\mathtt{AST}$, and since a similar observation holds for all $\R_{\tic}$-RSTs, $\R_{\tic}$ is $\mathtt{AST}$ (we will prove $\mathtt{iAST}$ of $\R_{\tic}$ with our new transformational processors in this section).
  In contrast, the DT framework from~\cite{kassinggiesl2023iAST} fails on this example, see~\cite{FLOPS2024} for details.
\end{exa}

\begin{figure}
  \scriptsize \center
  \begin{tikzpicture}
    \tikzstyle{adam}=[thick,draw=black!100,fill=white!100,minimum size=4mm,shape=rectangle split, rectangle split parts=2,rectangle split horizontal] \tikzstyle{adam2}=[thick,draw=red!100,fill=white!100,minimum size=4mm,shape=rectangle split, rectangle split parts=2,rectangle split horizontal] \tikzstyle{empty}=[rectangle,thick,minimum size=4mm]

    \node[adam] at (0, 0) (a) {$1$
      \nodepart{two} $\ta$};
    \node[adam] at (0, -0.7) (b) {$1$
      \nodepart{two} $\tf(\th(\tg),\tg)$};
    \node[adam] at (-2.6, -1.4) (c) {$\nicefrac{1}{2}$
      \nodepart{two} $\tf(\th(\tg),\tb_1)$};
    \node[adam] at (2.6, -1.4) (d) {$\nicefrac{1}{2}$
      \nodepart{two} $\tf(\th(\tg),\tb_2)$};
    \node[adam] at (-3.9, -2.1) (e1) {$\nicefrac{1}{4}$
      \nodepart{two} $\tf(\th(\tb_1),\tb_1)$};
    \node[adam2,label=below:{normal form}] at (-1.1, -2.1) (e2) {$\nicefrac{1}{4}$
        \nodepart{two} $\tf(\th(\tb_2),\tb_1)$};
    \node[adam] at (1.3, -2.1) (e3) {$\nicefrac{1}{4}$
      \nodepart{two} $\tf(\th(\tb_1),\tb_2)$};
    \node[adam] at (3.9, -2.1) (e4) {$\nicefrac{1}{4}$
      \nodepart{two} $\tf(\th(\tb_2),\tb_2)$};
    \node[adam] at (-3.9, -2.8) (e11) {$\nicefrac{1}{4}$
      \nodepart{two} $\tf(\ta,\tb_1)$};
    \node[adam] at (1.3, -2.8) (e33) {$\nicefrac{1}{4}$
      \nodepart{two} $\tf(\ta,\tb_2)$};
    \node[adam] at (3.9, -2.8) (e44) {$\nicefrac{1}{4}$
      \nodepart{two} $\ta$};
    \node[empty] at (-3.9, -3.5) (e111) {$\ldots$};
    \node[empty] at (1.3, -3.5) (e333) {$\ldots$};
    \node[empty] at (3.9, -3.5) (e444) {$\ldots$};

    \draw (a) edge[->] (b);
    \draw (b) edge[->] (c);
    \draw (b) edge[->] (d);
    \draw (c) edge[->] (e1);
    \draw (c) edge[->] (e2);
    \draw (d) edge[->] (e3);
    \draw (d) edge[->] (e4);
    \draw (e1) edge[->] (e11);
    \draw (e3) edge[->] (e33);
    \draw (e4) edge[->] (e44);
    \draw (e11) edge[->] (e111);
    \draw (e33) edge[->] (e333);
    \draw (e44) edge[->] (e444);
  \end{tikzpicture}
  \caption{$\R_{\tic}$-RST}
  \label{fig:running-3-RST}
\end{figure}

In the non-probabilistic DP framework, there exist several \emph{transformational} processors that increase the power of the framework substantially \cite{giesl2006mechanizing}.
To benefit from such transformations in the probabilistic setting as well, in \Cref{sec:Rewriting Processor} - \ref{sec:Rule Overlap Processor}, we introduce the \emph{rewriting}, \emph{instantiation}, \emph{forward instantiation}, and \emph{rule overlap instantiation} processors for our ADP framework.
The latter corresponds to a weaker version of the \emph{narrowing} processor from \cite{giesl2006mechanizing}, which is unsound in the probabilistic setting in general, as we will see in \cref{counterexample narrowing}.
These transformational processors had not been adapted in the probabilistic DT framework of \cite{kassinggiesl2023iAST}.
While one could also adapt these processors to the setting of \cite{kassinggiesl2023iAST}, the rewriting processor would be substantially weaker and it would fail in proving $\mathtt{iAST}$ of $\R_{\tic}$.

\subsection{Rewriting Processor}
\label{sec:Rewriting Processor}

\noindent
We start by considering the \emph{rewriting} processor.
Here, we restrict ourselves to innermost rewriting, as a corresponding processor for full rewriting is already unsound in the non-probabilistic setting.
For that reason, our novel processor can only be used to prove $\mathtt{iAST}$ (but not $\mathtt{AST}$) of $\R_{\tic}$.

In the non-probabilistic setting \cite{giesl2006mechanizing}, the rewriting processor may rewrite a redex in the right-hand side of a DP if this does not affect the construction of chains.
To ensure that, the usable rules for this redex must be non-overlapping (NO).
If the DP occurs in an innermost chain, then this redex is weakly innermost normalizing (or ``weakly innermost terminating''), hence by NO also terminating and confluent, and thus, it has a unique normal form \cite{Gramlich1995AbstractRB}.

In the probabilistic setting, to ensure that the probabilities for the normal forms stay the same, in addition to NO we require that the ADP used for the rewrite step is linear (L) (i.e., every variable occurs at most once in the left-hand side and in each term of the multi-distribution $\mu$ on the right-hand side) and non-erasing (NE) (i.e., each variable of the left-hand side occurs in each term of $\Supp(\mu)$).

\begin{defi}[Rewriting Processor]
  \label{def:RewritingProcessor}
  Let $\PP$ be an ADP problem with $\PP = \PP' \uplus \{\ell \to \{p_1:r_1, \ldots, p_k:r_k\}^{m}\}$.
  Let $\tau\!\in\!\posD(r_j)$ for some $1 \leq j \leq k$ such that $r_j|_{\tau} \in \TT$, i.e., there is no annotation below or at the position $\tau$.
  If $r_j \tored{}{}{\PP,\tau,\true} \{q_1\!:\!e_{1}, \ldots, q_h\!:\!e_{h}\}$, where $\tored{}{}{\PP,\tau,\true}$ denotes a rewrite step with $\tored{}{}{\PP}$ at position $\tau$ and the applied ADP from $\PP$ must have the flag $\true$, then we define

  \vspace*{-.6cm}

  {\small
    \[
      \begin{array}{rclll}
        \Proc_{\mathtt{r}}^{\mathbf{i}}(\PP) & = & \Bigl\{ \PP' \cup \{ & \ell \to & \{
        p_1:\flat(r_{1}), \ldots, p_k: \flat(r_{k})\}^{m},                                             \\
                                             &   &                      & \ell \to & \hspace*{-.175cm}
        \begin{array}[t]{l}
          \big( \{p_1:r_1, \ldots, p_k:r_k\} \setminus \{p_j:r_j\} \\
          \; \cup \; \{p_j \cdot q_1:e_1, \ldots, p_j \cdot q_h:e_h\}\big)^{m} \;\}\;\Bigr\}
        \end{array}
      \end{array}
    \]}
\end{defi}
\pagebreak[3]

In the non-probabilistic DP framework, when applying the rewriting processor to a DP problem $(\PP, \R)$, one only transforms the DPs $\PP$ by rewriting, but the rules $\R$ are left unchanged.
But since our ADPs represent both DPs and rules, when rewriting an ADP, we add a copy of the original ADP without any annotations (i.e., this corresponds to the original rule).
Another change to the rewriting processor in the non-probabilistic DP framework is the requirement that there exists no annotation below $\tau$.
Otherwise, rewriting would potentially remove annotations from $r_j$.
For the soundness of our processor, we have to ensure that this cannot happen.
Hence, the choice of the used VRF for this rewrite step is irrelevant.

\begin{restatable}[Soundness of the Rewriting Processor]{thm}{RewritingProc}
  \label{theorem:ptrs-rewriting-proc}
  $\Proc_{\mathtt{r}}^{\mathbf{i}}$ as in \Cref{def:RewritingProcessor} is \defemph{sound}\footnote{\label{RewritingComplete}
    For completeness in the non-probabilistic setting \cite{giesl2006mechanizing}, one uses a different definition of ``non-terminating'' (or ``infinite'') DP problems.
    In future work, we will examine if such a definition would also yield completeness of $\Proc_{\mathtt{r}}^{\mathbf{i}}$ in the probabilistic case.} for $\mathtt{iAST}$ if one of the following cases holds:
  \begin{enumerate}
    \item $\urules_{\PP}(r_j|_{\tau})$ is NO, and the rule used for rewriting $r_j|_{\tau}$ is L and NE.
    \item $\urules_{\PP}(r_j|_{\tau})$ is NO, and all its rules have the form $\ell' \to \{1:r'\}^{\true}$.
    \item $\urules_{\PP}(r_j|_{\tau})$ is NO, $r_j|_\tau$ is a ground term, and $r_j \tored{}{}{\PP,\tau,\true} \{q_1:e_{1}, \ldots, q_h:e_{h}\}$ is an innermost step.
  \end{enumerate}
\end{restatable}

\begin{myproofsketch} 
  Let
  $\overline{\PP'} =\Proc_{\mathtt{r}}^{\mathbf{i}}(\PP)$ and $\overline{\PP} = \overline{\PP'} \cup \{\ell \ruleArr{}{}{} \{ p_1:r_{1}, \ldots, p_k:r_{k}\}^{m}\}$.
  We call $\ell \ruleArr{}{}{} \{ p_1:r_{1}, \ldots, p_k: r_k\}^{m}$ the \emph{old} ADP, we call $\ell \to \left(\{p_1:r_1, \ldots, p_k:r_k\} \setminus \{p_j:r_j\} \cup \{p_j \cdot q_1:e_1, \ldots, p_j \cdot q_h:e_h\}\right)^{m}$ the \emph{new} ADP, and $\ell \ruleArr{}{}{} \{ p_1:\flat(r_{1}), \ldots, p_k: \flat(r_{k})\}^{m}$ is called the \emph{non-annotated old} ADP.

  For every $\PP$-CT $\F{T}$ we create a $\overline{\PP'}$-CT such that $|\F{T}'| \leq |\F{T}|$.
To this end, 
  we iteratively remove usages of the old ADP using a transformation $\Phi$ on CTs.
      The limit $\F{T}^{(\infty)}$ of this iteration is a $\overline{\PP'}$-CT 
      that converges with probability at most $|\F{T}|$. The transformation $\Phi$
      always removes a ``topmost'' usage 
 of the old ADP. There are two different cases, depending on whether the old ADP is
 applied
 at an annotated
      position:
      \begin{enumerate}
        \item[1.] If a $\overline{\PP}$-CT $\F{T}$ uses the old ADP in a node  $v$ at a position 
          $\pi \in \posT(t_v)$, then $\Phi(\F{T})$ is
          a corresponding $\overline{\PP}$-CT that uses the new ADP in node $v$.
          In the first case of \Cref{theorem:ptrs-rewriting-proc},
          the non-erasingness and non-overlappingness of $\PP$ guarantee that the
          rewrite step that we performed to obtain the new ADP
          would be performed later on in $\F{T}$ anyway.
          Hence, due to the linearity we can move this rewrite step to an earlier node
          without affecting the termination probability.
          The second case of \Cref{theorem:ptrs-rewriting-proc} corresponds to the original rewrite processor
          where all usable rules of $r_j|_{\tau}$ are non-probabilistic.
          The third case of \Cref{theorem:ptrs-rewriting-proc} is already an innermost rewrite step because of
          the restriction to ground terms, and moreover, it
          is the only possible rewrite step at this position.
        \item[2.] If a $\overline{\PP}$-CT $\F{T}$ uses the old ADP in a node  $v$ at a position 
          $\pi \notin \posT(t_v)$, then $\Phi(\F{T})$ is
          a corresponding $\overline{\PP}$-CT that uses the non-annotated old ADP at the root.
      \end{enumerate}
      \vspace*{-.3cm}
\end{myproofsketch}

In the last case of \Cref{theorem:ptrs-rewriting-proc}, for any instantiation only a single innermost rewrite step is possible for $r_j|_{\tau}$.
The restriction to innermost rewrite steps is only useful if $r_j|_\tau$ is ground.
Otherwise, an innermost step on $r_j|_\tau$ might become a non-innermost step when instantiating $r_j|_\tau$'s variables.

Next, we show why the rewriting processor needs the new requirement L in the first case that was not imposed in the non-probabilistic setting.
More precisely, we give counterexamples for soundness if the used rule is not left-linear, i.e., a variable occurs more than once in the left-hand side, and if the used rule is not right-linear, i.e., a variable occurs more than once in a term on the right-hand side.
The other new requirement NE is currently used in the soundness proof, but we were unable to find a counterexample to soundness if the used rule is not NE.
In fact, we conjecture that one can omit this requirement, but then one needs a much more complicated construction and estimation of the resulting termination probability in the soundness proof.
The reason is that with only L we can guarantee that performing this rewrite step at a (possibly) non-innermost redex can only increase the probability of innermost termination for the rewritten subterm but not decrease it.
Increasing the probability of termination for a proper subterm without any annotations means that we have a higher probability to apply a rewrite step at the position of an annotated symbol above it.
Recall that a CT requires that on each infinite path, redexes with annotated root symbol have to be rewritten infinitely often (they correspond to the rewrite steps with Case $(\mathbf{at})$ or $(\mathbf{af})$ of \cref{def:chaintree}).
Thus, a higher probability of termination of the proper subterm leads to a lower probability of the leaves in the CT as the probability to rewrite redexes with annotated root is higher.
However, proving this requires a much more involved approximation of the probability for termination than our current proof, where we additionally require NE.

\begin{exa}[Left-Linearity is Required for Soundness]
  \label{example:rew-proc-ll-for-soundness}
  To see why left-linearity is required for soundness of the rewriting processor in the probabilistic setting, consider the ADP problem $\PP_{\mathsf{ll}}$ with

    {\footnotesize \vspace*{-0.4cm}

      \hspace*{-.5cm}
      \begin{minipage}[t]{0.4\linewidth}
        \begin{align*}
          \tg(\tf(x,y)) & \to
          \{1:\td(\tG(\tf(\ta,\ta)),\tG(\tf(\ta,\ta)),\tG(\tf(\ta,\ta))) \}^{\tfalse} \\
        \end{align*}
      \end{minipage}
      \hspace*{-.1cm}
      \begin{minipage}[t]{0.41\linewidth}
        \begin{align}
          \tf(x,x) & \to \{1:\te(\tf(\ta,\ta))\}^{\ttrue} \label{Pll-f} \\
          \ta      & \to \{\nicefrac{1}{2}: \tb_1, \nicefrac{1}{2}: \tb_2\}^{\ttrue}\! \label{Pll-a}
        \end{align}
      \end{minipage}
    }

  \vspace*{.2cm}
  \noindent
  This example could also be made non-erasing by adding $x$ as an additional argument to $\te$ and by instantiating $x$ and $y$ by all possible values from $\{\tb_1, \tb_2 \}$ in the $\tg$-ADP.
  The ADP problem $\PP_{\mathsf{ll}}$ is not $\mathtt{iAST}$, as it allows for the following CT whose leaves have a probability $<1$.
  \begin{center}
    \scriptsize
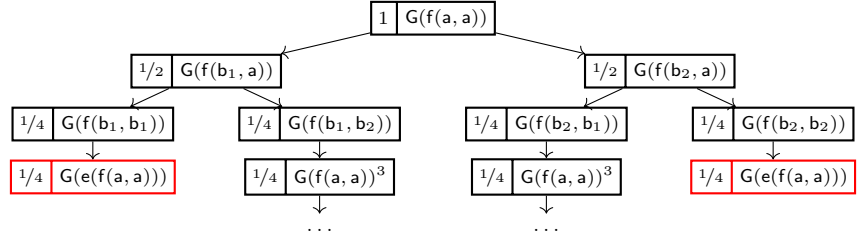
    \begin{tikzpicture}
      \tikzstyle{adam}=[thick,draw=black!100,fill=white!100,minimum size=4mm,shape=rectangle split, rectangle split parts=2,rectangle split horizontal] \tikzstyle{adam2}=[thick,draw=red!100,fill=white!100,minimum size=4mm,shape=rectangle split, rectangle split parts=2,rectangle split horizontal] \tikzstyle{empty}=[rectangle,thick,minimum size=4mm]

      \node[adam] at (0, -0.7) (b) {$1$
        \nodepart{two} $\tG(\tf(\ta,\ta))$};
      \node[adam] at (-3, -1.4) (c) {$\nicefrac{1}{2}$
        \nodepart{two} $\tG(\tf(\tb_1,\ta))$};
      \node[adam] at (3, -1.4) (d) {$\nicefrac{1}{2}$
        \nodepart{two} $\tG(\tf(\tb_2,\ta))$};
      \node[adam] at (-4.5, -2.1) (e1) {$\nicefrac{1}{4}$
        \nodepart{two} $\tG(\tf(\tb_1,\tb_1))$};
      \node[adam] at (-1.5, -2.1) (e2) {$\nicefrac{1}{4}$
        \nodepart{two} $\tG(\tf(\tb_1,\tb_2))$};
      \node[adam] at (1.5, -2.1) (e3) {$\nicefrac{1}{4}$
        \nodepart{two} $\tG(\tf(\tb_2,\tb_1))$};
      \node[adam] at (4.5, -2.1) (e4) {$\nicefrac{1}{4}$
        \nodepart{two} $\tG(\tf(\tb_2,\tb_2))$};
      \node[adam2] at (-4.5, -2.8) (e11) {$\nicefrac{1}{4}$
        \nodepart{two} $\tG(\te(\tf(\ta,\ta)))$};
      \node[adam] at (-1.5, -2.8) (e22) {$\nicefrac{1}{4}$
        \nodepart{two} $\tG(\tf(\ta,\ta))^3$};
      \node[adam] at (1.5, -2.8) (e33) {$\nicefrac{1}{4}$
        \nodepart{two} $\tG(\tf(\ta,\ta))^3$};
      \node[adam2] at (4.5, -2.8) (e44) {$\nicefrac{1}{4}$
        \nodepart{two} $\tG(\te(\tf(\ta,\ta)))$};
      \node[empty] at (-1.5, -3.5) (e222) {$\ldots$};
      \node[empty] at (1.5, -3.5) (e333) {$\ldots$};

      \draw (b) edge[->] (c);
      \draw (b) edge[->] (d);
      \draw (c) edge[->] (e1);
      \draw (c) edge[->] (e2);
      \draw (d) edge[->] (e3);
      \draw (d) edge[->] (e4);
      \draw (e1) edge[->] (e11);
      \draw (e2) edge[->] (e22);
      \draw (e3) edge[->] (e33);
      \draw (e4) edge[->] (e44);
      \draw (e22) edge[->] (e222);
      \draw (e33) edge[->] (e333);
    \end{tikzpicture}
  \end{center}
  Here, $\tG(\tf(\ta,\ta))^3$ denotes the term $\td(\tG(\tf(\ta,\ta)),\tG(\tf(\ta,\ta)),\tG(\tf(\ta,\ta)))$.
  Chain trees starting in $\tG(\te(\tf(\ta,\ta)))$ can never use a rewrite step with the $\tg$-ADP anymore and therefore, they converge with probability $1$.
  So we can rewrite a single $\tG$-term to a leaf with a probability of $\nicefrac{1}{2}$ and to three copies of itself with a probability of $\nicefrac{1}{2}$.
  Hence, the full CT corresponds to a random walk that terminates with probability $<1$.
  But without the restriction to left-linearity, we could apply the rewriting processor and replace the $\tg$-ADP by
  {\small$\tg(\tf(x,y)) \to \{1:\td(\tG(\te(\tf(\ta,\ta))),\tG(\te(\tf(\ta,\ta))),\tG(\te(\tf(\ta,\ta))))\}^{\tfalse}$}
  and several $\tg$-ADPs without annotations, since the usable rules $\urules_{\PP_{\mathsf{ll}}}(\tf(\ta,\ta)) = \{ \eqref{Pll-f}, \eqref{Pll-a} \}$ are NO.
  Now in every path of the CT this ADP can be used at most once and hence, the resulting ADP problem is $\mathtt{iAST}$, which shows unsoundness of the rewriting processor without left-linearity.
\end{exa}

\begin{exa}[Right-Linearity is Required for Soundness]
  \label{example:rew-proc-rl-for-soundness}
  To show the need for right-linearity, consider the ADP problem $\PP_{\mathsf{rl}}$ with

    {\footnotesize \vspace*{-0.4cm}

      \hspace*{-.6cm}
      \begin{minipage}[t]{0.4\linewidth}
        \begin{align*}
          \tf(\te(\tb_1,\tb_1)) & \to \{1:\th(\tF(\td(\ta)),\tF(\td(\ta)), \tF(\td(\ta)),\tF(\td(\ta)))\}^{\tfalse}\! \\
        \end{align*}
      \end{minipage}
      \hspace*{-.1cm}
      \begin{minipage}[t]{0.38\linewidth}
        \begin{align}
          \td(x) & \to \{1: \te(x,x)\}^{\ttrue} \label{Prl-d} \\
          \ta    & \to \{\nicefrac{1}{2}:\tb_1, \nicefrac{1}{2}:\tb_2\}^{\ttrue}\! \label{Prl-g}
        \end{align}
      \end{minipage}
    }

  \vspace*{.2cm}
  \noindent
  Note that for the term $\td(\ta)$ we have the following innermost $\PP_{\mathsf{rl}}$-CT:
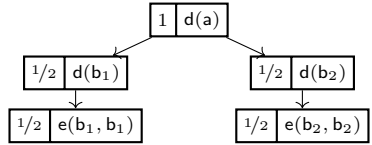
  \begin{center}
    \scriptsize
    \begin{tikzpicture}
      \tikzstyle{adam}=[thick,draw=black!100,fill=white!100,minimum size=4mm,shape=rectangle split, rectangle split parts=2,rectangle split horizontal] \tikzstyle{adam2}=[thick,draw=red!100,fill=white!100,minimum size=4mm,shape=rectangle split, rectangle split parts=2,rectangle split horizontal] \tikzstyle{empty}=[rectangle,thick,minimum size=4mm]

      \node[adam] at (0, 0) (a) {$1$
        \nodepart{two} $\td(\ta)$};
      \node[adam] at (-1.5, -0.7) (c) {$\nicefrac{1}{2}$
        \nodepart{two} $\td(\tb_1)$};
      \node[adam] at (1.5, -0.7) (d) {$\nicefrac{1}{2}$
        \nodepart{two} $\td(\tb_2)$};
      \node[adam] at (-1.5, -1.4) (c1) {$\nicefrac{1}{2}$
        \nodepart{two}
        $\te(\tb_1, \tb_1)$};
      \node[adam] at (1.5, -1.4) (d1) {$\nicefrac{1}{2}$
        \nodepart{two} $\te(\tb_2,\tb_2)$};

      \draw (a) edge[->] (c);
      \draw (a) edge[->] (d);
      \draw (c) edge[->] (c1);
      \draw (d) edge[->] (d1);
    \end{tikzpicture}
  \end{center}
  Hence, the term $\tF(\td(\ta))$ can rewrite to $\tF(\te(\tb_1,\tb_1))$ and then to four occurrences of $\tF(\td(\ta))$ again with probability $\nicefrac{1}{2}$, or it reaches a normal form with probability $\nicefrac{1}{2}$.
  This is again a random walk that terminates with probability $<1$, so that the ADP problem $\PP_{\mathsf{rl}}$ is not $\mathtt{iAST}$.
  But without the restriction to right-linearity of the used rule, it would be possible to apply the rewriting processor and replace the $\tf$-ADP by $\tf(\te(\tb_1,\tb_1)) \to \{1\!:\!\th(\tF(\te(\ta,\ta)),\tF(\te(\ta,\ta)),\tF(\te(\ta,\ta)),\tF(\te(\ta,\ta)))\}^{\tfalse}$ and several $\tf$-ADPs without annotations, as $\urules_{\PP_{\mathsf{rl}}}(\td(\ta)) = \{ \eqref{Prl-d}, \eqref{Prl-g} \}$ is NO.
  The term $\te(\ta,\ta)$ can now be rewritten to the term $\te(\tb_1,\tb_1)$ with probability $\nicefrac{1}{4}$, whereas one obtains a term of the form $\te(\tb_i,\tb_j)$ with $i \neq 1$ or $j \neq 1$ with probability $\nicefrac{3}{4}$.
  Hence, now $\tF(\te(\ta,\ta))$ can rewrite to a term with four subterms $\tF(\te(\ta,\ta))$ only with a probability of $\nicefrac{1}{4}$, and it reaches a leaf with probability $\nicefrac{3}{4}$.
  This is now a random walk that terminates with probability $1$ and the same happens for all possible CTs.
  Hence, the resulting ADP problem is $\mathtt{iAST}$, which shows unsoundness of the rewriting processor without right-linearity.
\end{exa}

As mentioned, while a rewriting processor could also be defined in the probabilistic DT framework of \cite{kassinggiesl2023iAST}, the rewriting processor benefits from our ADP framework, because it applies the rewrite relation $\tored{}{}{\PP}$ on ADPs.
In contrast, a rewriting processor in the DT framework of \cite{kassinggiesl2023iAST}
would have to replace a DT by \emph{multiple} new DTs, because the rewrite relation on DTs in \cite{kassinggiesl2023iAST} corresponds to an over-approximation of the actual rewrite relation on PTRSs.
Such a rewriting processor would fail for $\R_{\tic}$ whereas with the processor of \Cref{theorem:ptrs-rewriting-proc} we can now prove that $\R_{\tic}$ is $\mathtt{iAST}$.

\begin{exa}
  \label{leading ex rewriting}
  After applying the usable terms and the usable rules processor to $\DPair{\R_{\tic}}$, we obtain:

  \vspace*{-0.5cm}
  \begin{minipage}[t]{0.47\linewidth}
    \begin{align}
      \ta & \to \{1 : \tF(\tH(\tg),\tg)\}^{\ttrue} \label{Rtic-adp-1} \\
      \tg & \to\{\nicefrac{1}{2} : \tb_1, \nicefrac{1}{2} :
      \tb_2\}^{\ttrue} \label{Rtic-adp-2}
    \end{align}
  \end{minipage}
  \hfill
  \begin{minipage}[t]{0.47\linewidth}
    \begin{align}
      \th(\tb_1)   & \to \{1 : \tA\}^{\ttrue} \label{Rtic-adp-3} \\
      \tf(x,\tb_2) & \to \{1 : \tA\}^{\tfalse} \label{Rtic-adp-4}\!
    \end{align}
  \end{minipage}

  \vspace*{.2cm}

  Here, the usable terms processor has been used to replace $\tF(\tH(\tG),\tG)$ by $\tF(\tH(\tg),\tg)$ in the right-hand side of \eqref{Rtic-adp-1}.
  The reason is that $\tG$ is not usable as the $\tg$-ADP \eqref{Rtic-adp-2}
  does not contain any annotations.
  Moreover, the usable rules processor has been used to set the flag of \eqref{Rtic-adp-4} to $\false$, because neither $\tf$ nor $\tF$ occur below annotated symbols on right-hand sides.

  Next we can apply the rewriting processor on $(\ref{Rtic-adp-1})$ repeatedly until all $\tg$-symbols are rewritten, and replace it by the ADP $\ta \to \{\nicefrac{1}{4}:\tF(\tH(\tb_1),\tb_1), \nicefrac{1}{4}:\tF(\tH(\tb_2),\tb_1), \nicefrac{1}{4}:\tF(\tH(\tb_1),\tb_2), \nicefrac{1}{4}:\tF(\tH(\tb_2),\tb_2)\}^{\ttrue}$ as well as several ADPs $\ta \to \ldots$ without annotations in the right-hand side.
  Now in the subterms $\tF(\ldots,\tb_1)$ and $\tH(\tb_2)$, the annotations are removed from the roots by the usable terms processor, as these subterms cannot rewrite to annotated instances of left-hand sides of ADPs.
  So the $\ta$-ADP is changed to $\ta \to \{\nicefrac{1}{4}:\tf(\tH(\tb_1),\tb_1), \nicefrac{1}{4}:\tf(\th(\tb_2),\tb_1), \nicefrac{1}{4}:\tF(\tH(\tb_1),\tb_2), \nicefrac{1}{4}:\tF(\th(\tb_2),\tb_2)\}^{\ttrue}\; (\ref{Rtic-adp-1}')$.
  This ADP corresponds to the observations that explain why $\R_{\tic}$ is $\mathtt{iAST}$ in \Cref{ex:R-incomplete}: We have two terms in the right-hand side that correspond to one $\tA$ each, both with probability $\nicefrac{1}{4}$, one term that corresponds to a normal form with probability $\nicefrac{1}{4}$, and one that corresponds to two $\tA$'s with probability $\nicefrac{1}{4}$.
  So again this corresponds to a random walk where the number of $\tA$-symbols is decreased by 1 with probability $\nicefrac{1}{4}$, increased by 1 with probability $\nicefrac{1}{4}$, and kept the same with probability $\nicefrac{1}{2}$.

  Indeed, we can now use the reduction pair processor with the polynomial interpretation
  $\Pol$ that maps $\tA$, $\tF$, and $\tH$ to 1 and all other symbols to 0, to remove all
  annotations from the $\ta$-ADP $(\ref{Rtic-adp-1}')$, because it contains the right-hand
  side $\tf(\th(\tb_2), \tb_1)$ without annotations and thus,
  $\Pol_{sum}^{\sharp}(\tA) = \Pol(\tA) = 1 > \Pol_{sum}^{\sharp}(\tf(\th(\tb_2), \tb_1)) = 0$.
  Another application of the usable terms processor removes the remaining $\tA$-annotations from \eqref{Rtic-adp-3}
  and \eqref{Rtic-adp-4}.
  Since there are no more annotations left, this proves $\mathtt{iAST}$ of $\R_{\tic}$.
\end{exa}

\subsection{Instantiation Processor}
\label{sec:Instantiation Processor}

\noindent
In the non-probabilistic setting \cite{giesl2006mechanizing}, the idea of the \emph{instantiation} processor is to consider all possible \emph{predecessors} $s^{\sharp} \to t^\sharp$ of a dependency pair $u^\sharp \to v^\sharp$ in a chain and to compute the skeleton $\renterm(\capterm_\R(t^\sharp))$ of $t^\sharp$ (see \Cref{DP Framework}) that remains unchanged when we reduce $t^\sharp\sigma$ to $u^\sharp\sigma$ for some instantiation $\sigma$, i.e., when going from one DP in a chain to the next.
Then $\renterm(\capterm_\R(t^\sharp))$ and $u^\sharp$ must unify with some most general unifier (\emph{mgu}) $\delta$, and $t^\sharp\sigma \to_{\R}^* u^\sharp\sigma$ implies that $\sigma$ is an instance of $\delta$.
Hence, the instantiation processor replaces the DP $u^\sharp \to v^\sharp$ by $u^\sharp\delta \to v^\sharp\delta$.

To adapt the instantiation processor to ADPs and $\mathtt{AST}$, we consider all terms in the multi-distributions on the right-hand sides of all predecessors.
In the non-probabilistic DP framework, the instantiation processor only instantiates the DPs, but the rules are left unchanged.
Due to the duality of ADPs, similar to the rewriting processor, 
when instantiating an ADP, we add a copy of the original ADP without any annotations again (i.e., this corresponds to the original non-instantiated rule).
In the following, $\vr(\PP)$ is a variable-renamed copy of $\PP$ where all variables are replaced by fresh ones, and we use $\delta(\ell \ruleArr{}{}{} \{p_1:r_1, \ldots, p_k:r_k\}^{m}) = \ell \delta \ruleArr{}{}{} \{p_1:r_1 \delta, \ldots, p_k:r_k \delta\}^{m}$ for any substitution $\delta$ and ADP $\ell \ruleArr{}{}{} \{p_1:r_1, \ldots, p_k:r_k\}^{m}$.

\begin{restatable}[Instantiation Processor for $\mathtt{AST}$]{thm}{InstProc}
  \label{theorem:inst-proc-ast}
  Let $\PP$ be an ADP problem with $\PP = \PP' \uplus \{\ell \ruleArr{}{}{} \mu^{m}\}$.
  Then $\Proc_{\mathtt{ins}}$ is sound and complete for $\mathtt{AST}$, where $\Proc_{\mathtt{ins}}(\PP)\!=\!\{\PP' \cup N \cup \{\ell \ruleArr{}{}{} \flat(\mu)^{m}\}\}$ with

  \vspace*{-.2cm}

  {\footnotesize
    \[
      N = \Biggl\{ \delta(\ell \ruleArr{}{}{} \mu^{m})
      \Bigg|
      \begin{array}{c}
        \ell' \ruleArr{}{}{} \nu^{m'} \in \vr(\PP), r' \in \Supp(\nu), \\
        t \trianglelefteq_{\sharp} r', \delta =
        mgu(\renterm(\capterm_\PP(t^\sharp)), \ell^\sharp) \!
      \end{array}\Biggr\} \!
    \]}
\end{restatable}

\begin{myproofsketch} 
  Let $\overline{\PP} = \PP' \cup N \cup \{\ell \ruleArr{}{}{} \flat(\mu)^{m}\}$. We again
  use the Starting Lemma (\Cref{lemma:starting})
  to restrict the form of the CTs that have to be regarded.

  \textit{Soundness}:
  Every $\PP$-CT gives rise to a $\overline{\PP} \cup \{\ell \ruleArr{}{}{} \mu^{m}\}$-CT that converges with the
  same probability and uses the ADP $\ell \ruleArr{}{}{} \mu^{m}$ at most at the root.
  The core idea is that every rewrite step with $\ell \ruleArr{}{}{} \mu^{m}$ at a node
  below the root can also be done with a rule from $N$, 
  or we can use $\ell \ruleArr{}{}{} \flat(\mu)^m$
  if the annotations do not matter, e.g., we rewrite at a position that is not annotated.

  If there is a $\overline{\PP} \cup \{\ell \ruleArr{}{}{} \mu^{m}\}$-CT 
  which only uses $\ell \ruleArr{}{}{} \mu^{m}$ at the root and
  converges with a probability $< 1$, then this also holds for one of its direct subtrees,
  which is a $\overline{\PP}$-CT.

  \textit{Completeness}: 
  Every $\overline{\PP}$-CT  gives rise to a $\PP$-CT that converges with the
  same probability.
  The core idea is that every rewrite step with an ADP from $N$ 
  or the ADP $\ell \ruleArr{}{}{} \flat(\mu)^{m}$ 
  is also possible with the more general ADP $\ell \ruleArr{}{}{} \mu^{m}$ 
  that may also contain more annotations.
\end{myproofsketch}

For innermost rewriting, as in the estimation of the dependency graph, we do not need $\renterm$, and only consider those mgu's $\delta$ where the instantiated left-hand sides are in argument normal form.

\begin{restatable}[Instantiation Processor for $\mathtt{iAST}$]{thm}{InstProcInnermost}
  \label{theorem:inst-proc-iast}
  Let $\PP$ be an ADP problem with $\PP = \PP' \uplus \{\ell \ruleArr{}{}{} \mu^{m}\}$.
  Then $\Proc_{\mathtt{ins}}^{\mathbf{i}}$ is sound and complete for $\mathtt{iAST}$, where $\Proc_{\mathtt{ins}}^{\mathbf{i}}(\PP)\!=\!\{\PP' \cup N \cup \{\ell \ruleArr{}{}{} \flat(\mu)^{m}\}\}$ with

  \vspace*{-.2cm}

  {\footnotesize
    \[
      N = \Biggl\{ \delta(\ell \ruleArr{}{}{} \mu^{m})
      \Bigg|
      \begin{array}{c}
        \ell' \ruleArr{}{}{} \nu^{m'} \in \vr(\PP), r' \in \Supp(\nu), \\
        t \trianglelefteq_{\sharp} r', \delta =
        mgu(\capterm_\PP(t^\sharp), \ell^\sharp),
        \{\ell' \delta, \ell \delta\} \subseteq \ANF_{\PP} \!
      \end{array}\Biggr\} \!
    \]}
\end{restatable}

\begin{myproofsketch}
 Similar to the proof of \Cref{theorem:inst-proc-ast}, just considering innermost rewrite steps.
\end{myproofsketch}

\begin{exa}
  \label{example:inst}
  Consider the PTRS $\R_{\mathsf{ins}}$ with the rules \vspace*{-0.2cm}
  \begin{align*}
    \tc            & \to \{1 : \ta\} \qquad \tc \to \{1 : \tb\} \\
    \tf(x,y,z)     & \to \{1 : \tg(x,y,z)\} \\
    \tg(\ta,\tb,u) & \to \{\nicefrac{1}{2} : \tf(u,u,u), \nicefrac{1}{2} : \tg(\ta,\tb,u)\}\!
  \end{align*}
  $\R_{\mathsf{ins}}$ is not $\mathtt{AST}$, since $(1:\tg(\ta,\tb,\tc))$ starts an RST without leaves.
  However, $\R_{\mathsf{ins}}$ is $\mathtt{iAST}$.
  Its canonical ADPs $\DPair{\R_{\mathsf{ins}}}$ are \vspace*{-0.1cm}
  \begin{align*}
    \tc            & \to \{1 : \ta\}^{\ttrue} \qquad \tc \to \{1 : \tb\}^{\ttrue} \\
    \tf(x,y,z)     & \to \{1:\tG(x,y,z)\}^{\ttrue} \\
    \tg(\ta,\tb,u) & \to \{\nicefrac{1}{2}:\tF(u,u,u), \nicefrac{1}{2}:\tG(\ta,\tb,u)\}^{\ttrue}\!
  \end{align*}
  \vspace*{-0.5cm}

  \noindent
  Using only the processors of \Cref{The Probabilistic ADP Framework}, we cannot prove that $\R_{\mathsf{ins}}$ is $\mathtt{iAST}$.
  However, we can apply the instantiation processor on the $\tf$-ADP.

  As the term $t^\sharp = \tF(u,u,u)$ in the right-hand side of the $\tg$-ADP does not
  have any proper subterms with (possibly annotated) defined symbols, we have\linebreak $\capterm_{\DPair{\R_{\mathsf{ins}}}}(t^\sharp) = t^\sharp$.
  For the left-hand side $\ell^\sharp = \tF(x,y,z)$, $\delta = mgu(t^\sharp, \ell^\sharp)$ instantiates $x$, $y$, and $z$ by $u$.
  So the instantiation processor replaces the $\tf$-ADP by $\tf(u,u,u) \to \{1:\tG(u,u,u)\}^{\ttrue}$ (and moreover, we add $\tf(x,y,z) \to \{1:\tg(x,y,z)\}^{\ttrue}$).
  We can now remove the annotation in the transformed $\tf$-ADP by the dependency graph processor, and afterwards remove the annotation in the $\tg$-ADP by applying the reduction pair processor with the polynomial interpretation that maps $\tG$ to the constant $1$ and every other function symbol to $0$.
  As we removed all annotations, $\DPair{\R_{\mathsf{ins}}}$ and hence $\R_{\mathsf{ins}}$ must be $\mathtt{iAST}$.
\end{exa}

\subsection{Forward Instantiation Processor}
\label{sec:Forward Instantiation Processor}

\noindent
Next we adapt the \emph{forward instantiation} processor.
For non-probabilistic DPs, the idea of the forward instantiation processor \cite{giesl2006mechanizing}
is to consider all possible \emph{successors} $u^\sharp \to v^\sharp$ of a DP $s^\sharp \to t^\sharp$ in a chain, again, in order to find the skeleton that remains unchanged when rewriting $t^\sharp\sigma$ to $u^\sharp\sigma$ for some substitution $\sigma$.
To find this skeleton, one reverses the rules of the TRS and then proceeds as for the instantiation processor.
Note that these reversed rules might violate the variable conditions of TRSs, i.e., the right-hand side of a reversed rule may contain variables that do not occur in the left-hand side or the left-hand side may be a variable, i.e., we ignore these restrictions here.
We say that an ADP problem $\PP$ is \emph{collapsing} if it contains an ADP $\ell \to \mu^m$ such that $\Supp(\mu) \cap \VV \neq \emptyset$, i.e., one of the right-hand sides of an ADP is a variable.
For a non-collapsing ADP problem $\PP$ and a term $\ell' \in \TT$, let $\capterm_{\PP}^{-1}(\ell'^\sharp)$ result from $\ell'^\sharp$ by replacing all its subterms $f(\ldots)$ by fresh variables where $f$ occurs on the root position of some right-hand side in $\flat(\PP)$.
For a collapsing ADP problem $\PP$, $\capterm_{\PP}^{-1}(\ell'^\sharp)$ is simply a fresh variable.

\begin{restatable}[Forward Instantiation Processor for $\mathtt{AST}$]{thm}{ForwardInstProc}
  \label{theorem:forward-inst-proc-ast}
  Let $\PP$ be an ADP problem with $\PP = \PP' \uplus \{\ell \ruleArr{}{}{} \mu^{m}\}$.
  Then $\Proc_{\mathtt{fins}}$ is sound and complete for $\mathtt{AST}$, where $\Proc_{\mathtt{fins}}(\PP)\!=\!\{\PP' \cup N \cup \{\ell \ruleArr{}{}{} \flat(\mu)^{m}\}\}$.
  Here,
  {\footnotesize
      \[
        N = \Biggl\{ \delta(\ell \ruleArr{}{}{} \mu^{m})
        \Bigg|
        \begin{array}{c}
          \ell' \ruleArr{}{}{} \nu^{m'} \in \vr(\PP),
          r \in \Supp(\mu),
          t \trianglelefteq_{\sharp} r, \\
          \delta = mgu(t^\sharp, \renterm(\capterm_{\PP}^{-1}(\ell'^\sharp))) \!
        \end{array}\Biggr\} \!
      \]}
\end{restatable}

\begin{myproofsketch}
 Similar to the proof of \Cref{theorem:inst-proc-ast}.
\end{myproofsketch}

As in the non-probabilistic setting, the variant of this processor for innermost rewriting still has to use the renaming of all occurrences of variables via $\renterm$ since reversing the rules does not necessarily preserve the innermost evaluation strategy.
However, instead of considering all ADPs from $\PP$ in $\capterm_{\PP}^{-1}$ we can restrict ourselves to the usable rules of $t^\sharp$.

\begin{restatable}[Forward Instantiation Processor for $\mathtt{iAST}$]{thm}{ForwardInstProcInnermost}
  \label{theorem:forward-inst-proc-iast}
  Let $\PP$ be an ADP problem with $\PP = \PP' \uplus \{\ell \ruleArr{}{}{} \mu^{m}\}$.
  Then $\Proc_{\mathtt{fins}}^{\mathbf{i}}$ is sound and complete for $\mathtt{iAST}$, where $\Proc_{\mathtt{fins}}^{\mathbf{i}}(\PP)\!=\!\{\PP' \cup N \cup \{\ell \ruleArr{}{}{} \flat(\mu)^{m}\}\}$.
  Here,
  {\footnotesize
      \[
        N = \Biggl\{ \delta(\ell \ruleArr{}{}{} \mu^{m})
        \Bigg|
        \begin{array}{c}
          \ell' \ruleArr{}{}{} \nu^{m'} \in \vr(\PP),
          r \in \Supp(\mu),
          t \trianglelefteq_{\sharp} r, \\
          \delta = mgu(t^\sharp, \renterm(\capterm_{\urules_{\PP}(t^\sharp)}^{-1}(\ell'^\sharp))),
          \{\ell \delta, \ell' \delta \} \subseteq \ANF_{\PP} \!
        \end{array}\Biggr\} \!
      \]}
\end{restatable}

\begin{myproofsketch}
   Similar to the proof of \Cref{theorem:forward-inst-proc-ast}, just considering innermost rewrite steps.
\end{myproofsketch}

\begin{exa}
  \label{example:forward-inst}
  Consider the PTRS $\R_{\mathsf{fins}}$ with the rules

    {\small \vspace*{-0.5cm}
      \begin{minipage}[t]{0.47\linewidth}
        \begin{align*}
          \tf(x)   & \to \{\nicefrac{1}{2}:\tg(x), \nicefrac{1}{2}:\th(x)\} \\
          \tg(\ta) & \to \{1:\tf(\tq(\ta))\} \\
          \th(\tb) & \to \{1:\tf(\tq(\tb))\}\!
        \end{align*}
      \end{minipage}
      \hfill
      \begin{minipage}[t]{0.47\linewidth}
        \vspace*{0.3cm}
        \begin{align*}
          \tq(\ta) & \to \{1:\ta\} \\
          \tq(\tb) & \to \{1:\tb\}\!
        \end{align*}
      \end{minipage}
    }

  \vspace*{.2cm}
  \noindent
  When trying to prove $\mathtt{iAST}$, by the usable rules and the usable terms processor, we obtain the ADP problem $\PP_{\mathsf{fins}}$:

  {\small \vspace*{-0.5cm}
  \begin{minipage}[t]{0.47\linewidth}
    \begin{align*}
      \tf(x)   & \to \{\nicefrac{1}{2}:\tG(x), \nicefrac{1}{2}:\tH(x)\}^{\tfalse} \\
      \tg(\ta) & \to \{1:\tF(\tq(\ta))\}^{\tfalse} \\
      \th(\tb) & \to \{1:\tF(\tq(\tb))\}^{\tfalse}\!
    \end{align*}
  \end{minipage}
  \hfill
  \begin{minipage}[t]{0.47\linewidth}
    \vspace*{0.3cm}
    \begin{align*}
      \tq(\ta) & \to \{1:\ta\}^{\ttrue} \\
      \tq(\tb) & \to \{1:\tb\}^{\ttrue}\!
    \end{align*}
  \end{minipage}
  }

  \vspace*{.2cm}
  \noindent
  Here, the instantiation processor is useless, as both the mgu of
  $\capterm_{\PP_{\mathsf{fins}}}(\tF(\tq(\ta)))\linebreak =
    \tF(y)$ and $\tF(x)$, and the mgu of $\capterm_{\PP_{\mathsf{fins}}}(\tF(\tq(\tb))) = \tF(z)$ and $\tF(x)$ do not modify $\tF(x)$. However, the forward instantiation processor can replace the original $\tf$-ADP with $\tf(\ta) \to \{\nicefrac{1}{2}:\tG(\ta), \nicefrac{1}{2}:\tH(\ta)\}^{\tfalse}$ and $\tf(\tb) \to \{\nicefrac{1}{2}:\tG(\tb), \nicefrac{1}{2}:\tH(\tb)\}^{\tfalse}$. The reason is that the only possible successors of the $\tf$-ADP are the $\tg$- and the $\th$-ADP. For the $\tg$-ADP, we have $\urules_{\PP_{\mathsf{fins}}}(\tG(x)) = \emptyset$ and thus, $mgu(\tG(x), \renterm(\capterm_{\urules_{\PP_{\mathsf{fins}}}(\tG(x))}^{-1}(\tG(\ta)))) = mgu(\tG(x),\tG(\ta))$ instantiates $x$ with $\ta$. Similarly, for the $\th$-ADP, the mgu instantiates $x$ with $\tb$. (Note that here the restriction to the usable rules is crucial in the innermost forward instantiation processor, because $\capterm_{\PP_{\mathsf{fins}}}^{-1}$ would replace the symbol $\ta$ by a fresh variable, since it occurs on the root position in the right-hand side of an ADP from $\PP_{\mathsf{fins}}$. Hence, $mgu(\tG(x), \renterm(\capterm_{\PP_{\mathsf{fins}}}^{-1}(\tG(\ta)))) = mgu(\tG(x),\tG(y))$ would not modify $x$.)

  We can now remove the annotations of the normal forms $\tG(\tb)$ and $\tH(\ta)$ from the new instantiated $\tf$-ADPs with the usable terms processor, and apply the reduction pair processor with the polynomial interpretation that maps every function symbol to the constant $1$, in order to remove all annotations of both new $\tf$-ADPs.
  Finally, we can remove all remaining annotations and prove $\mathtt{iAST}$ using the dependency graph processor.
  Since the PTRS is non-overlapping and linear, this also implies $\mathtt{AST}$, see \cite{FOSSACS24}.
  Note that without the forward instantiation processor, we would have to find a polynomial interpretation that is at least linear.
\end{exa}

\cref{example:forward-inst} shows that the processors that instantiate a given ADP are not only sometimes needed for a successful termination proof (as in \cref{example:inst}), but sometimes they can also ease the search for algebras (e.g., polynomial or matrix interpretations) in the reduction pair processor, the most time-consuming part of the ADP framework.

\subsection{Rule Overlap Instantiation Processor}
\label{sec:Rule Overlap Processor}

\noindent
The \emph{narrowing} processor \cite{arts2000termination,giesl2006mechanizing}
can only be used in a weaker version in the probabilistic setting, and only for innermost rewriting.
Let $\PP = \PP' \uplus \{\ell \ruleArr{}{}{} \{ p_1:r_{1}, \ldots, p_k: r_k\}^{m}\}$ be an ADP problem.
For each $1 \leq j \leq k$ and each $t \trianglelefteq_{\sharp} r_j$, we define its \emph{narrowing substitutions} and its \emph{narrowing results} (as in \cite{noschinski2013analyzing}, where narrowing was adapted to dependency tuples for complexity analysis of ordinary term rewriting).
If we have to perform rewrite steps on (an instance of) $t$ in order to enable the next application of an ADP at an annotated position, then the idea of the narrowing processor is to perform the first step of this reduction already on the ADP $\ell \ruleArr{}{}{} \{ p_1:r_{1}, \ldots, p_k: r_k\}^{m}$.
So whenever there is a $t \trianglelefteq_{\sharp} r_j$ and a non-variable position $\tau$ in $t$ such that $t|_\tau$ unifies with the left-hand side $\ell'$ of some ADP $\ell' \ruleArr{}{}{} \{
  p_1':r_{1}', \ldots, p_{k'}': r_{k'}'\}^{m'} \in \vr(\PP)$ using an mgu $\delta$ (such that $\ell \delta, \ell' \delta \in \ANF_{\PP}$ if we consider innermost rewriting), then $\delta$ is a \emph{narrowing substitution} of $t$. While the corresponding \emph{narrowing result}
could also be defined for probabilistic rules, to simplify the presentation let us assume for the moment that the rule just has the form $\ell' \to \{1:r'\}$.
Then the corresponding \emph{narrowing result} is $s = t[r']_{\tau} \delta$ if we rewrite at a position of $t$ that used to be annotated in $r_j$, and $s = t[\flat(r')]_{\tau} \delta$ otherwise.

If $\delta_1, \ldots ,\delta_d$ are all narrowing substitutions of $t$ with the corresponding narrowing results $s_1,\ldots,s_d$, then one would like to define a narrowing processor that replaces $\ell \ruleArr{}{}{} \{ p_1:r_{1}, \ldots, p_k: r_k\}^{m}$ by $\ell \delta_e \to \{p_1:r_1 \delta_e, \ldots, p_j:s_e, \ldots, p_k:r_k \delta_e\}$, for all $1 \leq e \leq d$.
However, the main idea of the narrowing processor, i.e., performing the first rewrite step directly on the ADPs, is unsound for probabilistic ADP problems, as shown by the following example.

\begin{exa}
  \label{counterexample narrowing}
  Consider the following ADP problem:

  {\small \vspace*{-0.4cm}
  \begin{minipage}[t]{0.47\linewidth}
    \begin{align*}
      \tf(\tb_1,\td_1) & \to \{1: \tF(\ta,\te)\}^{\tfalse} \\
      \tf(\tb_2,\td_2) & \to \{1: \tF(\ta,\te)\}^{\tfalse} \\
      \ta              & \to \{\nicefrac{1}{2}:\tb_1, \nicefrac{1}{2}:\tb_2\}^{\ttrue}\!
    \end{align*}
  \end{minipage}
  \hfill
  \begin{minipage}[t]{0.47\linewidth}
    \vspace*{0.3cm}
    \begin{align*}
      \te & \to \{1:\td_1\}^{\ttrue} \\
      \te & \to \{1:\td_2\}^{\ttrue}\!
    \end{align*}
  \end{minipage}
  }

  \vspace*{.3cm}
  \noindent
  This ADP problem is not $\mathtt{iAST}$, because it allows for the CT below on the left without any leaves.
  Note that all occurring terms are ground terms, hence all narrowing substitutions are just the identity function.

  But if we apply the narrowing processor to the ADPs in order to rewrite $\te$, then we replace the two $\tf$-ADPs by the following four new ADPs.

    {\small \vspace*{-0.5cm}
      \begin{minipage}[t]{0.47\linewidth}
        \begin{align*}
          \tf(\tb_1,\td_1)\to \{1:\tF(\ta,\td_1)\}^{\tfalse} \\
          \tf(\tb_1,\td_1)\to \{1:\tF(\ta,\td_2)\}^{\tfalse}\!
        \end{align*}
      \end{minipage}
      \hfill
      \begin{minipage}[t]{0.47\linewidth}
        \begin{align*}
          \tf(\tb_2,\td_2)\to \{1:\tF(\ta,\td_1)\}^{\tfalse} \\
          \tf(\tb_2,\td_2)\to \{1:\tF(\ta,\td_2)\}^{\tfalse}\!
        \end{align*}
      \end{minipage}
    }

  \vspace*{.2cm}
  \noindent
  This new ADP problem is $\mathtt{iAST}$, as we reach a normal form with probability $\nicefrac{1}{2}$ after each application of an $\tf$-ADP.
  For example, if we use the first ADP, then we get the CT below on the right.
  There we reach the normal form $\tF(\tb_2,\td_1)$ with probability $\nicefrac{1}{2}$.
  \begin{center}
    \scriptsize
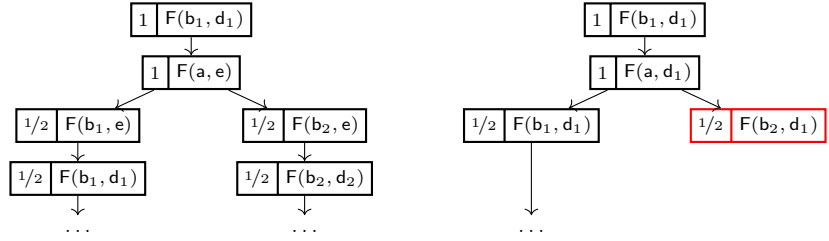
    \begin{tikzpicture}
      \tikzstyle{adam}=[thick,draw=black!100,fill=white!100,minimum size=4mm,shape=rectangle split, rectangle split parts=2,rectangle split horizontal] \tikzstyle{adam2}=[thick,draw=red!100,fill=white!100,minimum size=4mm,shape=rectangle split, rectangle split parts=2,rectangle split horizontal] \tikzstyle{empty}=[rectangle,thick,minimum size=4mm]

      \node[adam] at (-3, 0) (a) {$1$
        \nodepart{two} $\tF(\tb_1,\td_1)$};
      \node[adam] at (-3, -0.7) (b) {$1$
        \nodepart{two} $\tF(\ta,\te)$};
      \node[adam] at (-4.5, -1.4) (c) {$\nicefrac{1}{2}$
        \nodepart{two} $\tF(\tb_1,\te)$};
      \node[adam] at (-1.5, -1.4) (d) {$\nicefrac{1}{2}$
        \nodepart{two} $\tF(\tb_2,\te)$};
      \node[adam] at (-4.5, -2.1) (e1) {$\nicefrac{1}{2}$
        \nodepart{two} $\tF(\tb_1,\td_1)$};
      \node[adam] at (-1.5, -2.1) (e2) {$\nicefrac{1}{2}$
        \nodepart{two} $\tF(\tb_2,\td_2)$};
      \node[empty] at (-4.5, -2.8) (e222) {$\ldots$};
      \node[empty] at (-1.5, -2.8) (e333) {$\ldots$};

      \draw (a) edge[->] (b);
      \draw (b) edge[->] (c);
      \draw (b) edge[->] (d);
      \draw (c) edge[->] (e1);
      \draw (d) edge[->] (e2);
      \draw (e1) edge[->] (e222);
      \draw (e2) edge[->] (e333);

      \node[adam] at (3, 0) (a) {$1$
        \nodepart{two} $\tF(\tb_1,\td_1)$};
      \node[adam] at (3, -0.7) (b) {$1$
        \nodepart{two} $\tF(\ta,\td_1)$};
      \node[adam] at (1.5, -1.4) (c) {$\nicefrac{1}{2}$
        \nodepart{two} $\tF(\tb_1,\td_1)$};
      \node[adam2] at (4.5, -1.4) (d) {$\nicefrac{1}{2}$
        \nodepart{two} $\tF(\tb_2,\td_1)$};
      \node[empty] at (1.5, -2.8) (e222) {$\ldots$};

      \draw (a) edge[->] (b);
      \draw (b) edge[->] (c);
      \draw (b) edge[->] (d);
      \draw (c) edge[->] (e222);
    \end{tikzpicture}
  \end{center}
  The difference is that when narrowing the ADP, we have to decide how to rewrite $\te$ before we ``split'' the term into two different successors with a certain probability, i.e., before we rewrite $\ta$ to $\{ \nicefrac{1}{2}:\tb_1, \nicefrac{1}{2}:\tb_2 \}$.
\end{exa}

Thus, in the probabilistic setting, we can only transform the ADP by applying a rewrite rule if we can ensure the same conditions as for the rewriting processor in \Cref{sec:Rewriting Processor}.
So in the probabilistic setting, the narrowing processor can only instantiate the ADP by the narrowing substitutions, but it must not perform any rewrite step.
Instead, the rewrite steps have to be done via the rewriting processor afterwards.
Thus, instead of calling it \emph{narrowing processor}, we call it \emph{rule overlap instantiation processor} as it only instantiates the ADPs but does not perform any rewrite steps.

\subsubsection{Rule Overlap Instantiation Processor for $\mathtt{iAST}$}

\noindent
Unfortunately, 
considering only the narrowing substitutions $\delta_1, \ldots ,\delta_d$ of $t$ is not sufficient.
There could be another subterm $t' \trianglelefteq_{\sharp} r_j$ (with $t' \neq t$) 
which was involved in a CT (i.e., $t'^\sharp \sigma \to^*_{\nonprob(\PP)}
\tilde{\ell}\tilde{\sigma}$ for some substitutions $\sigma, \tilde{\sigma}$ and a left-hand side $\tilde{\ell}$ of an ADP, where in the innermost case we require $\ito^*_{\nonprob(\PP)}$ instead of $\to^*_{\nonprob(\PP)}$), but this CT is no longer possible when instantiating $t'$ to $t' \delta_1, \ldots, t' \delta_d$. 
We say that $t'$ is \emph{covered}\footnote{This was called ``captured'' in \cite{noschinski2013analyzing}.} by $\delta_1, \ldots, \delta_d$ if for each narrowing substitution $\rho$ of $t'$, there is a $\delta_e$ with $1 \leq e \leq d$ such that $\delta_e$ is more general than $\rho$, i.e., $\rho = \delta_e \rho'$ for some substitution $\rho'$. 
So the narrowing processor has to add another ADP $\ell \ruleArr{}{}{} \{ p_1:\anno_{\capt_1(\delta_1,\ldots,\delta_d)}(r_{1}), \ldots, p_k: \anno_{\capt_k(\delta_1,\ldots,\delta_d)}(r_k)\}^{m}$, where $\capt_i(\delta_1,\ldots,\delta_d)$ contains all positions of subterms $t' \trianglelefteq_{\sharp} r_j$ which are not covered by $\delta_1, \ldots, \delta_d$. 
(Therefore, in contrast to instantiation and forward instantiation, 
here we do not have to add another copy of the original rule without annotations,
because the original rule is represented by this additional ADP.)
Thus, for innermost rewriting, we obtain the following processor.

\begin{restatable}[Rule Overlap Instantiation Processor]{thm}{RuleOverlapInstProc}
  \label{theorem:rule-overlap-inst}
  Let $\PP$ be an ADP problem with $\PP = \PP' \uplus \{\ell \ruleArr{}{}{} \{ p_1:r_{1}, \ldots, p_k: r_k\}^{m}\}$, let $1 \leq j \leq k$, and let $t \trianglelefteq_{\sharp} r_j$.
  Let $\delta_1,\ldots,\delta_d$ be all narrowing substitutions of $t$, where $d \geq 0$.
  Then $\Proc_{\mathtt{roi}}^{\mathbf{i}}(\PP)\!=\!\{\PP' \cup N\}$ is sound and complete for $\mathtt{iAST}$, where
  \[
    \begin{array}{rcl}
      N & =    & \Bigl\{ \ell \delta_e \to \{p_1:r_1 \delta_e, \ldots,
      p_k:r_k \delta_e\}^{m} \Big| 1 \leq e \leq d \Bigr\}             \\
        & \cup & \Bigl\{
      \begin{array}{rcrl}
        \ell \ruleArr{}{}{} \{ p_1:\anno_{\capt_1(\delta_1,\ldots,\delta_d)}(r_{1}), \ldots, p_k: \anno_{\capt_k(\delta_1,\ldots,\delta_d)}(r_k)\}^{m}\!
      \end{array}\ \Bigr\} \!
    \end{array}
  \]
\end{restatable}

\begin{myproofsketch} Again, we 
  use the Starting Lemma (\Cref{lemma:starting})
  to restrict the form of the CTs that have to be regarded.
  
  \textit{Soundness}: 
  Every $\PP$-CT $\F{T}$ gives rise to a $(\PP' \cup N)$-CT  $\F{T}'$
  with $|\F{T}'| = |\F{T}|$.
  The core idea of this construction is that every rewrite step with $\ell \ruleArr{}{}{} \{ p_1:r_{1}, \ldots, p_k: r_k\}^{m}$ can also be done with a rule from $N$.
  If we use $\ell \ruleArr{}{}{} \{ p_1:\anno_{\capt_1(\delta_1,\ldots,\delta_d)}(r_{1}),
  \ldots, p_k: \anno_{\capt_k(\delta_1,\ldots,\delta_d)}(r_k)\}^{m} \in N$, we may create
  fewer annotations than with the old ADP $\ell \ruleArr{}{}{} \{ p_1:r_{1}, \ldots, p_k: r_k\}^{m}$.
  However,
in the $\PP$-CT $\F{T}$
  we never rewrite at the position of those annotations that do not get created in the CT
  $\F{T}'$, hence we can ignore them.

  \textit{Completeness}: 
  Similar to the completeness proof of \cref{theorem:inst-proc-ast}.
  We can replace each ADP $\ell \delta_e \ruleArr{}{}{} \{ p_1:r_{1}\delta_e, \ldots, p_k: r_k\delta_e\}^{m}$ 
  with the more general ADP $\ell \ruleArr{}{}{} \{ p_1:r_{1}, \ldots, p_k: r_k\}^{m}$, 
  and each ADP $\ell \ruleArr{}{}{} \{ p_1:\anno_{\capt_1(\delta_1,\ldots,\delta_d)}(r_{1}), 
  \ldots, p_k: \anno_{\capt_k(\delta_1,\ldots,\delta_d)}(r_k)\}^{m}$ 
  can be replaced by $\ell \ruleArr{}{}{} \{ p_1:r_{1}, \ldots, p_k: r_k\}^{m}$ as well, 
  which may lead to more annotations than before.
\end{myproofsketch}

\begin{exa}
  Consider the PTRS $\R_{\mathsf{roi}}$.

  \vspace*{-0.8cm}
  \begin{minipage}[t]{0.57\linewidth}
    \vspace*{0.3cm}
    \begin{align*}
      \tf(\td(x)) & \to \{ \nicefrac{3}{4}: \te(\tf(\tg(x)),\tf(\th(x))), \nicefrac{1}{4}: \ta \}\!
    \end{align*}
  \end{minipage}
  \hfill
  \begin{minipage}[t]{0.37\linewidth}
    \begin{align*}
      \tg(\ta) & \to \{1:\td(\ta)\} \\
      \th(\tb) & \to \{1:\td(\tb)\}\!
    \end{align*}
  \end{minipage}

  \vspace*{.2cm}

  \noindent
  The usable terms processor transforms the canonical ADP problem into the following ADP problem $\PP_{\mathsf{roi}}$.

  \vspace*{-0.8cm}
  \begin{minipage}[t]{0.57\linewidth}
    \vspace*{0.3cm}
    \begin{align*}
      \tf(\td(x)) & \to \{ \nicefrac{3}{4}: \te(\tF(\tg(x)),\tF(\th(x))), \nicefrac{1}{4}: \ta \}^{\ttrue}\!
    \end{align*}
  \end{minipage}
  \hfill
  \begin{minipage}[t]{0.37\linewidth}
    \begin{align*}
      \tg(\ta) & \to \{1:\td(\ta)\}^{\ttrue} \\
      \th(\tb) & \to \{1:\td(\tb)\}^{\ttrue}\!
    \end{align*}
  \end{minipage}

  \vspace*{.2cm}

  \noindent
  The ADP problem $\PP_{\mathsf{roi}}$ (and thus also the original PTRS $\R_{\mathsf{roi}}$) is $\mathtt{AST}$ and hence $\mathtt{iAST}$, because for every instantiation, at most one of the two ``recursive $\tF$-calls'' in the right-hand side of the $\tf$-ADP can be applied.
  The reason is that we can either use the $\tg$-rule if the variable $x$ is instantiated with $\ta$, or we can apply the $\th$-rule if the variable is instantiated with $\tb$, but not both.
  We apply $\Proc_{\mathtt{roi}}$ using the term $t^\sharp = \tF(\tg(x))$, 
  where the only narrowing substitution of $t = \tf(\tg(x))$ is $\delta = \{x/\ta\}$.
  For the other subterm $\tF(\th(x))$ with annotated root, 
  its flattened version $\tf(\th(x))$ is not covered by this substitution, 
  because
the substitution $\delta$ is not more general than
 the narrowing substitution $\delta' = \{x/\tb\}$ of $\tf(\th(x))$.
Hence, 
  we have to generate an additional ADP where this second subterm is annotated.
  Thus, we replace the former $\tf$-ADP by the following two new ADPs.
  \begin{align*}
    \tf(\td(\ta)) & \to \{ \nicefrac{3}{4}: \te(\tF(\tg(\ta)),\tF(\th(\ta))), \nicefrac{1}{4}: \ta \}^{\ttrue} \\
    \tf(\td(x))   & \to \{ \nicefrac{3}{4}: \te(\tf(\tg(x)),\tF(\th(x))), \nicefrac{1}{4}: \ta \}^{\ttrue}\!
  \end{align*}
  Now one can remove the annotation of $\tF(\th(\ta))$ from the first ADP by the usable terms processor and then apply the reduction pair processor with the polynomial interpretation that maps $\tF$ to 1 and all other symbols to 0 to remove all annotations, which proves $\mathtt{iAST}$.
  Again, proving $\mathtt{iAST}$ with such a simple polynomial interpretation would not be possible without the rule overlap instantiation processor.
\end{exa}

\subsubsection{Rule Overlap Instantiation Processor for $\mathtt{AST}$}

\noindent
For full rewriting, even a processor like the one of \Cref{theorem:rule-overlap-inst}
which only applies the narrowing substitutions but does not perform any rewrite steps is unsound, as shown by the following example.

\begin{exa}
  \label{example:ROI-unsound-AST}
  Reconsider the ADP problem from \Cref{example:usable-rules-ce-rules-full2}
  with the ADPs
  \[ \begin{array}{rcl}
      \ta                       & \to & \{ \nicefrac{5}{8}:\tc(\ta,\ta), \; \nicefrac{3}{8}:\tz\}^\true \\
      \tf(\tc(x_1,x_2)) & \to & \{ 1:\tc(\tF(x_1),\tF(x_2))\}^\true
    \end{array}\]
  As shown in \Cref{example:usable-rules-ce-rules-full2}, this ADP problem is not $\mathtt{AST}$.

  The only narrowing substitution for terms of the form $\tf(x_i)$ (i.e., for the two corresponding terms in the right-hand side of the $\tf$-ADP) is $\delta$ with $\delta(x_i) = \tc(x_1^i,x_2^i)$, 
  since $\delta$ is the mgu of $\tf(x_i)$ and the (variable-renamed) left-hand side $\tf(\tc(x_1^i,x_2^i))$.
  Applying the rule overlap instantiation processor for each $\tf(x_i) \trianglelefteq_{\sharp} \tc(\tF(x_1),\tF(x_2))$ results in the following ADP.
    \[ \begin{array}{l}
          \tf\left(\tc\left(\tc(x_1^1,x_2^1),\tc(x_1^2,x_2^2)\right)\right) \to \left\{
          1\!:\!\tc\left(\tF\left(\tc(x_1^1,x_2^1)\right),\tF\left(\tc(x_1^2,x_2^2)\right)\right) \right\}^\true
        \end{array}\]

  With the ADP problem of \Cref{example:usable-rules-ce-rules-full2}, 
  we could rewrite $\tF(\ta)$ to $\tF(\tc(\ta,\ta))$ with probability $\nicefrac{5}{8}$, 
  and afterwards create two copies of $\tF(\ta)$ using the original $\tf$-ADP, 
  leading to a random walk biased towards non-termination.

  With the new ADP problem, we can instead rewrite $\tF\big(\tc(\ta,\ta)\big)$ to two copies of itself.
  However, now this is not done with probability $\nicefrac{5}{8}$, but only with probability $(\nicefrac{5}{8})^2 = \nicefrac{25}{64}$.
  The reason is that we first have to rewrite each of the two different $\ta$ subterms 
  to $\tc(\ta,\ta)$, each with probability $\nicefrac{5}{8}$.
  Thus, we can only rewrite $\tF\big(\tc(\ta,\ta)\big)$ to the term
  \[\tF\big(\tc(\tc(\ta,\ta),\tc(\ta,\ta))\big)\]
  with probability $\nicefrac{25}{64}$.
  Now we can indeed create two copies of $\tF\big(\tc(\ta,\ta)\big)$ using the new $\tf$-ADP.
  However, this is now a random walk biased towards termination, since $\nicefrac{25}{64} < \nicefrac{1}{2}$.

  Indeed, the new ADP problem created by the rule overlap instantiation processor is $\mathtt{AST}$, 
  while the original ADP problem was not $\mathtt{AST}$, i.e., $\Proc_{\mathtt{roi}}$ is unsound for $\mathtt{AST}$.
\end{exa}

The see why $\Proc_{\mathtt{roi}}$ is unsound for $\mathtt{AST}$, note that after applying an ADP, the next rewrite step can be ``completely inside'' the used substitution if one does not impose an innermost rewriting strategy.
So for the ``non-$\mathtt{AST}$'' reduction in \Cref{example:ROI-unsound-AST}, after using
the (original) $\tf$-ADP to rewrite $\tF(\tc(\ta,\ta))$ to $\tc(\tF(\ta),\tF(\ta))$, the
next ``inner'' rewrite step rewrites the $\ta$-subterms which did not occur in the
right-hand side of the $\tf$-ADP but which were ``completely inside'' the substitution
used for $x_1$ and $x_2$.
However, if we only consider the right-hand side of the $\tf$-ADP, as it does not contain any $\ta$-symbol, we only note that after one application of the $\tf$-ADP we eventually have to use the $\tf$-ADP again, leading to the narrowing substitution $\delta$ in \Cref{example:ROI-unsound-AST}.
But when instantiating the $\tf$-ADP with this substitution, in order to simulate the previous reduction with the new $\tf$-ADP, we now need two instead of one ``inner'' rewrite step for the $\ta$-subterms.
In the probabilistic setting, this is a problem if the critical rewrite step (e.g., from $\ta$ to $\tc(\ta,\ta)$) is only chosen with a probability $< 1$, because the requirement of choosing the critical step two times has a lower probability than only choosing it once.

\section{Conclusion and Evaluation}
\label{Evaluation}

\noindent
In this work, we developed the ADP framework to prove $\mathtt{AST}$ and $\mathtt{iAST}$ of PTRSs completely automatically.
By using annotated DPs instead of dependency tuples as in \cite{kassinggiesl2023iAST}, we obtain \emph{complete} criteria for both $\mathtt{iAST}$ and $\mathtt{AST}$, which also simplifies the framework substantially.
As explained at the end of \Cref{Introduction}, this paper is based on our conference papers \cite{FLOPS2024,JPK60}
and it extends them by unifying the two separate frameworks for $\mathtt{AST}$ and $\mathtt{iAST}$ and by several new ADP processors, explanations, and remarks.

Compared to the non-probabilistic DP framework for termination of TRSs \cite{arts2000termination,gieslLPAR04dpframework,giesl2006mechanizing,hirokawa2005automating,DBLP:journals/iandc/HirokawaM07}, analyzing $\mathtt{(i)AST}$ automatically is significantly more difficult\linebreak
due to the lack of a ``minimality property'' in the probabilistic setting, which would allow several further processors.
Moreover, the ADP framework for PTRSs is restricted to reduction pairs based on barycentric, $\NN$-collapsible algebras (e.g., to multilinear polynomial interpretations).
The soundness proofs for our processors are much more involved than in the non-probabilistic setting, due to the more complex structure of chain trees.
However, the processors themselves are analogous to their non-probabilistic counterparts, and thus, existing implementations of the processors can easily be adapted to their new probabilistic versions.

Our novel ADP framework for full $\mathtt{AST}$ is particularly useful when analyzing overlapping PTRSs, as for such PTRSs we cannot use the criteria of \cite{FOSSACS24}
for classes of PTRSs where $\mathtt{iAST}$ implies $\mathtt{AST}$.
The frameworks for $\mathtt{AST}$ and $\mathtt{iAST}$ differ in power, since some processors are less powerful for $\mathtt{AST}$, and some are not even applicable for $\mathtt{AST}$.
\Cref{fig:diff-ADP} illustrates the differences between the ADP frameworks for $\mathtt{AST}$ and $\mathtt{iAST}$.
The parts highlighted in green
  show the differences to the non-probabilistic DP framework.
Here, ``($\lnot$)S'' and ``($\lnot$)C'' stand for ``(not) sound'' and ``(not) complete''.

\begin{table}
\smallskip
{\renewcommand\cellgape{}
\begin{center}
  \scriptsize
  \begin{tabular}{@{\;\;\;}lcc@{\;\;\;}}
    \toprule
    \textbf{Processor} & \textbf{ADPs for $\mathtt{AST}$} & \textbf{ADPs for $\mathtt{iAST}$} \\
    \midrule
    Chain Crit.\                                & \makecell{S \& C                                                     \\\textcolor{ForestGreen}{for non-duplicating}} & \makecell{S \& C \\\phantom{noop}} \\[1px]
    \hline
    Dep.\ Graph                                 & S \& C           & S \& C \\
    \hline
    \textcolor{ForestGreen}{Usable Terms}                         & S \& C           & S \& C \\
    \hline
    Usable Rules                                & \makecell{ $\lnot$ S                                                                           \\\textcolor{ForestGreen}{even with $C_\varepsilon$-Rules} \\ \textcolor{ForestGreen}{see \Cref{example:usable-rules-ce-rules-full2}}}& \makecell{\phantom{noop}\\S \& C\\\phantom{noop} } \\[1px]
    \hline
    Reduction Pairs                             & \makecell{S \& C                                                     \\\textcolor{ForestGreen}{barycentric, $\IN$-collapsible}\\\textcolor{ForestGreen}{algebras}}& \makecell{S \& C \\\textcolor{ForestGreen}{barycentric, $\IN$-collapsible}\\\textcolor{ForestGreen}{algebras}} \\
    \hline
    \textcolor{ForestGreen}{Probability Removal}                  & S \& C           & S \& C \\[1px]
    \hline
    Subterm Criterion                           & \makecell{\textcolor{ForestGreen}{$\lnot$ S}                                                                     \\\textcolor{ForestGreen}{see \Cref{ex:subterm-crit-unsound-AST}}} & \makecell{S \& C\\\textcolor{ForestGreen}{at most $1$ annotation}} \\[1px]
    \hline
    From Full to Innermost Rewriting & \makecell{\textcolor{ForestGreen}{$\lnot$ S}                                                                     \\\textcolor{ForestGreen}{see \Cref{example:from-f-to-i-unsound}}} & \dash \\[1px]
    \hline
    Rewriting Processor                         & \makecell{$\lnot$ S                                                                            \\ \phantom{noop}} & \makecell{S \& \textcolor{ForestGreen}{$\lnot$ C}\\ \textcolor{ForestGreen}{see \Cref{RewritingComplete}}} \\[1px]
    \hline
    Instantiation Processor                     & S \& C           & S \& C \\[1px]
    \hline
    Forward Inst.\ Processor                    & S \& C           & S \& C \\[1px]
    \hline
    Narrowing Processor                         & \makecell{\textcolor{ForestGreen}{$\lnot$ S}                                                                     \\\textcolor{ForestGreen}{see \Cref{counterexample narrowing}}} & \makecell{\textcolor{ForestGreen}{$\lnot$ S} \\\textcolor{ForestGreen}{see \Cref{counterexample narrowing}}} \\[1px]
    \hline
    \textcolor{ForestGreen}{Rule Overlap Inst.\ Processor}        & \makecell{\textcolor{ForestGreen}{$\lnot$ S}                                                                     \\\textcolor{ForestGreen}{see \Cref{example:ROI-unsound-AST}}} & \makecell{S \& C\\\phantom{blabla}} \\[1px]
    \bottomrule
  \end{tabular}
\end{center}}
  \caption{Differences between the ADP framework for $\mathtt{AST}$ and $\mathtt{iAST}$.}\label{fig:diff-ADP}
\end{table}

\smallskip

We implemented our new contributions in our termination prover
\textsf{AProVE} \cite{JAR-AProVE2017}
and compared the new version of \textsf{AProVE}
based on the probabilistic ADP framework
including all new processors and improvements
presented in this paper (\textsf{journal}) with
a previous version of \textsf{AProVE}
which only contains the processors
as presented
in our corresponding
conference papers \cite{FLOPS2024,JPK60} (\textsf{conf}). Moreover,
we also compare with a version of \textsf{AProVE} implementing our earlier framework from
\cite{kassinggiesl2023iAST} which used dependency tuples instead of ADPs
(\textsf{dt}).
As explained at the end of \Cref{Introduction},
compared to the conference papers of \textsf{conf},
in \textsf{journal}
we added the
subterm criterion processor, the instantiation, forward instantiation, and
rule overlap instantiation
processors,
and we
extended the
probabilistic reduction pair processors
to algebras, which allows the use of, e.g., matrix orders. Similarly, matrix
orders can now also be used in our corresponding criterion of
\cite{kassinggiesl2023iAST}
for a direct application of orders to prove $\mathtt{AST}$ (outside of the ADP
framework). This direct application of orders was only used with polynomial
interpretations in \textsf{dt} and \textsf{conf}, but in
\textsf{journal}, we also use it with matrix orders.
Furthermore, note that \textsf{dt} is restricted to the analysis of
innermost rewriting,
and cannot handle arbitrary rewrite strategies.

The main goal for probabilistic termination
analysis is to become as powerful as termination analysis in the non-probabilistic setting.
Therefore, in our first experiment, we
considered the non-probabilistic TRSs of the \emph{TPDB} \cite{TPDB} (the
benchmark set used in the annual \emph{Termination and Com\-plexity Competition (TermComp)}
\cite{TermComp}) and compared \textsf{journal}, \textsf{conf}, and \textsf{dt} with
\textsf{AProVE}'s tech\-niques for
ordinary non-probabilistic TRSs (\textsf{AProVE-NP}),
because at the current \emph{TermComp},
\textsf{AProVE-NP} was the
most powerful tool for termination of ordinary
non-probabilistic TRSs.
\textsf{AProVE-NP}
includes many addi\-tio\-nal DP processors and benefits from using separate
dependency pairs
for each occurrence of a defined symbol in the right-hand side of a rule
instead of ADPs.
Clearly, a TRS can be represented as a PTRS with trivial probabilities,
and then (innermost) $\mathtt{AST}$ is the same as (innermost) termination.
While all three variants \textsf{journal}, \textsf{conf}, and \textsf{dt}
have a probability removal processor to
switch to the classical DP framework for such problems, we disabled that
processor in this experiment.
As in \emph{TermComp},
we used a timeout of 300 seconds for each example.
The ``TRS Innermost'' and ``TRS Standard'' categories contain 366 and 1512 benchmarks and
the results regarding innermost rewriting are shown in \Cref{table:TRS-Innermost} and
\Cref{table:TRS-Standard}, respectively.
One can see that the transformations are very important for automatic termination proofs
as we get
around 5\% closer to \textsf{AProVE-NP}'s results
when using only the rewriting processor (in \textsf{conf}), and 12-16\% closer when also
using the remaining transformational processors that use instantiations (in \textsf{journal}).

\begin{table}[t]
  \caption{Evaluation Results}\label{table:Evaluation-Results}
  \begin{subtable}{1\linewidth}
    \centering
    \caption{TRS-Innermost}\label{table:TRS-Innermost}
    \begin{tabular}{@{\;\;\;}c@{\;\;}c@{\qquad}cccc@{\;\;\;}}
      \toprule
      \textbf{TRS-Innermost} & \#-Examples & \textsf{AProVE-NP} & \textsf{journal} & \textsf{conf} & \textsf{dt} \\
      \midrule
      & 366 & 293                & 170              & 148           & 134         \\
      \bottomrule
    \end{tabular}
  \end{subtable}
  
  \vspace*{0.2cm}
  \begin{subtable}{1\linewidth}
    \centering
    \caption{TRS-Standard}\label{table:TRS-Standard}
    \begin{tabular}{@{\;\;\;}c@{\;\;}c@{\qquad}cccc@{\;\;\;}}
      \toprule
      \textbf{TRS-Standard} & \#-Examples & \textsf{AProVE-NP} & \textsf{journal} & \textsf{conf} & \textsf{dt} \\
      \midrule
      (Innermost) & 1512 & 1114               & 791              & 667           & 612         \\
      (Full)      & 1512 & 1031               & 241              & 201           & \dash       \\
      \bottomrule
    \end{tabular}
  \end{subtable}

  \vspace*{0.2cm}
  \begin{subtable}{1\linewidth}
    \centering
    \caption{PTRS-Standard}\label{table:PTRS}
    \begin{tabular}{@{\;\;\;}c@{\;\;}c@{\qquad}cccc@{\;\;\;}}
      \toprule
      \textbf{PTRS-Standard} & \#-Examples & \textsf{AProVE-NP} & \textsf{journal} & \textsf{conf} & \textsf{dt} \\
      \midrule
      (Innermost) & 128 & \dash              & 107              & 88            & 77          \\
      (Full)      & 128 & \dash              & 58               & 56            & \dash       \\
      \bottomrule
    \end{tabular}
  \end{subtable}%
\end{table}

Furthermore, we also tried to prove termination w.r.t.\ an arbitrary evaluation strategy for the examples from the ``TRS Standard'' category.
Here, \textsf{AProVE-NP} can prove termination of 1031 examples, while \textsf{journal} and \textsf{conf} (using the ADP framework for full rewriting) can only show termination of 241 and 201 examples, respectively, see \Cref{table:TRS-Standard} again.
Recall that \textsf{dt} cannot handle arbitrary evaluation strategies.
Again, our new transformations clearly increase power, while
it becomes clear that proving $\mathtt{AST}$ is significantly harder than proving
$\mathtt{iAST}$.

In the next experiment, we used the PTRS benchmark set of the \emph{TPDB},
containing 128 typical probabilistic programs,
including examples with complicated probabilistic structure and probabilistic algorithms
on lists and trees.
For instance, it contains
the following PTRS $\R_{\tqs}$ for
probabilistic quicksort. Here,
lists are represented via the constructors
$\tnil$ (for the empty list) and $\tcons$, where, e.g.,
$\tcons(\ts(\tz), \tcons(\ts(\tz), \tcons(\tz, \tnil)))$ represents the list
$[1,1,0]$. To ease readability,
we write $r$
instead of $\{1:r\}$.

\vspace*{-.3cm}

{\small
  \[
    \begin{array}{rcl}
      \trotate(\tcons(x,\xs))    & \to & \{ \nicefrac{1}{2} : \tcons(x,\xs), \; \nicefrac{1}{2} :
      \trotate(\tapp(\xs, \tcons(x,\tnil)))\}                                                     \\
      \tqs(\xs)                  & \to & \tif(\tisempty(\xs),\;
      \tlow(\thd(\xs), \ttail(\xs)), \; \thd(\xs), \;
      \thigh(\thd(\xs), \ttail(\xs)))                                                             \\
      \tisempty(\tnil)           & \to & \ttrue                                                   \\
      \tisempty(\tcons(x,\xs))   & \to & \tfalse                                                  \\
      \thd(\tcons(x, \xs))       & \to & x                                                        \\
      \ttail(\tcons(x, \xs))     & \to & \xs                                                      \\
      \tif(\ttrue, \xs, x, \ys)  & \to & \tnil                                                    \\
      \tif(\tfalse, \xs, x, \ys) & \to & \tapp(\tqs(\trotate(\xs)),\; \tcons(x, \,
        \tqs(\trotate(\ys))))
    \end{array}
  \]}

The $\trotate$-rules rotate
a list randomly often (they are
$\mathtt{AST}$, but not termina\-ting). Thus, by choosing the first element
of the resulting list, one obtains random pivot elements for the recursive calls of
$\tqs$ in the second $\tif$-rule.
In addition to the rules above, $\R_{\tqs}$ contains rules for
list concatenation ($\tapp$), and rules such that
$\tlow(x,\xs)$ ($\thigh(x,\xs)$)
returns all elements of the list $\xs$ that are smaller (greater or equal) than
$x$, see App.\ \ref{Probabilistic Quicksort}.
Proving $\mathtt{iAST}$ of the
above rules requires transformational processors
to instantiate and rewrite the $\tisempty$-,
$\thd$-,
and $\ttail$-subterms in the
right-hand side of the $\tqs$-rule.
Thus, \textsf{dt} and \textsf{conf} fail for this example,
but \textsf{journal} can prove $\mathtt{iAST}$ of $\R_{\tqs}$.
(Our ADP framework for $\mathtt{AST}$ is not applicable here, since the PTRS is
duplicating.)

The \textsf{AProVE} variant
\textsf{dt} can prove $\mathtt{iAST}$ for 77 of the
128 PTRSs from the corresponding collection of the \emph{TPDB},
and \textsf{conf} can prove it for 88.
Adding the new and improved processors in \textsf{journal} increases this number to
107, which demonstrates their power for PTRSs with non-trivial probabilities.
When regarding full rewriting,
\textsf{conf} can prove $\mathtt{AST}$ for 56
examples and \textsf{journal} succeeds on 58 PTRSs, see \Cref{table:PTRS}.
For details on our experiments and for instructions on how to run our implementation
in \textsf{AProVE} via its \emph{web interface} or locally, see:
\[\mbox{\url{https://aprove-developers.github.io/ProbabilisticADPs/}}\]

On this website, we also performed experiments where we disabled individual transformational processors
of the ADP framework, which shows the usefulness of each processor.

In future work, we
will try to adapt the ADP framework in order to analyze stronger properties (like
$\mathtt{PAST}$) and we want to develop techniques (and processors for the ADP framework)
in order to \emph{disprove} $\mathtt{AST}$.

\bibliographystyle{plainnat}
\bibliography{biblioPaper.bib}

\appendix

\section{Examples}
\label{Examples}

\noindent
In this appendix, we present several examples to illustrate specific strengths and weaknesses of the ADP framework and the new transformational processors.
They show that in contrast to most other techniques for analyzing $\mathtt{AST}$, due to probabilistic term rewriting, our approach is also suitable for the analysis of algorithms on algebraic data structures other than numbers.\oldcomment{JG Maybe also of the RPP with matrix orders and of the Subterm Criterion Processor.}

\subsection{Creating and Using Lists}
\label{Examples-List}

\noindent
Similar to \Cref{alg1}, the following algorithm first creates a random list, filled with random numbers, and afterwards uses the list for further computations.
In general, our ADP framework is well suited to analyze algorithms that access or modify randomly generated lists.

The algorithm below computes the sum of all numbers in the generated list.
Here, the function $\tcreateL(\xs)$ adds a prefix of arbitrary length filled with natural numbers according to a geometrical distribution in front of the list $\xs$.
Moreover, $\tapp(\xs,\ys)$ concatenates the two lists $\xs$ and $\ys$.
Finally, for a non-empty list $\xs$ of numbers, $\tsum(\xs)$ computes a singleton list whose only element is the sum of all numbers in $\xs$.
So $\tsum(\tcons(\ts(\tz), \tcons(\ts(\tz), \tcons(\tz, \tnil))))$ evaluates to
$\tcons(\ts(\ts(\tz)), \tnil)$. \pagebreak[3]

\vspace*{-.2cm}

{\scriptsize
  \begin{align*}
    \tinit                                       & \to \{ 1 : \tsum(\tcreateL(\tnil))\} \\
    \taddNum(x, \xs)                             & \to \{ \nicefrac{1}{2} : \tcons(x, \xs) ,\nicefrac{1}{2} : \taddNum(\ts(x), \xs)\} \\
    \tcreateL(\xs)                               & \to \{ \nicefrac{1}{2} : \taddNum(\tz, \xs) ,\nicefrac{1}{2} : \tcreateL(\taddNum(\tz, \xs))\} \\
    \tplus(\tz, y)                               & \to \{ 1 : y\} \\
    \tplus(\ts(x), y)                            & \to \{ 1 : \ts(\tplus(x, y))\} \\
    \tsum(\tcons(x, \tnil))                      & \to \{ 1 : \tcons(x, \tnil)\} \\
    \tsum(\tcons(x, \tcons(y, \ys)))             & \to \{ 1 : \tsum(\tcons(\tplus(x, y), \ys))\} \\
    \tsum(\tapp(\xs, \tcons(x, \tcons(y, \ys)))) & \to \{ 1
    : \tsum(\tapp(\xs, \tsum(\tcons(x, \tcons(y, \ys)))))\} \\
    \tapp(\tcons(x, \xs), \ys)                   & \to \{ 1 : \tcons(x, \tapp(\xs, \ys))\} \\
    \tapp(\tnil, \ys)                            & \to \{ 1 : \ys\} \\
    \tapp(\xs, \tnil)                            & \to \{ 1 : \xs \}\!
  \end{align*}
}

Note that the left-hand sides of the two rules $\tapp(\tnil, \ys) \to \{ 1 : \ys \}$ and $\tapp(\xs, \tnil) \to \{ 1 : \xs \}$ overlap and moreover, the last $\tsum$-rule overlaps with the first $\tapp$-rule.
Hence, this PTRS does not belong to any known class where $\mathtt{iAST}$ implies full $\mathtt{AST}$ \cite{FOSSACS24}.
Furthermore, there exists no polynomial order that proves $\mathtt{AST}$ for this example directly (i.e., without the use of DPs), because the left-hand side of the last $\tsum$-rule is embedded in its right-hand side.
With our ADP framework for full $\mathtt{AST}$, \aprove{}
can prove $\mathtt{AST}$ (hence also $\mathtt{iAST}$) of this example automatically.

\subsection{Probabilistic Quicksort}
\label{Probabilistic Quicksort}

\noindent
In \Cref{Evaluation} we already presented the main rules of the PTRS $\R_{\tqs}$ implementing a probabilistic quicksort algorithm, where $\mathtt{iAST}$ can only be proved if we use the new transformational processors.
The following PTRS is the full implementation of this probabilistic quicksort algorithm.

\vspace*{-.3cm}

{\scriptsize
  \begin{align*}
    \trotate(\tnil)                   & \!\to\! \{ 1 : \tnil \} \\
    \trotate(\tcons(x,\xs))           & \!\to\! \{ \nicefrac{1}{2} : \tcons(x,\xs), \nicefrac{1}{2} : \trotate(\tapp(\xs, \tcons(x,\tnil)))\} \\
    \tisempty(\tnil)                  & \!\to\! \{ 1 : \ttrue\} \\
    \tisempty(\tcons(x, \xs))         & \!\to\! \{ 1 : \tfalse\} \\
    \tqs(\xs)                         & \!\to\! \{ 1 : \tif(\tisempty(\xs), \tlow(\thd(\xs), \ttail(\xs)), \thd(\xs), \thigh(\thd(\xs), \ttail(\xs)))\} \\
    \tif(\ttrue, \xs, x, \ys)         & \!\to\! \{ 1 : \tnil\} \\
    \tif(\tfalse, \xs, x, \ys)        & \!\to\! \{ 1 : \tapp(\tqs(\trotate(\xs)), \tcons(x, \tqs(\trotate(\ys))))\} \\
    \thd(\tcons(x, \xs))              & \!\to\! \{ 1 : x\} \\
    \ttail(\tcons(x, \xs))            & \!\to\! \{ 1 : \xs\} \\
    \tlow(x,\tnil)                    & \!\to\! \{ 1 : \tnil\} \\
    \tlow(x,\tcons(y,\ys))            & \!\to\! \{ 1 : \tiflow(\tleq(x,y),x,\tcons(y,\ys))\} \\
    \tiflow(\ttrue,x,\tcons(y,\ys))   & \!\to\! \{ 1 : \tlow(x,\ys)\} \\
    \tiflow(\tfalse,x,\tcons(y,\ys))  & \!\to\! \{ 1 : \tcons(y,\tlow(x,\ys))\} \\
    \thigh(x,\tnil)                   & \!\to\! \{ 1 : \tnil\} \\
    \thigh(x,\tcons(y,\ys))           & \!\to\! \{ 1 : \tifhigh(\tleq(x,y),x,\tcons(y,\ys))\} \\
    \tifhigh(\ttrue,x,\tcons(y,\ys))  & \!\to\! \{ 1 : \tcons(y,\thigh(x,\ys))\} \\
    \tifhigh(\tfalse,x,\tcons(y,\ys)) & \!\to\! \{ 1 : \thigh(x,\ys)\} \\
    \tleq(\tz,x)                      & \!\to\! \{ 1 : \ttrue\} \\
    \tleq(\ts(x),\tz)                 & \!\to\! \{ 1 : \tfalse\} \\
    \tleq(\ts(x),\ts(y))              & \!\to\! \{ 1 : \tleq(x,y)\} \\
    \tapp(\tnil,\ys)                  & \!\to\! \{ 1 : \ys\} \\
    \tapp(\tcons(x,\xs),\ys)          & \!\to\! \{ 1 : \tcons(x,\tapp(\xs,\ys))\}\!
  \end{align*}
}
As already briefly explained in \Cref{Evaluation}, this quicksort algorithm searches for a random pivot element using the first two $\trotate$ rules.
Here, we rotate the list and always move the head element to the end of the list.
With probability $\nicefrac{1}{2}$ we stop this iteration and use the current head element as the next pivot element.
The remaining rules implement the classical quicksort algorithm without any probabilities.
As mentioned in \Cref{Evaluation}, $\tapp$ computes list concatenation, $\tlow(x,\xs)$ returns all elements of the list $\xs$ that are smaller than $x$, and $\thigh$ works analogously.
Furthermore, $\thd$ returns the head element of a list and $\ttail$ returns the rest of the list without the head.
Finally, $\tisempty$ checks whether the list is empty.

Using our ADP framework, \aprove{} can automatically prove that this PTRS is $\mathtt{iAST}$.
(As mentioned in \Cref{Evaluation}, our ADP framework for $\mathtt{AST}$ is not applicable here, since the PTRS is duplicating.) This example illustrates the need for the new transformational processors: The rewriting processor (\Cref{theorem:ptrs-rewriting-proc}) is required to evaluate the functions $\thd$, $\ttail$, and $\tisempty$, and the rule overlap instantiation processor (\Cref{theorem:rule-overlap-inst}) is required to determine all possible terms that these functions can actually be applied on.
So for example, we need to detect that if we have a term of the form $\tisempty(x)$ for a variable $x$, then in order to apply any rewrite step, the variable $x$ must be instantiated with either $\tnil$ or a term of the form $\tcons(y,\ys)$ for some new variables $y$ and $\ys$.

\subsection{Moving Elements in Lists Probabilistically}

\noindent
Another interesting probabilistic algorithm that deals with lists is the following: We are given two lists $L_1$ and $L_2$.
If one of the two lists is empty, then the algorithm terminates.
Otherwise, we either move the head of list $L_1$ to $L_2$ or vice versa, both with probability $\nicefrac{1}{2}$, and then we repeat this procedure.
This algorithm is represented by the following PTRS.

  {\scriptsize
    \begin{align*}
      \tor(\tfalse,\tfalse)     & \!\to\! \{ 1 : \tfalse\} \\
      \tor(\ttrue,x)            & \!\to\! \{ 1 : \ttrue\} \\
      \tor(x,\ttrue)            & \!\to\! \{ 1 : \ttrue\} \\
      \tmoveelements(\xs,\ys)   & \!\to\! \{ 1 : \tif(\tor(\tisempty(\xs),\tisempty(\ys)), \xs, \ys)\} \\
      \tif(\ttrue, \xs, \ys)    & \!\to\! \{ 1 : \xs\} \\
      \tif(\tfalse, \xs, \ys)   & \!\to\!
      \{ \nicefrac{1}{2} : \tmoveelements(\ttail(\xs),\tcons(\thd(\xs),\ys)), \\
                                & \phantom{ \!\to\! \{\,} \nicefrac{1}{2} : \tmoveelements(\tcons(\thd(\ys),\xs), \ttail(\ys))\} \\
      \tisempty(\tnil)          & \!\to\! \{ 1 : \ttrue\} \\
      \tisempty(\tcons(x, \xs)) & \!\to\! \{ 1 : \tfalse\} \\
      \thd(\tcons(x, \xs))      & \!\to\! \{ 1 : x\} \\
      \ttail(\tcons(x, \xs))    & \!\to\! \{ 1 : \xs\} \\
    \end{align*}
  }

This algorithm is $\mathtt{iAST}$ (and even $\mathtt{AST}$) because we can view this as a classical random walk on the number of elements in the first list that is both bounded from below by $0$ and from above by the sum of the lengths of both lists.
In order to prove this automatically, we again have to use some kind of instantiation processor, e.g., the rule overlap instantiation processor, to find all possible terms that the functions $\thd$, $\ttail$, and $\tisempty$ can actually be applied on.
In addition, we also need the rewriting processor again to evaluate these functions afterwards.
Once the $\tmoveelements$-ADP contains the symbols $\tcons$ and $\tnil$ in the left-hand side, we can detect the structure of the random walk using an application of the reduction pair processor that removes all annotations of this ADP.
After that, there is no SCC in the dependency graph left, and we have proven $\mathtt{iAST}$.
Again, our ADP framework for $\mathtt{AST}$ is not applicable, since the PTRS is duplicating.

\subsection{Trees}
\label{Examples-Tree}

\noindent
This example illustrates the use of probabilistic term rewriting for algorithms on trees.
In the following algorithm (adapted from \cite{AG01}), we consider binary trees represented via $\tleaf$ and $\ttree(x,y)$, where $\tconcat(x,y)$ replaces the rightmost leaf of the tree $x$ by $y$.
The algorithm first creates two random trees and then checks whether the first tree has less leaves than the second one.

\vspace*{-.2cm}

{\scriptsize
  \begin{align*}
    \tinit                                   & \to \{ 1 : \tlessleaves(\tcreateT(\tleaf), \tcreateT(\tleaf)) \} \\
    \tconcat(\tleaf, y)                      & \to \{ 1 : y \} \\
    \tconcat(\ttree(u, v), y)                & \to \{ 1 : \ttree(u, \tconcat(v, y)) \} \\
    \tlessleaves(x, \tleaf)                  & \to \{ 1 : \tfalse \} \\
    \tlessleaves(\tleaf, \ttree(x, y))       & \to \{ 1 : \ttrue \} \\
    \tlessleaves(\ttree(u, v), \ttree(x, y)) & \to \{ 1 : \tlessleaves(\tconcat(u, v), \tconcat(x, y)) \} \\
    \tcreateT(\xs)                           & \to \{ 1 : \xs \} \\
    \tcreateT(\xs)                           & \to \{ \nicefrac{1}{3} : \xs ,\nicefrac{1}{3} : \tcreateT(\ttree(\xs, \tleaf)) ,\nicefrac{1}{3} : \tcreateT(\ttree(\tleaf, \xs)) \} \!
  \end{align*}
}

Note that the last two rules are overlapping and thus, there is no known criterion \cite{FOSSACS24} which would allow us to conclude $\mathtt{AST}$ from a proof of $\mathtt{iAST}$.
In contrast, our ADP framework is able to prove full $\mathtt{AST}$ (hence $\mathtt{iAST}$) for this example, while the direct application of polynomial interpretations fails.

\subsection{Conditions on Numbers}

\noindent
Another important task of termination analysis is to handle conditions on numbers.
Such conditions occur in almost every program and often have an impact on the termination behavior.
This is also true for probabilistic programs.
For a successful proof of $\mathtt{iAST}$ without transformations, the rules of the PTRS have to express these conditions via suitable ``patterns'' in their left-hand sides, i.e., if one wants to check whether a number is zero or not, then one needs two different rules $\tf(\tz) \to \ldots$ and $\tf(\ts(x)) \to \ldots$ to perform this case analysis.
However, in programs, conditions are usually implemented via an $\tif$-construct, where one could use, e.g., an additional function $\tgt$ to check whether a number is greater than another.
The same is true for conditions on other data structures than numbers, as we have seen in the previous examples.
If one wants to check whether a list is empty, then without transformations, one needs two rules (for $\tnil$ and $\tcons$), whereas with transformations, one can use conditions and auxiliary functions like $\tisempty$.

The following PTRS implements the classical random walk, but we check the condition $x > 0$ not directly in the left-hand side of the rule but with an additional function $\tgt(x,y)$ which determines whether $x > y$ holds.
In addition, to decrease the value of a number by one, we use the predecessor function $\tp(x) = x-1$.

  {\scriptsize
    \begin{align*}
      \tgt(\tz,\tz)     & \!\to\! \{ 1 : \tfalse \} \\
      \tgt(\ts(x),\tz)  & \!\to\! \{ 1 : \ttrue \} \\
      \tgt(\tz,\ts(y))  & \!\to\! \{ 1 : \tfalse \} \\
      \tgt(\ts(x),s(y)) & \!\to\! \{ 1 : \tgt(x,y) \} \\
      \tp(\tz)          & \!\to\! \{ 1 : \tz \} \\
      \tp(\ts(x))       & \!\to\! \{ 1 : x \} \\
      \tloop(x)         & \!\to\! \{ 1 : \tif(\tgt(x,\tz), x) \} \\
      \tif(\tfalse, x)  & \!\to\! \{ 1 : \tstop \} \\
      \tif(\ttrue, x)   & \!\to\! \{ \nicefrac{1}{2} : \tloop(\tp(x)), \nicefrac{1}{2} : \tloop(\ts(x)) \}\!
    \end{align*}
  }

In this case, we need the rewriting processor to evaluate the functions $\tgt$ and $\tp$ and again we need the rule overlap instantiation processor to check for all possible terms that these functions can actually be applied on.
Thus, the ADP framework proves $\mathtt{iAST}$ for this example.
Since the PTRS is duplicating, the ADP framework for $\mathtt{AST}$ cannot be applied.

\subsection{Limits of the Instantiation Processors}

\noindent
Whenever we have an ADP where the left-hand side of the rule is also contained in the support of the right-hand side, then instantiations become useless, because we will always have at least the same ADP again after applying the processor.
For example, we need to apply one of the instantiation processors in order to prove innermost termination for the TRS with the rules $\tc \to \ta$, $\tc \to \tb$, $\tf(x,y,z) \to \tg(x,y,z)$, and $\tg(\ta,\tb,z) \to \tf(z,z,z)$.
If we make the $\tf$- and $\tg$-rules probabilistic by adding the possibility to do nothing in a rewrite step with the probability $\nicefrac{1}{2}$, then we result in the following PTRS.

\vspace*{-.3cm}

{\small
  \begin{align*}
    \tc            & \to \{1 : \ta\} \qquad \tc \to \{1 : \tb\} \\
    \tf(x,y,z)     & \!\to\! \{\nicefrac{1}{2} : \tg(x,y,z), \nicefrac{1}{2} : \tf(x,y,z)\} \\
    \tg(\ta,\tb,z) & \!\to\! \{\nicefrac{1}{2} : \tf(z,z,z), \nicefrac{1}{2} : \tg(\ta,\tb,z)\}\!
  \end{align*}
}

\noindent
This PTRS is $\mathtt{iAST}$ (but not $\mathtt{AST}$, since the original TRS was also just innermost terminating but not terminating), but we are unable to show this using the instantiation processor, because if one tries to instantiate any of the rules, this will result in at least the same rule after the processor.
In contrast, in \cref{example:inst} we had almost the same PTRS but the $\tf$-rule remained non-probabilistic.
There, we were able to apply the instantiation processor and prove $\mathtt{iAST}$ using the ADP framework.
(However, while instantiation does not help in our example above, we can prove $\mathtt{iAST}$ using the rule overlap instantiation processor.)

\subsection{Transformations do not Suffice for Inductive Reasoning}

\noindent
Transformational processors are useful to perform a case analysis, but they do not suffice for PTRSs where one needs inductive reasoning for the termination analysis.
For example, we cannot show $\mathtt{AST}$ (not even $\mathtt{iAST}$) of the following PTRS \cite{Niederhauser23}
even though the $\tevenif$-structure seems similar to the one of the probabilistic quicksort example.

\vspace*{-.5cm}

{\small
  \begin{align*}
    \teven(\tz)          & \!\to\! \{ 1 : \ttrue \}    & \teven(\ts(\tz))    & \!\to\! \{ 1 : \tfalse \} \\
    \teven(\ts(\ts(x)))  & \!\to\! \{ 1 : \teven(x) \} & \tloop(x)           & \!\to\! \{ 1 : \tevenif(\teven(x), x) \} \\
    \tevenif(\tfalse, x) & \!\to\! \{ 1 : \tstop \}    & \tevenif(\ttrue, x) & \!\to\! \{ \nicefrac{1}{2} : \tloop(x), \nicefrac{1}{2} : \tloop(\ts(x)) \}\!
  \end{align*}
}

\noindent
The idea here is that the recursion of $\tloop$ stops if its argument is an even number.
If it is not even, then we either increase the value by 1 or use the same value again.
Here, a simple case analysis does not suffice, but we actually have to show (inductively), that if a number $x$ is odd, then $x+1$ is even.
This is not possible with the new transformational processors but needs other types of processors for inductive reasoning (e.g., as in \cite{DPInductionJAR2011} for the non-probabilistic DP framework).

\section{Proofs}
\label{Proofs}

\noindent
In this section, we prove the chain criterion and the soundness and completeness of our processors of the ADP framework.

To this end, we first give some additional definitions and notations, 
and then prove the lemmas about ADP problems mentioned in \Cref{The Probabilistic ADP Framework}
that we will need throughout the rest of this appendix.
We start by defining the notion of a \emph{subtree}.

\begin{defi}[Subtree]
  \label{def:chain-tree-induced-sub}
  Let $\PP$ be an ADP problem and let $\F{T} = (V,E,L)$ be a tree that satisfies
  Conditions
  (1)-(3) of a $\PP$-CT.
  Let $W \subseteq V$ be non-empty, weakly connected, and for all $v \in W$ we have $vE \cap W = \emptyset$ or $vE \cap W = vE$.
  Then, we define the \emph{subtree} $\F{T}[W]$ by $\F{T}[W] = (W,E \cap (W \times W),L^W)$.
  Let $w \in W$ be the root of $G^{\F{T}[W]}$.
  To ensure that the root of our subtree has the probability $1$ again, we use the labeling $L^W(v) = (\frac{p_v^{\F{T}}}{p_w^{\F{T}}}: t_v^{\F{T}})$ for all nodes $v \in W$.
  If $W$ contains the root of $(V, E)$, then we call the subtree \emph{grounded}.
\end{defi}

The property of being non-empty and weakly connected ensures that the resulting graph $G^{\F{T}[W]}$ is a tree again.
The property that we have $vE \cap W = \emptyset$ or $vE \cap W = vE$ ensures that the sum of probabilities for the successors of a node $v$ is equal to the probability for the node $v$ itself.

\begin{exa}
  Reconsider the PTRS $\R_1$ with the rule $\tg \to \{\nicefrac{3}{4}:\td(\tg), \, \nicefrac{1}{4}:\tz\}$.
  Below one can see the $\R_1$-RST from \Cref{Probabilistic Rewriting} (on the left), and the subtree that starts at the node of the term $\td(\tg)$ (on the right).
  Note that the probabilities are normalized such that the root has the probability $1$ again.
  \begin{center}
    \scriptsize
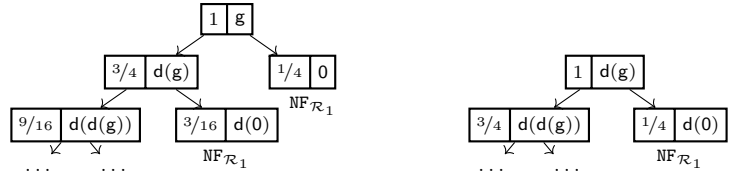
    \begin{tikzpicture}
      \tikzstyle{adam}=[thick,draw=black!100,fill=white!100,minimum size=4mm, shape=rectangle split, rectangle split parts=2,rectangle split horizontal] \tikzstyle{empty}=[rectangle,thick,minimum size=4mm]

      \node[adam] at (-3, 0) (a) {$1$
        \nodepart{two}$\tg$};
      \node[adam] at (-4, -0.7) (b) {$\nicefrac{3}{4}$
        \nodepart{two}$\td(\tg)$};
      \node[adam,label=below:{$\quad \mathtt{NF}_{\R_{1}}$}] at (-2, -0.7) (c) {$\nicefrac{1}{4}$
          \nodepart{two}$\tz$};
      \node[adam] at (-5, -1.4) (d) {$\nicefrac{9}{16}$
        \nodepart{two}$\td(\td(\tg))$};
      \node[adam,label=below:{$\mathtt{NF}_{\R_{1}}$}] at (-3, -1.4) (e) {$\nicefrac{3}{16}$
          \nodepart{two}$\td(\tz)$};
      \node[empty] at (-5.5, -2) (f) {$\ldots$};
      \node[empty] at (-4.5, -2) (g) {$\ldots$};

      \draw (a) edge[->] (b);
      \draw (a) edge[->] (c);
      \draw (b) edge[->] (d);
      \draw (b) edge[->] (e);
      \draw (d) edge[->] (f);
      \draw (d) edge[->] (g);

      \node[adam] at (2, -0.7) (b2) {$1$
        \nodepart{two}$\td(\tg)$};
      \node[adam] at (1, -1.4) (d2) {$\nicefrac{3}{4}$
        \nodepart{two}$\td(\td(\tg))$};
      \node[adam,label=below:{$\mathtt{NF}_{\R_{1}}$}] at (3, -1.4) (e2) {$\nicefrac{1}{4}$
          \nodepart{two}$\td(\tz)$};
      \node[empty] at (0.5, -2) (f2) {$\ldots$};
      \node[empty] at (1.5, -2) (g2) {$\ldots$};

      \draw (b2) edge[->] (d2);
      \draw (b2) edge[->] (e2);
      \draw (d2) edge[->] (f2);
      \draw (d2) edge[->] (g2);
    \end{tikzpicture}
  \end{center}
  \vspace*{-10px}
\end{exa}

We divide the inner nodes of a $\PP$-CT into $V \setminus \ctleaf = \aV \uplus \nV$ such that the rewrite step at each node from $\aV$ is done with Case $(\mathbf{at})$ or $(\mathbf{af})$, and the rewrite step at each node from $\nV$ is done with Case $(\mathbf{nt})$ or $(\mathbf{nf})$.
Note that $\aV$ contains the ``important'' nodes that count for the global Case
(4) of a $\PP$-CT, i.e., every infinite path must contain infinitely many nodes from $\aV$.

Now we start by proving the \emph{$\aV$-Partition Lemma}.
Compared to the $\P$-Partition Lemma (\Cref{lemma:p-partition-main} from \Cref{The Probabilistic ADP Framework}), 
the lemma presented (and proven) here is a bit more detailed and technical.
Instead of simply partitioning the set of ADPs, we partition the set of nodes $\Gamma$ within a chain tree.
In this way, we obtain an even more versatile lemma that we can use throughout our proofs.
Furthermore, note that this lemma does not rely on the structure of ADPs or rewriting,
but instead, it works for arbitrary infinite trees labeled by probabilities.

\begin{lem}[$\aV$-Partition Lemma]
  \label{lemma:p-partition}
  Let $\PP$ be an ADP problem and let $\F{T} = (V,E,L)$ be a $\PP$-CT that converges with probability $<1$.
  Assume that we can partition $\aV = \aV_1 \uplus \aV_2$ such that every subtree that only contains inner nodes from $\nV \cup \aV_1$ converges with probability $1$.
  Then there is a grounded subtree $\F{T}'$ that converges with probability $<1$ such that
  every infinite path has an infinite number of nodes from $\aV_2$.
\end{lem}

\begin{myproof}
  Let $\F{T} = (V,E,L)$ be a $\PP$-CT with $|\F{T}| = c <1$ for some $c \in \IR$.
  Since we have $0 \leq c <1$, there is an $\varepsilon > 0$ such that $c + \varepsilon <1$.
  Remember that the formula for the geometric series is
  \[
    \sum_{n = 1}^{\infty} \left(\frac{1}{d}\right)^n = \frac{1}{d-1}, \text{ for all } d \in \IR \text{ such that } \frac{1}{|d|} <1.
  \]
  Let $d = \frac{1}{\varepsilon} + 2$.
  Now, we have $\frac{1}{d} = \frac{1}{\frac{1}{\varepsilon} + 2} <1$ and
  \begin{equation}\label{eq:setting-d}
    \frac{1}{\varepsilon} + 1 < \frac{1}{\varepsilon} + 2 \Leftrightarrow \frac{1}{\varepsilon} + 1 < d \Leftrightarrow \frac{1}{\varepsilon} < d-1 \Leftrightarrow \frac{1}{d-1} < \varepsilon \Leftrightarrow \sum_{n = 1}^{\infty} \left(\frac{1}{d}\right)^n < \varepsilon.
  \end{equation}
  We will now construct a subtree $\F{T}' = (V',E',L')$ such that every infinite path has an infinite number of $\aV_2$ nodes and such that
  \begin{equation}\label{eq:sum-after-all-cuts}
    |\F{T}'| \leq |\F{T}| + \sum_{n = 1}^{\infty} \left(\frac{1}{d}\right)^n
  \end{equation}
  and then, we finally have
  \[
    |\F{T}'| \stackrel{\eqref{eq:sum-after-all-cuts}}{\leq} |\F{T}| + \sum_{n = 1}^{\infty} \left(\frac{1}{d}\right)^n = c + \sum_{n = 1}^{\infty} \left(\frac{1}{d}\right)^n \stackrel{\eqref{eq:setting-d}}{<} c + \varepsilon <1.
  \]

  The idea of this construction is that we cut infinite subtrees of pure $\nV \cup \aV_1$ nodes as soon as the probability for normal forms is high enough.
  In this way, one obtains paths where after finitely many $\nV \cup \aV_1$ nodes, there is either a node from $\aV_2$ or a leaf.

  For any node $v \in V$, let $\ctlevelTwo(v)$ be the number of $\aV_2$ nodes in the path from the root to $v$.
  Furthermore, for any set $W \subseteq V$ and $k \in \IN$, let $\ctlevelTwowithborder(W,k) = \{v \in W \mid \ctlevelTwo(v) \leq k \lor (v \in \aV_2 \land \ctlevelTwo(v) \leq k+1)\}$ be the set of all nodes in $W$ that have at most $k$ nodes from $\aV_2$ in the path from the root to its predecessor.
  So if $v \in \ctlevelTwowithborder(W,k)$ is not in $\aV_2$, then we have at most $k$ nodes from $\aV_2$ in the path from the root to $v$ and if $v \in \ctlevelTwowithborder(W,k)$ is in $\aV_2$, then we have at most $k+1$ nodes from $\aV_2$ in the path from the root to $v$.
  We will inductively define a set $U_k \subseteq V$ such that $U_k \subseteq \ctlevelTwowithborder(V,k)$ and then define the subtree as $\F{T}' = \F{T}[\bigcup_{k \in \IN} U_k]$.

  We start by considering the subtree $\F{T}_0 = \F{T}[\ctlevelTwowithborder(V,0)]$.
  This tree only contains inner nodes from $\nV \cup \aV_1$.
  While the node set $\ctlevelTwowithborder(V,0)$ itself may contain nodes from $\aV_2$, they can only occur at the leaves of $\F{T}_0$.
  Therefore, we get $|\F{T}_0|=1$ by the prerequisite of the lemma.
  In \cref{Possibilities for Te} one can see the different possibilities for $\F{T}_0$.
  Either $\F{T}_0$ is finite or $\F{T}_0$ is infinite.
  In the first case, we can add all the nodes to $U_0$ since there is no infinite path of pure $\nV \cup \aV_1$ nodes.
  Hence, we define $U_0 = \ctlevelTwowithborder(V,0)$.
  In the second case, we have to cut the tree at a specific depth once the probability of leaves is high enough.
  Let $\ctdepth_{0}(w)$ be the depth of the node $w$ in the tree $\F{T}_0$.
  Moreover, let $D_{0}(k) = \{v \in \ctlevelTwowithborder(V,0) \mid \ctdepth_{0}(w) \leq k\}$ be the set of nodes in $\F{T}_0$ that have a depth of at most $k$.
  Since $|\F{T}_0|=1$ and $|\cdot|$ is monotonic w.r.t.\ the depth of the tree $\F{T}_0$, we can find an $M_{0} \in \IN$ such that
  \[
    \sum_{v \in \ctleaf^{\F{T}_0}, d_{0}(v) \leq M_{0}} p_v^{\F{T}_0} \geq 1 - \frac{1}{d}
  \]
  Here, $\ctleaf^{\F{T}}$ and $p_v^{\F{T}}$ denote the set of leaves and the probability of the node $v$ in the tree $\F{T}$, respectively.

  We include all nodes from $D_{0}(M_{0})$ in $U_0$ and delete every other node of $\F{T}_0$.
  In other words, we cut the tree after depth $M_{0}$.
  This cut can be seen in \cref{Possibilities for Te}, indicated by the dotted line.
  We now know that this cut may increase the probability of leaves by at most $\frac{1}{d}$.
  Therefore, we define $U_0 = D_{0}(M_{0})$.

  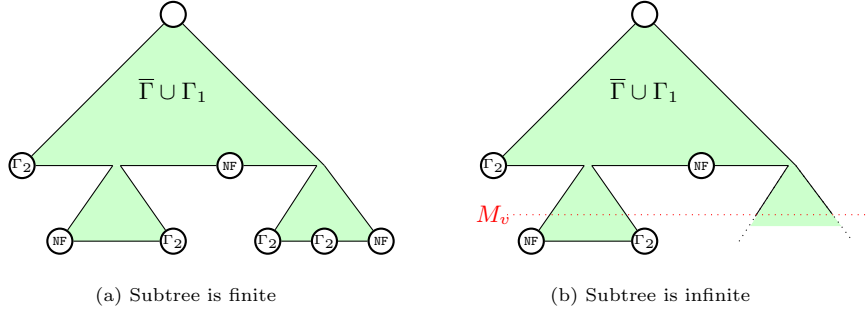
\begin{figure}
    \centering
    \begin{subfigure}[b]{0.4\textwidth}
      \centering
      \begin{tikzpicture}
        \tikzstyle{adam}=[circle,thick,draw=black!100,fill=white!100,minimum size=3mm] \tikzstyle{empty}=[circle,thick,minimum size=3mm]

        \node[adam] at (0, -1) (a) {};
        \node[empty] at (2, -3) (b) {};
        \node[adam, label=center:{\tiny $\aV_2$}] at (-2, -3) (c) {};
        \node[adam, label=center:{\tiny $\NF$}] at (0.75, -3) (d) {};
        \node[empty] at (-0.75, -3) (e) {};
        \node[empty, label=center:{\small $\nV \cup \aV_1$}] at (0, -2) (middleA) {};

        \node[adam, label=center:{\tiny $\NF$}] at (-1.5, -4) (nf1) {};
        \node[adam, label=center:{\tiny $\aV_2$}] at (0, -4) (nf2) {};

        \node[adam, label=center:{\tiny $\NF$}] at (2.75, -4) (bb) {};
        \node[adam, label=center:{\tiny $\aV_2$}] at (1.25, -4) (bc) {};
        \node[adam, label=center:{\tiny $\aV_2$}] at (2, -4) (bd) {};
        \node[empty, label=center:{\small $ $}] at (2, -3.5) (middleA) {};

        \node[empty, label=center:{\small $ $}] at (-0.75, -3.5) (middleA) {};

        \node[empty] at (0, -6) (stretch) {};

        \draw (a) edge[-] (2, -3);
        \draw (1.9, -3) edge[-] (d);
        \draw (d) edge[-] (-0.7, -3);
        \draw (-0.8, -3) edge[-] (c);
        \draw (a) edge[-] (c);

        \draw (2, -3) edge[-] (bb);
        \draw (bb) edge[-] (bd);
        \draw (bd) edge[-] (bc);
        \draw (1.9, -3) edge[-] (bc);

        \draw (-0.8, -3) -- (nf1) -- (nf2) -- (-0.7, -3);

        \begin{scope}[on background layer]
          \fill[green!20!white,on background layer] (0, -1) -- (-2, -3) -- (2, -3); \fill[green!20!white,on background layer] (2, -3) -- (1.25, -4) -- (2.75, -4); \fill[green!20!white,on background layer] (1.9, -3) -- (1.95, -4) -- (2, -3); \fill[green!20!white,on background layer] (1.9, -3) -- (1.25, -4) -- (2.75, -4); \fill[green!20!white,on background layer] (-0.8, -3) -- (-1.5, -4) -- (0, -4); \fill[green!20!white,on background layer] (-0.7, -3) -- (-0.8, -3) -- (-0.75, -4); \fill[green!20!white,on background layer] (-0.7, -3) -- (-1.5, -4) -- (0, -4);
        \end{scope}
      \end{tikzpicture}
      \vspace{-2.2cm}
      \caption{Subtree is finite}
    \end{subfigure}
    \hspace{30px}
    \begin{subfigure}[b]{0.4\textwidth}
      \centering
      \begin{tikzpicture}
        \tikzstyle{adam}=[circle,thick,draw=black!100,fill=white!100,minimum size=3mm] \tikzstyle{empty}=[circle,thick,minimum size=3mm]

        \node[adam] at (0, -1) (a) {};
        \node[empty] at (2, -3) (b) {};
        \node[adam, label=center:{\tiny $\aV_2$}] at (-2, -3) (c) {};
        \node[adam, label=center:{\tiny $\NF$}] at (0.75, -3) (d) {};
        \node[empty] at (-0.75, -3) (e) {};
        \node[empty, label=center:{\small $\nV \cup \aV_1$}] at (0, -2) (middleA) {};

        \node[adam, label=center:{\tiny $\NF$}] at (-1.5, -4) (nf1) {};
        \node[adam, label=center:{\tiny $\aV_2$}] at (0, -4) (nf2) {};

        \node[empty] at (2, -3) (l1) {};
        \node[empty] at (1.9, -3) (l2) {};

        \node[empty] at (2.75, -4) (bb) {};
        \node[empty] at (1.25, -4) (bc) {};
        \node[empty] at (2, -3.5) (middleA) {};

        \node[empty] at (-0.75, -3.5) (middleA) {};

        \node[empty, label=center:{\small \textcolor{red}{$M_v$}}] at (-2, -3.65) (cut) {};

        \node[empty] at (0, -6) (stretch) {};

        \draw (a) edge[-] (2, -3);
        \draw (1.9, -3) edge[-] (d);
        \draw (d) edge[-] (-0.7, -3);
        \draw (-0.8, -3) edge[-] (c);
        \draw (a) edge[-] (c);

        \draw (-0.8, -3) -- (nf1) -- (nf2) -- (-0.7, -3);

        \draw[] (2, -3) edge ($(l1)!0.8cm!(bb)$) edge [dotted] ($(l1)!1.2cm!(bb)$);
        \draw[] (1.9, -3) edge ($(l2)!0.8cm!(bc)$) edge [dotted] ($(l2)!1.2cm!(bc)$);

        \draw[] (3, -3.65) edge [red, dotted] (cut);

        \begin{scope}[on background layer]
          \fill[green!20!white,on background layer] (0, -1) -- (-2, -3) -- (2, -3); \fill[green!20!white,on background layer] (2, -3) -- (1.4, -3.8) -- (2.6, -3.8); \fill[green!20!white,on background layer] (1.9, -3) -- (1.95, -3.8) -- (2, -3); \fill[green!20!white,on background layer] (1.9, -3) -- (1.4, -3.8) -- (2.6, -3.8); \fill[green!20!white,on background layer] (-0.8, -3) -- (-1.5, -4) -- (0, -4); \fill[green!20!white,on background layer] (-0.7, -3) -- (-0.8, -3) -- (-0.75, -4); \fill[green!20!white,on background layer] (-0.7, -3) -- (-1.5, -4) -- (0, -4);
        \end{scope}
      \end{tikzpicture}
      \vspace{-2.2cm}
      \caption{Subtree is infinite}
    \end{subfigure}
    \caption{Possibilities for $\F{T}_0$ and $\F{T}_v$}\label{Possibilities for Te}
  \end{figure}

  For the induction step, assume that we have already defined a subset $U_i \subseteq \ctlevelTwowithborder(V,i)$.
  Let $H_i = \{v \in U_i \mid v \in \aV_2, \ctlevelTwo(v) = i+1\}$ be the set of leaves in $\F{T}[U_i]$ that are in $\aV_2$.
  For each $v \in H_i$, we consider the subtree that starts at $v$ until we reach the next node from $\aV_2$, including the node itself.
  Everything below such a node will be cut.
  To be precise, we regard the tree $\F{T}_v = (V_v,E_v,L_v) = \F{T}[\ctlevelTwowithborder(vE^*,i+1)]$.
  Here, $vE^*$ is the set of all nodes that are reachable from $v$.

  First, we show that $|\F{T}_v| = 1$.
  For every direct successor $w$ of $v$, the subtree $\F{T}_w = \F{T}_v[w E_v^*]$ of $\F{T}_v$ that starts at $w$ does not contain any inner nodes from $\aV_2$.
  Hence, we have $|\F{T}_w| = 1$ by the prerequisite of the lemma again, and hence
  \[
    |\F{T}_v| = \sum_{w \in vE} p_w \cdot |\F{T}_w| = \sum_{w \in vE} p_w \cdot 1 = \sum_{w \in vE} p_w = 1.
  \]
  For the construction of $U_{i+1}$, we have the same cases as before, see \Cref{Possibilities for Te}.
  Either $\F{T}_v$ is finite or $\F{T}_v$ is infinite.
  Let $Z_v$ be the set of nodes that we want to add to our node set $U_{i+1}$ from the tree $\F{T}_v$.
  In the first case we can add all the nodes again and set $Z_v = V_v$.
  In the second case, we once again cut the tree at a specific depth once the probability for leaves is high enough.
  Let $\ctdepth_v(z)$ be the depth of the node $z$ in the tree $\F{T}_v$.
  Moreover, let $D_v(k) = \{v \in V_v \mid \ctdepth_v(z) \leq k\}$ be the set of nodes in $\F{T}_v$ that have a depth of at most $k$.
  Since $|\F{T}_v|=1$ and $|\cdot|$ is monotonic w.r.t.\ the depth of the tree $\F{T}_v$, we can find an $M_v \in \IN$ such that
  \[
    \sum_{w \in \ctleaf^{\F{T}_v}, d_v(w) \leq M_v} p_w^{\F{T}_v} \geq 1 - \left(\frac{1}{d}\right)^{i+1} \cdot \frac{1}{|H_i|}
  \]
  We include all nodes from $D_v(M_v)$ in $U_{i+1}$ and delete every other node of $\F{T}_v$.
  In other words, we cut the tree after depth $M_v$.
  We now know that this cut may increase the probability of leaves by at most $\left(\frac{1}{d}\right)^{i+1} \cdot \frac{1}{|H_i|}$.
  Therefore, we set $Z_v = D_v(M_v)$.

  We do this for each $v \in H_i$ and in the end, we set $U_{i+1} = U_i \cup \bigcup_{v \in H_i} Z_v$.

  It is straightforward to see that $\bigcup_{k \in \IN} U_k$ satisfies the conditions of \Cref{def:chain-tree-induced-sub}, as we only cut after certain nodes in our construction.
  Hence, $\bigcup_{k \in \IN} U_k$ is non-empty and weakly connected, and for each of its nodes, it either contains no or all successors.
  Furthermore, $\F{T}' = \F{T}[\bigcup_{k \in \IN} U_k]$ is a subtree which does not contain an infinite path of pure $\nV \cup \aV_1$ nodes as we cut every such path after a finite depth.

  It remains to prove that $|\F{T}'| \leq |\F{T}| + \sum_{n = 1}^{\infty} \left(\frac{1}{d}\right)^n$ holds.
  During the $i$-th iteration of the construction, we may increase the value of $|\F{T}|$ by the sum of all probabilities corresponding to the new leaves resulting from the cuts.
  As we cut at most $|H_i|$ trees in the $i$-th iteration and for each such tree, we added at most a total probability of $\left(\frac{1}{d}\right)^{i+1} \cdot \frac{1}{|H_i|}$ for the new leaves, the value of $|\F{T}|$ might increase by
  \[
    |H_i| \cdot \left(\frac{1}{d}\right)^{i+1} \cdot \frac{1}{|H_i|} = \left(\frac{1}{d}\right)^{i+1}
  \]
  in the $i$-th iteration, and hence in total, we then get
  \[
    |\F{T}'| \leq |\F{T}| + \sum_{n = 1}^{\infty} \left(\frac{1}{d}\right)^n,
  \]
  as desired (see \eqref{eq:sum-after-all-cuts}).
\end{myproof}

Next, we prove the \emph{Starting Lemma}. 
The proof nicely illustrates how to use the $\aV$-Partition Lemma.

\StartingLemma*

\begin{myproof}
  Assume that every $\PP$-CT $\F{T}$ converges with probability $1$ if it starts with $(1:t)$ and $\posT(t) = \{\varepsilon\}$.
  We now prove that then also every $\PP$-CT $\F{T} = (V,E,L)$ that starts with $(1:t)$ for some arbitrary annotated term $t$ converges with probability $1$, and thus $\PP$ is $\mathtt{AST}$\@.
  We prove the claim by induction on the number of annotations in the initial term $t$.

  If $t$ contains no annotation, then $\F{T}$ is trivially finite (it cannot contain an infinite path, since there are no nodes in $\aV$) and hence, it converges with probability $1$.
  Next, if $t$ contains exactly one annotation at position $\pi$, then we can ignore everything above the annotation, as we will never use an $(\mathbf{at})$ or $(\mathbf{af})$ step above the annotated position, and we cannot duplicate or change annotations by rewriting above them, since we use VRFs and not GVRFs.
  For $t|_{\pi}$ with $\posT(t|_{\pi}) = \{\varepsilon\}$, we know by our assumption that such a CT converges with probability~$1$.

  Next, we regard the induction step, and assume for a contradiction that for a term $t$ with $n > 1$ annotations, there is a CT $\F{T} = (V,E,L)$ that converges with probability $<1$.
  Here, our induction hypothesis is that every $\PP$-CT $\F{T}$ that starts with $(1:t')$, where $t'$ contains $m$ annotations for some $1 \leq m < n$ converges with probability~$1$.
  Let $\Pi_1 = \{\tau\}$ and $\Pi_2 = \{\chi \in \posT(t) \mid \chi \neq \tau\}$ for some $\tau \in \posT(t)$ and consider the two terms $\anno_{\Pi_1}(t)$ and $\anno_{\Pi_2}(t)$, which contain both strictly less than $n$ annotations.
  By our induction hypothesis, we know that every $\PP$-CT that starts with $(1:\anno_{\Pi_1}(t))$ or $(1:\anno_{\Pi_2}(t))$ converges with probability $1$.
  Let $\F{T}_1 = (V, E, L_1)$ be the tree that starts with $(1:\anno_{\Pi_1}(t))$ and uses the same rules, the same positions, and the same VRFs as in $\F{T}$.

  We can partition the $\aV^{\F{T}}$-nodes (i.e., the $\aV$-nodes of $\F{T}$) into $\aV_1^{\F{T}} = \aV^{\F{T}_1}$ and $\aV_2^{\F{T}} = \aV^{\F{T}} \setminus \aV_1^{\F{T}}$ (i.e., the $\aV$-nodes of $\F{T}_1$, and the rest).
  Note that $\F{T}_1$ itself may not be a $\PP$-CT again, since there might exist paths without an infinite number of $\aV^{\F{T}_1}$-nodes, but obviously every subtree $\F{T}'_1$ of $\F{T}_1$ such that every infinite path has an infinite number of $\aV^{\F{T}_1}$-nodes is a $\PP$-CT again.
  Moreover, by extending such a subtree to be grounded, i.e., adding the initial path from the root of $\F{T}_1$ to $\F{T}'_1$, we created a $\PP$-CT that starts with $\anno_{\Pi_1}(t)$, and hence by our induction hypothesis, converges with probability $1$.
  Thus, this also holds for $\F{T}'_1$.

  We want to use the $\aV$-Partition Lemma (\Cref{lemma:p-partition}) for the tree $\F{T}$.
  For this, we have to show that every subtree $\F{T}'_1$ of $\F{T}$ that only contains inner nodes from $\nV^{\F{T}} \cup \aV_1^{\F{T}}$ converges with probability $1$.
  But since $\F{T}'_1$ only contains inner nodes from $\nV^{\F{T}} \cup \aV_1^{\F{T}}$ it must either contain infinitely many $\aV_1^{\F{T}}$-nodes (and by the previous paragraph it converges with probability $1$), or it contains only finitely many $\aV_1^{\F{T}}$-nodes, hence must be finite itself, and converges with probability $1$.

  We have shown that the conditions for the $\aV$-Partition Lemma (\cref{lemma:p-partition}) are satisfied.
  Thus, we can apply the $\aV$-Partition Lemma to obtain a grounded subtree $\F{T}'$ of $\F{T}$ with $|\F{T}'| <1$ such that on every infinite path, we have an infinite number of $\aV_2^{\F{T}}$ nodes.
  Let $\F{T}_2$ be the tree that starts with $\anno_{\Pi_2}(t)$ and uses the same rules, the same positions, and the same VRFs as in $\F{T}'$.
  Again, all local properties for a $\PP$-CT are satisfied for $\F{T}_2$.
  Additionally, this time we know that every infinite path has an infinite number of nodes from $\aV_2^{\F{T}}$ in $\F{T}'$, hence we also know that the global property of a CT is satisfied.
  This means that $\F{T}_2$ is a $\PP$-CT that starts with $\anno_{\Pi_2}(t)$ and with $|\F{T}_2| <1$.
  This is our desired contradiction, which proves the induction step.

  So we have proven that we can restrict ourselves to CTs that start with $(1:t)$ for a term $t \in \TT^\sharp$ where $\posT(t) = \{\varepsilon\}$ when analyzing $\mathtt{AST}$ and $\mathtt{iAST}$.
  Finally, since we eventually have to rewrite at the root of our term $t$ (that only contains an annotation at the root), we can assume that $\flat(t) = s \theta$ for a substitution $\theta$ and an ADP $s \to \mu^m \in \PP$.
  Note that for $\mathtt{iAST}$, we can only perform such a rewrite step if $\flat(t) = s \theta \in \ANF_{\PP}$.
\end{myproof}

Next, we prove the chain criterion for full rewriting.
In the following,
we use the shorthand notations $\annoD(t) = \anno_{\posD(t)}(t)$ and $\annoEps(t) = \anno_{\{\varepsilon\}}(t)$.

\ProbChainCriterionFull*

\begin{myproof}
  \smallskip

  \noindent
  \underline{\emph{Soundness:}} Assume that $\R$ is not $\mathtt{AST}$.
  Then, there exists an $\R$-RST $\F{T}=(V,E,L)$ whose root is labeled with $(1:t)$ for some term $t \in \TT$ that converges with probability $<1$.
  We will construct a $\DPair{\R}$-CT $\F{T}' = (V,E,L')$ with the same underlying tree
  structure and an adjusted labeling such that $p_v^{\F{T}} = p_v^{\F{T}'}$ for all $v \in
  V$,
  where all the inner nodes are in $\aV$.
  Since the tree structure and the probabilities are the same, we then get $|\F{T}| = |\F{T}'|$.
    So there exists a $\DPair{\R}$-CT $\F{T}'$ that converges with probability $<1$ and $\DPair{\R}$ is not $\mathtt{AST}$ either.
  \vspace*{-10px}
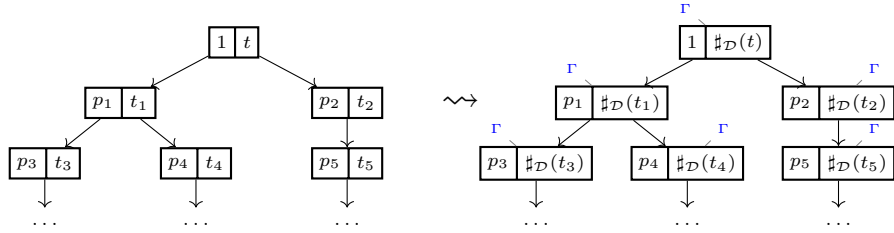
  \begin{center}
    \scriptsize
    \begin{tikzpicture}
      \tikzstyle{adam}=[thick,draw=black!100,fill=white!100,minimum size=4mm, shape=rectangle split, rectangle split parts=2,rectangle split horizontal] \tikzstyle{empty}=[rectangle,thick,minimum size=4mm]

      \node[adam] at (-3, 0) (a) {$1$
        \nodepart{two} $t$};
      \node[adam] at (-4.5, -0.8) (b) {$p_1$
        \nodepart{two} $t_{1}$};
      \node[adam] at (-1.5, -0.8) (c) {$p_2$
        \nodepart{two} $t_{2}$};
      \node[adam] at (-5.5, -1.6) (d) {$p_3$
        \nodepart{two} $t_3$};
      \node[adam] at (-3.5, -1.6) (e) {$p_4$
        \nodepart{two} $t_4$};
      \node[adam] at (-1.5, -1.6) (f) {$p_5$
        \nodepart{two} $t_5$};
      \node[empty] at (-5.5, -2.4) (g) {$\ldots$};
      \node[empty] at (-3.5, -2.4) (h) {$\ldots$};
      \node[empty] at (-1.5, -2.4) (i) {$\ldots$};

      \node[empty] at (0, -0.8) (arrow) {\Large $\leadsto$};

      \node[adam,pin={[pin distance=0.1cm, pin edge={,-}] 140:\tiny \textcolor{blue}{$\aV$}}] at (3.5, 0) (a2) {$1$
          \nodepart{two} $\annoD(t)$};
      \node[adam,pin={[pin distance=0.1cm, pin edge={,-}] 140:\tiny \textcolor{blue}{$\aV$}}] at (2, -0.8) (b2) {$p_1$
          \nodepart{two} $\annoD(t_1)$};
      \node[adam,pin={[pin distance=0.1cm, pin edge={,-}] 45:\tiny \textcolor{blue}{$\aV$}}] at (5, -0.8) (c2) {$p_2$
          \nodepart{two} $\annoD(t_2)$};
      \node[adam,pin={[pin distance=0.1cm, pin edge={,-}] 140:\tiny \textcolor{blue}{$\aV$}}] at (1, -1.6) (d2) {$p_3$
          \nodepart{two} $\annoD(t_3)$};
      \node[adam,pin={[pin distance=0.1cm, pin edge={,-}] 45:\tiny \textcolor{blue}{$\aV$}}] at (3, -1.6) (e2) {$p_4$
          \nodepart{two} $\annoD(t_4)$};
      \node[adam,pin={[pin distance=0.1cm, pin edge={,-}] 45:\tiny \textcolor{blue}{$\aV$}}] at (5, -1.6) (f2) {$p_5$
          \nodepart{two} $\annoD(t_5)$};
      \node[empty] at (1, -2.4) (g2) {$\ldots$};
      \node[empty] at (3, -2.4) (h2) {$\ldots$};
      \node[empty] at (5, -2.4) (i2) {$\ldots$};

      \draw (a) edge[->] (b);
      \draw (a) edge[->] (c);
      \draw (b) edge[->] (d);
      \draw (b) edge[->] (e);
      \draw (c) edge[->] (f);
      \draw (d) edge[->] (g);
      \draw (e) edge[->] (h);
      \draw (f) edge[->] (i);

      \draw (a2) edge[->] (b2);
      \draw (a2) edge[->] (c2);
      \draw (b2) edge[->] (d2);
      \draw (b2) edge[->] (e2);
      \draw (c2) edge[->] (f2);
      \draw (d2) edge[->] (g2);
      \draw (e2) edge[->] (h2);
      \draw (f2) edge[->] (i2);
    \end{tikzpicture}
  \end{center}
  \vspace*{-10px}
  We label all nodes $v \in V$ in $\F{T}'$ with $\annoD(t_v)$, where $t_v$ is the term for the node $v$ in $\F{T}$.
  The annotations ensure that we rewrite with Case $(\mathbf{at})$ so that the node $v$ is contained in $\aV$.
  We only have to show that the edge relation represents valid rewrite steps with $\tored{}{}{\DPair{\R}}$.
  Let $v \in V \setminus \ctleaf$ and $vE = \{w_1, \ldots, w_k\}$ be the set of its successors.
  Since $v$ is not a leaf, we have $t_v \to_{\R} \{\tfrac{p_{w_1}}{p_v}:t_{w_1}, \ldots, \tfrac{p_{w_k}}{p_v}:t_{w_k}\}$.
  This means that there is a rule $\ell \to \{p_1:r_1, \ldots, p_k:r_k\} \in \R$, a position $\pi$, and a substitution $\sigma$ such that ${t_v}|_\pi = \ell\sigma$.
  Furthermore, we have $t_{w_j} = t_v[r_j \sigma]_{\pi}$ for all $1 \leq j \leq k$.

  The corresponding ADP for the rule is $\ell \to \{ p_1 : \annoD(r_1), \ldots, p_k : \annoD(r_k) \}^{\ttrue}$.
  Furthermore, $\pi \in \posT(\annoD(t_v))$ as all defined symbols are annotated in $\annoD(t_v)$.
  Hence, we can rewrite $\annoD(t_v)$ with $\ell \to \{ p_1 : \annoD(r_1), \ldots, p_k : \annoD(r_k) \}^{\ttrue}$, using the position $\pi$, the substitution $\sigma$, and Case $(\mathbf{at})$ applies.
  Furthermore, we take some VRF $(\varphi_j)_{1 \leq j \leq k}$ that is surjective on the positions of the variables in the right-hand side, i.e., for all $1 \leq j \leq k$ and all positions $\tau \in \pos_{\VSet}(r_j)$ there exists a $\tau' \in \pos_{\VSet}(\ell)$ such that $\varphi_j(\tau') = \tau$.
  Such a VRF must exist, since $\R$ is non-duplicating.
  We have $\annoD(t_v) \tored{}{}{\DPair{\R}} \{p_1: \annoD(t_{w_1}), \ldots, p_k: \annoD(t_{w_k})\}$ since by rewriting with Case $(\mathbf{at})$ we get $\annoD(t_v)[\anno_{\posT(r_j) \cup   \Psi_j}(r_j \sigma)]_{\pi} = \annoD(t_{w_j})$ with $\Psi_j$ defined as in \Cref{def:ADPs-and-Rewriting-full-main}.
  Note that since the used VRF is surjective on the variable positions of the right-hand side, we do not remove any annotation in the substitution.
  Furthermore, we annotated all defined symbols in $r_j$.
  Thus, we result in $\annoD(t_{w_j})$ where all defined symbols are annotated again.
  \smallskip

  \noindent
  \underline{\emph{Completeness:}} Assume that $\DPair{\R}$ is not $\mathtt{AST}$.
  Then, there exists a $\DPair{\R}$-CT $\F{T} = (V,E,L)$ whose root is labeled with $(1:t)$ for some annotated term $t \in \TT^{\sharp}$ that converges with probability $<1$.
  We will construct an $\R$-RST $\F{T}' = (V,E,L')$ with the same underlying tree structure and an adjusted labeling such that $p_v^{\F{T}} = p_v^{\F{T}'}$ for all $v \in V$.
  Since the tree structure and the probabilities are the same, we then get $|\F{T}'| = |\F{T}|$.
  Therefore, there exists an $\R$-RST $\F{T}'$ that converges with probability $<1$.
  Hence, $\R$ is not $\mathtt{AST}$ either.

  \begin{center}
    \scriptsize
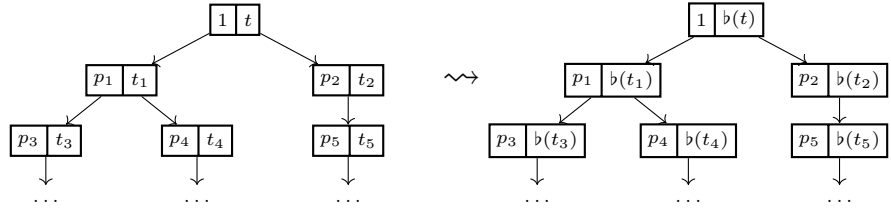
    \begin{tikzpicture}
      \tikzstyle{adam}=[thick,draw=black!100,fill=white!100,minimum size=4mm, shape=rectangle split, rectangle split parts=2,rectangle split horizontal] \tikzstyle{empty}=[rectangle,thick,minimum size=4mm]

      \node[adam] at (-3, 0) (a) {$1$
        \nodepart{two} $t$};
      \node[adam] at (-4.5, -0.8) (b) {$p_1$
        \nodepart{two} $t_{1}$};
      \node[adam] at (-1.5, -0.8) (c) {$p_2$
        \nodepart{two} $t_{2}$};
      \node[adam] at (-5.5, -1.6) (d) {$p_3$
        \nodepart{two} $t_3$};
      \node[adam] at (-3.5, -1.6) (e) {$p_4$
        \nodepart{two} $t_4$};
      \node[adam] at (-1.5, -1.6) (f) {$p_5$
        \nodepart{two} $t_5$};
      \node[empty] at (-5.5, -2.4) (g) {$\ldots$};
      \node[empty] at (-3.5, -2.4) (h) {$\ldots$};
      \node[empty] at (-1.5, -2.4) (i) {$\ldots$};

      \node[empty] at (0, -0.8) (arrow) {\Large $\leadsto$};

      \node[adam] at (3.5, 0) (a2) {$1$
        \nodepart{two} $\flat(t)$};
      \node[adam] at (2, -0.8) (b2) {$p_1$
        \nodepart{two} $\flat(t_1)$};
      \node[adam] at (5, -0.8) (c2) {$p_2$
        \nodepart{two} $\flat(t_2)$};
      \node[adam] at (1, -1.6) (d2) {$p_3$
        \nodepart{two} $\flat(t_3)$};
      \node[adam] at (3, -1.6) (e2) {$p_4$
        \nodepart{two} $\flat(t_4)$};
      \node[adam] at (5, -1.6) (f2) {$p_5$
        \nodepart{two} $\flat(t_5)$};
      \node[empty] at (1, -2.4) (g2) {$\ldots$};
      \node[empty] at (3, -2.4) (h2) {$\ldots$};
      \node[empty] at (5, -2.4) (i2) {$\ldots$};

      \draw (a) edge[->] (b);
      \draw (a) edge[->] (c);
      \draw (b) edge[->] (d);
      \draw (b) edge[->] (e);
      \draw (c) edge[->] (f);
      \draw (d) edge[->] (g);
      \draw (e) edge[->] (h);
      \draw (f) edge[->] (i);

      \draw (a2) edge[->] (b2);
      \draw (a2) edge[->] (c2);
      \draw (b2) edge[->] (d2);
      \draw (b2) edge[->] (e2);
      \draw (c2) edge[->] (f2);
      \draw (d2) edge[->] (g2);
      \draw (e2) edge[->] (h2);
      \draw (f2) edge[->] (i2);
    \end{tikzpicture}
  \end{center}
  \vspace*{-10px}
  We label all nodes $v \in V$ in $\F{T}'$ with $\flat(t_v)$, where $t_v$ is the term for the node $v$ in $\F{T}$, i.e., we remove all annotations.
  We only have to show that $\F{T}'$ is indeed a valid RST, i.e., that the edge relation represents valid rewrite steps with $\to_{\R}$, but this follows directly from the fact that if we remove all annotations in \Cref{def:ADPs-and-Rewriting-full-main}, then we get the ordinary probabilistic term rewriting relation again.
\end{myproof}

Next, we define the set $\PosDPoss(t)$ of all positions of subterms of $t$
that may be used as a redex now or in future rewrite steps, 
because the subterm has a defined root symbol and is not in normal form w.r.t.\ $\R$.
This set is needed, as we cannot simply annotate every defined symbol in the soundness proof of the innermost chain criterion.
Instead,   we can only guarantee that all the symbols in $\PosDPoss(t)$ are annotated.

\begin{defi}[{\normalfont{$\PosDPoss$}}]
  \label{def:prop-important-sets}
  Let $\R$ be a PTRS.
  For a term $t \in \TT$ we define $\normalfont{\PosDPoss}(t) = \{\pi \mid \pi \in \posD(t), t|_\pi \notin \NF_{\R}\}$.
\end{defi}

\begin{exa}
  \label{example:important-sets}
  Consider the following PTRS $\R$ over a signature with $\SignatureD = \{\tf,\tg\}$ and $\SignatureC = \{\ta,\ts\}$ with the rules $\tf(\ta,\ta) \to \{ 1: \ts(\tf(\tg,\tg))\}$ and $\tg \to \{ 1: \ta\}$.
  For the term $t = \ts(\tf(\tg,\tg))$ we have $\PosDPoss(t) = \{1, 1.1, 1.2\}$.
\end{exa}

Finally, for two (possibly annotated) terms $s,t$ we define $s \flateq t$ if $\flat(s) = \flat(t)$.
Remember that for innermost rewriting, we do not use any VRFs, hence, we can ignore them in the proofs for innermost rewriting.

\ProbChainCriterionInnermost*

\begin{myproof}
  In the following, we will often implicitly use that for an annotated term $t \in \TT^\sharp$, we have $\flat(t) \in \ANF_{\R}$ iff $t \in \ANF_{\DPair{\R}}$ since a rewrite rule and its corresponding canonical annotated dependency pair have the same left-hand side.
  \smallskip

  \noindent
  \underline{\emph{Soundness:}} Assume that $\R$ is not $\mathtt{iAST}$.
  Then, there exists an innermost $\R$-RST $\F{T}=(V,E,L)$ whose root is labeled with $(1:t)$ for some term $t \in \TT$ that converges with probability $<1$.
  We use almost the same construction as in the proof of soundness for $\mathtt{AST}$, however, instead of annotating all defined symbols, we only ensure that all defined symbols from $\PosDPoss(t)$ are annotated.

  \begin{center}
    \scriptsize
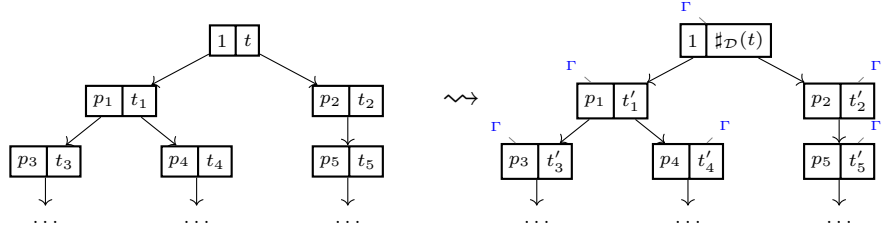
    \begin{tikzpicture}
      \tikzstyle{adam}=[thick,draw=black!100,fill=white!100,minimum size=4mm, shape=rectangle split, rectangle split parts=2,rectangle split horizontal] \tikzstyle{empty}=[rectangle,thick,minimum size=4mm]

      \node[adam] at (-3, 0) (a) {$1$
        \nodepart{two} $t$};
      \node[adam] at (-4.5, -0.8) (b) {$p_1$
        \nodepart{two} $t_{1}$};
      \node[adam] at (-1.5, -0.8) (c) {$p_2$
        \nodepart{two} $t_{2}$};
      \node[adam] at (-5.5, -1.6) (d) {$p_3$
        \nodepart{two} $t_3$};
      \node[adam] at (-3.5, -1.6) (e) {$p_4$
        \nodepart{two} $t_4$};
      \node[adam] at (-1.5, -1.6) (f) {$p_5$
        \nodepart{two} $t_5$};
      \node[empty] at (-5.5, -2.4) (g) {$\ldots$};
      \node[empty] at (-3.5, -2.4) (h) {$\ldots$};
      \node[empty] at (-1.5, -2.4) (i) {$\ldots$};

      \node[empty] at (0, -0.8) (arrow) {\Large $\leadsto$};

      \node[adam,pin={[pin distance=0.1cm, pin edge={,-}] 140:\tiny \textcolor{blue}{$\aV$}}] at (3.5, 0) (a2) {$1$
          \nodepart{two} $\annoD(t)$};
      \node[adam,pin={[pin distance=0.1cm, pin edge={,-}] 140:\tiny \textcolor{blue}{$\aV$}}] at (2, -0.8) (b2) {$p_1$
          \nodepart{two} $t'_1$};
      \node[adam,pin={[pin distance=0.1cm, pin edge={,-}] 45:\tiny \textcolor{blue}{$\aV$}}] at (5, -0.8) (c2) {$p_2$
          \nodepart{two} $t'_2$};
      \node[adam,pin={[pin distance=0.1cm, pin edge={,-}] 140:\tiny \textcolor{blue}{$\aV$}}] at (1, -1.6) (d2) {$p_3$
          \nodepart{two} $t'_3$};
      \node[adam,pin={[pin distance=0.1cm, pin edge={,-}] 45:\tiny \textcolor{blue}{$\aV$}}] at (3, -1.6) (e2) {$p_4$
          \nodepart{two} $t'_4$};
      \node[adam,pin={[pin distance=0.1cm, pin edge={,-}] 45:\tiny \textcolor{blue}{$\aV$}}] at (5, -1.6) (f2) {$p_5$
          \nodepart{two} $t'_5$};
      \node[empty] at (1, -2.4) (g2) {$\ldots$};
      \node[empty] at (3, -2.4) (h2) {$\ldots$};
      \node[empty] at (5, -2.4) (i2) {$\ldots$};

      \draw (a) edge[->] (b);
      \draw (a) edge[->] (c);
      \draw (b) edge[->] (d);
      \draw (b) edge[->] (e);
      \draw (c) edge[->] (f);
      \draw (d) edge[->] (g);
      \draw (e) edge[->] (h);
      \draw (f) edge[->] (i);

      \draw (a2) edge[->] (b2);
      \draw (a2) edge[->] (c2);
      \draw (b2) edge[->] (d2);
      \draw (b2) edge[->] (e2);
      \draw (c2) edge[->] (f2);
      \draw (d2) edge[->] (g2);
      \draw (e2) edge[->] (h2);
      \draw (f2) edge[->] (i2);
    \end{tikzpicture}
  \end{center}
  \vspace*{-10px}
  We construct the new labeling $L'$ for the $\DPair{\R}$-CT inductively such that for all inner nodes $v \in V \setminus \ctleaf$ with children nodes $vE = \{w_1,\ldots,w_k\}$ we have $t_v' \itored{}{}{\DPair{\R}} \{\tfrac{p_{w_1}}{p_v}:t_{w_1}', \ldots, \tfrac{p_{w_k}}{p_v}:t_{w_k}'\}$ and use Case $(\mathbf{at})$ so that all nodes are in $\aV$.
  Let
    $W \subseteq V$ be the set of nodes $v$ where we have already defined the labeling $L'(v)$.
  During our construction, we ensure that for all $v \in W$ we have
  \begin{equation} \label{chain-crit-1-soundness-induction-hypothesis}
    t_v \flateq t_v' \text{ and } \PosDPoss(t_v) \subseteq \posT(t_v').
  \end{equation}
  This means that the corresponding term $t_v$ for the node $v$ in $\F{T}$ has the same structure as the term $t_v'$ in $\F{T}'$, and additionally, all the possible redexes in $t_v$ are annotated in $t_v'$.
  We label the root of $\F{T}'$ with $\annoD(t)$.
  Here, we have $t \flateq \annoD(t)$ and $\PosDPoss(t) \subseteq \posD(t) = \posT(\annoD(t))$.
  As long as there is still an inner node $v \in W$ such that its successors are not contained in $W$, we do the following.
  Let $vE = \{w_1, \ldots, w_k\}$ be the set of its successors.
  We need to define the corresponding terms $t_{w_1}', \ldots, t_{w_k}'$ for the nodes $w_1, \ldots, w_k$.
  Since $v$ is not a leaf, we have $t_v \itor \{\tfrac{p_{w_1}}{p_v}:t_{w_1}, \ldots, \tfrac{p_{w_k}}{p_v}:t_{w_k}\}$.
  This means that there is a rule $\ell \to \{p_1:r_1, \ldots, p_k:r_k\} \in \R$, a position $\pi$, and a substitution $\sigma$ such that ${t_v}|_\pi = \ell\sigma \in \ANF_{\R}$.
  Furthermore, we have $t_{w_j} = t_v[r_j \sigma]_{\pi}$ for all $1 \leq j \leq k$.
  So the labeling of the successor $w_j$ in $\F{T}$ is $L(w_j) = (p_v \cdot p_j: t_v[r_j\sigma]_\pi)$ for all $1 \leq j \leq k$.

  The corresponding ADP for the rule is $\ell \to \{ p_1 : \annoD(r_1), \ldots, p_k : \annoD(r_k) \}^{\ttrue}$.
  Furthermore, $\pi \in \PosDPoss(t_v) \subseteq_{(IH)} \posT(t_v')$ and $t_v \flateq_{(IH)} t_v'$.
  Hence, we can rewrite $t_v'$ with $\ell \to \{ p_1 : \annoD(r_1), \ldots, p_k : \annoD(r_k) \}^{\ttrue}$, using the position $\pi$ and the substitution $\sigma$, and Case $(\mathbf{at})$ applies.
  We get $t_v' \itored{}{}{\DPair{\R}} \{p_1: t_{w_1}', \ldots, p_k: t_{w_k}'\}$ with $t_{w_j}' = t_v'[\annoD(r_j) \sigma]_{\pi}$.
  This means that we have $t_{w_j} \flateq t_{w_j}'$.
  It remains to prove $\PosDPoss(t_{w_j}) \subseteq \posT(t_{w_j}')$ for all $1 \leq j \leq k$.
  For all positions $\tau \in \PosDPoss(t_{w_j}) = \PosDPoss(t_v[r_j \sigma]_{\pi})$ that are orthogonal or above $\pi$, we have $\tau \in \PosDPoss(t_v,\R) \subseteq_{(IH)} \posT(t_v')$, and all annotations orthogonal or above $\pi$ remain in $t_{w_j}'$ as they were in $t_v'$.
  For all positions $\tau \in \PosDPoss(t_{w_j}) = \PosDPoss(t_v[r_j \sigma]_{\pi})$ that are below $\pi$, we know that, due to innermost evaluation, at least the defined root symbol of a term that is not in normal form must be inside $r_j$, and thus $\tau \in \posT(t_{w_j}')$, as all defined symbols of $r_j$ are annotated in $t_{w_j}' = t_v'[\annoD(r_j) \sigma]_{\pi}$.
  \smallskip

  \noindent
  \underline{\emph{Completeness:}}
  Analogous to completeness for $\mathtt{AST}$.
\end{myproof}

Next, we prove soundness and completeness of the ADP processors.
Recall that most processors are complete by \Cref{thm:proc-complete}.

\ProbDepGraphProc*

\begin{myproof}
  Let $\overline{X} = X \cup \flat(\PP \setminus X)$ for $X \subseteq \PP$.
  \smallskip

  \noindent
  \underline{\emph{Completeness:}} By \Cref{thm:proc-complete}.
  \medskip

  \noindent
  \underline{\emph{Soundness:}} Let $\F{G}$ be the $\PP$-dependency graph.
  Suppose that every $\overline{\PP_i}$-CT converges with probability $1$ for all $1 \leq i \leq n$.
  We prove that then also every $\PP$-CT converges with probability 1.
  Let $\F{W} = \{\PP_1, \ldots, \PP_n\} \cup \{\{\alpha\} \subseteq \PP \mid \alpha$ is not in an SCC of $\F{G}\}$ be the set of all SCCs and all singleton sets of nodes that do not belong to any SCC\@.
  For two $X_1,X_2 \in \F{W}$ we say that $X_2$ is a \emph{direct successor} of $X_1$ (denoted $X_1 >_{\F{G}} X_2$) 
  if there exist nodes $\alpha \in X_1$ and $\beta \in X_2$ such that there is an edge from $\alpha$ to $\beta$ in $\F{G}$.
  The core steps of this proof are the following:
  \begin{enumerate}
    \item[1.] We show that every ADP problem $\overline{X}$ with $X \in \F{W}$ is $\mathtt{AST}$\@.
    \item[2.] We show that composing SCCs maintains the $\mathtt{AST}$ property.
    \item[3.] We show that for every $X \in \F{W}$, the ADP problem $\overline{\bigcup_{X >_{\F{G}}^* Y}Y}$ is $\mathtt{AST}$ by induction on $>_{\F{G}}$.
    \item[4.] We conclude that $\PP$ must be $\mathtt{AST}$\@.
  \end{enumerate}

  \medskip

  \noindent
  \textbf{\underline{1.
      Every ADP problem $\overline{X}$ with $X \in \F{W}$ is $\mathtt{AST}$\@.}}

  \noindent
  We start by proving the following:
  \begin{equation}
    \label{W is AST}
    \mbox{Every ADP problem $\overline{X}$ with $X \in \F{W}$ is $\mathtt{AST}$\@.}
  \end{equation}
  To prove~\eqref{W is AST}, note that if $X$ is an SCC, then it follows from our assumption that $\overline{X}$ is $\mathtt{AST}$\@.
  If $X$ is a singleton set of a node that does not belong to any SCC, then assume for a contradiction that $\overline{X}$ is not $\mathtt{AST}$\@.
  By \cref{lemma:starting} there exists an $\overline{X}$-CT $\F{T} = (V,E,L)$ that converges with probability $<1$ and starts with $(1:t)$ where $\posT(t) = \{\varepsilon\}$ and $\flat(t) = s\theta$ for a substitution $\theta$ and some ADP $\alpha = s \to \{ p_1:r_1, \ldots, p_k:r_k \}^m \in \overline{X}$.
  If $\alpha \notin X$, then the resulting terms after the first rewrite step contain no annotations anymore and this cannot start a CT that converges with probability $<1$.
  Hence, we have $\alpha \in X$ and thus, $X = \{\alpha \}$, since $X$ is a singleton set.
  Assume for a contradiction that there exists a node $v \in \aV$ in $\F{T}$ that is not the root and introduces new annotations.
  W.l.o.G., let $v$ be reachable from the root without traversing any other node that introduces new annotations.
  This means that for the corresponding term $t_v$ for node $v$ there is a $t' \trianglelefteq_{\sharp} t_v$ at position $\tau$ such that $t' = s \sigma'$ for some substitution $\sigma'$ (since $s \to \dots$ is the only ADP in $\overline{X}$ that contains any annotations in the right-hand side).
  Let $(z_0, \ldots, z_m)$ with $z_m = v$ be the path from the root to $v$ in $\F{T}$.
  The first rewrite step at the root must be $s \theta \tored{}{}{\overline{X}} \{p_1:r_1 \theta, \ldots, p_k:r_k \theta\}$.
  After that, we only use ADPs with the flag $\ttrue$ below the annotated position that will be used for the rewrite step at node $v$, as otherwise, the position $\tau$ would not be annotated in $t_v$.
  Therefore, we must have an $1 \leq j \leq k$ and a $t'' \trianglelefteq_{\sharp} r_j$ such that $t''^\sharp \theta \to_{\nonprob(\PP)}^* s^\sharp \sigma'$, which means that there must be a self-loop for the only ADP in $X$, which is a contradiction to our assumption that $X$ is a singleton consisting of an ADP that is not in any SCC of $\F{G}$.\linebreak
  \indent Now, we have proven that the $\overline{X}$-CT $\F{T}$ does not introduce new annotations.
  By definition of a $\PP$-CT, every infinite path must contain an infinite number of nodes in $\aV$, i.e., nodes where we rewrite at an annotation.
  Thus, every path in $\F{T}$ must be finite, which means that $\F{T}$ is finite itself, as the tree is finitely branching.
  But every finite CT converges with probability $1$, which is a contradiction to our assumption that $\F{T}$ converges with probability $<1$.

  \medskip

  \noindent
  \textbf{\underline{2.
      Composing SCCs maintains the $\mathtt{AST}$ property.}}

  \noindent
  In the second step, we show that composing SCCs maintains the $\mathtt{AST}$ property.
  More precisely, we prove the following:
  \begin{equation}
    \label{Composing AST}
    \parbox{.9\textwidth}{Let $\hat{X} \subseteq \F{W}$ and $\hat{Y} \subseteq \F{W}$
      such that there are no $X_1,X_2 \!\in\! \hat{X}$ and $Y \!\in\! \hat{Y}$ which
      satisfy both $X_1 >_{\F{G}}^* Y >_{\F{G}}^* X_2$ and $Y \not\in \hat{X}$, and such that there are no $Y_1,Y_2 \!\in\! \hat{Y}$ and $X \!\in\! \hat{X}$ which
      satisfy both $Y_1 >_{\F{G}}^* X >_{\F{G}}^* Y_2$ and $X \not\in \hat{Y}$.
      If both $ \overline{\bigcup_{X \in \hat{X}} X} $ and $ \overline{\bigcup_{Y \in \hat{Y}} Y} $ are $\mathtt{AST}$, then $ \overline{\bigcup_{X \in \hat{X}} X \cup \bigcup_{Y \in \hat{Y}} Y} $ is $\mathtt{AST}$.}
  \end{equation}
  To show~\eqref{Composing AST}, we assume that both $\overline{\bigcup_{X \in \hat{X}} X}$ and $\overline{\bigcup_{Y \in \hat{Y}} Y}$ are $\mathtt{AST}$\@.
  Let $\overline{Z} = \overline{\bigcup_{X \in \hat{X}} X \cup \bigcup_{Y \in \hat{Y}} Y}$.
  The property in~\eqref{Composing AST} for $\hat{X}$ and $\hat{Y}$ says that a path between two nodes from $\bigcup_{X \in \hat{X}} X$ that only traverses nodes from $Z$ must also be a path that only traverses nodes from $\bigcup_{X \in \hat{X}} X$, so that $\bigcup_{Y \in \hat{Y}} Y$ cannot be used to ``create'' new paths between two nodes from $\bigcup_{X \in \hat{X}} X$, and vice versa.
  Assume for a contradiction that $\overline{Z}$ is not $\mathtt{AST}$\@.
  By \cref{lemma:starting} there exists a $\overline{Z}$-CT $\F{T} = (V,E,L)$ that converges with probability $<1$ and starts with $(1:t)$ where $\posT(t) = \{\varepsilon\}$ and $\flat(t) = s \theta$ for a substitution $\theta$ and an ADP $s \to \ldots \in \overline{Z}$.

  If $s \to \ldots \notin \bigcup_{X \in \hat{X}} X \cup \bigcup_{Y \in \hat{Y}} Y$, then the resulting terms contain no annotations anymore and this cannot start a CT that converges with probability $<1$.
  W.l.o.g., we may assume that the ADP that is used for the rewrite step at the root is in $\bigcup_{X \in \hat{X}} X$.
  Otherwise, we simply swap $\bigcup_{X \in \hat{X}} X$ with $\bigcup_{Y \in \hat{Y}} Y$ in the following.

  We can partition the set $\aV$ of our $\overline{Z}$-CT $\F{T}$ into the sets
  \begin{itemize}
    \item[$\bullet$] $\aV_1 := \{v \in \aV \mid v$ together with the labeling and its successors represents a step with an ADP from $\bigcup_{X \in \hat{X}} X\}$
    \item[$\bullet$] $\aV_2 := \aV \setminus \aV_1$
  \end{itemize}
  Note that the labeling represents a step with an ADP from $\PP \setminus \bigcup_{X \in \hat{X}} X$ at nodes $v \in \aV_2$.
  Every $\overline{\bigcup_{Y \in \hat{Y}} Y}$-CT converges with probability $1$, since $\overline{\bigcup_{Y \in \hat{Y}} Y}$ is $\mathtt{AST}$\@.
  Thus, also every $\overline{\bigcup_{Y \in \hat{Y}} Y \setminus \bigcup_{X \in \hat{X}} X}$-CT converges with probability $1$ (as it contains fewer annotations than $\overline{\bigcup_{Y \in \hat{Y}} Y}$).
  Furthermore, we have $|\F{T}| <1$ by our assumption.
  By the $\aV$-Partition Lemma (\cref{lemma:p-partition}) we can find a grounded sub $\overline{Z}$-CT $\F{T}' = (V',E',L')$ with $|\F{T}'| <1$ such that every infinite path has an infinite number of $\aV_1$-nodes.

  We now construct a $\overline{\bigcup_{X \in \hat{X}} X}$-CT $\F{T}'' = (V',E',L'')$ with $\aV_1 \cap \aV^{\F{T}'} \subseteq \aV^{\F{T}''}$ that has the same underlying tree structure and adjusted labeling such that all nodes get the same probabilities as in $\F{T}'$.
  Hence, we then obtain $|\F{T}'| = |\F{T}''|$.
  Moreover, every infinite path in $\F{T}''$ contains infinitely many $\aV^{\F{T}''}$-nodes, since every path in $\F{T}'$ contains infinitely many $\aV_1$-nodes and we have $\aV_1 \cap \aV^{\F{T}'} \subseteq \aV^{\F{T}''}$.
  But this implies that $\F{T}''$ is indeed a $\overline{\bigcup_{X \in \hat{X}} X}$-CT with $|\F{T}''| <1$, which is a contradiction to our assumption that $\overline{\bigcup_{X \in \hat{X}} X}$ is $\mathtt{AST}$\@.

  \vspace*{-10px}
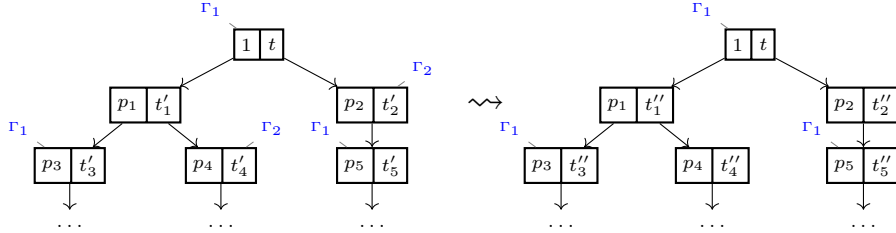
  \begin{center}
    \scriptsize
    \begin{tikzpicture}
      \tikzstyle{adam}=[thick,draw=black!100,fill=white!100,minimum size=4mm, shape=rectangle split, rectangle split parts=2,rectangle split horizontal] \tikzstyle{empty}=[rectangle,thick,minimum size=4mm]

      \node[adam,pin={[pin distance=0.1cm, pin edge={,-}] 145:\tiny \textcolor{blue}{$\aV_1$}}] at (-3, 0) (a) {$1$
          \nodepart{two} $t$};
      \node[adam] at (-4.5, -0.8) (b) {$p_1$
        \nodepart{two} $t_{1}'$};
      \node[adam,pin={[pin distance=0.1cm, pin edge={,-}] 35:\tiny \textcolor{blue}{$\aV_2$}}] at (-1.5, -0.8) (c) {$p_2$
          \nodepart{two} $t_{2}'$};
      \node[adam,pin={[pin distance=0.1cm, pin edge={,-}] 145:\tiny \textcolor{blue}{$\aV_1$}}] at (-5.5, -1.6) (d) {$p_3$
          \nodepart{two} $t_3'$};
      \node[adam,pin={[pin distance=0.1cm, pin edge={,-}] 35:\tiny \textcolor{blue}{$\aV_2$}}] at (-3.5, -1.6) (e) {$p_4$
          \nodepart{two} $t_4'$};
      \node[adam,pin={[pin distance=0.1cm, pin edge={,-}] 145:\tiny \textcolor{blue}{$\aV_1$}}] at (-1.5, -1.6) (f) {$p_5$
          \nodepart{two} $t_5'$};
      \node[empty] at (-5.5, -2.4) (g) {$\ldots$};
      \node[empty] at (-3.5, -2.4) (h) {$\ldots$};
      \node[empty] at (-1.5, -2.4) (i) {$\ldots$};

      \node[empty] at (0, -0.8) (arrow) {\Large $\leadsto$};

      \node[adam,pin={[pin distance=0.1cm, pin edge={,-}] 145:\tiny \textcolor{blue}{$\aV_1$}}] at (3.5, 0) (a2) {$1$
          \nodepart{two} $t$};
      \node[adam] at (2, -0.8) (b2) {$p_1$
      \nodepart{two} $t''_{1}$};
      \node[adam] at (5, -0.8) (c2) {$p_2$
      \nodepart{two} $t''_{2}$};
      \node[adam,pin={[pin distance=0.1cm, pin edge={,-}] 145:\tiny \textcolor{blue}{$\aV_1$}}] at (1, -1.6) (d2) {$p_3$
          \nodepart{two} $t''_3$};
      \node[adam] at (3, -1.6) (e2) {$p_4$
        \nodepart{two} $t''_4$};
      \node[adam,pin={[pin distance=0.1cm, pin edge={,-}] 145:\tiny \textcolor{blue}{$\aV_1$}}] at (5, -1.6) (f2) {$p_5$
          \nodepart{two} $t''_5$};
      \node[empty] at (1, -2.4) (g2) {$\ldots$};
      \node[empty] at (3, -2.4) (h2) {$\ldots$};
      \node[empty] at (5, -2.4) (i2) {$\ldots$};

      \draw (a) edge[->] (b);
      \draw (a) edge[->] (c);
      \draw (b) edge[->] (d);
      \draw (b) edge[->] (e);
      \draw (c) edge[->] (f);
      \draw (d) edge[->] (g);
      \draw (e) edge[->] (h);
      \draw (f) edge[->] (i);

      \draw (a2) edge[->] (b2);
      \draw (a2) edge[->] (c2);
      \draw (b2) edge[->] (d2);
      \draw (b2) edge[->] (e2);
      \draw (c2) edge[->] (f2);
      \draw (d2) edge[->] (g2);
      \draw (e2) edge[->] (h2);
      \draw (f2) edge[->] (i2);
    \end{tikzpicture}
  \end{center}
  \vspace*{-10px}
  The core idea of this construction is that annotations introduced by rewrite steps at a node $v \in \aV_2$ are not important for our computation.
  The reason is that if annotations are introduced using an ADP from $\bigcup_{Y \in \hat{Y}} Y$ that is not contained in $\bigcup_{X \in \hat{X}} X$, then by the prerequisite of~\eqref{Composing AST}, we know that such an ADP has no path in the dependency graph to an ADP in $\bigcup_{X \in \hat{X}} X$.
  Hence, by definition of the dependency graph, we are never able to use these terms for a rewrite step with an ADP from $\bigcup_{X \in \hat{X}} X$ to introduce new annotations.
  We can therefore apply the non-annotated ADP from $\bigcup_{Y \in \hat{Y}} Y$ instead.

  We now construct the new labeling $L''$ for the $\overline{\bigcup_{X \in \hat{X}} X}$-CT $\F{T}''$ recursively.
  Let $W \subseteq V$ be the set of nodes where we have already defined the labeling $L''$.
  Furthermore, for any term $t'_v$, let $\Junk_{\hat{X}}(t'_v)$ denote the positions of all annotated subterms $s \trianglelefteq_{\sharp} t'_v$ that can never be used for a rewrite step with an ADP from $\hat{X}$, as indicated by the dependency graph.
  To be precise, we define $\pi \in \Junk_{\hat{X}}(t'_v)$:$\Leftrightarrow$ there is no $B \in \F{W}$ with $B >_{\F{G}}^* X$ for some $X \in \hat{X}$ such that there is an ADP $\ell \to \{p_1: r_1, \ldots, p_k: r_k\}^{m} \in B$, and a substitution $\sigma$ with $\annoEps(t_v'|_{\pi}) \to_{\nonprob(\PP)}^* \ell^\sharp \sigma$.
  During our construction, we ensure that for every $v \in W$ we have
  \begin{equation}\label{dep-graph-construction-induction-hypothesis}
     t'_v \flateq t''_v \text{ and } \posT(t'_v) \setminus \Junk_{\hat{X}}(t'_v) \subseteq \posT(t''_v).
  \end{equation}

  We start by setting $t''_v = t'_v = t$ for the root $v$ of $\F{T}'$.
  Here, our property~\eqref{dep-graph-construction-induction-hypothesis} is clearly satisfied.
  As long as there is still an inner node $v \in W$ such that its successors are not contained in $W$, we do the following.
  Let $vE = \{w_1, \ldots, w_k\}$ be the set of its successors.
  We need to define the corresponding terms for the nodes $w_1, \ldots, w_k$ in $\F{T}''$.
  Since $v$ is not a leaf and $\F{T}'$ is a $\overline{Z}$-CT, we have $t'_v \tored{}{}{\overline{Z}} \{\tfrac{p_{w_1}}{p_v}:t'_{w_1}, \ldots, \tfrac{p_{w_k}}{p_v}:t'_{w_k}\}$, and hence, we have to deal with the following two cases:\linebreak
  \vspace*{-0.4cm}
  \begin{enumerate}
    \item If we use an ADP from $\bigcup_{X \in \hat{X}} X$ in $\F{T}'$, then we perform the rewrite step with the same ADP, the same VRF $(\varphi_j)_{1 \leq j \leq k}$, the same position $\pi$, and the same substitution in $\F{T}''$.
          Since we have $t'_v \flateq_{(IH)} t''_v$, we also get $t'_{w_j} \flateq t''_{w_j}$ for all $1 \leq j \leq k$.
          Furthermore, since we rewrite at position $\pi$ it cannot be in $\Junk_{\hat{X}}(t'_v)$, and hence, if $\pi \in \posT(t'_v)$, then also $\pi \in \posT(t''_v)$ by \eqref{dep-graph-construction-induction-hypothesis}.
          Thus, whenever we create annotations in the rewrite step in $\F{T}'$ (a step with $(\mathbf{af})$ or $(\mathbf{at})$), then we do the same in $\F{T}''$ (the step is also an $(\mathbf{af})$ or $(\mathbf{at})$ step, respectively), and whenever we remove annotations in the rewrite step in $\F{T}''$ (a step with $(\mathbf{af})$ or $(\mathbf{nf})$), then the same happened in $\F{T}'$ (the step is also an $(\mathbf{af})$ or $(\mathbf{nf})$ step).
          Therefore, we also get $\posT(t'_{w_j}) \setminus \Junk_{\hat{X}}(t'_{w_j}) \subseteq \posT(t''_{w_j})$ for all $1 \leq j \leq k$ and \eqref{dep-graph-construction-induction-hypothesis} is again satisfied.
    \item If we use an ADP from $\PP \setminus \bigcup_{X \in \hat{X}} X$ in $\F{T}'$, and we use the ADP $\ell \to \{p_1:r_1, \ldots, p_k:r_k\}^{m}$, then we can use $\ell \to \{p_1:\flat(r_1), \ldots, p_k:\flat(r_k)\}^{m}$ instead, with the same VRF $(\varphi_j)_{1 \leq j \leq k}$, the same position $\pi$, and the same substitution.
          Note that if $\pi \in \posT(t'_v)$, then all positions of annotations introduced by the ADP are in 
          $\Junk_{\hat{X}}(t'_{w_j})$ for all $1 \leq j \leq k$, 
          since the used ADP is not in $\bigcup_{X \in \hat{X}} X$ and by \eqref{Composing AST} 
          we cannot use another ADP from $\bigcup_{Y \in \hat{Y}} Y \setminus \bigcup_{X \in \hat{X}} X$ 
          to create a path in the dependency graph to a node in $\bigcup_{X \in \hat{X}} X$ again.
          Otherwise, we remove the annotations during the application of the rule anyway.
          Again, \eqref{dep-graph-construction-induction-hypothesis} is satisfied.
  \end{enumerate}
  We have now shown that~\eqref{Composing AST} holds.

  \medskip

  \noindent
  \textbf{\underline{3.
      For every $X \in \F{W}$, the ADP problem $\overline{\bigcup_{X >_{\F{G}}^* Y}Y}$ is $\mathtt{AST}$\@.}}

  \noindent
  Using~\eqref{W is AST} and~\eqref{Composing AST}, by induction on $>_{\F{G}}$ we now prove that
  \begin{equation}
    \label{SCC induction} \mbox{for every $X \in \F{W}$, the ADP problem $\overline{\bigcup_{X >_{\F{G}}^* Y}Y}$ is $\mathtt{AST}$\@.}
  \end{equation}
  Note that $>_{\F{G}}$ is well founded, since $\F{G}$ is finite.

  For the base case, we consider an $X \in \F{W}$ that is minimal w.r.t.\ $>_{\F{G}}$.
  Hence, we have $\bigcup_{X >_{\F{G}}^* Y} Y = X$.
  By~\eqref{W is AST}, $\overline{X}$ is $\mathtt{AST}$\@.

  For the induction step, we consider an $X \in \F{W}$ and assume that $\overline{\bigcup_{Y >_{\F{G}}^* Z} Z}$ is $\mathtt{AST}$ for every $Y \in \F{W}$ with $X >_{\F{G}}^+ Y$.
  Let $\mathtt{Succ}(X) = \{Y \in \F{W} \mid X >_{\F{G}} Y\} = \{Y_1, \ldots Y_m\}$ be the set of all direct successors of $X$.
  The induction hypothesis states that $\overline{\bigcup_{Y_u >_{\F{G}}^* Z} Z}$ is $\mathtt{AST}$ for all $1 \leq u \leq m$.
  We first prove by induction that for all $1 \leq u \leq m$, $\overline{\bigcup_{1 \leq i \leq u} \bigcup_{Y_i >_{\F{G}}^* Z} Z}$ is $\mathtt{AST}$\@.

  In the inner induction base, we have $u = 1$ and hence $\overline{\bigcup_{1 \leq i \leq u} \bigcup_{Y_i >_{\F{G}}^* Z} Z} = \overline{\bigcup_{Y_1 >_{\F{G}}^* Z} Z}$.
  By our outer induction hypothesis we know that $\overline{\bigcup_{Y_1 >_{\F{G}}^* Z} Z}$ is $\mathtt{AST}$\@.

  In the inner induction step, assume that the claim holds for some $1 \leq u < m$.
  Then $\overline{\bigcup_{Y_{u+1} >_{\F{G}}^* Z} Z}$ is $\mathtt{AST}$ by our outer induction hypothesis and\linebreak
  $\overline{\bigcup_{1 \leq i \leq u} \bigcup_{Y_{i} >_{\F{G}}^* Z} Z}$ is $\mathtt{AST}$ by our inner induction hypothesis.
  By~\eqref{Composing AST}, we know that then $\overline{\bigcup_{1 \leq i \leq u+1} \bigcup_{Y_{i} >_{\F{G}}^* Z} Z}$ is $\mathtt{AST}$ as well.
  The conditions for~\eqref{Composing AST} are clearly satisfied, as we use the reflexive, transitive closure $>_{\F{G}}^*$ of the direct successor relation in both $\bigcup_{1 \leq i \leq u} \bigcup_{Y_{i} >_{\F{G}}^* Z} Z$ and $\bigcup_{Y_{u+1} >_{\F{G}}^* Z} Z$.

  Now we have shown that $\overline{\bigcup_{1 \leq i \leq m} \bigcup_{Y_i >_{\F{G}}^* Z} Z}$ is $\mathtt{AST}$\@.
  We know that $\overline{X}$ is $\mathtt{AST}$ by our assumption and that $\overline{\bigcup_{1 \leq i \leq m} \bigcup_{Y_i >_{\F{G}}^* Z} Z}$ is $\mathtt{AST}$\@.
  Hence, by~\eqref{Composing AST} we obtain that $\overline{\bigcup_{X >_{\F{G}}^* Y} Y}$ $\mathtt{AST}$\@.
  Again, the conditions of~\eqref{Composing AST} are satisfied, since $X$ is strictly greater w.r.t.\ $>_{\F{G}}^+$ than all $Z$ with $Y_i >_{\F{G}}^* Z$ for some $1 \leq i \leq m$.

  \medskip

  \noindent
  \textbf{\underline{4.
      $\PP$ is $\mathtt{AST}$\@.}}

  \noindent
  In~\eqref{SCC induction} we have shown that $\overline{\bigcup_{X >_{\F{G}}^* Y} Y}$ for every $X \in \F{W}$ is $\mathtt{AST}$\@.
  Let $X_1, \ldots, X_m\linebreak
    \in \F{W}$ be the maximal elements of $\F{W}$ w.r.t.\ $>_{\F{G}}$. By induction, one can prove that $\overline{\bigcup_{1 \leq i \leq u} \bigcup_{X_i >_{\F{G}}^* Y}
      Y}$ is $\mathtt{AST}$ for all $1 \leq u \leq m$ by~\eqref{Composing AST}, analogous to the previous induction. Again, the conditions of~\eqref{Composing AST} are satisfied as we use the reflexive, transitive closure of $>_{\F{G}}$. In the end, we know that $\overline{\bigcup_{1 \leq i \leq m} \bigcup_{X_i >_{\F{G}}^* Y} Y} = \PP$ is $\mathtt{AST}$ and this ends the proof.
\end{myproof}

\ProbDepGraphProcInnermost*

\begin{myproof}
  Let $\overline{X} = X \cup \flat(\PP \setminus X)$ for $X \subseteq \PP$.
  \smallskip

  \noindent
  \underline{\emph{Completeness:}} By \Cref{thm:proc-complete}.
  \medskip

  \noindent
  \underline{\emph{Soundness:}}
  The proof is analogous to the proof for $\mathtt{AST}$ but uses the innermost dependency graph $\F{G}$ instead.
  Only Step 1.\ and 2.\ change slightly due to innermost rewriting.

  \underline{Step 1.\ } Reusing the notation from the proof for $\mathtt{AST}$, existence of a node $v \in \aV$ in $\F{T}$ that is not the root and where we introduce new annotations leads to a term $t' \trianglelefteq_{\sharp} t_v$ such that $t' = s \sigma' \in \ANF_{\PP}$ for some substitution $\sigma'$ and the ADP $s \to \ldots \in X$ where $s \sigma' \in \ANF_{\PP}$.
  The first rewrite step at the root is $s \theta \itored{}{}{\overline{X}} \{p_1:r_1 \theta, \ldots, p_k:r_k \theta\}$.
  Therefore, we must have a $1 \leq j \leq k$ and a $t'' \trianglelefteq_{\sharp} r_j$ such that $t''^\sharp \theta \ito_{\nonprob(\PP)}^* s^\sharp \sigma'$ and $s \theta, s \sigma' \in \ANF_{\PP}$, which means that there must be a self-loop for the only ADP in $X$, which is the desired contradiction.

  \underline{Step 2.\ } The construction in this step now considers innermost CTs.
  Hence, the definition of $\Junk_{\hat{X}}(t'_v)$ uses the innermost dependency graph and innermost rewriting.
  To be precise, we define $\pi \in \Junk_{\hat{X}}(t'_v)$:$\Leftrightarrow$ there is no $B \in \F{W}$ with $B >_{\F{G}}^* X$ for some $X \in \hat{X}$ such that there is an ADP $\ell \to \{p_1: r_1, \ldots, p_k: r_k\}^{m} \in B$, and a substitution $\sigma$ with $\annoEps(t_v'|_{\pi}) \ito_{\nonprob(\PP)}^* \ell^\sharp \sigma$ and $\ell^\sharp \sigma \in \ANF_{\nonprob(\PP)}$.
  Due to the definition of the innermost dependency graph, we can still perform each rewrite step from $\F{T}'$ analogously in $\F{T}''$ with $\aV_1 \cap \aV^{\F{T}'} \subseteq \aV^{\F{T}''}$.
\end{myproof}

\UsableTermsProc*

\begin{myproof}

  \noindent
  \underline{\emph{Completeness:}} By \Cref{thm:proc-complete}.
  \medskip

  \noindent
  \underline{\emph{Soundness:}}
  Let $\PP$ be not $\mathtt{AST}$.
  Then by \Cref{lemma:starting} there exists a $\PP$-CT $\F{T} = (V,E,L)$ that converges with probability $<1$ whose root is labeled with $(1: t)$ and $\posT(t) = \{\varepsilon\}$.
  We will now create a $\mathcal{T}_\mathtt{UT}(\PP)$-CT $\F{T}' = (V,E,L')$, with the same underlying tree structure, and an adjusted labeling such that $p_v^{\F{T}} = p_v^{\F{T}'}$ for all $v \in V$.
  Hence, we get $|\F{T}'| = |\F{T}| <1$, and thus $\mathcal{T}_\mathtt{UT}(\PP)$ is not $\mathtt{AST}$ either.

  We construct the new labeling $L'$ for the $\mathcal{T}_\mathtt{UT}(\PP)$-CT $\F{T}'$ recursively.
  Let $W \subseteq V$ be the set of nodes where we have already defined the labeling $L'$.
  During our construction, we ensure that for every node $v \in W$ we have
  \begin{equation}\label{usable-terms-soundness-induction-hypothesis}
    t_v \flateq t'_v \text{ and } \posT(t_v) \setminus \Junk(t_v) \subseteq \posT(t'_v).
  \end{equation}
  Here, for any term $t_v$, let $\Junk(t_v)$ be the set of positions that can never be used for a rewrite step with an ADP that contains annotations.
  To be precise, we define $\pi \in \Junk(t_v)$:$\Leftrightarrow$ there is no ADP $\ell \to \{p_1: r_1, \ldots, p_k: r_k\}^{m} \in \PP$ with annotations and no substitution $\sigma$ such that $\annoEps(t_v|_{\pi}) \to_{\nonprob(\PP)}^* \ell^\sharp \sigma$.

  We start with the same term $t$ at the root.
  Here, our property~\eqref{usable-terms-soundness-induction-hypothesis} is clearly satisfied.
  As long as there is still an inner node $v \in W$ such that its successors are not contained in $W$, we do the following.
  Let $vE = \{w_1, \ldots, w_k\}$ be the set of its successors.
  We need to define the terms for the nodes $w_1, \ldots, w_k$ in $\F{T}'$.
  Since $v$ is not a leaf and $\F{T}$ is a $\PP$-CT, we have $t_v \tored{}{}{\PP} \{\tfrac{p_{w_1}}{p_v}:t_{w_1}, \ldots, \tfrac{p_{w_k}}{p_v}:t_{w_k}\}$.
  If we performed a step with $\tored{}{}{\PP}$ using the ADP $\ell \to \{ p_1: r_1, \ldots, p_k:r_k\}^{m}$, the VRF $(\varphi_j)_{1 \leq j \leq k}$, the position $\pi$, and the substitution $\sigma$ in $\F{T}$, then we can use the ADP $\ell \!\to\! \{ p_1: \sharp_{\Delta_{\PP}(r_1)}(r_1), \ldots, p_k:\sharp_{\Delta_{\PP}(r_k)}(r_k)\}^{m}$ with the same VRF $(\varphi_j)_{1 \leq j \leq k}$, the same position $\pi$, and the same substitution $\sigma$.
  Now, we directly get $t_{w_j} \flateq t'_{w_j}$ for all $1 \leq j \leq k$.
  To prove $\posT(t_{w_j}) \setminus \Junk(t_{w_j}) \subseteq \posT(t'_{w_j})$, note that if $\pi \in \posT(t_v) \cap \Junk(t_v)$, then $\ell \to \{ p_1: r_1, \ldots, p_k:r_k\}^{m}$ contains no annotations by definition of $\Junk(t_v)$.
  Therefore, it does not matter whether we rewrite with case $(\mathbf{at})$ or $(\mathbf{nt})$ ($(\mathbf{af})$ or $(\mathbf{nf})$).
  Otherwise, if $\pi \in \posT(t_v) \setminus \Junk(t_v)$, then the original rule contains the same terms with possibly more annotations, but all missing annotations are in $\Junk(t_v)$ by definition of $\sharp_{\Delta_{\PP}(r_j)}(r_j)$.
  Thus, we get $\posT(t_{w_j}) \setminus \Junk(t_{w_j}) \subseteq \posT(t'_{w_j})$ for all $1 \leq j \leq k$.

  Finally, in order to see that $\F{T}'$ is indeed a $\mathcal{T}_\mathtt{UT}(\PP)$-CT, we have to make sure that on every infinite path there are still an infinite number of rewrite steps with Case $(\mathbf{at})$ or $(\mathbf{af})$.
  Assume otherwise, i.e., that there exists an infinite path with no such rewrite steps.
  The same path exists in $\F{T}$, which is a $\PP$-CT, so there are infinitely many rewrite steps with Case $(\mathbf{at})$ or $(\mathbf{af})$ in $\F{T}$.
  This means that every rewrite step with Case $(\mathbf{at})$ or $(\mathbf{af})$ at a node $v$ in this path is performed at a position in $\Junk(t_{w_j})$.
  But since such rewrite steps decrease the number of annotations in the term, we cannot have infinitely many of them without rewriting at a position $\posT(t_v) \setminus \Junk(t_v)$, which is a contradiction.
\end{myproof}

\UsableTermsProcInnermost*

\begin{myproof}
  \smallskip

  \noindent
  \underline{\emph{Completeness:}} By \Cref{thm:proc-complete}.
  \medskip

  \noindent
  \underline{\emph{Soundness:}}
  We do the same construction as for $\mathtt{AST}$ but the definition of $\Junk(t_v)$ changes according to innermost rewriting.
  Here, we define $\pi \in \Junk(t_v)$:$\Leftrightarrow$ there is no ADP $\ell \to \{p_1: r_1, \ldots, p_k: r_k\}^{m} \in \PP$ with annotations and no substitution $\sigma$ such that $\annoEps(t_v|_{\pi}) \ito_{\nonprob(\PP)}^* \ell^\sharp \sigma$ and $\ell^\sharp \sigma \in \ANF_{\PP}$.
\end{myproof}

\UsableRulesProcInnermost*

\begin{myproof}
  Let $\overline{\PP} = \urules(\PP) \cup \{\ell \to \mu^{\tfalse} \mid \ell \to \mu^{m} \in \PP \setminus \urules(\PP)\}$.
  \smallskip

  \noindent
  \underline{\emph{Completeness:}} By \Cref{thm:proc-complete}.

  \smallskip

  \noindent
  \underline{\emph{Soundness:}} Assume that $\PP$ is not $\mathtt{iAST}$.
  Then by \Cref{lemma:starting} there exists a $\PP$-CT $\F{T} = (V,E,L)$ that converges with probability $<1$ whose root is labeled with $(1: t)$ and $\flat(t) = s \theta \in \ANF_{\PP}$ for a substitution $\theta$ and an ADP $s \to \ldots \in \PP$, and $\posT(t) = \{\varepsilon\}$.
  By the definition of usable rules, as in the non-probabilistic case, rules $\ell \to \mu \in \PP$ that are not usable (i.e., $\ell \to \mu \not\in \overline{\PP}$) will never be used below an annotated symbol in such a $\PP$-CT.
  Hence, we can also view $\F{T}$ as a $\overline{\PP}$-CT that converges with probability $<1$ and thus $\overline{\PP}$ is not $\mathtt{iAST}$.
\end{myproof}

The proof for the reduction pair processor is split into two parts.
First, we consider the technical part of the proof, and then we conclude the soundness and completeness of the processor.
We use a slightly more complicated lemma here instead of \Cref{lemma:prob-RPP} presented
in \Cref{The Probabilistic ADP Framework}.
While \Cref{lemma:prob-RPP}  imposed conditions on the rewrite relation $\tored{}{}{\PP}$,
we only impose conditions on the ADPs and lift them to $\tored{}{}{\PP}$-steps in the
proof of the following lemma.
Moreover, in \Cref{lemma:prob-RPP}, we considered the case where one term in the support
decreases 
strictly in every rewrite step. Instead, we now want to regard
CTs where such ``strictly decreasing''  ADPs are used
infinitely often in each infinite path of the CT.

Again, let $\mathcal{A}_{sum}^{\sharp}(r) = \sum_{t \trianglelefteq_{\sharp} r} \AA(t^\sharp)$.
Moreover, we use the \emph{prefix order}
($\pi \leq \tau \Leftrightarrow \text{ there exists } \chi \in \IN^* \text{ such that } \pi.\chi = \tau$) to compare positions.

\begin{lem}[Proving $\mathtt{AST}$ and $\mathtt{iAST}$ on CTs with Algebras]
  \label{lemma:prob-RPP-CT-lemma}
  Let  
    $(\AA,\succ)$ be a weakly monotonic, $\NN$-collapsible, barycentric $\Sigma^\sharp$-algebra.
  Let $\PP = \PP_{\succcurlyeq} \uplus \PP_{\succ}$ with $\PP_{\succ} \neq \emptyset$ where:
  {\small
  \[\begin{array}{cclll}
      (1) &     & \forall \ell \to \mu^{m} \in \PP                                    & : & \mathcal{A}_{sum}^{\sharp}(\ell^\sharp) \succcurlyeq \IE(\mathcal{A}_{sum}^{\sharp}(\mu))     \\
      (2) & (a) & \forall \ell \to \mu^{m} \in \PP_{\succ} : \exists r \in \Supp(\mu) & : & \mathcal{A}_{sum}^{\sharp}(\ell^\sharp) \succ \mathcal{A}_{sum}^{\sharp}(r)                   \\
          & (b) & \text{If } m = \ttrue,\text{ then we additionally have }            & : & \mathcal{A}(\ell) \succcurlyeq \mathcal{A}(\flat(r)) \\
      (3) &     & \forall \ell \to \mu^{\ttrue} \in \PP                               & : &
      \mathcal{A}(\ell) \succcurlyeq \IE(\mathcal{A}(\flat(\mu)))\!
    \end{array}\]}

  \noindent
  Let $\F{T} = (V,E,L)$ be a $\PP$-CT.
  We can partition $\aV = \aV_{\succcurlyeq} \uplus \aV_{\succ}$ according to $\PP = \PP_{\succcurlyeq} \uplus \PP_{\succ}$.
  If
  \begin{itemize}
    \item[\normalfont{(+)}] every infinite path in $\F{T}$ has an infinite number of $\aV_{\succ}$ nodes,
  \end{itemize}
  then $|\F{T}| = 1$.
\end{lem}

\begin{myproof}
  This proof uses the proof idea for $\mathtt{AST}$ from~\cite{mciver2017new}.
  For a $\PP$-CT $\F{T}$ that satisfies (+), the core steps of the proof are the following:
  \begin{enumerate}
    \item[I.] We extend the conditions (1), (2), and (3) to rewrite steps instead of just rules (and thus, to edges of a CT).
    \item[II.] We create a CT $\F{T}^{\leq N}$ for any $N \in \IN$.
    \item[III.] We prove that $|\F{T}^{\leq N}| \geq p_{min}^{N}$ for any $N \in \IN$.
    \item[IV.] We prove that $|\F{T}^{\leq N}|=1$ for any $N \in \IN$.
    \item[V.] Finally, we prove that $|\F{T}|=1$.
  \end{enumerate}
  Here, let $p_{min}>0$ be the minimal probability that occurs in the ADPs from $\PP$.
  As $\PP$ consists of only finitely many ADPs and all occurring multi-distributions are finite as well, this minimum is well defined.
  We have $p_{min}^{N} = \underbrace{p_{min} \cdot \ldots \cdot p_{min}}_\text{$N$ times}$.

  \medskip

  \noindent
  \textbf{\underline{I.
      We extend the conditions to rewrite steps instead of just rules}}

  \noindent
  We show that the conditions (1), (2), and (3) of the lemma extend to rewrite steps instead of just rules:
  \begin{enumerate}
    \item[(a)] If $s \to \{ p_1:t_1, \ldots, p_k:t_k \}$ using a rewrite rule $\ell \to \{ p_1:r_1, \ldots, p_k:r_k \}$\linebreak
      with $\AA(\ell) \succcurlyeq \AA(r_j)$ for some $1 \leq j \leq k$, then we have $\AA(s) \succcurlyeq \AA(t_j)$.
    \item[(b)] If $a \tored{}{}{\PP} \{ p_1:b_1, \ldots, p_k:b_k \}$ using $\ell \to \{ p_1:r_1, \ldots, p_k:r_k \}^{m} \in \PP_{\succ}$ at a position $\pi \in \posT(a)$, then $\mathcal{A}_{sum}^{\sharp}(a) \succ \mathcal{A}_{sum}^{\sharp}(b_j)$ for some $1 \leq j \leq k$.
    \item[(c)] If $s \to \{ p_1:t_1, \ldots, p_k:t_k \}$ using a rewrite rule $\ell \to \{ p_1:r_1, \ldots, p_k:r_k \}$\linebreak
      with $\AA(\ell) \succcurlyeq \sum_{1 \leq j \leq k} p_j \cdot \AA(r_j)$, then $\AA(s) \succcurlyeq \sum_{1 \leq j \leq k} p_j \cdot \AA(t_j)$.
    \item[(d)] If $a \tored{}{}{\PP} \{ p_1:b_1, \ldots, p_k:b_k \}$ using $\ell \to \{ p_1:r_1, \ldots, p_k:r_k \}^{m} \in \PP$, then $\mathcal{A}_{sum}^{\sharp}(a) \succcurlyeq \sum_{1 \leq j \leq k} p_j \cdot \mathcal{A}_{sum}^{\sharp}(b_j)$.
  \end{enumerate}

  \noindent
  \underline{Proofs:}

  \begin{itemize}
    \item[(a)] In this case, there exist a rule $\ell \to \{ p_1:r_1, \ldots, p_k:r_k \}$ with $\AA(\ell) \succcurlyeq \AA(r_j)$ for some $1 \leq j \leq k$, a substitution $\sigma$, and a position $\pi$ of $s$ such that $s|_\pi =\ell\sigma$ and $t_h = s[r_h \sigma]_\pi$ for all $1 \leq h \leq k$.

      We perform structural induction on $\pi$.
      So in the induction base, let $\pi = \varepsilon$.
      Hence, we have $s = \ell\sigma \to \{ p_1:r_1 \sigma, \ldots, p_k:r_k \sigma\}$.
      By assumption, we have $\AA(\ell) \succcurlyeq \AA(r_j)$ for some $1 \leq j \leq k$.
      As these inequations hold for all instantiations of the occurring variables, for $t_j = r_j\sigma$ we have
      \[ \AA(s) = \AA(\ell\sigma) \succcurlyeq \AA(r_j\sigma) = \AA(t_j). \]

      In the induction step, we have $\pi = i.\pi'$, $s = f(s_1,\ldots,s_i,\ldots,s_n)$, $f \in \Sigma^\sharp$, $s_i \to \{ p_1:t_{i,1}, \ldots, p_k:t_{i,k} \}$, and $t_j = f(s_1,\ldots,t_{i,j},\ldots,s_n)$ with $t_{i,j} = s_i[r_j\sigma]_{\pi'}$ for all $1 \leq j \leq k$.
      Then by the induction hypothesis we have $\AA(s_i) \succcurlyeq \AA(t_{i,j})$.
      For $t_j = f(s_1,\ldots,t_{i,j},\ldots,s_n)$ we obtain
      \[
        \begin{array}{l@{\;}l@{\;}l}
          
          \AA(s) & =            & \AA(f(s_1,\ldots,s_i,\ldots,s_n))                                                                                    \\
                 & =            & f_{\AA}(\AA(s_1),\ldots,\AA(s_i),\ldots,\AA(s_n))                                                                    \\
                 & \succcurlyeq & f_{\AA}(\AA(s_1),\ldots,\AA(t_{i,j}),\ldots,\AA(s_n))                                                                \\
                 &              & \hspace*{1cm} \text{ \textcolor{blue}{(by weak monotonicity of $f_{\AA}$ and $\AA(s_i) \succcurlyeq \AA(t_{i,j})$)}} \\
                 & =            & \AA(f(s_1,\ldots,t_{i,j},\ldots,s_n))                                                                                \\
                 & =            & \AA(t_j).
        \end{array}
      \]

    \item[(b)] In this case, there exist an ADP $\ell \to \{ p_1:r_1, \ldots, p_k:r_k \}^{m} \in \PP_{\succ}$, a VRF $(\varphi_j)_{1 \leq j \leq k}$, a substitution $\sigma$, and position $\pi \in \posT(a)$ with $\flat(a|_{\pi}) = \ell \sigma$ and $b_j \flateq a[r_j \sigma]_{\pi}$.
      First, assume that $m = \ttrue$.
      Let $I_1 = \{\tau \in \posT(a) \mid \tau < \pi\}$ be the set of positions of all annotations strictly above $\pi$, $I_2 = \{\tau \in \posT(a) \mid \gamma \in \pos_{\VSet}(\ell), \pi < \tau \leq \pi.\gamma\}$ be the set of positions of all annotations inside the left-hand side $\ell$ of the used redex $\ell \sigma$ (but not on the root of $\ell$), $I_3 = \{\tau \in \posT(a) \mid \gamma \in \pos_{\VSet}(\ell), \pi.\gamma < \tau\}$ be the set of positions of all annotations inside the substitution, and let $I_4 = \{\tau \in \posT(a) \mid \tau \bot \pi\}$ be the set of positions of all annotations orthogonal to $\pi$.
      Furthermore, for each $\tau \in I_1$ let $\kappa_\tau$ be the position such that $\tau.\kappa_\tau = \pi$, and for each $\tau \in I_3$ let $\gamma_\tau$ and $\rho_\tau$ be the positions such that $\gamma_\tau \in \pos_{\VSet}(\ell)$ and $\pi.\gamma_\tau.\rho_\tau = \tau$.
      By Requirement (2), there exists a $1 \leq j \leq k$ with $\mathcal{A}(\ell^\sharp) = \mathcal{A}_{sum}^{\sharp}(\ell^\sharp) \succ \mathcal{A}_{sum}^{\sharp}(r_j) = \sum_{t \trianglelefteq_{\sharp} r_j} \AA(t^\sharp)$ and, additionally, $\AA(\ell) \succcurlyeq \AA(\flat(r_j))$ since $m = \ttrue$.
      As these inequations hold for all instantiations of the occurring variables, we have

        {\scriptsize
          \begin{longtable}{L@{\;}L}
                         &\mathcal{A}_{sum}^{\sharp}(a)\\[6pt]
            =            & \sum_{t \trianglelefteq_{\sharp} a} \AA(t^\sharp) \\[6pt]
            =            & \AA(\annoEps(a|_{\pi})) + \sum_{\tau \in I_1} \AA(\annoEps(a|_{\tau})) + \sum_{\tau \in I_2} \AA(\annoEps(a|_{\tau})) \\
                         & \; + \sum_{\tau \in I_3} \AA(\annoEps(a|_{\tau})) + \sum_{\tau \in I_4} \AA(\annoEps(a|_{\tau})) \\[6pt]
            \succcurlyeq & \AA(\annoEps(a|_{\pi})) + \sum_{\tau \in I_1} \AA(\annoEps(a|_{\tau})) \\
                         & \; + \sum_{\tau \in I_3} \AA(\annoEps(a|_{\tau})) + \sum_{\tau \in I_4} \AA(\annoEps(a|_{\tau})) \\
                         & \hspace*{.5cm} \text{\textcolor{blue}{(removed $I_2$)}} \\[6pt]
            =            & \AA(\annoEps(\ell) \sigma) + \sum_{\tau \in I_1} \AA(\annoEps(a|_{\tau})) \\
                         & \; + \sum_{\tau \in I_3} \AA(\annoEps(a|_{\tau})) + \sum_{\tau \in I_4} \AA(\annoEps(a|_{\tau})) \\
                         & \hspace*{.5cm} \text{\textcolor{blue}{(as $\annoEps(a|_{\pi}) = \annoEps(\ell) \sigma$)}} \\[6pt]
            \succ        & \sum_{t \trianglelefteq_{\sharp} r_j} \AA(\annoEps(t)\sigma) + \sum_{\tau \in I_1} \AA(\annoEps(a|_{\tau})) \\
                         & \; + \sum_{\tau \in I_3} \AA(\annoEps(a|_{\tau})) + \sum_{\tau \in I_4} \AA(\annoEps(a|_{\tau}))  \\
                         & \hspace*{.5cm} \text{\textcolor{blue}{(as $\AA(\annoEps(\ell)) \succ \sum_{t \trianglelefteq_{\sharp} r_j} \AA(\annoEps(t))$, hence $\AA(\annoEps(\ell)\sigma) \succ \sum_{t \trianglelefteq_{\sharp} r_j} \AA(\annoEps(t)\sigma)$)}} \\[6pt]
            \succcurlyeq & \sum_{t \trianglelefteq_{\sharp} r_j\sigma} \AA(\annoEps(t)) + \sum_{\tau \in I_1} \AA(\annoEps(a|_{\tau}[r_j \sigma]_{\kappa_{\tau}}))  \\
                         & \; + \sum_{\tau \in I_3} \AA(\annoEps(a|_{\tau})) + \sum_{\tau \in I_4} \AA(\annoEps(a|_{\tau})) \\
                         & \hspace*{.5cm} \text{\textcolor{blue}{(by $\AA(\ell) \succcurlyeq \AA(r_j)$ and (a))}} \\[6pt]
            \succcurlyeq & \sum_{t \trianglelefteq_{\sharp} r_j\sigma} \AA(\annoEps(t)) + \sum_{\tau \in I_1} \AA(\annoEps(a|_{\tau}[r_j \sigma]_{\kappa_{\tau}})) \\
                         & \; + \sum_{\tau \in I_3, \varphi_j(\gamma_\tau) \neq \bot} \AA(\annoEps(b_j|_{\pi.\varphi_j(\gamma_\tau).\rho_\tau})) + \sum_{\tau \in I_4} \AA(\annoEps(a|_{\tau})) \\
                         & \hspace*{.5cm} \text{\textcolor{blue}{(moving $\tau = \pi.\gamma_\tau.\rho_\tau \in I_3$ via the VRF)}} \\[6pt]
            =            & \sum_{t \trianglelefteq_{\sharp} b_j} \AA(t^\sharp) \\[6pt]
            =            & \mathcal{A}_{sum}^{\sharp}(b_j)\!
          \end{longtable}
        }

      \noindent
      In the case $m = \tfalse$, we additionally remove $\sum_{\tau \in I_1}
      \AA(\annoEps(a|_{\tau}[r_j \sigma]_{\kappa_\tau}))$, and the inequation remains correct.

    \item[(c)] In this case, there exist a rule $\ell \to \{ p_1:r_1, \ldots, p_k:r_k \}$ with $\AA(\ell) \succcurlyeq \sum_{1 \leq j \leq k} p_j \cdot \AA(r_j)$, a substitution $\sigma$, and a position $\pi$ of $s$ such that $s|_\pi =\ell\sigma$, and $t_j = s[r_j \sigma]_\pi$ for all $1 \leq j \leq k$.

      We perform structural induction on $\pi$.
      So in the induction base $\pi = \varepsilon$ we have $s = \ell\sigma \to \{ p_1:r_1\sigma, \ldots, p_k:r_k\sigma \}$.
      As $\AA(\ell) \succcurlyeq \sum_{1 \leq j \leq k} p_j \cdot \AA(r_j)$ holds for all instantiations of the occurring variables, for $t_j = r_j\sigma$ we obtain
      \[
        \AA(s) \;=\; \AA(\ell\sigma) \;\succcurlyeq\;\sum_{1 \leq j \leq k} p_j \cdot \AA(r_j\sigma) \;=\; \sum_{1 \leq j \leq k} p_j \cdot \AA(t_j).
      \]

      In the induction step, we have $\pi = i.\pi'$, $s = f(s_1,\ldots,s_i,\ldots,s_n)$, $s_i \to \{ p_1:t_{i,1}, \ldots, p_k:t_{i,k} \}$, and $t_j = f(s_1,\ldots,t_{i,j},\ldots,s_n)$ with $t_{i,j} = s_i[r_j\sigma]_{\pi'}$ for all $1 \leq j \leq k$.
      Then by the induction hypothesis we have $\AA(s_i) \succcurlyeq \sum_{1 \leq j \leq k} p_j \cdot \AA(t_{i,j})$.
      Thus, we have

      \begin{longtable}{L@{\;}L@{\;}L}
        \AA(s) & =            & \AA(f(s_1,\ldots,s_i,\ldots,s_n))                                                                                                         \\
               & =            & f_{\AA}(\AA(s_1),\ldots,\AA(s_i),\ldots,\AA(s_n))                                                                                         \\
               & \succcurlyeq & f_{\AA}(\AA(s_1),\ldots,\sum_{1 \leq j \leq k} p_j \cdot \AA(t_{i,j}),\ldots,\AA(s_n))                                                    \\
               &              & \; \text{\textcolor{blue}{(by weak monotonicity of $f_{\AA}$ and $\AA(s_i) \succcurlyeq \sum_{1 \leq j \leq k} p_j \cdot \AA(t_{i,j})$)}} \\
               & \succcurlyeq & \sum_{1 \leq j \leq k} p_j \cdot f_{\AA}(\AA(s_1),\ldots,\AA(t_{i,j}),\ldots,\AA(s_n))                                                    \\
               &              & \; \text{\textcolor{blue}{(as $f_{\AA}$ is concave)}}                                                                                     \\
               & =            & \sum_{1 \leq j \leq k} p_j \cdot \AA(f(s_1,\ldots,t_{i,j},\ldots,s_n))                                                                    \\
               & =            & \sum_{1 \leq j \leq k} p_j \cdot \AA(t_j).
      \end{longtable}

    \item[(d)] In this case, there exist an ADP $\ell \to \{ p_1:r_1, \ldots, p_k:r_k \}^{m} \in \PP$, a substitution $\sigma$, and position $\pi$ with $\flat(a|_{\pi}) = \ell \sigma$ and $b_j \flateq a[r_j \sigma]_{\pi}$.
      First, assume that $m = \ttrue$ and $\pi \in \posT(a)$.
      Let $I_1 = \{\tau \in \posT(a) \mid \tau < \pi\}$ be the set of positions of all annotations strictly above $\pi$, $I_2 = \{\tau \in \posT(a) \mid \gamma \in \pos_{\VSet}(\ell), \pi < \tau \leq \pi.\gamma\}$ be the set of positions of all annotations inside the left-hand side $\ell$ of the used redex $\ell \sigma$ (but not on the root of $\ell$), $I_3 = \{\tau \in \posT(a) \mid \gamma \in \pos_{\VSet}(\ell), \pi.\gamma < \tau\}$ be the set of positions of all annotations inside the substitution, and let $I_4 = \{\tau \in \posT(a) \mid \tau \bot \pi\}$ be the set of positions of all annotations orthogonal to $\pi$.
      Furthermore, for each $\tau \in I_1$ let $\kappa_\tau$ be the position such that $\tau.\kappa_\tau = \pi$, and for each $\tau \in I_3$ let $\gamma_\tau$ and $\rho_\tau$ be the positions such that $\gamma_\tau \in \pos_{\VSet}(\ell)$ and $\pi.\gamma_\tau.\rho_\tau = \tau$.
      By Requirement (1), we have $\mathcal{A}(\annoEps(\ell)) = \mathcal{A}_{sum}^{\sharp}(\annoEps(\ell)) \succcurlyeq \sum_{1 \leq j \leq k} p_j \cdot \sum_{t \trianglelefteq_{\sharp} r_j} \AA(\annoEps(t))$ and by (3) we have $\AA(\ell) \succcurlyeq \sum_{1 \leq j \leq k} p_j \cdot \AA(\flat(r_j))$.
      As these inequations hold for all instantiations of the occurring variables, we have

        {\scriptsize
          \begin{longtable}{L@{\;}L}
            & \mathcal{A}_{sum}^{\sharp}(a)\\[6pt]
            =            & \sum_{t \trianglelefteq_{\sharp} a} \AA(t^\sharp) \\[6pt]
            =            & \AA(\annoEps(a|_{\pi})) + \sum_{\tau \in I_1} \AA(\annoEps(a|_{\tau})) + \sum_{\tau \in I_2} \AA(\annoEps(a|_{\tau})) \\
                         & \; + \sum_{\tau \in I_3} \AA(\annoEps(a|_{\tau})) + \sum_{\tau \in I_4} \AA(\annoEps(a|_{\tau})) \\[6pt]
            \succcurlyeq & \AA(\annoEps(a|_{\pi})) + \sum_{\tau \in I_1} \AA(\annoEps(a|_{\tau})) \\
                         & \; + \sum_{\tau \in I_3} \AA(\annoEps(a|_{\tau})) + \sum_{\tau \in I_4} \AA(\annoEps(a|_{\tau})) \\
                         & \hspace*{2cm} \text{\textcolor{blue}{(removed $I_2$)}} \\[6pt]
            =            & \AA(\annoEps(\ell)\sigma) + \sum_{\tau \in I_1} \AA(\annoEps(a|_{\tau})) \\
                         & \; + \sum_{\tau \in I_3} \AA(\annoEps(a|_{\tau})) + \sum_{\tau \in I_4} \AA(\annoEps(a|_{\tau})) \\
                         & \hspace*{2cm} \text{\textcolor{blue}{(as $a|_{\pi} = \annoEps(\ell) \sigma$)}} \\[6pt]
            \succcurlyeq & \sum_{1 \leq j \leq k} p_j \cdot \sum_{t \trianglelefteq_{\sharp} r_j \sigma} \AA(\annoEps(t)) + \sum_{\tau \in I_1} \AA(\annoEps(a|_{\tau})) \\
                         & \; + \sum_{\tau \in I_3} \AA(\annoEps(a|_{\tau})) + \sum_{\tau \in I_4} \AA(\annoEps(a|_{\tau})) \\
                         & \hspace*{2cm} \text{\textcolor{blue}{(by $\AA(\annoEps(\ell)) \succcurlyeq \sum_{1 \leq j \leq k} p_j \cdot \sum_{t \trianglelefteq_{\sharp} r_j} \AA(\annoEps(t))$,}} \\
                         & \hspace*{3cm} \text{\textcolor{blue}{hence $\AA(\annoEps(\ell)\sigma) \succcurlyeq \sum_{1 \leq j \leq k} p_j \cdot \sum_{t \trianglelefteq_{\sharp} r_j \sigma} \AA(\annoEps(t))$)}} \\[6pt]
            \succcurlyeq & \sum_{1 \leq j \leq k} p_j \cdot \sum_{t \trianglelefteq_{\sharp} r_j \sigma} \AA(\annoEps(t)) + \sum_{\tau \in I_1} \sum_{1 \leq j \leq k} p_j \cdot \AA(\annoEps(a|_{\tau}[r_j \sigma]_{\kappa_\tau})) \\
                         & + \sum_{\tau \in I_3} \AA(\annoEps(a|_{\tau})) + \sum_{\tau \in I_4} \AA(\annoEps(a|_{\tau}))   \\
                         & \hspace*{2cm} \text{\textcolor{blue}{(by $\AA(\ell) \succcurlyeq \sum_{1 \leq j \leq k} p_j \cdot \AA(r_j)$ and (c))}}  \\[6pt]
            =            & \sum_{1 \leq j \leq k} p_j \cdot \sum_{t \trianglelefteq_{\sharp} r_j \sigma} \AA(\annoEps(t)) + \sum_{1 \leq j \leq k} \sum_{\tau \in I_1} p_j \cdot \AA(\annoEps(a|_{\tau}[r_j \sigma]_{\kappa_\tau})) \\
                         & + \sum_{\tau \in I_3} \AA(\annoEps(a|_{\tau})) + \sum_{\tau \in I_4} \AA(\annoEps(a|_{\tau})) \\[6pt]
            =            & \sum_{1 \leq j \leq k} p_j \cdot \sum_{t \trianglelefteq_{\sharp} r_j \sigma} \AA(\annoEps(t)) + \sum_{1 \leq j \leq k} p_j \cdot \sum_{\tau \in I_1} \AA(\annoEps(a|_{\tau}[r_j \sigma]_{\kappa_\tau})) \\
                         & + \sum_{\tau \in I_3} \AA(\annoEps(a|_{\tau})) + \sum_{\tau \in I_4} \AA(\annoEps(a|_{\tau}))   \\[6pt]
            =            & \sum_{1 \leq j \leq k} p_j \cdot \sum_{t \trianglelefteq_{\sharp} r_j \sigma} \AA(\annoEps(t)) + \sum_{1 \leq j \leq k} p_j \cdot \sum_{\tau \in I_1} \AA(\annoEps(a|_{\tau}[r_j \sigma]_{\kappa_\tau})) \\
                         & + \sum_{1 \leq j \leq k} p_j \cdot \sum_{\tau \in I_3} \AA(\annoEps(a|_{\tau})) + \sum_{1 \leq j \leq k} p_j \cdot \sum_{\tau \in I_4} \AA(\annoEps(a|_{\tau})) \\[6pt]
            =            & \sum_{1 \leq j \leq k} p_j \cdot \big(\sum_{t \trianglelefteq_{\sharp} r_j \sigma} \AA(\annoEps(t)) + \sum_{\tau \in I_1} \AA(\annoEps(a|_{\tau}[r_j \sigma]_{\kappa_\tau})) \\
                         & \; + \sum_{\tau \in I_3} \AA(\annoEps(a|_{\tau})) + \sum_{\tau \in I_4} \AA(\annoEps(a|_{\tau}))\big) \\[6pt]
            \succcurlyeq & \sum_{1 \leq j \leq k} p_j \cdot \big(\sum_{t \trianglelefteq_{\sharp} r_j \sigma} \AA(\annoEps(t)) + \sum_{\tau \in I_1} \AA(\annoEps(a|_{\tau}[r_j \sigma]_{\kappa_\tau})) \\
                         & \; + \sum_{\tau \in I_3, \varphi_j(\gamma_\tau) \neq \bot} \AA(\annoEps(b_j|_{\pi.\varphi_j(\gamma_\tau).\rho_\tau})) + \sum_{\tau \in I_4} \AA(\annoEps(a|_{\tau}))\big) \\
                         & \hspace*{2cm} \text{\textcolor{blue}{(moving $\tau = \pi.\gamma_\tau.\rho_\tau \in I_3$ via the VRF)}}\\[6pt]
            =            & \sum_{1 \leq j \leq k} p_j \cdot \sum_{t \trianglelefteq_{\sharp} b_j} \AA(t^\sharp) \\[6pt]
            =            & \mathcal{A}_{sum}^{\sharp}(b_j) \!
          \end{longtable}
        }

      \noindent
      In the case $\pi \notin \posT(a)$, we need to remove $\AA(\annoEps(\ell)\sigma)$ as this annotated subterm does not exist in $a$, 
      and therefore also $\sum_{t \trianglelefteq_{\sharp} r_j \sigma} \AA(t^\sharp)$ in the end, leading to the same result.
      In the case $m = \tfalse$, we additionally remove $\sum_{i \in I_1} \AA(\annoEps(a|_{\tau}[r_j \sigma]_{\kappa_\tau}))$ in the end.
  \end{itemize}

  \noindent
  \textbf{\underline{II.
      We create a CT $\F{T}^{\leq N}$ for any $N \in \IN$}}

  \noindent
  Let $\F{T} = (V,E,L)$ be a $\PP$-CT that satisfies (+).
  Since $\AA$ is $\IN$-collapsible, there exists a concave embedding $g_{\AA \to \IN}: A \to \IN$.
  We define the \emph{tree value} of any node $v \in V$ in our CT by
  \[
    \tval: V \to \IN, \qquad v \mapsto
    \begin{cases}
      0,                                          & \text{ if } v \in \ctleaf \\
      g_{\AA \to \IN}(\mathcal{A}_{sum}^{\sharp}(t_v \delta_{\ta})) + 1, & \text{ otherwise }
    \end{cases}
  \]
  Remember that $t_v$ denotes the term in the labeling of the node $v$.
  The substitution $\delta_{\ta}$ maps every occurring variable to some constant $\ta$.
  By definition of $\succ$, we know that $\mathcal{A}_{sum}^{\sharp}(t) \succ \mathcal{A}_{sum}^{\sharp}(s)$ 
  implies $\mathcal{A}_{sum}^{\sharp}(t \delta_{\ta}) \succ \mathcal{A}_{sum}^{\sharp}(s \delta_{\ta})$.
  The substitution ensures that we are only working with ground terms, so that $\tval$ really maps each term to a natural number.
  Furthermore, note that we have $\tval(v) = 0$ if and only if $v$ is a leaf in $\F{T}$.

  For any $N \in \IN$, we create a modified tree $\F{T}^{\leq N}$, where we cut everything below a node $v$ of the tree with $\tval(v) \geq N+1$.
  Let $C = \ctleaf^{\F{T}^{\leq N}} \setminus \ctleaf^{\F{T}}$ be the set of all new leaves in $\F{T}^{\leq N}$ due to the cut.
  So for all $v \in C$ we have $\tval(v) \geq N+1$.

  Our goal is to prove that we have $|\F{T}| = 1$.
  First, we prove that $|\F{T}^{\leq N}| = 1$ for any $N \in \IN$.
  However, this does not yet prove that $|\F{T}| = 1$, which we will show afterwards.

  \medskip

  \noindent
  \textbf{\underline{III.
      We prove that $|\F{T}^{\leq N}| \geq p_{min}^{N}$ for any $N \in \IN$}}

  \noindent
  Note that in a rewrite step $a \tored{}{}{\PP} \{p_1:b_1, \ldots, p_k:b_k\}$ with Case $(\mathbf{nt})$ or $(\mathbf{nf})$ or in a rewrite step with an ADP from $\PP_{\succcurlyeq}$, we cannot guarantee that there exists a $b_j$ that is strictly decreasing in value.
  It is also possible that all values stay the same.
  Thus, there does not have to be a single witness path of length $N$ in $\F{T}^{\leq N}$ that shows termination with a probability of at least $p_{min}^{N}$.
  Instead, we have to use multiple witness paths of finite length to ensure that we have $|\F{T}^{\leq N}| \geq p_{min}^{N}$\footnote{$\IN$-collapsibility ensures that $p_{min}^{N} > 0$ is a lower bound on the termination probability. If one wants to adapt this theorem to more complex algebras (that are not $\IN$-collapsible), then one needs to find different ways to obtain a constant lower bound $> 0$ for every $N \in \IN$.}.
  We first explain how to find such a set of witness paths in a finite sub $\PP$-CT that starts with a node from $\PP_{\succ}$, and then we prove by induction that we have $|\F{T}^{\leq N}| \geq p_{min}^{N}$ in a second step.

  \medskip

  \noindent
  \textbf{\underline{III.1.
      We find witnesses in subtrees of $\F{T}^{\leq N}$}}

  \noindent
  In this part, we prove a first observation regarding the existence of certain witnesses that shows a guaranteed decrease of values.
  Let $\aV$ now refer to $\F{T}^{\leq N}$,

  \noindent
  \begin{minipage}{.55\textwidth}
    i.e., $\aV$ only contains inner nodes of $\F{T}^{\leq N}$ and thus, $\aV \cap C = \emptyset$.
    Moreover, let $\aV_{\succ}$ and $\aV_{\succcurlyeq}$ correspond to the rewrite steps at nodes in $\aV$ with $\PP_{\succ}$ and $\PP_{\succcurlyeq}$, respectively.
    For every $v \in \aV_{\succ}$, let $\F{T}_v$ be the subtree that starts at $v$ and where we cut every edge after the second node from $\aV_{\succ}$.
    This is illustrated in \Cref{Tx figure}, where an empty node stands for a leaf that already existed in $\F{T}^{\leq N}$ (i.e., it is in $\ctleaf^{\F{T}}$ or in $C$).
    Since $\F{T}$ satisfies (+) and is finitely branching, we know that $\F{T}_v$ must
  \end{minipage}
  \begin{minipage}{.45\textwidth}
    \centering
    \begin{tikzpicture}
      \tikzstyle{adam}=[circle,thick,draw=black!100,fill=white!100,minimum size=4mm] \tikzstyle{empty}=[circle,thick,minimum size=4mm]

      \node[adam, label=center:{\tiny $\aV_{\succ}$}] at (0, 0) (a) {};
      \node[adam, label=center:{\tiny $\aV_{\succcurlyeq}$}] at (2, -2) (b) {};
      \node[adam, label=center:{\tiny }] at (-2, -2) (c) {};
      \node[adam, label=center:{\tiny $\aV_{\succ}$}] at (0.75, -2) (d) {};
      \node[adam, label=center:{\tiny $\aV_{\succcurlyeq}$}] at (-0.75, -2) (e) {};
      \node[empty, label=center:{\small $\nV$}] at (0, -1) (middleA) {};

      \node[adam, label=center:{\tiny }] at (-1.5, -3.5) (nf1) {};
      \node[adam, label=center:{\tiny }] at (0, -3.5) (nf2) {};

      \node[adam, label=center:{\tiny $\aV_{\succ}$}] at (2.75, -3.5) (bb) {};
      \node[adam, label=center:{\tiny }] at (1.25, -3.5) (bc) {};
      \node[adam, label=center:{\tiny $\aV_{\succ}$}] at (2, -3.5) (bd) {};
      \node[empty, label=center:{\small $\nV$}] at (2, -2.8) (middleA) {};

      \node[empty, label=center:{\small $\nV$}] at (-0.75, -2.8) (middleA) {};

      \draw (a) edge[-] (b);
      \draw (b) edge[-] (d);
      \draw (d) edge[-] (e);
      \draw (e) edge[-] (c);
      \draw (a) edge[-] (c);

      \draw (b) edge[-] (bb);
      \draw (bb) edge[-] (bd);
      \draw (bd) edge[-] (bc);
      \draw (b) edge[-] (bc);

      \draw (e) -- (nf1) -- (nf2) -- (e);

      \begin{scope}[on background layer]
        \fill[green!20!white,on background layer] (0, 0) -- (-2, -2) -- (2, -2); \fill[green!20!white,on background layer] (2, -2) -- (1.25, -3.5) -- (2.75, -3.5); \fill[green!20!white,on background layer] (-0.75, -2) -- (-1.5, -3.5) -- (0, -3.5);
      \end{scope}
    \end{tikzpicture}
    \captionof{figure}{Subtree $\F{T}_v$}\label{Tx figure}
  \end{minipage}
  be finite.
  We want to prove that for such a tree $\F{T}_v$, we have a set of leaves (a set of witnesses) that show a certain decrease of values compared to the root value $\tval(v)$.
  To be precise, we want to prove that there exists a set $W^v \subseteq \ctleaf^{\F{T}_v}$ of leaves in $\F{T}_v$ with the following two properties:
  \begin{enumerate}
    \item[](W-1) For all $w \in W^v$ we have $\adtval(w) \leq \adtval(v) - 1$.
    \item[](W-2) $\sum_{w \in W^v} p_w^{\F{T}_v} \cdot p_{min}^{\adtval(w)} \geq p_{min}^{\adtval(v)}$.
  \end{enumerate}
  Here, we use an \emph{adjusted tree value} function $\adtval : V^{\F{T}^{\leq N}} \to \IN$, such that for every node $v \in V^{\F{T}^{\leq N}}$ we have
  \[
    \adtval(v) =
    \begin{cases}
      0,        & \text{if } v \text{ is a leaf in } \F{T}^{\leq N} \\
      \tval(v), & \text{otherwise}
    \end{cases}
  \]
  This is the same tree value function as before, except for the nodes in $C$.
  For all $v \in C$ we have $\tval(v) \geq N+1$ and $\adtval(v) = 0$.
  Again, we have $p_{min}^{n} = \underbrace{p_{min} \cdot \ldots \cdot p_{min}}_\text{$n$ times}$ for any natural number $n \in \IN$, e.g., $\adtval(v) \in \IN$.
  The first property (W-1) states that all of our witnesses have a strictly smaller value than the root.
  Furthermore, we have to be careful that the probabilities for our witnesses are not too low.
  The second property (W-2) states that the sum of all probabilities for the witnesses is still big enough.
  The additional $p_{min}^{\adtval(w)}$ is used to allow smaller probabilities for our witnesses if the value decrease is high enough.

  In order to show the existence of such a set $W^v$, we prove by induction on the height $H$ of $\F{T}_v$ that there exists a set of nodes $W^v_i$ such that for all $1 \leq i \leq H$ we have
  \begin{enumerate}
    \item[] (W-1!) For all $w \in W^v_i$ we have $\adtval(w) \leq \adtval(v) - 1$.
    \item[] (W-2!) $\sum_{w \in W^v_i} p_w^{\F{T}_v} \cdot p_{min}^{\adtval(w)} \geq p_{min}^{\adtval(v)}$.
    \item[] (W-3!) Every $w \in W^v_i$ is a leaf of $\F{T}_v$ before or at the $i$-th depth of $\F{T}_v$.
  \end{enumerate}
  Then in the end if $i = H$ is the height of the tree $\F{T}_v$, we get a set $W^v = W_H^v$ such that every node in $W_H^v$ is a leaf in $\F{T}_v$ (i.e., $W_H^v \subseteq \ctleaf^{\F{T}_v}$) and both (W-1) and (W-2) are satisfied.

  In the induction base, we consider depth $i = 1$ and look at the rewrite step at the root.
  The first edge represents a rewrite step with $\PP_{\succ}$ at a position $\pi \in \posT(t_v)$.
  Let $vE = \{w_1, \ldots, w_k\}$ be the set of the successors of $v$.
  We have $t_v \tored{}{}{\PP} \{p_1:t_{w_1}, \ldots, p_k:t_{w_k}\}$ using a rule from $\PP_{\succ}$.
  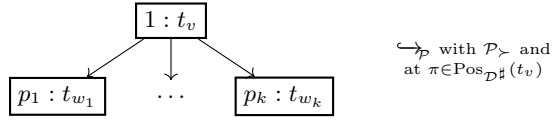
\begin{figure}[H]
    \centering \small
    \begin{tikzpicture}
      \tikzstyle{adam}=[rectangle,thick,draw=black!100,fill=white!100,minimum size=4mm] \tikzstyle{empty}=[rectangle,thick,minimum size=4mm]

      \node[adam] at (0, 0) (a) {$1:t_v$};
      \node[adam] at (-1.5, -1) (b) {$p_{1}:t_{w_1}$};
      \node[adam] at (1.5, -1) (c) {$p_{k}:t_{w_k}$};
      \node[empty] at (0, -1) (d) {$\ldots$};
      \node[empty] at (4, -0.5) (e) {$\substack{\tored{}{}{\PP} \text{ with } \PP_{\succ} \text{ and}\\
            \text{at } \pi \in \posT(t_v)}$};
      \node[empty] at (-4, -0.5) (f) {\textcolor{white}{$\substack{\tored{}{}{\PP} \text{ with } \PP_{\succ} \text{ and}\\
              \text{at } \pi \in \posT(t_v)}$}};

      \draw (a) edge[->] (b);
      \draw (a) edge[->] (c);
      \draw (a) edge[->] (d);
    \end{tikzpicture}
    \caption{Nodes in the induction base}
  \end{figure}
  Due to (b) there is a $1 \leq j \leq k$ with $\mathcal{A}_{sum}^{\sharp}(t_v) \succ \mathcal{A}_{sum}^{\sharp}(t_{w_j})$.
  This means $g_{\AA \to \IN}(\mathcal{A}_{sum}^{\sharp}(t_v \delta_{\ta})) \geq g_{\AA \to \IN}(\mathcal{A}_{sum}^{\sharp}(t_{w_j} \delta_{\ta})) + 1$, 
  and hence $\tval(v) \geq \tval(w_j) + 1$, which also implies that $\adtval(v) \geq \adtval(w_j) + 1$.
  Since $0 < p_{min} \leq 1$, we therefore have $p_{min}^{\adtval(v)} \leq p_{min}^{\adtval(w_j) + 1}$.
  Thus we can set $W_0^v = \{w_j\}$ and have (W-1!) satisfied, since $\adtval(v) \geq \adtval(w_j) + 1$ so $\adtval(w_j) \leq \adtval(v) - 1$.
  Property (W-3!) is clearly satisfied and (W-2!) holds as well since we have
  \[
  \mbox{\footnotesize
$\sum_{w \in W_1^v} p_{w}^{\F{T}_v} \cdot p_{min}^{\adtval(w)} = p_j \cdot p_{min}^{\adtval(w_j)} \!\stackrel{\bigl(\substack{
        p_j \geq p_{min} \land\\ \adtval(w_j) \leq \adtval(v) - 1}\bigr)}{\geq}\!\! p_{min} \cdot p_{min}^{\adtval(v) - 1} = p_{min}^{\adtval(v)}.$}
  \]

  In the induction step, we consider depth $i > 1$.
  Due to the induction hypothesis, there is a set $W_{i-1}^v$ that satisfies (W-1!), (W-2!), and (W-3!).
  For every node $w \in W_{i-1}^v$ that is not a leaf of $\F{T}_v$, we proceed as follows: Let $wE = \{w_1, \ldots, w_k\}$ be the set of its successors.
  We rewrite $t_w$ either with $\tored{}{}{\PP}$ and an ADP from $\PP_{\succcurlyeq}$ or with $\tored{}{}{\PP}$ and a position $\pi \notin \posT(t_w)$, which rewrites $t_w$ to a multi-distribution $\{p_1:t_{w_1}, \ldots,p_k:t_{w_k}\}$.
  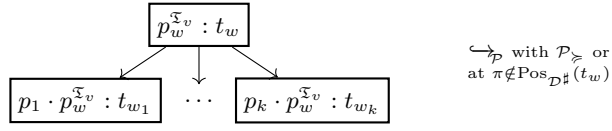
\begin{figure}[H]
    \centering \small \vspace*{-.2cm}
    \begin{tikzpicture}
      \tikzstyle{adam}=[rectangle,thick,draw=black!100,fill=white!100,minimum size=4mm] \tikzstyle{empty}=[rectangle,thick,minimum size=4mm]

      \node[adam] at (0, 0) (a) {$p_w^{\F{T}_v}:t_w$};
      \node[adam] at (-1.5, -1) (b) {$p_{1} \cdot p_w^{\F{T}_v}:t_{w_1}$};
      \node[adam] at (1.5, -1) (c) {$p_{k} \cdot p_w^{\F{T}_v}:t_{w_k}$};
      \node[empty] at (0, -1) (d) {$\ldots$};
      \node[empty] at (4.5, -0.5) (e) {$\substack{\tored{}{}{\PP} \text{ with } \PP_{\succcurlyeq} \text{ or}\\
            \text{at } \pi \notin \posT(t_w)}$};
      \node[empty] at (-4.5, -0.5) (f) {\textcolor{white}{$\substack{\tored{}{}{\PP} \text{ with } \PP_{\succcurlyeq} \text{ or}\\
              \text{at } \pi \notin \posT(t_w)}$}};

      \draw (a) edge[->] (b);
      \draw (a) edge[->] (c);
      \draw (a) edge[->] (d);
    \end{tikzpicture}
    \caption{Induction step}
    \vspace*{-.2cm}
  \end{figure}
  Due to (d), we have $\mathcal{A}_{sum}^{\sharp}(t_w) \succcurlyeq \sum_{1 \leq j \leq k} p_{j} \cdot \mathcal{A}_{sum}^{\sharp}(t_{w_j})$.
  Since $g_{\AA \to \IN}$ is a concave embedding we get
  \begin{align*}
                          g_{\AA \to \IN}(\mathcal{A}_{sum}^{\sharp}(t_w \delta_{\ta})) & \geq g_{\AA \to \IN}\big(\sum_{1 \leq j \leq k} p_{j} \cdot \mathcal{A}_{sum}^{\sharp}(t_{w_j} \delta_{\ta})\big) \\
                           & \geq \sum_{1 \leq j \leq k} p_{j} \cdot g_{\AA \to \IN}(\mathcal{A}_{sum}^{\sharp}(t_{w_j} \delta_{\ta})) \\
    \Longleftrightarrow \adtval(w) & \geq \sum_{1 \leq j \leq k} p_{j} \cdot \adtval(w_j)
  \end{align*}
  Hence, we either have $\adtval(w) = \adtval(w_j)$ for all $1 \leq j \leq k$ or there exists at least one $1 \leq j \leq k$ with $\adtval(w) > \adtval(w_j)$.

  We partition the set $W_{i-1}^v$ into the disjoint subsets $W_{i-1}^{v(1)}$, $W_{i-1}^{v(2)}$, and $W_{i-1}^{v(3)}$, where
  \begin{itemize}
    \item[$\bullet$] $w \in W_{i-1}^{v(1)}$:$\Leftrightarrow$ $w \in W_{i-1}^v$ and $w$ is a leaf of $\F{T}_v$.
    \item[$\bullet$] $w \in W_{i-1}^{v(2)}$:$\Leftrightarrow$ $w \in W_{i-1}^v$ and $\adtval(w) = \adtval(w_j)$ for all $1 \leq j \leq k$.
    \item[$\bullet$] $w \in W_{i-1}^{v(3)}$:$\Leftrightarrow$ $w \in W_{i-1}^v$ and there exists a $1 \leq j \leq k$ with $\adtval(w) > \adtval(w_j)$.
      We denote this node $w_j$ by $w^+$.
  \end{itemize}
  In the first case, $w$ is already contained in $W_{i-1}^v$.
  In the second case, we get $\adtval(w) = \adtval(w_j)$ for all $1 \leq j \leq k$.
  And in the third case, we have $\adtval(w) > \adtval(w^+)$, hence $\adtval(w^+) \leq \adtval(w) - 1$.
  So for all of these nodes, we can be sure that the value does not increase.
  We can now define $W_{i}^v$ as:
  \[
    \begin{array}{lcl}
      W_{i}^v & =    & W_{i-1}^{v(1)}                    \\
              & \cup & \bigcup_{w \in W_{i-1}^{v(2)}} wE \\
              & \cup & \{w^+ \mid w \in W_{i-1}^{v(3)}\} \\
    \end{array}
  \]
  Intuitively, this means that every leaf remains inside the set of witnesses ($W_{i-1}^{v(1)}$) and for every inner node $w$ we have two cases.
  If there exists a successor $w^+$ with a strictly smaller value, then we replace the node $w$ by this successor $w^+$ in our set of witnesses ($\{w^+ \mid w \in W_{i-1}^{v(3)}\}$).
  Otherwise, all the successors $w_1, \ldots, w_k$ of the node $w$ have the same value as the node $w$ itself, so that we have to replace $w$ with all of its successors in our set of witnesses as there is no single node with a guaranteed value decrease ($\bigcup_{w \in W_{i-1}^{v(2)}} wE$).

  It remains to show that our properties (W-1!), (W-2!), and (W-3!) are still satisfied for $W_{i}^v$.
  In order to see that (W-1!) is satisfied, note that we have $\adtval(w) \leq \adtval(v) - 1$, for all $w \in W_{i-1}^v$ by our induction hypothesis.
  Thus, for all $w' \in W_{i-1}^{v(1)} \subseteq W_{i-1}^v$, we also obtain $\adtval(w')\leq \adtval(v) - 1$.
  For all $w' \in \bigcup_{w \in W_{i-1}^{v(2)}} wE$, we have $\adtval(w') = \adtval(w)$ for some $w \in W_{i-1}^v$ and thus $\adtval(w') = \adtval(w) \leq \adtval(v) - 1$.
  Finally, for all $w' \in \{w^+ \mid w \in W_{i-1}^{v(3)}\}$, we have $\adtval(w') \leq \adtval(w) - 1$ for some $w \in W_{i-1}^v$ and thus $\adtval(w') \leq \adtval(w) - 1 \leq \adtval(v) - 2 \leq \adtval(v) - 1$.

  Property (W-3!) holds as well, since every node $w \in W_{i-1}$ that is not a leaf of $\F{T}_v$ is at depth $i-1$ of the tree $\F{T}_v$ by our induction hypothesis.
  We exchange each such node with one or all of its successors.
  The leaves in $W_{i-1}$ are at a depth of at most $i$ by induction hypothesis and remain in $W_{i}$.
  Hence, all the nodes of $W_{i}$ are at a depth of at most $i$, and the nodes that are no leaves are at a depth of precisely $i$.

  Finally, we regard (W-2!).
  For $\bigcup_{w \in W_{i-1}^{v(2)}} wE$ we have:

  \begin{longtable}{C@{\;}L}
       & \sum_{w' \in \bigcup_{w \in W_{i-1}^{v(2)}} wE} p_{w'}^{\F{T}_v} \cdot p_{min}^{\adtval(w')}  \\[5pt]
     = & \sum_{w \in W_{i-1}^{v(2)}} \sum_{w' \in wE}p_{w'}^{\F{T}_v} \cdot p_{min}^{\adtval(w')}      \\[5pt]
     = & \sum_{w \in W_{i-1}^{v(2)}, wE = \{w_1,\ldots,w_k\}} \sum_{1 \leq j \leq k} p_{w_j}^{\F{T}_v} \cdot p_{min}^{\adtval(w_j)}  \\[5pt]
     = & \sum_{w \in W_{i-1}^{v(2)}, wE = \{w_1,\ldots,w_k\}} \sum_{1 \leq j \leq k} p_{w_j}^{\F{T}_v} \cdot p_{min}^{\adtval(w)}    \\
       & \qquad \text{(as $\adtval(w_j) = \adtval(w)$ for all $1 \leq j \leq k$)}                      \\[5pt]
     = & \sum_{w \in W_{i-1}^{v(2)}, wE = \{w_1,\ldots,w_k\}} \sum_{1 \leq j \leq k} p_j \cdot p_w^{\F{T}_v} \cdot p_{min}^{\adtval(w)}  \\
       & \qquad \text{(as $p_{w_j}^{\F{T}_v} = p_j \cdot p_w^{\F{T}_v}$ for all $1 \leq j \leq k$)}    \\[5pt]
     = & \sum_{w \in W_{i-1}^{v(2)}, wE = \{w_1,\ldots,w_k\}} p_w^{\F{T}_v} \cdot p_{min}^{\adtval(w)} \cdot \sum_{1 \leq j \leq k} p_j \\[5pt]
     = & \sum_{w \in W_{i-1}^{v(2)}} p_w^{\F{T}_v} \cdot p_{min}^{\adtval(w)} \cdot 1 \qquad\qquad \text{(as $\sum_{1 \leq j \leq k} p_j = 1$)} \\[5pt]
     = & \sum_{w \in W_{i-1}^{v(2)}} p_w^{\F{T}_v} \cdot p_{min}^{\adtval(w)}.
  \end{longtable}

  For $\{w^+ \mid w \in W_{i-1}^{v(3)}\}$ we have:

  \begin{longtable}{C@{\;}LL}
          & \sum_{w' \in \{w^+ \mid w \in W_{i-1}^{v(3)}\}} p_{w'}^{\F{T}_v} \cdot p_{min}^{\adtval(w')} \\
     =    & \sum_{w \in W_{i-1}^{v(3)}} p_{w^+}^{\F{T}_v} \cdot p_{min}^{\adtval(w^+)}   \\
     \geq & \sum_{w \in W_{i-1}^{v(3)}} p_{min} \cdot p_w^{\F{T}_v} \cdot p_{min}^{\adtval(w^+)}         & \text{(as $p_{w^+}^{\F{T}_v} \geq p_{min} \cdot p_w^{\F{T}_v}$)} \\
     \geq & \sum_{w \in W_{i-1}^{v(3)}} p_{min} \cdot p_w^{\F{T}_v} \cdot p_{min}^{\adtval(w)-1}         & \text{(as $\adtval(w^+) \leq \adtval(w) - 1$)}                   \\
     =    & \sum_{w \in W_{i-1}^{v(3)}} p_w^{\F{T}_v} \cdot p_{min}^{\adtval(w)}.
  \end{longtable}

  To summarize, we have
  \[
    \begin{array}{cl}
                           & \sum_{w' \in W_{i}^v} p_{w'}^{\F{T}_v} \cdot p_{min}^{\adtval(w')}                                      \\[5pt]
       =                   & \sum_{w' \in W_{i-1}^{v(1)}} p_{w'}^{\F{T}_v} \cdot p_{min}^{\adtval(w')}                               \\[-2pt]
                           & + \; \sum_{w' \in \bigcup_{w \in W_{i-1}^{v(2)}} wE} p_{w'}^{\F{T}_v} \cdot p_{min}^{\adtval(w')}       \\[-2pt]
                           & \quad + \; \sum_{w' \in \{w^+ \mid w \in W_{i-1}^{v(3)}\}} p_{w'}^{\F{T}_v} \cdot p_{min}^{\adtval(w')} \\[10pt]
       \geq                & \sum_{w \in W_{i-1}^{v(1)}} p_{w}^{\F{T}_v} \cdot p_{min}^{\adtval(w)}                                  \\[-2pt]
                           & + \; \sum_{w \in W_{i-1}^{v(2)}} p_{w}^{\F{T}_v} \cdot p_{min}^{\adtval(w)}                             \\[-2pt]
                           & \quad + \; \sum_{w \in W_{i-1}^{v(3)}} p_{w}^{\F{T}_v} \cdot p_{min}^{\adtval(w)}                       \\[5pt]
       \stackrel{(IH)}{\geq} & p_{min}^{\adtval(v)}.
    \end{array}
  \]

  \medskip

  \noindent
  \textbf{\underline{III.2.
      We prove that $|\F{T}^{\leq N}| \geq p_{min}^{N}$ for any $N \in \IN$ by induction}}

  \noindent
  We now prove that $|\F{T}^{\leq N}| \geq p_{min}^{N}$ holds for any $N \in \IN$.
  Let $Z_{k}$ denote the set of nodes in $\F{T}^{\leq N}$ from $\aV_{\succ}$ that have precisely $k-1$ nodes from $\aV_{\succ}$ above them (so they are themselves the $k$-th node from $\aV_{\succ}$).
  Here, $\aV$, $\aV_{\succcurlyeq}$, and $\aV_{\succ}$ again refer to $\F{T}^{\leq N}$, i.e., we again have $\aV \cap C = \emptyset$.
  Let $\ctleaf_k$ denote the set of all leaves in $\F{T}^{\leq N}$ that are reachable by a path that uses less than $k$ nodes from $\aV_{\succ}$.
  We show by induction that for all $1 \leq k \leq N+1$, we have
  \[
    \sum_{v \in Z_{k} \cup \ctleaf_{k}, 0 \leq \adtval(v) \leq N+1-k} p_v^{\F{T}^{\leq N}} \cdot p_{min}^{\adtval(v)} \geq p_{min}^{N}
  \]
  Then, for $k = N+1$ we finally have:

  \begin{longtable}{L@{\;}C@{\;}L}
    p_{min}^{N} & \leq & \sum_{v \in Z_{N+1} \cup \ctleaf_{N+1}, 0 \leq \adtval(v) \leq N+1-(N+1)} p_v^{\F{T}^{\leq N}} \cdot p_{min}^{\adtval(v)} \\
                & =    & \sum_{v \in Z_{N+1} \cup \ctleaf_{N+1}, 0 \leq \adtval(v) \leq 0} p_v^{\F{T}^{\leq N}} \cdot p_{min}^{\adtval(v)}         \\
                & =    & \sum_{v \in Z_{N+1} \cup \ctleaf_{N+1}, \adtval(v) = 0} p_v^{\F{T}^{\leq N}} \cdot p_{min}^{\adtval(v)}                   \\
                & =    & \sum_{v \in Z_{N+1} \cup \ctleaf_{N+1}, \adtval(v) = 0} p_v^{\F{T}^{\leq N}} \cdot 1                                      \\
                & =    & \sum_{v \in Z_{N+1} \cup \ctleaf_{N+1}, \adtval(v) = 0} p_v^{\F{T}^{\leq N}}                                              \\
                & =    & \sum_{v \in \ctleaf_{N+1}} p_v^{\F{T}^{\leq N}} \qquad \text{(as $\adtval(v) = 0$ iff $v$ is a leaf of $\F{T}^{\leq N}$)} \\
                & \leq & \sum_{v \in \ctleaf^{\F{T}^{\leq N}}} p_v^{\F{T}^{\leq N}}              \\
                & =    & |\F{T}^{\leq N}|
  \end{longtable}

  In the induction base, we have $k = 1$, and thus

  \begin{longtable}{L@{\;}C@{\;}L}
     &      & \sum_{v \in Z_{1} \cup \ctleaf_{1}, 0 \leq \adtval(v) \leq N+1-1} p_v^{\F{T}^{\leq N}} \cdot p_{min}^{\adtval(v)} \\
     & =    & \sum_{v \in Z_{1} \cup \ctleaf_{1}, 0 \leq \adtval(v) \leq N} p_v^{\F{T}^{\leq N}} \cdot p_{min}^{\adtval(v)}     \\
     & \geq & \sum_{v \in Z_{1} \cup \ctleaf_{1}, 0 \leq \adtval(v) \leq N} p_v^{\F{T}^{\leq N}} \cdot p_{min}^{N} \quad \text{(since $\adtval(v) \leq N$)} \\
     & =    & p_{min}^{N} \cdot \sum_{v \in Z_{1} \cup \ctleaf_{1}, 0 \leq \adtval(v) \leq N} p_v^{\F{T}^{\leq N}}              \\
     & =    & p_{min}^{N} \cdot \sum_{v \in Z_{1} \cup \ctleaf_{1}} p_v^{\F{T}^{\leq N}}                                        \\
     & =    & p_{min}^{N} \cdot 1 \quad \text{(since $\sum_{v \in Z_{1} \cup \ctleaf_{1}} p_v^{\F{T}^{\leq N}} = 1$)}           \\
     & =    & p_{min}^{N}
  \end{longtable}

  Here, we have $\sum_{v \in Z_{1} \cup \ctleaf_{1}} p_v^{\F{T}^{\leq N}} = 1$, since $Z_{1} \cup \ctleaf_{1}$ are the leaves of the finite subtree where we cut everything below the first node of $\PP_{\succ}$ (i.e., we cut directly after the nodes in $Z_{1}$).
  This tree is finite, because by (+) there is no infinite path without $\aV_{\succ}$ nodes.
  All finite CTs converge with probability~$1$.

  In the induction step, we assume that the statement holds for some $1 \leq k \leq N$.
  Then we have

    {\small
      \begin{longtable}{C@{\;}L}
                            & p_{min}^{N}                                            \\
        \stackrel{(IH)}{\leq} & \sum_{v \in Z_{k} \cup \ctleaf_{k}, 0 \leq \adtval(v) \leq N+1-k} p_v^{\F{T}^{\leq N}} \cdot p_{min}^{\adtval(v)}                      \\
        =                   & \sum_{v \in \ctleaf_{k}} p_v^{\F{T}^{\leq N}} \cdot p_{min}^{\adtval(v)} \; + \; \sum_{v \in Z_{k}, 1 \leq \adtval(v) \leq N+1-k} p_v^{\F{T}^{\leq N}} \cdot p_{min}^{\adtval(v)}                                      \\
        \leq                & \sum_{v \in \ctleaf_{k}} p_v^{\F{T}^{\leq N}} \cdot p_{min}^{\adtval(v)} \; + \; \sum_{v \in Z_{k}, 1 \leq \adtval(v) \leq N+1-k} p_v^{\F{T}^{\leq N}} \cdot \sum_{w \in W^v} p_w^{\F{T}_v} \cdot p_{min}^{\adtval(w)} \\
                            & \hspace*{2cm} \text{(existence of the set $W^v$ by the previous Step 3.1 and (W-2))} \\
        =                   & \sum_{v \in \ctleaf_{k}} p_v^{\F{T}^{\leq N}} \cdot p_{min}^{\adtval(v)} \; + \; \sum_{v \in Z_{k}, 1 \leq \adtval(v) \leq N+1-k} \sum_{w \in W^v} p_v^{\F{T}^{\leq N}} \cdot p_w^{\F{T}_v} \cdot p_{min}^{\adtval(w)} \\
        =                   & \sum_{v \in \ctleaf_{k}} p_v^{\F{T}^{\leq N}} \cdot p_{min}^{\adtval(v)} \; + \; \sum_{v \in Z_{k}, 1 \leq \adtval(v) \leq N+1-k} \sum_{w \in W^v} p_w^{\F{T}^{\leq N}} \cdot p_{min}^{\adtval(w)}                     \\
                            & \hspace*{2cm} \text{(as $p_v^{\F{T}^{\leq N}} \cdot p_w^{\F{T}_v} = p_w^{\F{T}^{\leq N}}$).}
      \end{longtable}
    }

  Every node in $W^v$ is either contained in $\ctleaf_{k+1}$ or contained in $Z_{k+1}$.
  The reason for that is that $W^v$ only contains leaves of $\F{T}_v$ and a leaf in $\F{T}_v$ is either also a leaf in $\F{T}^{\leq N}$ so that it is contained in $\ctleaf_{k+1}$, or contained in $\PP_{\succ}$ and thus in $Z_{k+1}$, since $v$ is contained in $Z_{k}$ and there is no other inner node from $\PP_{\succ}$ in $\F{T}_v$.
  Furthermore, we know that $\adtval(w) \leq \adtval(v) - 1$ for all $w \in W^v$ by (W-1).
  Moreover, we have $\ctleaf_{k} \subseteq \ctleaf_{k+1}$.
  Thus, we get
  \[
    \begin{array}{c@{\;}l}
      & \sum_{v \in \ctleaf_{k}} p_v^{\F{T}^{\leq N}} \cdot p_{min}^{\adtval(v)} \; + \; \sum_{v \in Z_{k}, 1 \leq \adtval(v) \leq N+1-k} \sum_{w \in W^v} p_w^{\F{T}^{\leq N}} \cdot p_{min}^{\adtval(w)} \\
      \leq & \sum_{w \in Z_{k+1} \cup \ctleaf_{k+1}, 0 \leq \adtval(w) \leq N+1-(k+1)} p_w^{\F{T}^{\leq N}} \cdot p_{min}^{\adtval(w)}.
    \end{array}
  \]
  To summarize, we have shown that $|\F{T}^{\leq N}| \geq p_{min}^{N}$.

  \medskip

  \noindent
  \textbf{\underline{IV.
      We prove that $|\F{T}^{\leq N}|=1$ for any $N \in \IN$}}

  \noindent
  We have proven that $|\F{T}^{\leq N}| \ge p_{min}^N$ holds for all $\PP$-CTs $\F{T}$ that satisfy (+).
  Hence, for any $N \in \IN$, we have
  \begin{equation}
    \label{RPP-part-4-inf-definition}
    p^{\star}_{N}:=\inf_{\F{T} \text{ is a $\PP$-CT satisfying (+)}}
    (|\F{T}^{\leq N}|) \geq p_{min}^{N} > 0.
  \end{equation}
  We now prove by contradiction that this is enough to ensure $p^{\star}_N = 1$.
  So assume that $p^{\star}_N <1$.
  Then we define $\epsilon := \frac{p^{\star}_N \cdot (1-p^{\star}_N)}{2}>0$.
  By definition of the infimum, $p^{\star}_N + \epsilon$ is not a lower bound of $|\F{T}^{\leq N}|$ for all $\PP$-CTs $\F{T}$ that satisfy (+).
  Hence, there must exist a $\PP$-CT $\F{T}$ satisfying (+) such that
  \begin{equation}\label{RPP-part-4-pstarEpsilon}
    p^{\star}_N\leq |\F{T}^{\leq N}| < p^{\star}_N + \epsilon.
  \end{equation}
  For readability, let $Z = \ctleaf^{\F{T}^{\leq N}}$ be the set of leaves of the tree $\F{T}^{\leq N}$ and let $\overline{Z} = V^{\F{T}^{\leq N}} \setminus \ctleaf^{\F{T}^{\leq N}}$ be the set of inner nodes of the tree $\F{T}^{\leq N}$.
  Since $p_v^{\F{T}^{\leq N}} = p_v^{\F{T}}$ for all $v \in V^{\F{T}^{\leq N}}$, in the following we just write $p_v$.
  By the monotonicity of $|\cdot|$ w.r.t.\ the depth of the tree $\F{T}^{\leq N}$, there must exist a natural number $m^{\star} \in \IN$ such that
  \begin{equation}
    \label{RPP-part-4-p-star-half}
    \sum_{v \in Z, \ctdepth(v) \leq m^*} p_v > \tfrac{p^{\star}_N}{2}.
  \end{equation}
  Here, $\ctdepth(v)$ denotes the depth of node $v$ in $\F{T}^{\leq N}$.
  For every $v \in V^{\F{T}^{\leq N}}$ with $\ctdepth(v) = m^{\star}$ and $v \not\in Z$, we define the sub $\PP$-CT of $\F{T}^{\leq N}$ starting at node $v$ by $\F{T}^{\leq N}(v) = \F{T}^{\leq N}[v(E^{\F{T}^{\leq N}})^*]$.
  Then we have
  \begin{align}\label{RPP-part-4-limit}
    |\F{T}^{\leq N}| = \sum_{v \in Z, \ctdepth(v) \leq m^*} p_v + \sum_{v \in \overline{Z}, \ctdepth(v) = m^*} p_v \cdot |\F{T}^{\leq N}(v)|.
  \end{align}
  Furthermore, we have
  \begin{equation}
    \label{RPP-part-4-first-observation}
    \sum_{v \in \overline{Z}, \ctdepth(v) = m^*} p_v = 1 - \sum_{v \in Z, \ctdepth(v) \leq m^*} p_v,
  \end{equation}
  since $\sum_{v \in \overline{Z}, \ctdepth(v) = m^*} p_v + \sum_{v \in Z, \ctdepth(v) \leq m^*} p_v = 1$, as the nodes $v \in \overline{Z}$ with $d(v) = m^*$ are the leaves of the finite, grounded subtree of $\F{T}^{\leq N}$ where we cut every edge after the nodes of depth $m^*$.
  We obtain
  \begin{longtable}{RL}
             & p^{\star}_N+\epsilon     \\
    {}>{}    & |\F{T}^{\leq N}| \qquad \text{(by~\eqref{RPP-part-4-pstarEpsilon})}
    \\
    {}={}    & \sum_{v \in Z, \ctdepth(v) \leq m^*} p_v + \sum_{v \in \overline{Z}, \ctdepth(v) = m^*} p_v \cdot \underbrace{|\F{T}^{\leq N}(v)|}_{\geq \, p^{\star}_N} \qquad \text{(by~\eqref{RPP-part-4-limit} and~\eqref{RPP-part-4-inf-definition})}
    \\
    {}\geq{} & \sum_{v \in Z, \ctdepth(v) \leq m^*} p_v + \sum_{v \in \overline{Z}, \ctdepth(v) = m^*} p_v \cdot p^{\star}_N                                              \\
    {}={}    & \sum_{v \in Z, \ctdepth(v) \leq m^*} p_v + p^{\star}_N \cdot \sum_{v \in \overline{Z}, \ctdepth(v) = m^*} p_v                                              \\
    {}={}    & \sum_{v \in Z, \ctdepth(v) \leq m^*} p_v + p^{\star}_N \cdot (1 - \sum_{v \in Z, \ctdepth(v) \leq m^*} p_v) \qquad \text{(by~\eqref{RPP-part-4-first-observation})}
    \\
    {}={}    & p^{\star}_N + \sum_{v \in Z, \ctdepth(v) \leq m^*} p_v - p^{\star}_N \cdot \sum_{v \in Z, \ctdepth(v) \leq m^*} p_v                                        \\
    {}={}    & p^{\star}_N + \sum_{v \in Z, \ctdepth(v) \leq m^*} p_v \cdot (1-p^{\star}_N)                             \\
    {}>{}    & p^{\star}_N + \left(1-p^{\star}_N\right)\cdot \tfrac{p^{\star}_N}{2} \qquad \text{(by~\eqref{RPP-part-4-p-star-half})}                                     \\
    {}={}    & p^{\star}_N + \epsilon, \quad \textcolor{red}{\lightning}
  \end{longtable}
  \noindent
  which is a contradiction.
  So $p^{\star}_N=1$.
  In particular, this means that for every $N \in \IN$ and every $\PP$-CT $\F{T}$ satisfying (+), we have
  \begin{equation}
    \label{RPP-part-4-limitMuLeqNIs1}
    |\F{T}^{\leq N}| = 1.
  \end{equation}

  \noindent
  \textbf{\underline{V.
      Finally, we prove that $|\F{T}|=1$}}

  \noindent
  We adjust the tree value function for this new tree $\F{T}^{\leq N}$ once again and define:
  \[
    \adtvaltwo: V^{\F{T}^{\leq N}} \to \IN, \qquad v \mapsto
    \begin{cases}
      N+1,                                     & \text{ if } v \in C               \\
      0,                                       & \text{ if } v \in \ctleaf^{\F{T}} \\
      g_{\AA \to \IN}(\mathcal{A}_{sum}^{\sharp}(t_v\delta_\ta)) + 1, & \text{ otherwise }
    \end{cases}
  \]

  Now for a node $v \in V^{\F{T}^{\leq N}}$ we have $0 \leq \adtvaltwo(v) \leq N+1$.
  Furthermore, we have $\tval(v) \geq \adtvaltwo(v)$ for all $v \in V^{\F{T}^{\leq N}}$.
  Let $|\F{T}^{\leq N}|_{\adtvaltwo} = \sum_{v \in \ctleaf^{\F{T}^{\leq N}}} p_v \cdot \adtvaltwo(v)$.
  This (possibly infinite) sum is well defined (i.e., it is a convergent series), because all addends are non-negative and the sum is bounded from above by $N+1$, since
  \[\mbox{\small $\sum\limits_{v \in \ctleaf^{\F{T}^{\leq N}}} \hspace*{-.3cm} p_v
        \cdot \adtvaltwo(v) \leq \hspace*{-.3cm} \sum\limits_{v \in \ctleaf^{\F{T}^{\leq
              N}}} \hspace*{-.3cm} p_v \cdot (N+1) = (N+1) \cdot \hspace*{-.3cm} \sum\limits_{v \in
          \ctleaf^{\F{T}^{\leq N}}} \hspace*{-.3cm} p_v \leq (N+1) \cdot 1 = N+1$.}\]

  For the root $\ctroot$ of $\F{T}$ and $\F{T}^{\leq N}$, we have $\adtvaltwo(\ctroot) \geq |\F{T}^{\leq N}|_{\adtvaltwo}$ because due to (d), we know that the (expected) value is non-increasing.

  Now we fix $N \in \IN$ and a CT $\F{T}$ satisfying (+), and obtain the corresponding transformed tree $\F{T}^{\leq N}$.
  Note that by~\eqref{RPP-part-4-limitMuLeqNIs1} we have $|\F{T}^{\leq N}| = 1 = q_N + q_N'$, where $q_N := \sum_{v \in \ctleaf^{\F{T}^{\leq N}}, \adtvaltwo(v) = 0} p_v$ and $q_N' = \sum_{v \in \ctleaf^{\F{T}^{\leq N}}, \adtvaltwo(v) > 0} p_v = \sum_{v \in C} p_v$.
  Now we can determine $|\F{T}^{\leq N}|_{\adtvaltwo}$.
  The probabilities of zero-valued nodes (i.e., leaves in the original tree) add up to $q_N$, while the probabilities of new leaves due to a cut add up to probability $q_N' = 1-q_N$.
  So $|\F{T}^{\leq N}|_{\adtvaltwo} = q_N \cdot 0 + (1-q_N) \cdot (N+1) = (1-q_N)\cdot (N+1)$.
  Thus,
  \[
    \tval(\ctroot) \geq \adtvaltwo(\ctroot) \geq |\F{T}^{\leq N}|_{\adtvaltwo}
    \geq (1-q_N) \cdot (N+1),
  \]
  which implies $q_N \geq 1-\tfrac{\tval(\ctroot)}{N+1}$.
  Note that $q_N$ is weakly monotonically increasing and bounded from above by 1 for $N \to \infty$.
  Hence, $q := \lim_{N \to \infty} q_N$ exists and $1 \geq q \geq \lim_{N \to \infty} (1-\tfrac{\tval(\ctroot^{\F{T}})}{N+1}) = 1$, i.e., $q = 1$.
  Hence, we obtain $|\F{T}| = \lim_{N \to \infty} q_N = q = 1$.
\end{myproof}

By \Cref{lemma:prob-RPP-CT-lemma}, we have finally proven the main technical part for the reduction pair processor in the probabilistic setting.
Next, we want to prove the soundness and completeness of the actual processor, which is fairly easy using \Cref{lemma:prob-RPP-CT-lemma} in addition to the $\aV$-Partition Lemma.

\ProbRedPairProc*

\begin{myproof}
  Let $\overline{\PP} = \PP_{\succcurlyeq} \cup \flat(\PP_{\succ})$.
  \smallskip

  \noindent
  \underline{\emph{Completeness:}} By \Cref{thm:proc-complete}.
  \medskip

  \noindent
  \underline{\emph{Soundness:}}
  Suppose that every (innermost) $\overline{\PP}$-CT converges with probability $1$.
  Assume for a contradiction that there exists an (innermost) $\PP$-CT $\F{T} = (V,E,L)$ that converges with probability $<1$.
  Let us partition $\aV = \aV_{\succcurlyeq} \uplus \aV_{\succ}$ according to the used ADPs for the rewrite steps.
  Since every $\overline{\PP}$-CT converges with probability $1$, so does every subtree of $\F{T}$ that only contains $\aV$-nodes from $\aV_{\succcurlyeq}$.
  By the $\aV$-Partition Lemma (\cref{lemma:p-partition}), there must be an (innermost) $\PP$-CT $\F{T}'$ with $|\F{T}'| <1$ such that every infinite path has an infinite number of edges corresponding to $\PP_{\succ}$ steps.
  But this is a contradiction to \cref{lemma:prob-RPP-CT-lemma}.
\end{myproof}

\ProbRemProc*

\begin{myproof}
  Let $\PP$ be an ADP problem such that every ADP in $\PP$ has the form $\ell \to \{1:r\}^{m}$.
  Note that every $\PP$-CT is a single (not necessarily finite) path.
  For such a CT $\F{T}$ that is only a single path, we have only two possibilities for $|\F{T}|$.
  If the path is finite, then $|\F{T}| = 1$, since we have a single leaf in this tree with probability $1$.
  Otherwise, we have an infinite path, which means that there is no leaf at all and hence $|\F{T}| = 0$.

  \medskip

  \noindent
  \underline{\emph{``only if''}}

  \noindent
  We first consider $\mathtt{AST}$.
  Assume that $(\nonprobDP(\PP),\nonprob(\PP))$ is not terminating.
  Then there exists an infinite $(\nonprobDP(\PP),\nonprob(\PP))$-chain
  \[t_0^\sharp \epsto_{\nonprobDP(\PP)} \circ \to_{\nonprob(\PP)}^* t_1^\sharp \epsto_{\nonprobDP(\PP)} \circ \to_{\nonprob(\PP)}^* t_2^\sharp \epsto_{\nonprobDP(\PP)} \circ \to_{\nonprob(\PP)}^* ...\]
  such that for all $i \in \IN$ we have $t_i^\sharp = \ell_i^\sharp \sigma_i$ for some $\ell_i^\sharp \to r_i^\sharp \in \nonprobDP(\PP)$ and some substitution $\sigma_i$.
  From this infinite $(\nonprobDP(\PP),\nonprob(\PP))$-chain, we will now construct an infinite $\PP$-CT $\F{T} = (V,E,L)$.
  As explained above, we then know that this infinite $\PP$-CT must be an infinite path, and thus $|\F{T}| = 0$.
  This means that $\PP$ is not $\mathtt{AST}$, and thus, a processor with $\Proc_{\mathtt{PR}}(\PP) = \emptyset$ would be unsound.
  \begin{center}
    \small
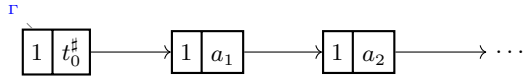
    \begin{tikzpicture}
      \tikzstyle{adam}=[thick,draw=black!100,fill=white!100,minimum size=4mm, shape=rectangle split, rectangle split parts=2,rectangle split horizontal] \tikzstyle{empty}=[rectangle,thick,minimum size=4mm]

      \node[adam,pin={[pin distance=0.1cm, pin edge={,-}] 135:\tiny \textcolor{blue}{$\aV$}}] at (-2, 0) (a2) {$1$
          \nodepart{two} $t_0^\sharp$};
      \node[adam] at (0, 0) (b2) {$1$
        \nodepart{two} $a_{1}^{\textcolor{white}{!}}$};
      \node[adam] at (2, 0) (c2) {$1$
        \nodepart{two} $a_{2}^{\textcolor{white}{!}}$};
      \node[empty] at (4, 0) (d2) {$\ldots$};

      \draw (a2) edge[->] (b2);
      \draw (b2) edge[->] (c2);
      \draw (c2) edge[->] (d2);
    \end{tikzpicture}
  \end{center}
  We start our CT with $(1:t_0^\sharp)$.
  In the non-probabilistic rewrite sequence, we have $t_0^\sharp \epsto_{\nonprobDP(\PP)}
    \circ \to_{\nonprob(\PP)}^* t_1^\sharp$, so there exists a natural number $k \geq 1$ such that
  \[\mbox{\small $t_0^\sharp = \ell_0^\sharp \sigma_0 \to_{\nonprobDP(\PP)} r_0^\sharp \sigma_0 = s_1^\sharp
        \to_{\nonprob(\PP)} s_2^\sharp \to_{\nonprob(\PP)}
        \ldots \to_{\nonprob(\PP)} s_{k}^\sharp = t_1^\sharp = \ell_1^\sharp \sigma_1$} \]
  Performing the same rewrite steps with $\PP$ yields terms $a_1,\ldots,a_k$ such that $s_i \trianglelefteq_{\sharp} a_i$ for all $1 \leq i \leq k$.
  Here, one needs that all ADPs that yield the rules in $\nonprob(\PP)$ have the flag $\ttrue$ and thus, the annotations above the redex are not removed.
  For all $1 \leq i \leq k$, we now construct $a_i$ inductively.

  In the induction base ($i = 1$), let $\ell_0 \to \{1: r_0'\}^{m} \in \PP$ be the ADP that was used to create the dependency pair $\ell_0^\sharp \to r_0^\sharp$ in $\nonprobDP(\PP)$.
  This means that we have $r_0 \trianglelefteq_{\sharp} r_0'$.
  Since we have $t_0^\sharp = \ell_0^\sharp \sigma_0$, we can also rewrite $t_0^\sharp$ with the ADP $\ell_0 \to \{1:r_0'\}^{m} \in \PP$ and the substitution $\sigma_0$ at the root position.
  We result in $a_1 = r_0'\sigma_0$ and thus we have $s_1 = r_0 \sigma_0 \trianglelefteq_{\sharp} r_0' \sigma_0 = a_1$.

  In the induction step, we assume that $s_i \trianglelefteq_{\sharp} a_i$ for some $1 \leq i < k$.
  Let $\pi$ be the annotated position of $a_i$ where $s_i = \flat(a_i)|_\pi$.
  In our non-probabilistic rewrite sequence we have $s_i^\sharp \to_{\nonprob(\PP)} s_{i+1}^\sharp$ using a rule $\ell' \to \flat(r') \in \nonprob(\PP)$ and substitution $\delta_i$ at a position $\tau \in \IN^+$ such that $s_i^\sharp|_{\tau} = \ell' \delta_i$ and $s_{i+1}^\sharp = s_i^\sharp[\flat(r') \delta_i]_{\tau}$.
  We can mirror this rewrite step with the ADP $\ell' \to \{1:r'\}^{\ttrue} \in \PP$, since by construction we have $s_i \trianglelefteq_{\sharp} a_i$ and $s_i = \flat(a_i)|_\pi$.
  We obtain $a_i \tored{}{}{\PP} a_{i+1} = a_i[\anno_{\Phi}(r' \delta_i)]_{\pi.\tau}$ with $\Phi = \posT(r_j) \cup \Psi_j$ (step with $(\mathbf{at})$) or $\Phi = \Psi_1$ (step with $(\mathbf{nt})$) by rewriting the subterm of $a_i$ at position $\pi.\tau$, which implies $s_{i+1} = s_i[\flat(r') \delta_i]_{\tau} \trianglelefteq_{\sharp} a_i[\anno_{\Phi}(r' \delta_i)]_{\pi.\tau} = a_{i+1}$.

  At the end of this induction, we result in $a_{k}$.
  Next, we can mirror the step $t_1^\sharp \epsto_{\nonprobDP(\PP)} \circ \to_{\nonprob(\PP)}^* t_2^\sharp$ from our non-probabilistic rewrite sequence with the same construction, etc.
  This results in an infinite $\PP$-CT.
  To see that this is indeed a $\PP$-CT, note that all the local properties are satisfied since every edge represents a rewrite step with $\tored{}{}{\PP}$.
  The global property is also satisfied since, in an infinite $(\nonprobDP(\PP),\nonprob(\PP))$-chain, we use an infinite number of steps with $\to_{\nonprobDP(\PP)}$ so that our resulting CT has an infinite number of $\aV$-nodes.

  For $\mathtt{iAST}$ we just have to consider innermost rewriting, while the construction remains the same.
  More precisely, we start with an innermost $(\nonprobDP(\PP),\nonprob(\PP))$-chain
  \[t_0^\sharp \iepsto_{\nonprobDP(\PP)} \circ \ito_{\nonprob(\PP)}^* t_1^\sharp \iepsto_{\nonprobDP(\PP)} \circ \to_{\nonprob(\PP)}^* t_2^\sharp \iepsto_{\nonprobDP(\PP)} \circ \ito_{\nonprob(\PP)}^* ...\]
  and construct an innermost $\PP$-CT.\@ But since all the steps in the chain are innermost, so are the steps in the CT.
  \medskip

  \noindent
  \underline{\emph{``if''}}

  \noindent
  Assume that $\PP$ is not $\mathtt{AST}$, i.e., that the processor $\Proc_{\mathtt{PR}}(\PP) = \emptyset$ is unsound.
  By \cref{lemma:starting}, there exists a $\PP$-CT $\F{T} = (V,E,L)$ that converges with probability $<1$ and starts with $(1:t^\sharp)$ such that $t^\sharp = \ell^\sharp \sigma_0$ for some substitution $\sigma_0$ and an ADP $\ell \to \{1:r\}^{m} \in \PP$.
  As explained above, this tree must be an infinite path.

  From $\F{T}$, we will now construct an infinite $(\nonprobDP(\PP),\nonprob(\PP))$-chain, which shows that $(\nonprobDP(\PP),\nonprob(\PP))$ is not terminating.
  We start our infinite chain with the term $t_0^\sharp = t^\sharp$.
  We have $t^\sharp \tored{}{}{\PP} \{1:a_1\}$, where $a_{1} = \anno_{\Phi_1}(r\sigma_0) = r \sigma_0$.

  There must be a term $r_0 \trianglelefteq_{\sharp} r$ (i.e., $r_0^\sharp \sigma_0 \trianglelefteq r \sigma_0$) such that if we replace $a_{1} = r \sigma_0$ with $r_0^\sharp \sigma_0$ and obtain the same $\PP$-CT except when we would rewrite terms that do not exist anymore (i.e., we ignore these rewrite steps), then we still end up in an infinite number of nodes in $\aV$ (otherwise, $\F{T}$ would not have an infinite number of nodes in $\aV$).

  Let $\pi$ be the annotated position of $r \sigma_0$ where $r_0 \sigma_0 = \flat(r \sigma_0)|_\pi$.
  We can rewrite the term $t_0^\sharp$ with the dependency pair $\ell^\sharp \to r_0^\sharp \in \nonprobDP(\PP)$, using the substitution $\sigma_0$ since $t_0 = t = \ell \sigma_0$.
  Hence, we result in $t_0^\sharp \to_{\nonprobDP(\PP)} r_0^\sharp \sigma_0 = s_1^\sharp$.
  Next, we mirror the rewrite steps from the $\PP$-CT that are performed strictly below the root of $r_0^\sharp \sigma_0$ with $\nonprob(\PP)$ until we would rewrite at the root of $r_0^\sharp \sigma_0$.
  With this construction, we ensure that each the term $s_i^\sharp$ in our $(\nonprobDP(\PP),\nonprob(\PP))$-chain satisfies $s_i \trianglelefteq_{\sharp} a_i$ and $s_i = \flat(a_i)|_{\pi}$.
  A rewrite step at position $\pi.\tau$ in $a_i$ with $\PP$ is then mirrored with $\nonprob(\PP)$ in $s_i^\sharp$ at position $\tau$.
  Note that we only use a finite number of $\nonprob(\PP)$ steps until we rewrite at the root.

  So eventually, we result in a term $t_1^\sharp = s_{k}^\sharp$ with $s_k = \flat(a_k)|_{\pi}$, and we rewrite at position $\pi$ in the $\PP$-CT.
  We mirror the step $a_k \tored{}{}{\PP} a_{k+1}$ with $\nonprobDP(\PP)$ and then use the same construction again until we reach term $t_2^\sharp$, etc.
  This construction creates a sequence $t_0^\sharp, t_1^\sharp, \ldots$ of terms such that
  \[t_0^\sharp \; \to_{\nonprobDP(\PP)} r_0^\sharp\sigma_0 \; \to_{\nonprob(\PP)}^* \; t_1^\sharp \;
    \to_{\nonprobDP(\PP)} r_1^\sharp\sigma_1\; \to_{\nonprob(\PP)}^*\; \ldots \]
  Therefore, $(\nonprobDP(\PP),\nonprob(\PP))$ is not terminating.

  Again, the proof works analogously for $\mathtt{iAST}$ just considering innermost rewriting.
\end{myproof}

\SubtermCriterionInnermost*

\begin{myproof}
  Consider the polynomial interpretation that maps each $f(x_1, \ldots, x_n)$ with $f \in \Sigma$ to $x_1 + \ldots + x_n + 1$, and each $f^\sharp(x_1, \ldots, x_n)$ with $f^\sharp \in \Sigma^\sharp$ to $x_{\proj(f^\sharp)}$, i.e., we map each annotated symbol to its projected argument by the simple projection $\proj$, and all other symbols are used to calculate the number of symbols in a term.

  When choosing $\PP_\succ$ to be $\PP_\triangleright$, then the conditions of the reduction pair processor that consider the $\sharp$-sums (Case (1) and (2a)) are both satisfied with this interpretation.
  On the other hand, the conditions on the flattened ADPs without annotations (Case (2b) and (3)) may not be satisfied.
  However, due to the restriction to innermost rewriting and the restriction that ADPs only contain at most a single annotated symbol in each term on the right-hand side, those cases are irrelevant, because of the following observation:

  If we perform an innermost rewrite step with an ADP $\ell \to \mu^m \in \PP$ at an annotated position $\tau$ with the substitution $\sigma$, then $\ell^\sharp \sigma \in \ANF_{\PP}$.
  Let $r \in \Supp(\mu)$ and $t \trianglelefteq_{\sharp} r$ if $r$ contains an annotated symbol.
  Note that there exists at most one such $t$ for each $r$, as there is at most one annotation in $r$.
  Since $\proj(\ell^\sharp) \trianglerighteq \proj(t^\sharp)$ by definition of the subterm criterion, and all proper subterms of $\ell^\sharp \sigma$ are in $\NF_{\PP}$, we have $\proj(t^\sharp) \in \NF_{\PP}$.
  Hence, we cannot rewrite below the projected position before the next $(\mathbf{at})$- or $(\mathbf{af})$-step at the annotated position.
  More precisely, this means that the $\sharp$-sum of the terms cannot change by rewrite steps at non-annotated positions.
  Thus, in this case the reduction pair processor works without requiring Case (2b) and (3), as these requirements are never used in its proof if the rules have no impact on the $\sharp$-sum of a term.
\end{myproof}

\RewritingProc*

\begin{myproof}
  Let $\overline{\PP'} =\Proc_{\mathtt{r}}^{\mathbf{i}}(\PP)$ and $\overline{\PP} = \overline{\PP'} \cup \{\ell \ruleArr{}{}{} \{ p_1:r_{1}, \ldots, p_k:r_{k}\}^{m}\}$.
  We call $\ell \ruleArr{}{}{} \{ p_1:r_{1}, \ldots, p_k: r_k\}^{m}$ the \emph{old} ADP,
  we call $\ell \to
  (\{p_1:r_1, \ldots, p_k:r_k\}\linebreak \setminus \{p_j:r_j\} \cup \{p_j \cdot q_1:e_1, \ldots, p_j \cdot q_h:e_h\})^{m}$ the \emph{new} ADP, and $\ell \ruleArr{}{}{} \{ p_1:\flat(r_{1}), \ldots, p_k: \flat(r_{k})\}^{m}$ is called the \emph{non-annotated old} ADP.

  \medskip

  \noindent
  \underline{\textbf{First Case}}

  \noindent
  We start with the case where $\urules_{\PP}(r_j|_{\tau})$ is NO and the used rule $\hat{\ell} \to \{ \hat{p}_1:\hat{r}_1, \ldots, \hat{p}_h:\hat{r}_h\}$ is L and NE.
  First, note that the rule used for rewriting $r_j|_{\tau}$ in the rewrite processor is contained in $\urules_{\PP}(r_j|_{\tau})$.\footnote{The rules that are applicable at position $\tau$ do not have to be applicable in an innermost RST, see \cite[Sect.\ 6]{AAECC01} and \cite[Ex.\ 5.14]{thiemanndiss2007}.}
  Since $\urules_{\PP}(r_j|_{\tau})$ is NO, we can therefore be sure that there is only a single rule applicable at position $\tau$.

  \smallskip

  \noindent
  Let $\PP$ be not $\mathtt{iAST}$.
  Then there exists a $\PP$-CT $\F{T} = (V,E,L)$ that converges with probability $c <1$.
  We will now create a $\overline{\PP'}$-CT $\F{T}' = (V',E',L')$ such that $|\F{T}'| \leq |\F{T}| <1$, and hence $\overline{\PP'}$ is not $\mathtt{iAST}$ either.

  The core steps of this construction are the following:
  \begin{enumerate}
    \item[1.] We iteratively remove usages of the old ADP using a transformation $\Phi$ on CTs.
      The limit $\F{T}^{(\infty)}$ of this iteration is a $\overline{\PP'}$-CT 
      that converges with probability at most $|\F{T}|$. The transformation $\Phi$
      always removes a ``topmost'' usage of the old ADP at some node $x \in V$. 
      There are two different cases, depending on whether the old ADP is
      applied at an annotated position:
      \begin{enumerate}
        \item[1.1.] For a $\PP$-CT $\F{T}_x$ that uses the old ADP at the root $x$ at a position \linebreak
          $\pi \in \posT(t_x)$, we create a new $\overline{\PP}$-CT $\Phi(\F{T}_x)$ that uses the new ADP at the root.
        \item[1.2.] For a $\PP$-CT $\F{T}_x$ that uses the old ADP at the root $x$ at a position \linebreak
          $\pi \not\in \posT(t_x)$, we create a new $\overline{\PP}$-CT $\Phi(\F{T}_x)$
          that uses the non-annotated old ADP at the root.
      \end{enumerate}
  \end{enumerate}
  This proof structure will also be used in the other cases for soundness.

  \smallskip

  \noindent
  \textbf{\underline{1.
      We iteratively remove usages of the old ADP}}

  \noindent
  W.l.o.g., in $\F{T}$ there exists at least one rewrite step performed at some node $x$ with the old ADP (otherwise, $\F{T}$ would already be a $\overline{\PP'}$-CT).
  Furthermore, we can assume that this is the first such rewrite step in the path from the root to the node $x$ and that $x$ is a node of minimum depth with this property.
  We will now replace this rewrite step with a rewrite step using the new ADP such that we result in a CT $\F{T}^{(1)}$ with the following connections between $\F{T}$ and $\F{T}^{(1)}$.
  Here, for an infinite path $p = v_1, v_2, \ldots$ in a CT $\F{T} = (V,E,L)$, by $\mathcal{C}(p) = w_1, w_2, \ldots$ we denote the sequence of $\{\aV,\nV\}$-labels for these nodes, i.e., we have $w_i = \aV$ if $v_i \in \aV$ or $w_i = \nV$ otherwise for all $i$.
  \begin{itemize}
    \item[(a)] $|\F{T}^{(1)}| \leq |\F{T}| = c$, and
    \item[(b)] for every infinite path $p$ in $\F{T}^{(1)}$ we can find an infinite path $p'$ in $\F{T}$ such that $\mathcal{C}(p)$ and $\mathcal{C}(p')$ only differ in a finite number of $\nV$-nodes, i.e., we can remove and add a finite number of $\nV$-nodes from $\mathcal{C}(p')$ to get $\mathcal{C}(p)$.
  \end{itemize}
  This construction only works because $\urules_{\PP}(r_j|_{\tau})$ is NO and there exists no annotation below $\tau$, and the resulting CT has only at most the same probability of termination as the original one due to the fact that $\hat{\ell} \to \{ \hat{p}_1:\hat{r}_1, \ldots, \hat{p}_h:\hat{r}_h\}^{m'}$ is also L and NE.
  Let $\F{T}_x$ be the induced subtree that starts at node $x$, i.e.
  $\F{T}_x = \F{T}[xE^*]$.
  The construction defines a new tree $\Phi(\F{T}_x)$ such that (a) and (b) w.r.t.\ $\F{T}_x$ and $\Phi(\F{T}_x)$ holds, and where we use the new ADP at the root node $x$ instead of the old one (i.e., we pushed the first use of the old ADP deeper into the tree).
  Then, by replacing the subtree $\F{T}_x$ with the new tree $\Phi(\F{T}_x)$ in $\F{T}$, we get a $\PP$-CT $\F{T}^{(1)}$, with (a) and (b) w.r.t.\ $\F{T}$ and $\F{T}^{(1)}$, and where we use the new ADP at node $x$ instead of the old one.
  We can then do this replacement iteratively for every use of the old ADP, i.e., we again replace the first use of the old ADP in $\F{T}^{(1)}$ to get $\F{T}^{(2)}$ with (a) and (b) w.r.t.\ $\F{T}^{(1)}$ and $\F{T}^{(2)}$, and so on.
  The limit of all these CTs $\lim_{i \to \infty} \F{T}^{(i)}$ is a $\overline{\PP'}$-CT, that we denote by $\F{T}^{(\infty)}$ and that converges with probability at most $c <1$, and hence, $\overline{\PP'}$ is not $\mathtt{iAST}$.

  To see that $\F{T}^{(\infty)}$ is indeed a valid $\overline{\PP'}$-CT, note that in every iteration of the construction we turn a use of the old ADP at minimum depth into a use of the new one.
  Hence, for every depth $H$ of the tree, we eventually turned every use of the old ADP up to depth $H$ into a use of the new one so that the construction will not change the tree above depth $H$ anymore, i.e., there exists an $m_H$ such that $\F{T}^{(\infty)}$ and $\F{T}^{(m_H)}$ are the same trees up to depth $H$.
  This means that the sequence $\lim_{i \to \infty} \F{T}^{(i)}$ really converges into a tree that satisfies the first three conditions of a $\overline{\PP'}$-CT.
  We only have to show that the last condition of a CT, namely that every infinite path in $\lim_{i \to \infty} \F{T}^{(i)}$ contains infinitely many nodes from $\aV$, holds as well.
  First, by induction on $n$ one can prove that all trees $\F{T}^{(i)}$ for all $1 \leq i \leq n$ satisfy Condition (4), because due to (b) we can find for each infinite path $p \in \F{T}^{(i)}$ an infinite path $p'$ in $\F{T}$ such that $\mathcal{C}(p)$ and $\mathcal{C}(p')$ only differ in a finite number of $\nV$-labels.
  Hence, if $p$ contains no nodes from $\aV$, then $p'$ contains no nodes from $\aV$, which is a contradiction to $\F{T}$ satisfying Condition (4).
  Moreover, we only replace subtrees after a node in $\aV$, the node itself remains in $\aV$, and after we replaced the subtree at a node $v$, we will never replace a predecessor of $v$ anymore.
  This means that the subsequence between the $i$-th and $(i+1)$-th occurrence of $\aV$ in $p$ and the subsequence between the $i$-th and $(i+1)$-th occurrence of $\aV$ in $p'$ only differ in a finite number of $\nV$ nodes again, for every $i$.
  Now, let $p = v_1, v_2, \ldots$ be an infinite path that starts at the root in $\F{T}^{(\infty)}$ and only contains finitely many nodes from $\aV$, i.e., there exists an $1 \leq i$ such that $v_i, v_{i+1}, \ldots$ contains no node from $\aV$.
  Again, let $m_H \in \IN$ such that $\F{T}^{(\infty)}$ and $\F{T}^{(m_H)}$ are the same trees up to depth $H$.
  The path $v_1, \ldots, v_H$ must be a path in $\F{T}^{(m_H)}$ as well.
  For two different $H_1, H_2$ with $i < H_1 < H_2$ we know that since the path $v_i, \ldots, v_{H_2}$ contains no node from $\aV$, the path must also exist in $\F{T}^{(m_{H_1})}$.
  We can now construct an infinite path in $\F{T}^{(m_i)}$ that contains no nodes from $\aV$, which is a contradiction.

  Next, we want to prove that we really have $|\F{T}^{(\infty)}| \leq c$.
  Again, by induction on $n$ one can prove that $|\F{T}^{(i)}| \leq c$ for all $1 \leq i \leq n$.
  Assume for a contradiction that $\F{T}^{(\infty)}$ converges with probability greater than $c$, i.e.
  $|\F{T}^{(\infty)}| > c$.
  Then there exists an $H \in \IN$ for the depth such that $\sum_{x \in \ctleaf^{\F{T}^{(\infty)}}, d(x) \leq H} p_x > c$.
  Here, $d(x)$ denotes the depth of node $x$.
  Again, let $m_H \in \IN$ such that $\F{T}^{(\infty)}$ and $\F{T}^{(m_H)}$ are the same trees up to depth $H$.
  But this would mean that $|\F{T}^{(m_H)}| \geq \sum_{x \in \ctleaf^{\F{T}^{(m_H)}}, d(x) \leq H} p_x = \sum_{x \in \ctleaf^{\F{T}^{(\infty)}}, d(x) \leq H} p_x > c$, which is a contradiction to $|\F{T}^{(m_H)}| \leq c$.

  It remains to show the mentioned construction $\Phi(\circ)$.

  \smallskip

  \noindent
  \textbf{\underline{1.1.
      Construction of $\Phi(\circ)$ if $\pi \in \posT(t_x)$}}

  \noindent
  Let $\F{T}_x$ be a $\PP$-CT that uses the old ADP at the root node $x$, i.e, $t_x \itored{}{}{\PP} \{p_{y_1}:t_{y_1}, \ldots, p_{y_k}:t_{y_k}\}$ using the ADP $\ell \ruleArr{}{}{} \{ p_1:r_{1}, \ldots, p_k: r_k\}^{m}$, the position $\pi$ with $\pi \in \posT(t_x)$, and a substitution $\sigma$ such that $\flat(t_x|_{\pi}) = \ell \sigma \in \ANF_{\PP}$.
  Then $t_{y_j} = t_x[r_j\sigma]_{\pi}$ if $m = \ttrue$, or $t_{y_j} = \disannoPos{\pi}(t_x[r_j\sigma]_{\pi})$, otherwise.

  \smallskip

  \noindent
  \textbf{\underline{1.1.1 General construction of $\Phi(\F{T}_x)$}}

  \noindent
  Instead of applying only the old ADP at the root $x$
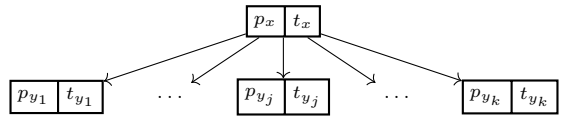
  \begin{center}
    \scriptsize
    \begin{tikzpicture}
      \tikzstyle{adam}=[thick,draw=black!100,fill=white!100,minimum size=4mm, shape=rectangle split, rectangle split parts=2,rectangle split horizontal] \tikzstyle{empty}=[rectangle,thick,minimum size=4mm]

      \node[adam] at (1.5, 2) (1) {$p_x$
        \nodepart{two} $t_x$};
      \node[adam] at (-1.5, 1) (11) {$p_{y_1}$
        \nodepart{two} $t_{y_1}$};
      \node[empty] at (0, 1) (12) {$\ldots$};
      \node[adam] at (1.5, 1) (13) {$p_{y_j}$
        \nodepart{two} $t_{y_j}$};
      \node[empty] at (3, 1) (14) {$\ldots$};
      \node[adam] at (4.5, 1) (15) {$p_{y_k}$
        \nodepart{two} $t_{y_k}$};

      \draw (1) edge[->] (11);
      \draw (1) edge[->] (12);
      \draw (1) edge[->] (13);
      \draw (1) edge[->] (14);
      \draw (1) edge[->] (15);
    \end{tikzpicture}
  \end{center}
  where we use an arbitrary rewrite step at node $y_j$ afterwards, we want to directly apply the rewrite rule at position $\pi.\tau$ of the term $t_{y_j}$ in our CT, which we performed on $r_j$ at position $\tau$ to transform the old into the new ADP, to get
  \begin{center}
    \scriptsize
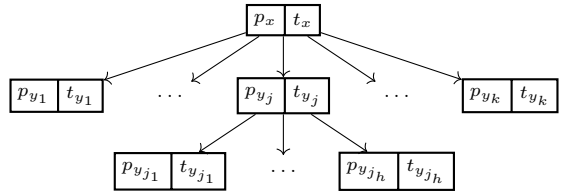
    \begin{tikzpicture}
      \tikzstyle{adam}=[thick,draw=black!100,fill=white!100,minimum size=4mm, shape=rectangle split, rectangle split parts=2,rectangle split horizontal] \tikzstyle{empty}=[rectangle,thick,minimum size=4mm]

      \node[adam] at (1.5, 2) (1) {$p_x$
        \nodepart{two} $t_x$};
      \node[adam] at (-1.5, 1) (11) {$p_{y_1}$
        \nodepart{two} $t_{y_1}$};
      \node[empty] at (0, 1) (12) {$\ldots$};
      \node[adam] at (1.5, 1) (13) {$p_{y_j}$
        \nodepart{two} $t_{y_j}$};
      \node[empty] at (3, 1) (14) {$\ldots$};
      \node[adam] at (4.5, 1) (15) {$p_{y_k}$
        \nodepart{two} $t_{y_k}$};
      \node[adam] at (0, 0) (131) {$p_{y_{j_1}}$
        \nodepart{two} $t_{y_{j_1}}$};
      \node[empty] at (1.5, 0) (132) {$\ldots$};
      \node[adam] at (3, 0) (133) {$p_{y_{j_h}}$
        \nodepart{two} $t_{y_{j_h}}$};

      \draw (1) edge[->] (11);
      \draw (1) edge[->] (12);
      \draw (1) edge[->] (13);
      \draw (1) edge[->] (14);
      \draw (1) edge[->] (15);
      \draw (13) edge[->] (131);
      \draw (13) edge[->] (132);
      \draw (13) edge[->] (133);
    \end{tikzpicture}
  \end{center}
  Then, we can contract the edge $(x,y_j)$ to get
  \begin{center}
    \scriptsize
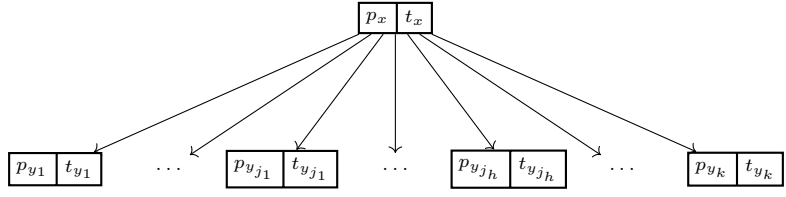
    \begin{tikzpicture}
      \tikzstyle{adam}=[thick,draw=black!100,fill=white!100,minimum size=4mm, shape=rectangle split, rectangle split parts=2,rectangle split horizontal] \tikzstyle{empty}=[rectangle,thick,minimum size=4mm]

      \node[adam] at (3, 3) (1) {$p_x$
        \nodepart{two} $t_x$};
      \node[adam] at (-1.5, 1) (11) {$p_{y_1}$
        \nodepart{two} $t_{y_1}$};
      \node[empty] at (0, 1) (12) {$\ldots$};
      \node[adam] at (1.5, 1) (13) {$p_{y_{j_1}}$
        \nodepart{two} $t_{y_{j_1}}$};
      \node[empty] at (3, 1) (14) {$\ldots$};
      \node[adam] at (4.5, 1) (15) {$p_{y_{j_h}}$
        \nodepart{two} $t_{y_{j_h}}$};
      \node[empty] at (6, 1) (16) {$\ldots$};
      \node[adam] at (7.5, 1) (17) {$p_{y_{k}}$
        \nodepart{two} $t_{y_{k}}$};

      \draw (1) edge[->] (11);
      \draw (1) edge[->] (12);
      \draw (1) edge[->] (13);
      \draw (1) edge[->] (14);
      \draw (1) edge[->] (15);
      \draw (1) edge[->] (16);
      \draw (1) edge[->] (17);
    \end{tikzpicture}
  \end{center}
  and this is equivalent to applying the new ADP.
  Note that the rewrite step at position $\pi.\tau$ may not be an innermost rewrite step in the CT.

  The subtrees that start at the nodes $y_1, \ldots, y_{j-1}, y_{j+1}, \ldots, y_{k}$ remain completely the same.
  We only have to construct a new subtree for node $y_j$, i.e.,
  \begin{wrapfigure}[10]{r}{0.37\textwidth}
    \vspace*{-0.9cm}
    \begin{center}
      \begin{tikzpicture}[scale=0.5]
        \begin{pgfonlayer}{nodelayer}
          \node [style=target,pin={[pin distance=0.05cm, pin edge={,-}] 140:\tiny \textcolor{blue}{$x$}}] (0) at (0, 6) {};
          \node [style=none] (1) at (0, 3) {};
          \node [style=target] (2) at (-3, 3) {};
          \node [style=target,pin={[pin distance=0.05cm, pin edge={,-}] 140:\tiny \textcolor{blue}{$y_j$}}] (3) at (0, 3) {};
          \node [style=none] (6) at (1.5, 0) {};
          \node [style=none] (7) at (-1.5, 0) {};
          \node [style=none] (9) at (-2, -1) {};
          \node [style=none] (10) at (2, -1) {};
          \node [style=moveBlock] (12) at (-0.75, 0.75) {};
          \node [style=moveBlock] (13) at (0, 0) {};
          \node [style=moveBlock] (14) at (0.75, 0.25) {};
          \node [style=moveBlock] (15) at (-0.75, 0.75) {};
          \node [style=none] (16) at (-0.5, 1.5) {};
          \node [style=none] (17) at (0, 1.25) {};
          \node [style=none] (18) at (0.5, 1.5) {};
          \node [style=none] (19) at (-0.75, 0) {};
          \node [style=none] (20) at (0, -0.75) {};
          \node [style=none] (21) at (0.75, -0.5) {};
          \node [style=target] (22) at (3, 3) {};
        \end{pgfonlayer}
        \begin{pgfonlayer}{edgelayer}
          \draw (0) to (1.center);
          \draw (0) to (2);
          \draw (3) to (6.center);
          \draw (3) to (7.center);
          \draw [style=dotWithoutHead] (7.center) to (9.center);
          \draw [style=dotWithoutHead] (6.center) to (10.center);
          \draw [style=dotWithoutHead, in=15, out=-105, looseness=0.50] (3) to (16.center);
          \draw [style=dotWithoutHead, in=120, out=-90, looseness=0.75] (3) to (17.center);
          \draw [style=dotWithoutHead, in=135, out=-75] (3) to (18.center);
          \draw [style=dotHead, in=90, out=-30, looseness=0.75] (18.center) to (14);
          \draw [style=dotHead, in=90, out=-150, looseness=0.75] (16.center) to (15);
          \draw [style=dotHead, in=90, out=-45] (17.center) to (13);
          \draw [style=dashHead, bend right=75, looseness=2.00] (14) to (3);
          \draw [style=dashHead, bend left=75, looseness=1.75] (15) to (3);
          \draw [style=dashHead, bend right=105, looseness=2.75] (13) to (3);
          \draw [style=dotWithoutHead] (15) to (19.center);
          \draw [style=dotWithoutHead] (13) to (20.center);
          \draw [style=dotWithoutHead] (14) to (21.center);
          \draw (0) to (22);
        \end{pgfonlayer}
      \end{tikzpicture}
    \end{center}
  \end{wrapfigure}
  the node that really changed when applying the rewriting processor.
  To be precise, let $\F{T}_{y_j} = \F{T}_x[y_jE^*]$ be the subtree starting at node $y_j$.
  The construction first creates a new subtree $\Psi(\F{T}_{y_j})$ such that (a) and (b) hold w.r.t.\ $\F{T}_{y_j}$ and $\Psi(\F{T}_{y_j})$, and that directly performs the first rewrite step at position $\pi.\tau$ at the root of the tree, by pushing it from the original position in the tree $\F{T}_{y_j}$ to the root.
  This can be seen in the diagram on the side.
  Again, this push only results in the exact same termination probability due to our restriction that $\hat{\ell} \to \{ \hat{p}_1:\hat{r}_1, \ldots, \hat{p}_h:\hat{r}_h\}$ is L and NE.
  Then, by replacing $\F{T}_{y_j}$ by $\Psi(\F{T}_{y_j})$ in $\F{T}_x$ we result in a tree $\F{T}_x'$ such that (a) and (b) hold w.r.t.\ $\F{T}_x$ and $\F{T}_x'$, and such that we perform the desired rewrite step at node $y_j$.
  Finally, we contract the edge $(x,y_j)$ in $\F{T}_x'$, in order to get $\Phi(\F{T}_x)$.
  Again, (a) and (b) hold w.r.t.\ $\F{T}_x$ and $\Phi(\F{T}_x)$, and we use the new ADP at the root $x$ in $\Phi(\F{T}_x)$.
  It only remains to explain the construction of $\Psi(\F{T}_{y_j})$.
  \smallskip

  \noindent
  \textbf{\underline{1.1.2.
      Construction of $\Psi(\F{T}_{y_j})$}}

  \noindent
  We will move the first rewrite step that takes place at position $\pi.\tau$ from the original tree $\F{T}_{y_j}$ (example on the left below) to the top of the new tree $\Psi(\F{T}_{y_j})$ (example on the right below) and show that (a) and (b) both hold after this construction.
  Below, the circled nodes represent the nodes where we perform a rewrite step at position $\pi.\tau$.
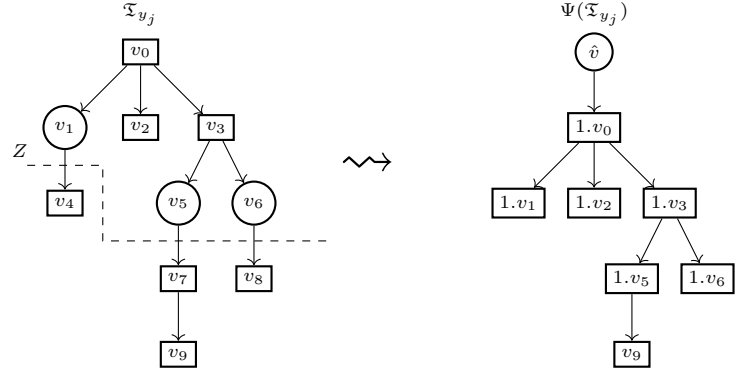
  \begin{center}
    \centering \scriptsize
    \begin{tikzpicture}
      \tikzstyle{adam}=[rectangle,thick,draw=black!100,fill=white!100,minimum size=3mm] \tikzstyle{empty}=[shape=circle,thick,minimum size=8mm] \tikzstyle{circle}=[shape=circle,draw=black!100,fill=white!100,thick,minimum size=3mm]

      \node[empty] at (-3.5, 0.5) (name) {$\F{T}_{y_j}$};
      \node[adam] at (-3.5, 0) (la) {$v_0$};

      \node[circle] at (-4.5, -1) (lb1) {$v_1$};
      \node[adam] at (-3.5, -1) (lb2) {$v_2$};
      \node[adam] at (-2.5, -1) (lb3) {$v_3$};

      \node[adam] at (-4.5, -2) (lc1) {$v_4$};

      \node[circle] at (-3, -2) (ld1) {$v_5$};
      \node[circle] at (-2, -2) (ld3) {$v_6$};

      \node[adam] at (-3, -3) (lf1) {$v_7$};
      \node[adam] at (-2, -3) (lf3) {$v_8$};

      \node[adam] at (-3, -4) (lg1) {$v_9$};

      \draw (la) edge[->] (lb1);
      \draw (la) edge[->] (lb2);
      \draw (la) edge[->] (lb3);
      \draw (lb1) edge[->] (lc1);
      \draw (lb3) edge[->] (ld1);
      \draw (lb3) edge[->] (ld3);
      \draw (ld1) edge[->] (lf1);
      \draw (ld3) edge[->] (lf3);
      \draw (lf1) edge[->] (lg1);

      \node[empty] at (-5.1,-1.3) (Z) {$Z$};

      \draw [dashed] (-5,-1.5) -- (-4,-1.5) -- (-4,-2.5) -- (-1,-2.5);

      \node[empty] at (-0.5, -1.5) (lead) {\huge $\leadsto$};

      \node[empty] at (2.5, 0.5) (name2) {$\Psi(\F{T}_{y_j})$};
      \node[circle] at (2.5, 0) (a) {$\hat{v}$};

      \node[adam] at (2.5, -1) (b1) {$1.v_0$};

      \node[adam] at (1.5, -2) (c1) {$1.v_1$};
      \node[adam] at (2.5, -2) (c2) {$1.v_2$};
      \node[adam] at (3.5, -2) (c3) {$1.v_3$};

      \node[adam] at (3, -3) (d1) {$1.v_5$};
      \node[adam] at (4, -3) (d2) {$1.v_6$};

      \node[adam] at (3, -4) (f1) {$v_9$};

      \draw (a) edge[->] (b1);
      \draw (b1) edge[->] (c1);
      \draw (b1) edge[->] (c2);
      \draw (b1) edge[->] (c3);
      \draw (c3) edge[->] (d1);
      \draw (c3) edge[->] (d2);
      \draw (d1) edge[->] (f1);
    \end{tikzpicture}
  \end{center}
  We will define the $\PP$-CT $\Psi(\F{T}_{y_j})$ that satisfies the properties (a) and (b) w.r.t.\ $\F{T}_{y_j}$ and $\Psi(\F{T}_{y_j})$, and that directly performs the rewrite step $t_{y_j} \itored{}{}{\PP, \pi.\tau, \ttrue} \{\hat{p}_{1}:t_{y_{j_1}}, \ldots, \hat{p}_{h}:t_{y_{j_h}}\}$, with the rule $\hat{\ell} \to \{ \hat{p}_1:\hat{r}_1, \ldots, \hat{p}_h:\hat{r}_h\}^{\ttrue} \in \PP$, a substitution $\hat{\sigma}$, and the position $\pi.\tau$, at the new root $\hat{v}$.
  Here, we have $\flat(t_{y_j}|_{\pi.\tau}) = \flat(r_j\sigma|_{\tau}) = \flat(r_j|_{\tau} \sigma) = \hat{\ell} \hat{\sigma}$.
  Let $Z$ be the set of all nodes $v$ of $\F{T}_{y_j}$ where we did not perform a rewrite step at position $\pi.\tau$ in the path from the root $x$ to the node $v$, or $v$ is the first node in the path that performs a rewrite step at position $\pi.\tau$.
  In the example we have $Z = \{v_0, \ldots, v_6\} \setminus \{v_4\}$.
  For each of these nodes $z \in Z$ and each $1 \leq e \leq h$, we create a new node $e.z \in V'$ with edges as in $\F{T}_{y_j}$ for the nodes in $Z$, e.g., for the node $1.v_3$ we create an edge to $1.v_5$ and $1.v_6$.
  Furthermore, we add the edges from the new root $\hat{v}$ to the nodes $e.{y_j}$ for all $1 \leq e \leq h$.
  Remember that $y_j$ was the root in the tree $\F{T}_{y_j}$ and has to be contained in $Z$.
  For example, for the node $\hat{v}$ we create an edge to $1.v_0$.
  For all these new nodes in $Z$, we show that the following holds:
  \begin{itemize}
    \item[] (T1) $t_z[\hat{r}_e \gamma]_{\pi.\tau} \flateq t'_{e.z}$ for the substitution $\gamma$ such that $\flat(t_z|_{\pi.\tau}) = \hat{\ell} \gamma$
    \item[] (T2) $\posT(t_z[\flat(\hat{r}_e) \gamma]_{\pi.\tau}) \subseteq \posT(t'_{e.z})$.
    \item[] (T3) $p_{e.z}^{\Psi(\F{T}_{y_j})} = p_z^{\F{T}_{y_j}} \cdot \hat{p}_e$
  \end{itemize}
  Note that we only regard the subtree of the $j$-th child of the root.
  By the prerequisite in the definition of the rewrite processor (there is no annotation below or at position $\tau$), there are no annotations on or below $\pi.\tau$ for all nodes in $Z$.

  Now, for a leaf $e.z \in V'$ either $z \in V$ is also a leaf (e.g., node $v_2$) or we rewrite the position $\pi.\tau$ at node $z$ in $\F{T}_{y_j}$ (e.g., node $v_1$).
  If we rewrite $t_{z} \itored{}{}{\PP, \pi.\tau, \ttrue} \{\hat{p}_{1}:t_{w_1}, \ldots, \hat{p}_{h}:t_{w_h}\}$, then we have $t_{w_e} = t_z[\flat(\hat{r}_e)\gamma]_{\pi.\tau}$.
  Here, we get $t_{w_e} \flateq t_z[\flat(\hat{r}_e)\gamma]_{\pi.\tau} \flateq_{(T1)} t'_{e.z}$, $\posT(t_{w_e}) \subseteq \posT(t_z[\flat(\hat{r}_e) \gamma]_{\pi.\tau}) \subseteq_{(T2)} \posT(t'_{e.z})$ and $p_{e.z}^{\Psi(\F{T}_{y_j})} = p_z^{\F{T}_{y_j}} \cdot \hat{p}_e =_{(T3)} p_{w_e}^{\F{T}_{y_j}}$, and we can again copy the rest of this subtree of $\F{T}_{y_j}$ in our newly generated tree $\Psi(\F{T}_{y_j})$.
  In our example, $v_1$ has the only successor $v_4$, hence we can copy the subtree starting at node $v_4$, which is only the node itself, to the node $1.v_1$ in $\Psi(\F{T}_{y_j})$.
  For $v_5$ we have the only successor $v_7$, hence we can copy the subtree starting at node $v_7$, which is the node itself together with its successor $v_9$, to the node $1.v_5$ in $\Psi(\F{T}_{y_j})$.
  So essentially, we just have to define the part of the tree before we reach the rewrite step in $\F{T}_{y_j}$, and then, we have to show that (a) and (b) for $\F{T}_{y_j}$ and $\Psi(\F{T}_{y_j})$ are satisfied.
  We show the latter first, and then explain the proof that this label gives us indeed a valid $\PP$-CT.

  We start by showing (a) for $\F{T}_{y_j}$ and $\Psi(\F{T}_{y_j})$.
  Let $u$ be a leaf in $\Psi(\F{T}_{y_j})$.
  If $u = e.v$ for some node $v \in Z$ that is a leaf in $\F{T}_{y_j}$ (e.g., node $1.v_2$), then also $e.v$ must be a leaf in $\Psi(\F{T}_{y_j})$ for every $1 \leq e \leq h$.
  Here, we get $\sum_{1 \leq e \leq h} p_{e.v}^{\Psi(\F{T}_{y_j})} \stackrel{\text{(T3)}}{=} \sum_{1 \leq e \leq h} p_v^{\F{T}_{y_j}} \cdot \hat{p}_e = p_v^{\F{T}_{y_j}} \cdot \sum_{1 \leq e \leq h} \hat{p}_e = p_v^{\F{T}_{y_j}} \cdot 1 = p_v^{\F{T}_{y_j}}$.
  If $u = e.v$ for some node $v \in Z$ that is not a leaf in $\F{T}_{y_j}$ (e.g., node $1.v_1$), then we know by construction that all successors of $v$ in $\F{T}_{y_j}$ are not contained in $Z$ and are leaves.
  Here, we get $p_{e.v}^{\Psi(\F{T}_{y_j})} \stackrel{\text{(T3)}}{=} p_v^{\F{T}_{y_j}} \cdot \hat{p}_e = p_w^{\F{T}_{y_j}}$ for the (unique) $e$-th successor $w$ of $v$.
  Finally, if $u$ does not have the form $u = e.v$, then $u$ is also a leaf in $\F{T}_{y_j}$ with $p_{u}^{\Psi(\F{T}_{y_j})} = p_{u}^{\F{T}_{y_j}}$.
  Note that these cases cover no leaf of $\F{T}_{y_j}$ twice.
  This implies that we have
    {\small \allowdisplaybreaks
      \begin{align*}
             & |\Psi(\F{T}_{y_j})| \\
        =    & \sum_{v \in \ctleaf^{\Psi(\F{T}_{y_j})}} p_{e.v}^{\Psi(\F{T}_{y_j})} \\
        =    & \sum_{\substack{e.v \in \ctleaf^{\Psi(\F{T}_{y_j})} \\ v \in Z}} p_{e.v}^{\Psi(\F{T}_{y_j})} + \sum_{\substack{e.v \in \ctleaf^{\Psi(\F{T}_{y_j})}\\ v \in wE, v \not\in Z, w \in Z}} p_{e.v}^{\Psi(\F{T}_{y_j})} + \sum_{\substack{v \in \ctleaf^{\Psi(\F{T}_{y_j})}\\v \in \ctleaf^{\F{T}_{y_j}}, v \in wE, w \not\in Z}} p_v^{\Psi(\F{T}_{y_j})} \\
        \leq & \sum_{\substack{v \in \ctleaf^{\F{T}_{y_j}} \\ v \in Z}} \left( \sum_{1 \leq e \leq h} p_{e.v}^{\Psi(\F{T}_{y_j})} \right) + \sum_{\substack{v \in \ctleaf^{\F{T}_{y_j}}, 1 \leq e \leq h\\ v \in wE, v \not\in Z, w \in Z}} p_{e.v}^{\Psi(\F{T}_{y_j})} + \sum_{\substack{v \in \ctleaf^{\F{T}_{y_j}}\\ v \in wE, w \not\in Z}} p_v^{\Psi(\F{T}_{y_j})} \\
        \leq & \sum_{\substack{v \in \ctleaf^{\F{T}_{y_j}} \\ v \in Z}} \!\! p_v^{\F{T}_{y_j}} + \sum_{\substack{v \in \ctleaf^{\F{T}_{y_j}}\\ v \in wE, v \not\in Z, w \in Z}} p_v^{\F{T}_{y_j}} + \sum_{\substack{v \in \ctleaf^{\F{T}_{y_j}}\\ v \in wE, w \not\in Z}} p_v^{\F{T}_{y_j}} \\
        =    & \sum_{v \in \ctleaf^{\F{T}_{y_j}}} p_v^{\F{T}_{y_j}} \\
        =    & |\F{T}_{y_j}|
      \end{align*}
    }
  Next, we show (b) for $\F{T}_{y_j}$ and $\Psi(\F{T}_{y_j})$.
  Let $p = u_0, u_1, \ldots$ be an infinite path in $\Psi(\F{T}_{y_j})$ that starts at the root $\hat{v}$.
  If for all $1 \leq i$ we have $u_i = e.v_i$ for some node $v_i \in Z$ and $1 \leq e \leq h$, then $p = v_1, \ldots$ is our desired path in $\F{T}_{y_j}$.
  Otherwise, there is a maximal $1 \leq o$ such that for all $1 \leq i \leq o$ we have $u_i = e.v_i$ for some node $v_i \in Z$ and $1 \leq e \leq h$.
  Then our desired path is $v_1, \ldots, v_o, w, u_{o+1}, \ldots$.
  Here, $w$ is the $e$-th successor of $v_o$ in $\F{T}_{y_j}$.
  Note that in the first case, we remove one $\nV$-label at the start of our path, while in the other case, we just move an $\nV$-label from the first position to a later one in the path.
  This shows that (b) is satisfied.

  Finally, we prove that $\Psi(\F{T}_{y_j})$ is a valid $\PP$-CT.
  We only need to prove that the construction satisfies all of our conditions for the nodes in $Z$.
  As the rest of the tree is copied, we can be sure that all Conditions (1)-(3) of a $\PP$-CT are satisfied.
  Additionally, due to (b) w.r.t.\ $\F{T}_{y_j}$ and $\Psi(\F{T}_{y_j})$, we get (4) as well, because $\F{T}_{y_j}$ satisfies (4).

  \smallskip

  At the root, after applying the old ADP, we directly perform the rewrite step $t_{y_j} \itored{}{}{\PP} \{\hat{p}_{1}:t'_{1.y_{j}}, \ldots, \hat{p}_{h}:t'_{h.y_{j}}\}$, with the rule $\hat{\ell} \to \{ \hat{p}_1:\hat{r}_1, \ldots, \hat{p}_h:\hat{r}_h\}^{\ttrue} \in \PP$, a substitution $\hat{\sigma}$, and the position $\tau$.
  Then, $t_{e.y_{j}} = t_{y_j}[\flat(\hat{r}_e) \hat{\sigma}]_{\pi.\tau}$.
  Here, the conditions (T1)-(T3) are clearly satisfied.
  We now construct the rest of this subtree by mirroring the rewrite steps in the original tree (which is always possible due to the fact that $\urules_{\PP}(r_j|_{\tau})$ is NO), and once we encounter the rewrite step that we moved to the top, i.e., once we use a rewrite step at position $\pi.\tau$, we skip this rewrite step and directly go on with the $e$-th successor if we are in a path that went to the $e$-th successor in the initial rewrite step, as described above.
  In the following, we distinguish between two different cases for a rewrite step at a node $u$:
  \begin{enumerate}
    \item[(A)] We use a step with $\itored{}{}{\PP}$ in $\F{T}_{y_j}$ at a position orthogonal to $\pi.\tau$.
    \item[(B)] We use a step with $\itored{}{}{\PP}$ in $\F{T}_{y_j}$ at a position below $\pi.\tau$.
      Note that this is the more interesting case, where we need to use the properties L and NE.
  \end{enumerate}
  Note that we cannot rewrite above $\pi.\tau$ before rewriting at position $\pi.\tau$ due to the innermost restriction.

  \noindent
  \textbf{(A) If we have} $t_u \itored{}{}{\PP} \{\tfrac{p_{q_1}}{p_u}:t_{q_1}, \ldots, \tfrac{p_{q_{\overline{h}}}}{p_u}:t_{q_{\overline{h}}}\}$, then there is a rule $\bar{\ell} \to \{ \bar{p}_1:\bar{r}_1, \ldots, \bar{p}_{{\overline{h}}}:\bar{r}_{{\overline{h}}}\}^{\overline{m}} \in \PP$, a substitution $\delta$, and a position $\zeta \in \IN^+$ with $\flat(t_u|_{\zeta}) = \bar{\ell} \delta \in \ANF_{\PP}$.
  Furthermore, let $\zeta \bot \pi.\tau$.
  Then, we have $t'_{e.u}|_{\zeta} \flateq_{(T1)} t_u[\hat{r}_e \gamma]_{\pi.\tau}|_{\zeta} = t_{u}|_{\zeta}$ for the substitution $\gamma$ such that $t_u|_{\pi.\tau} = \hat{\ell} \gamma$, and we can rewrite $t'_{e.u}$ using the same rule, same substitution, and same position.
  Then (T3) is again satisfied.
  Furthermore, we have $t'_{e.{q_j}} = t'_{e.u}[\bar{r}_e \delta]_{\zeta}$ if $\zeta \in \posT(t'_{e.{u}})$ and $m = \ttrue$, $t'_{e.{q_j}} = \disannoPos{\zeta}(t'_{e.u}[\bar{r}_e \delta]_{\zeta})$ if $\zeta \in \posT(t'_{e.{u}})$ and $m = \tfalse$, $t'_{e.{q_j}} = t'_{e.u}[\flat(\bar{r}_e) \delta]_{\zeta}$ if $\zeta \notin \posT(t'_{e.{u}})$ and $m = \ttrue$, or $t'_{e.{q_j}} = \disannoPos{\zeta}(t'_{e.u}[\flat(\bar{r}_e) \delta]_{\zeta})$ if $\zeta \notin \posT(t'_{e.{u}})$ and $m = \tfalse$.
  In all four cases we get (T1) (using the same substitution $\gamma$) and (T2) as well.

  \noindent
  \textbf{(B) If we have} $t_u \itored{}{}{\PP} \{\tfrac{p_{q_1}}{p_u}:t_{q_1}, \ldots, \tfrac{p_{q_{\overline{h}}}}{p_u}:t_{q_{\overline{h}}}\}$, then there is a rule $\bar{\ell} \to \{ \bar{p}_1:\bar{r}_1, \ldots, \bar{p}_{{\overline{h}}}:\bar{r}_{{\overline{h}}}\}^{\overline{m}} \in \PP$, a substitution $\delta$, and a position $\zeta \in \IN^+$ with $\flat(t_u|_{\zeta}) = \bar{\ell} \delta \in \ANF_{\PP}$.
  Furthermore, let $\zeta > \pi.\tau$.
  Since $\hat{\ell} \to \{ \hat{p}_1:\hat{r}_1, \ldots, \hat{p}_h:\hat{r}_h\}$ is L and NE, we know that $\hat{\ell}$ contains exactly the same variables as $\hat{r}_e$ and all of them exactly once.
  Furthermore, since $\urules_{\PP}(r_j|_{\tau})$ is non-overlapping, we know that the rewriting must be completely inside the substitution $\gamma$ for the substitution $\gamma$ such that $t_u|_{\pi.\tau} = \hat{\ell} \gamma$, i.e., there is a position $\eta_c$ of a variable $c$ in $\hat{\ell}$ and another position $\chi_c$ with $\pi.\tau.\eta_c.\chi_c = \zeta$.
  Let $\varphi_e(c)$ be the (unique) variable position of $c$ in $\hat{r}_e$.
  Then, we have $t'_{e.u}|_{\pi.\tau.\varphi_e(c).\chi_c} \flateq_{(T1)} t_u[\hat{r}_e \gamma]_{\pi.\tau}|_{\pi.\tau.\varphi_e(c).\chi_c} = \hat{r}_e \gamma|_{\varphi_e(c).\chi_c} = \gamma(c)|_{\chi_c} = \hat{\ell} \hat{\sigma}|_{\eta_c.\chi_c} = t_{u}|_{\pi.\tau.\eta_c.\chi_c}$, and we can rewrite $t'_{e.u}$ using the same rule, same substitution, and position $\pi.\tau.\varphi_e(c).\chi_c$.
  Again, in all four cases (T1)-(T3) are satisfied.

  \smallskip

  \noindent
  \textbf{\underline{1.2.
      $\pi \notin \posT(t_x)$}}

  \noindent
  If we have $\pi \notin \posT(t_x)$, then we can simply use the non-annotated old ADP instead of the old one.
  Since we would remove all annotations in the right-hand side of the rule anyway, due to $\pi \notin \posT(t_x)$, this leads to the same labels in the resulting $\overline{\PP}$-CT.

  \medskip

  \noindent
  \underline{\textbf{Second Case}}

  \noindent
  Next, we prove the theorem in the case where $\urules_{\PP}(r_j|_{\tau})$ is NO and all rules in $\urules_{\PP}(r_j|_{\tau})$ have the form $\ell' \to \{1:r'\}$ for some terms $\ell'$ and $r'$.
  \smallskip

  \noindent
  The proof uses the same idea as in the first case but the construction of $\Phi(\circ)$ is easier since $\urules_{\PP}(r_j|_{\tau})$ is non-probabilistic.
  First assume that $\urules_{\PP}(r_j|_{\tau})$ is not weakly innermost terminating.
  This means that after using the ADP $\ell \ruleArr{}{}{} \{ p_1:r_{1}, \ldots, p_k: r_k\}^{m}$ as we did in the soundness proof for the first case, every subterm at a position above $\pi.\tau$ will never be in $\NF_{\PP}$ in the subtree starting at the $j$-th successor $y_j$.
  Furthermore, since all rules in $\urules_{\PP}(r_j|_{\tau})$ have the form $\ell' \to \{1:r'\}$, we can simply remove all nodes that perform a rewrite step below $\pi.\tau$.
  To be precise, if there is a node $v$ that performs a rewrite step below position $\pi.\tau$, then we have $t_v \itored{}{}{\PP} \{1 : t_w\}$ for the only successor $w$ of $v$.
  Here, we have $t_w \flateq t_v[r' \sigma]_{\zeta}$ for the used substitution $\sigma$ and position $\zeta$ below $\pi.\tau$.
  The construction $\Psi(\F{T}_{y_j})$ contracts all edges $(x,y)$ where we use a rewrite step at a position below $\pi.\tau$.
  This only removes $\nV$-nodes, as there is no annotation below or at position $\pi.\tau$.
  Furthermore, we adjust the labeling such that the subterm at position $\pi.\tau$ remains the same for the whole CT.
  Finally, we exchange the rewrite step at the root $x$ from using the old ADP to using the new ADP.
  Since all subterms at a position above $\pi.\tau$ will never be reduced to normal forms in the original CT, it does not matter which subterms really occur.
  It is easy to see that $\Phi(\F{T}_x)$ is a valid $\PP$-CT and that (a) and (b) hold w.r.t.\ $\F{T}_x$ and $\Phi(\F{T}_x)$, and this ends the proof if $\urules_{\PP}(r_j|_{\tau})$ is not weakly innermost terminating.

  If $\urules_{\PP}(r_j|_{\tau})$ is weakly innermost terminating, then it follows directly that it is also confluent and terminating \cite[Thm.\ 3.2.11]{GramlichDiss}.
  We can use the construction $\Psi(\circ)$ from the first case to iteratively push the next innermost rewrite step that is performed below position $\pi.\tau$ to a higher position in the tree, until we reach the node that performs the rewrite step at position $\pi.\tau$.
  Note that we do not need the conditions L or NE for the used rule here, because only Case (A) of the construction can happen.
  There is no rewrite step below possible (Case (B)), since we move an innermost rewrite step further up and there is no rewrite step above possible, since before doing this construction it was a valid innermost $\PP$-CT.
  Since $\urules_{\PP}(r_j|_{\tau})$ is terminating, this construction ends after a finite number of $\Psi(\circ)$-applications in a tree $\F{T}_{y_j}^{(\infty)}$.
  Now $\F{T}_{y_j}^{(\infty)}$ first rewrites below or at position $\pi.\tau$ until it is a normal form.
  Since all rules in $\urules_{\PP}(r_j|_{\tau})$ have the form $\ell' \to \{1:r'\}$, these rewrite steps are a single path in $\F{T}_{y_j}^{(\infty)}$, and it does not matter how long the path is.
  Furthermore, it does not matter which rewrite strategy we use, since $\urules_{\PP}(r_j|_{\tau})$ is confluent.
  We will always reach the same normal form at the end of this path.
  Hence, we can replace the steps with the old ADP and the corresponding innermost rewrite steps in the CT by a step with the new ADP (where the rewriting does not necessarily correspond to the innermost strategy), i.e., we move this non-innermost step directly to the point where the ADP is applied.
  We will reach the same normal form and can copy the rest of the tree again.

  \medskip

  \noindent
  \underline{\textbf{Third Case}}

  \noindent
  Finally, we prove the theorem in the case where $\urules_{\PP}(r_j|_{\tau})$ is NO, $r_j|_{\tau}$ is a ground term, and we have $D_j \itored{}{}{\PP, \pi.\tau, \ttrue} \{q_1:E_{1}, \ldots, q_h:E_{h}\}$, i.e., it is an innermost step.
  \smallskip

  \noindent
  Once again, we use the same idea as in the proof for the second case but the construction of $\Phi(\circ)$ is, again, easier.
  Note that if $\urules_{\PP}(r_j|_{\tau})$ is non-overlapping, $r_j$ contains no variable below position $\tau$, and we perform an innermost rewrite step, then this is always an innermost rewrite step in every possible CT and this is the only possible rewrite step at this position.
  Hence, we can move this innermost step directly after the use of the ADP using the construction $\Psi(\circ)$.
  Again, here only Case (A) can happen.
  The reason is that we have to perform this rewrite step eventually if we want to rewrite above position $\tau$, and all other rewrite steps that we can perform in such a situation would be at orthogonal positions.
  So we get the same normal forms in the leaves with the same probability.
\end{myproof}

For the proofs of the instantiation processors, we first show some basic properties of $\capterm_\PP$ and $\renterm$,
similar to the known properties of the non-probabilistic versions of $\capterm_\PP$ and $\renterm$.

\begin{lem}[Property of $\capterm_\PP$ and $\renterm$]
  \label{lemma:cap-props}
  Let $t,s \in \TT^\sharp$.
  \begin{enumerate}
    \item If $t \sigma \ito_{\normalfont{\nonprob}(\PP)}^* s$ for some substitution $\sigma$ that maps each variable to a normal form, i.e., $\sigma(x) \in \NF_{\PP}$ for all $v \in \VSet$, then $s = \capterm_\PP(t) \delta$ for some substitution $\delta$ which only differs from $\sigma$ on the fresh variables that are introduced by $\capterm_\PP$.
    \item If $t \sigma \to_{\normalfont{\nonprob}(\PP)}^* s$ for some substitution $\sigma$, then $s = \renterm(\capterm_\PP(t)) \delta$ for some substitution $\delta$ which only differs from $\sigma$ on the fresh variables that are introduced by $\renterm$.
  \end{enumerate}
\end{lem}

\begin{myproof}
  \begin{enumerate}
    \item Let $t \sigma \ito_{\nonprob(\PP), \rho_1} s_1 \ito_{\nonprob(\PP), \rho_2}
            \ldots \ito_{\nonprob(\PP), \rho_n} s_n = s$, where $\ito_{\nonprob(\PP), \rho}$ again denotes a rewrite step at position $\rho$. Let $\{\pi_1, \ldots , \pi_m\}$ be the set of positions where $\capterm_\PP$ replaces the subterms of $t$ by corresponding fresh variables $x_1,\ldots,x_m$. Note that the rewrite steps cannot be ``completely inside'' the used substitution $\sigma$, as it maps every variable to a normal form. Hence, by the definition of $\capterm_\PP$, for each $\rho_i$ there is a higher position $\pi_j \leq \rho_i$. Thus, $s$ can at most differ from $t \sigma$ on positions below a $\pi_j$. We define $\delta$ to be like $\sigma$ but on the fresh variables $x_1,\ldots, x_m$ we define $\delta(x_j) = s|_{\pi_j}$. Then by construction $\capterm_\PP(t) \delta = s$ and $\delta$ and $\sigma$ differ only on the fresh variables.

    \item Let $t \sigma \to_{\nonprob(\PP), \rho_1} s_1 \to_{\nonprob(\PP), \rho_2} \ldots \to_{\nonprob(\PP), \rho_n} s_n = s$.
          Let $\{\pi_1, \ldots , \pi_m\}$ be the set of positions where $\renterm \circ \capterm_\PP$ replaces the subterms of $t$ by corresponding fresh variables $x_1,\ldots,x_m$.
          By the definition of $\renterm$ and $\capterm_\PP$, for each $\rho_i$ there is a higher position $\pi_j \leq \rho_i$.
          (If the rewrite step is ``completely inside'' the substitution $\sigma$, then $\renterm$ replaces the corresponding variable in $t$ by a fresh one, otherwise $\capterm_\PP$ replaces the subterm in $t$ by a new variable, which is again replaced by $\renterm$ afterwards.) Hence, $s$ can at most differ from $t \sigma$ on positions below a $\pi_j$.
          We define $\delta$ to be like $\sigma$ but on the fresh variables $x_1,\ldots, x_m$ we define $\delta(x_j) = s|_{\pi_j}$.
          Then by construction $\renterm(\capterm_\PP(t)) \delta = s$, and $\delta$ and $\sigma$ differ only on the fresh variables.
  \end{enumerate}
\end{myproof}

Next, we prove soundness and completeness of the instantiation processor for both innermost and full rewriting.
There are only small changes in the proof due to the rewrite strategy that we highlight once they occur.
The same holds for the following forward instantiation processor.

\InstProc* \InstProcInnermost*

\begin{myproof}[of both \Cref{theorem:inst-proc-ast} and \Cref{theorem:inst-proc-iast}]
  We will use the following two observations.
  As in the proof of \Cref{theorem:ptrs-rewriting-proc}, let $\mu = \{ p_1:r_1, \ldots, p_k:r_k \}$, $\overline{\PP'} = \PP' \cup N \cup \{\ell \ruleArr{}{}{} \{ p_1:\flat(r_{1}), \ldots, p_k: \flat(r_{k})\}^{m}\}$, and $\overline{\PP} = \overline{\PP'} \cup \{\ell \ruleArr{}{}{} \{ p_1:r_{1}, \ldots, p_k:r_{k}\}^{m}\}$.
  First, note that $\ANF_{\PP} = \ANF_{\overline{\PP}} = \ANF_{\overline{\PP'}}$, since the left-hand sides in $\overline{\PP}$ and $\overline{\PP'}$ are either already from rules in $\PP$ or instantiated left-hand sides from $\PP$.
  So it suffices to consider only $\ANF_{\PP}$.
  Second, assume that there exists a $\overline{\PP}$-CT $\F{T}$ that converges with probability $<1$ whose root is labeled with $(1 : t)$ and $\flat(t) = s \theta$ for a substitution $\theta$ and an ADP $s \to \ldots \in \overline{\PP}$, and $\posT(t) = \{\varepsilon\}$.
  If in addition, the ADP $\ell \ruleArr{}{}{} \{ p_1:r_{1}, \ldots, p_k: r_k\}^{m}$ may only be used at the root, then we know that not only $\overline{\PP}$ is not $\mathtt{AST}$ but also that $\overline{\PP'}$ is not $\mathtt{AST}$.
  The reason is that if only the root $v$ uses the ADP $\ell \ruleArr{}{}{} \{ p_1:r_{1}, \ldots, p_k: r_k\}^{m}$ then all the subtrees starting at one of its direct successors $vE = \{w_1, \ldots, w_k\}$ are $\overline{\PP'}$-CTs.
  Furthermore, since $\F{T}$ converges with probability $<1$, there must be at least one subtree $\F{T}[w_iE^*]$ that starts at the node $w_i$ for some $1 \leq i \leq k$ and also converges with probability $<1$.

  \smallskip

  \noindent
  \underline{\emph{Soundness:}} Let $\PP$ be not $\mathtt{(i)AST}$.
  Then by \Cref{lemma:starting} there exists an (innermost) $\PP$-CT $\F{T} = (V,E,L)$ that converges with probability $<1$ whose root is labeled with $(1: t)$ and $\flat(t) = s \theta$ for a substitution $\theta$ and an ADP $s \to \ldots \in \PP$, and $\posT(t) = \{\varepsilon\}$.
  We will now create an (innermost) $\overline{\PP}$-CT $\F{T}' = (V,E,L')$, with the same underlying tree structure, and an adjusted labeling such that $p_v^{\F{T}} = p_v^{\F{T}'}$ for all $v \in V$.
  Furthermore, we will at most use the ADP $\ell \ruleArr{}{}{} \{ p_1:r_{1}, \ldots, p_k: r_k\}^{m}$ at the root.
  Since the tree structure and the probabilities are the same, we then get $|\F{T}'| = |\F{T}| <1$, and hence, using our previous discussion, $\overline{\PP'}$ is not $\mathtt{(i)AST}$ either.

  The core idea is that every rewrite step with $\ell \ruleArr{}{}{} \{ p_1:r_{1}, \ldots, p_k: r_k\}^{m}$ at a node $v$ that is not the root can also be done with a rule from $N$, or we can use $\ell \ruleArr{}{}{} \{ p_1:\flat(r_{1}), \ldots, p_k: \flat(r_{k})\}^{m}$ if the annotations do not matter, e.g., we rewrite at a position that is not annotated.

  We only consider $\mathtt{AST}$ in the following construction.
  If not mentioned otherwise, the proof for $\mathtt{iAST}$ works analogously just by considering innermost rewrite steps.
  We construct the new labeling $L'$ for the $\overline{\PP}$-CT $\F{T}'$ inductively such that for all nodes $v \in V \setminus \ctleaf$ with children nodes $vE = \{w_1,\ldots,w_h\}$ we have $t'_v \tored{}{}{\overline{\PP}} \{\tfrac{p_{w_1}}{p_v}:t'_{w_1}, \ldots, \tfrac{p_{w_h}}{p_v}:t'_{w_h}\}$ and for all non-root nodes in $\aV$ we even have $t'_v \tored{}{}{\overline{\PP'}} \{\tfrac{p_{w_1}}{p_v}:t'_{w_1}, \ldots, \tfrac{p_{w_h}}{p_v}:t'_{w_h}\}$.
  Let $W \subseteq V$ be the set of nodes $v$ where we have already defined the labeling $L'(x)$.
  During our construction, we ensure that for every node $v \in W$ we have
  \begin{equation} \label{soundness-inst-induction-hypothesis}
    t_v \flateq t_v' \text{ and } \posT(t_v) \subseteq \posT(t_v').
  \end{equation}
  This means that the corresponding term $t_v$ for the node $v$ in $\F{T}$ has the same structure as the term $t_v'$ in $\F{T}'$, and additionally, every annotation in $t_v$ also exists in $t_v'$.
  The second condition ensures that if we rewrite using Case $(\mathbf{af})$ or $(\mathbf{at})$ in $\F{T}$, we do the same in $\F{T}'$, i.e., the corresponding node $v$ remains in $\aV$ in $\F{T}'$.
  We label the root of $\F{T}'$ by $(1:t)$.
  Here, \eqref{soundness-inst-induction-hypothesis} obviously holds.
  As long as there is still an inner node $v \in W$ such that its successors are not contained in $W$, we do the following.
  Let $vE = \{w_1, \ldots, w_h\}$ be the set of its successors.
  We need to define the corresponding terms $t_{w_1}', \ldots, t_{w_m}'$ for the nodes $w_1, \ldots, w_h$.
  Since $v$ is not a leaf and $\F{T}$ is a $\PP$-CT, we have $t_v \tored{}{}{\PP} \{\tfrac{p_{w_1}}{p_v}:t_{w_1}, \ldots, \tfrac{p_{w_h}}{p_v}:t_{w_h}\}$.
  We have the following three different cases:

  \smallskip

  \noindent
  \textbf{(A) We have} $t_v \tored{}{}{\PP} \{\tfrac{p_{w_1}}{p_v}:t_{w_1}, \ldots, \tfrac{p_{w_h}}{p_v}:t_{w_h}\}$ 
  with an ADP $\ell' \ruleArr{}{}{} \{ p_1:r_{1}', \ldots, p_h: r_h'\}^{m'}$ 
  that is either different to $\ell \ruleArr{}{}{} \{ p_1:r_{1}, \ldots, p_k: r_k\}^{m}$ or the node $v$ is the root, using the position $\pi$, 
  a VRF $(\varphi_j)_{1 \leq j \leq k}$, and a substitution $\sigma$ such that $\flat(t_v|_\pi)=\ell\sigma$.
  Then, $t_v \flateq_{(IH)} t_v'$, and hence also $\flat(t_v'|_\pi)=\ell\sigma$.
  Thus, we can rewrite the term $t'_v$ using the same ADP, the same position, the same substitution, and the same VRF.
  This means that we have $t'_v \tored{}{}{\overline{\PP}} \{\tfrac{p_{w_1}}{p_v}:t'_{w_1}, \ldots, \tfrac{p_{w_h}}{p_v}:t'_{w_h}\}$.
  Let $1 \leq j \leq h$.
  If $\pi \in \posT(t_v)$, then also $\pi \in \posT(t'_v)$ by \eqref{soundness-inst-induction-hypothesis}.
  Whenever we create annotations in the rewrite step in $\F{T}$ (a step with $(\mathbf{af})$ or $(\mathbf{at})$), then we do the same in $\F{T}'$ (the step is also an $(\mathbf{af})$- or $(\mathbf{at})$-step, respectively), and whenever we remove annotations in the rewrite step in $\F{T}'$ (a step with $(\mathbf{nt})$ or $(\mathbf{nf})$), then we do the same in $\F{T}$ (the step is also either a $(\mathbf{nt})$- or $(\mathbf{nf})$-step).
  Therefore, we also get $\posT(t_{w_j}) \subseteq \posT(t'_{w_j})$ for all $1 \leq j \leq k$ and \eqref{soundness-inst-induction-hypothesis} is again satisfied.

  \smallskip

  \noindent
  \textbf{(B) We have} $t_v \tored{}{}{\PP} \{\tfrac{p_{w_1}}{p_v}:t_{w_1}, \ldots, \tfrac{p_{w_k}}{p_v}:t_{w_k}\}$ 
  using the ADP $\ell \ruleArr{}{}{} \{ p_1:r_{1}, \ldots, p_k: r_k\}^{m}$, 
  the position $\pi$, a VRF $(\varphi_j)_{1 \leq j \leq k}$, and a substitution $\sigma$ 
  such that $\flat(t_v|_\pi)=\ell\sigma$, and $\pi \not\in \posT(t_v)$.
  Since $t_v \flateq_{(IH)} t_v'$, we can rewrite the term $t'_v$ using the ADP $\ell \ruleArr{}{}{} \{ p_1:\flat(r_{1}), \ldots, p_k: \flat(r_{k})\}^{m}$, the same position, the same substitution, and the same VRF.
  This means that we have $t'_v \tored{}{}{\overline{\PP}} \{\tfrac{p_{w_1}}{p_v}:t'_{w_1}, \ldots, \tfrac{p_{w_k}}{p_v}:t'_{w_k}\}$ and \eqref{soundness-inst-induction-hypothesis} is again satisfied.
  In order to prove this, one can do a similar analysis as above.
  Note that only cases $(\mathbf{nf})$ and $(\mathbf{nt})$ can be applied in $\F{T}$ (since $\pi \not\in \posT(t_v)$), and we would remove the annotations of the terms $r_j$ anyway in those cases.

  \smallskip

  \noindent
  \textbf{(C) Finally, we have} $t_v \tored{}{}{\PP} \{\tfrac{p_{w_1}}{p_v}:t_{w_1}, \ldots, \tfrac{p_{w_k}}{p_v}:t_{w_k}\}$ 
  using the ADP $\ell \ruleArr{}{}{} \{ p_1:r_{1}, \ldots, p_k: r_k\}^{m}$, the position $\pi$, 
  a VRF $(\varphi_j)_{1 \leq j \leq k}$, and a substitution $\sigma$ 
  such that $\flat(t_v|_\pi)=\ell\sigma$, $\pi \in \posT(t_v)$, and $v$ is not the root.
  Then $t_{w_j} = \disannoPos{\pi}(t_v[\sharp_{\posT(r_j) \cup \Psi_j}(r_j\sigma)]_{\pi})$ if $m = \tfalse$ 
  and $t_{w_j} = t_v[\sharp_{\posT(r_j) \cup \Psi_j}(r_j\sigma)]_{\pi}$, otherwise.

  We now look at the (not necessarily direct) predecessor $z$ of $v$ that is in $\aV$, where an ADP is applied on a position above or equal to $\pi$, 
  and where in the path from $z$ to $v$, no ADP is applied on a position on or above $\pi$.
  There is always such a node $z$.
  The reason is that $v$ is not the first node in $\aV$ and by the Starting Lemma (\Cref{lemma:starting}) 
  we can assume that a step at the root of the term takes place at the root of the CT, i.e., $t_v$ results from right-hand sides of $\PP$.
  Furthermore, we only use rules with the flag $m = \ttrue$ as otherwise, the position $\pi$ would not be annotated in $t_v$.
  We show that the path from $z$ to $v$ can also be taken when using one of the new instantiations of 
  $\ell \ruleArr{}{}{} \{ p_1:r_{1}, \ldots, p_k: r_k\}^{m}$ instead.

  Let this predecessor $z \in \aV$ use the ADP $\ell' \to \{p_1':r_1', \ldots, p_h':r_h'\}^{m}$, 
  the position $\pi'$, and the substitution $\sigma'$.
  Furthermore, let $1 \leq j \leq h$ and $\tau$ be the position of $r_j'$ such that
  $\annoEps(r_j'|_{\tau}) \sigma' \to_{\nonprob(\PP)}^* \annoEps(t_v|_{\pi}) = \annoEps(\ell) \sigma$,
  which exists by the previous paragraph.

  \vspace{-0.2cm}
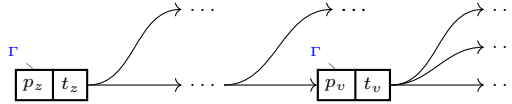
  \begin{center}
    \scriptsize
    \begin{tikzpicture}
      \tikzstyle{adam}=[thick,draw=black!100,fill=white!100,minimum size=4mm, shape=rectangle split, rectangle split parts=2,rectangle split horizontal] \tikzstyle{empty}=[rectangle,thick,minimum size=4mm]

      \node[adam,pin={[pin distance=0.1cm, pin edge={,-}] 140:\tiny \textcolor{blue}{$\aV$}}] at (0, 0) (a) {$p_z$
          \nodepart{two} $t_z$};
      \node[empty] at (2.0, 1) (b) {$\ldots$};
      \node[empty] at (2.0, 0) (c) {$\ldots$};
      \node[empty] at (4.0, 1) (d) {$\ldots$};
      \node[adam,pin={[pin distance=0.1cm, pin edge={,-}] 140:\tiny \textcolor{blue}{$\aV$}}] at (4.0, 0) (e) {$p_v$
          \nodepart{two} $t_v$};
      \node[empty] at (4.0, 1) (f) {$\ldots$};
      \node[empty] at (6.0, 0) (g) {$\ldots$};
      \node[empty] at (6.0, 0.5) (h) {$\ldots$};
      \node[empty] at (6.0, 1.0) (i) {$\ldots$};

      \draw (a) edge[->, in = 180, out = 0] (b);
      \draw (a) edge[->, in = 180, out = 0] (c);
      \draw (c) edge[->, in = 180, out = 0] (d);
      \draw (c) edge[->, in = 180, out = 0] (e);
      \draw (e) edge[->, in = 180, out = 0] (g);
      \draw (e) edge[->, in = 180, out = 0] (h);
      \draw (e) edge[->, in = 180, out = 0] (i);
    \end{tikzpicture}
  \end{center}
  \vspace{-0.2cm}

  The following consideration now differs for $\mathtt{iAST}$ and $\mathtt{AST}$ due to the different approximations ($\mathtt{iAST}$ uses only $\capterm_{\PP}$ and $\mathtt{AST}$ uses $\renterm \circ \capterm_{\PP}$).

  \begin{itemize}
    \item \underline{$\mathtt{AST}$:} From $\annoEps(r_j'|_{\tau}) \sigma' \to_{\nonprob(\PP)}^* \annoEps(\ell) \sigma$ together with \cref{lemma:cap-props} 
    we get $\annoEps(\ell) \sigma = \renterm(\capterm_\PP(\annoEps(r_j'|_{\tau}))) \delta$ for some substitution $\delta$ 
    that differs from $\sigma'$ at most on the variables that are introduced by $\renterm$.
    W.l.o.g., we can assume that $\sigma'$ is equal to $\delta$ on all these fresh variables, and since the ADPs are variable-renamed, 
    we can also assume that $\sigma$ is equal to $\delta$ on all the fresh variables and all the variables from $\ell'$.
    Hence, $\annoEps(\ell) \delta = \renterm(\capterm_\PP(\annoEps(r_j'|_{\tau}))) \delta$ implies that we can find an mgu $\gamma$ of $\annoEps(\ell)$ 
    and $\renterm(\capterm_\PP(\annoEps(r_j'|_{\tau})))$ with $\sigma = \gamma \zeta$ for some substitution $\zeta$.
    Hence, we can apply the new ADP $\ell \gamma \ruleArr{}{}{} \{ p_1:r_{1}\gamma, \ldots, p_k: r_k\gamma\}^{m} \in N$ 
    with the position $\pi$, the same VRF, and the substitution $\zeta$.
    This means that we have $t'_v \tored{}{}{\PP} \{\tfrac{p_{w_1}}{p_v}:t'_{w_1}, \ldots, \tfrac{p_{w_k}}{p_v}:t'_{w_j}\}$ 
    with $t_{w_j}' = t_v'[\sharp_{\posT(r_j) \cup \Psi_j}(r_j\gamma \zeta)]_{\pi}$ if $m = \ttrue$, 
    or $t_{w_j}' = \disannoPos{\pi}(t_v'[\sharp_{\posT(r_j) \cup \Psi_j}(r_j\gamma \zeta)]_{\pi})$, otherwise.
    Since $\sigma = \gamma \zeta$ we directly get $t_{w_j} \flateq t_{w_j}'$ 
    and $\posT(t_{w_j}) \subseteq \posT(t_{w_j}')$ so that \eqref{soundness-inst-induction-hypothesis} is satisfied again.

    \item \underline{$\mathtt{iAST}$:} From $\annoEps(r_j'|_{\tau}) \sigma' \ito_{\nonprob(\PP)}^* \annoEps(\ell) \sigma$ 
    and the fact that $\sigma'$ maps every variable to a normal form due to innermost rewriting, 
    \cref{lemma:cap-props} implies $\annoEps(\ell) \sigma = \capterm_\PP(\annoEps(r_j'|_{\tau})) \delta$ 
    for some substitution $\delta$ that differs from $\sigma'$ at most on the variables that are introduced by $\capterm_\PP$.
    W.l.o.g., we can assume that $\sigma'$ is equal to $\delta$ on all these fresh variables, and since the ADPs are variable-renamed, 
    we can also assume that $\sigma$ is equal to $\delta$ on all the fresh variables and all the variables from $\ell'$.
    Hence, $\annoEps(\ell) \delta = \capterm_\PP(\annoEps(r_j'|_{\tau})) \delta$ shows there is an mgu $\gamma$ of $\annoEps(\ell)$ 
    and $\capterm_\PP(\annoEps(r_j'|_{\tau}))$ with $\sigma = \gamma \zeta$ for some substitution $\zeta$.
    Moreover, the property $\{\ell' \sigma', \ell \sigma\} \subseteq \ANF_{\PP}$ must remain true when replacing $\sigma$ and $\sigma'$ 
    by the more general substitution $\gamma$, i.e., $\{\ell' \gamma, \ell \gamma\} \subseteq \ANF_{\PP}$.
    The rest is analogous to the $\mathtt{AST}$ case.
  \end{itemize}

  \medskip

  \noindent
  \underline{\emph{Completeness:}} Let $\overline{\PP'}$ be not $\mathtt{(i)AST}$.
  Then there exists an (innermost) $\overline{\PP'}$-CT $\F{T} = (V,E,L)$ that converges with probability $<1$.
  We will now create an (innermost) $\PP$-CT $\F{T}' = (V,E,L')$, with the same underlying tree structure, and an adjusted labeling such that $p_v^{\F{T}} = p_v^{\F{T}'}$ for all $v \in V$.
  Since the tree structure and the probabilities are the same, we then get $|\F{T}'| = |\F{T}|$, and thus $\PP$ is not $\mathtt{(i)AST}$ either.

  The core idea of this construction is that every rewrite step with an ADP from $N$ or the ADP $\ell \ruleArr{}{}{} \{ p_1:\flat(r_{1}), \ldots, p_k: \flat(r_k)\}^{m}$ is also possible with the more general ADP $\ell \ruleArr{}{}{} \{ p_1:r_{1}, \ldots, p_k: r_k\}^{m}$ that may also contain more annotations.
  As for soundness, we focus on $\mathtt{AST}$, but the proof is analogous for $\mathtt{iAST}$.
  We construct the new labeling $L'$ for the $\PP$-CT $\F{T}'$ inductively such that for all inner nodes $v \in V \setminus \ctleaf$ with children nodes $vE = \{w_1,\ldots,w_m\}$ we have $t'_v \tored{}{}{\PP} \{\tfrac{p_{w_1}}{p_v}:t'_{w_1}, \ldots, \tfrac{p_{w_m}}{p_v}:t'_{w_h}\}$.
  Let $W \subseteq V$ be the set of nodes $v$ where we have already defined the labeling $L'(x)$.
  During our construction, we ensure that for every $v \in W$ (analogous to the soundness construction) we have
  \begin{equation} \label{completeness-inst-induction-hypothesis}
    t_v \flateq t_v' \text{ and } \posT(t_v) \subseteq \posT(t_v').
  \end{equation}
  We label the root of $\F{T}'$ exactly as the root of $\F{T}$.
  Here, \eqref{completeness-inst-induction-hypothesis} obviously holds.
  As long as there is still an inner node $v \in W$ such that its successors are not contained in $W$, we do the following.
  Let $vE = \{w_1, \ldots, w_h\}$ be the set of its successors.
  We need to define the corresponding terms $t_{w_1}', \ldots, t_{w_h}'$ for the nodes $w_1, \ldots, w_h$.
  Since $v$ is not a leaf and $\F{T}$ is a $\overline{\PP'}$-CT, we have $t_v \tored{}{}{\overline{\PP'}} \{\tfrac{p_{w_1}}{p_v}:t_{w_1}, \ldots, \tfrac{p_{w_h}}{p_v}:t_{w_h}\}$.
  We have the following three different cases:
  \begin{enumerate}
    \item[(A)] If it is a step with $\tored{}{}{\PP}$ using an ADP from $\PP'$ in $\F{T}$, 
    then we perform a rewrite step with the same ADP, the same position, the same substitution, and the same VRF in $\F{T}'$.
    This is analogous to Case (A) of the soundness proof.
    \item[(B)] If it is a step with $\tored{}{}{\PP}$ using the ADP $\ell \ruleArr{}{}{} \{ p_1:\flat(r_{1}), \ldots, p_k: \flat(r_k)\}^{m}$ in $\F{T}$, 
    then we use the ADP $\ell \ruleArr{}{}{} \{ p_1:r_{1}, \ldots, p_k: r_k\}^{m}$ that contains more annotations in $\F{T}'$, 
    at the same position, using the same substitution, and the same VRF.
    Since we use the same rule but with more annotations, we end up with $t_{w_j} \flateq t_{w_j}'$ and $\posT(t_{w_j}) \subseteq \posT(t_{w_j}')$ again.
    \item[(C)] If it is a step with $\tored{}{}{\PP}$ using an ADP 
    $\ell \gamma \ruleArr{}{}{} \{ p_1:r_{1}\gamma, \ldots, p_k: r_k\gamma\}^{m}$ from $N$ in $\F{T}$ and a substitution $\delta$, 
    then we use the more general ADP $\ell \ruleArr{}{}{} \{ p_1:r_{1}, \ldots, p_k: r_k\}^{m}$ in $\F{T}'$, at the same position, and the same VRF.
    Furthermore, we use the substitution $\gamma \delta$.
  \end{enumerate}
\end{myproof}

\ForwardInstProc* \ForwardInstProcInnermost*

\begin{myproof}[of both \Cref{theorem:forward-inst-proc-ast} and \Cref{theorem:forward-inst-proc-iast}]
  As in the proof of \Cref{theorem:inst-proc-ast}, let $\mu = \{ p_1:r_1, \ldots, p_k:r_k \}$, $\overline{\PP'} = \PP' \cup N \cup \{\ell \ruleArr{}{}{} \{ p_1:\flat(r_{1}), \ldots, p_k: \flat(r_{k})\}^{m}\}$, and $\overline{\PP} = \overline{\PP'} \cup \{\ell \ruleArr{}{}{} \{ p_1:r_{1}, \ldots, p_k:r_{k}\}^{m}\}$.
  \smallskip

  \noindent
  \underline{\emph{Soundness:}} The core proof idea, the construction itself, and even the induction hypothesis are completely the same as for the soundness proof of \cref{theorem:inst-proc-ast}, only the third case (C) changes.

  \smallskip

  \noindent
  \textbf{(C) If we have} $t_v \tored{}{}{\PP} \{\tfrac{p_{w_1}}{p_v}:t_{w_1}, \ldots, \tfrac{p_{w_k}}{p_v}:t_{w_k}\}$ 
  using the ADP $\ell \ruleArr{}{}{} \{ p_1:r_{1}, \ldots, p_k: r_k\}^{m}$, 
  the position $\pi \in \posT(s)$, and a substitution $\sigma$ such that $\flat(t_v|_\pi)=\ell\sigma$ 
  (where $\ell\sigma \in \ANF_\PP$ in the innermost case), and $v$ is not the root, 
  then $t_{w_j} = \disannoPos{\pi}(t_v[\sharp_{\posT(r_j) \cup \Psi_j}(r_j\sigma)]_{\pi})$ if $m = \tfalse$ 
  and $t_{w_j} = t_v[\sharp_{\posT(r_j) \cup \Psi_j}(r_j\sigma)]_{\pi}$, otherwise.

  First consider the case, where there exists no successor $z$ of $v$ where an ADP is applied at an annotated subterm introduced by 
  some right-hand side $r_j$ for some $1 \leq j \leq k$.
  Then, we can use $\ell \ruleArr{}{}{} \{ p_1:\flat(r_{1}), \ldots, p_k: \flat(r_{k})\}^{m}$ instead, 
  because the annotations will never be used.

  Otherwise, there exists a successor $z$ of $v$ where an ADP is applied at an annotated subterm introduced by 
  the right-hand side $r_j$ for some $1 \leq j \leq k$.
  We show that the path to this successor can also be taken when using one of the new instantiations of 
  $\ell \ruleArr{}{}{} \{ p_1:r_{1}, \ldots, p_k: r_k\}^{m}$ instead.\footnote{It 
    suffices to consider only \emph{one} of those successors $v$ of $v$ to find an instantiation 
    of the ADP that could be used instead of the old ADP in order to reach \emph{all} such successors.
    The reason is that this instantiation is equal or more general than the actual concrete instantiation 
    of the ADP that was used to perform the ADP-step in the actual CT.
    Hence, every other such successor can also be used with the more general instantiation of the ADP.}
  At node $z$ we use an ADP $\ell' \ruleArr{}{}{} \{ p_1':r_{1}', \ldots, p_h': r_h'\}^{m'}$ and the substitution $\sigma'$.
  Moreover, let $\tau$ be the position of $r_j$ with $\annoEps(r_j|_{\tau}) \sigma \to_{\nonprob(\PP)}^* \annoEps(\ell') \sigma'$.
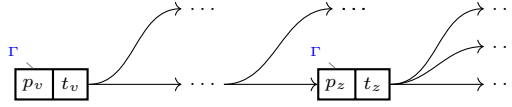
  \begin{center}
    \scriptsize
    \begin{tikzpicture}
      \tikzstyle{adam}=[thick,draw=black!100,fill=white!100,minimum size=4mm, shape=rectangle split, rectangle split parts=2,rectangle split horizontal] \tikzstyle{empty}=[rectangle,thick,minimum size=4mm]

      \node[adam,pin={[pin distance=0.1cm, pin edge={,-}] 140:\tiny \textcolor{blue}{$\aV$}}] at (0, 0) (a) {$p_v$
          \nodepart{two} $t_v$};
      \node[empty] at (2.0, 1) (b) {$\ldots$};
      \node[empty] at (2.0, 0) (c) {$\ldots$};
      \node[empty] at (4.0, 1) (d) {$\ldots$};
      \node[adam,pin={[pin distance=0.1cm, pin edge={,-}] 140:\tiny \textcolor{blue}{$\aV$}}] at (4.0, 0) (e) {$p_z$
          \nodepart{two} $t_z$};
      \node[empty] at (4.0, 1) (f) {$\ldots$};
      \node[empty] at (6.0, 0) (g) {$\ldots$};
      \node[empty] at (6.0, 0.5) (h) {$\ldots$};
      \node[empty] at (6.0, 1.0) (i) {$\ldots$};

      \draw (a) edge[->, in = 180, out = 0] (b);
      \draw (a) edge[->, in = 180, out = 0] (c);
      \draw (c) edge[->, in = 180, out = 0] (d);
      \draw (c) edge[->, in = 180, out = 0] (e);
      \draw (e) edge[->, in = 180, out = 0] (g);
      \draw (e) edge[->, in = 180, out = 0] (h);
      \draw (e) edge[->, in = 180, out = 0] (i);
    \end{tikzpicture}
  \end{center}
  Again we have two different cases:

  \begin{itemize}
    \item \underline{$\mathtt{AST}$:} Let $\QQ' = \nonprob(\PP)$ and let $\QQ = \QQ'^{-1}$.
          Note that $\annoEps(r_j|_{\tau}) \sigma \to_{\QQ'}^* \annoEps(\ell') \sigma'$ and hence, $\annoEps(\ell') \sigma' \to_{\QQ}^* \annoEps(r_j|_{\tau}) \sigma$.
          By \cref{lemma:cap-props} one directly obtains $\annoEps(r_j|_{\tau}) \sigma = \renterm(\capterm_{\QQ}(\annoEps(\ell'))) \delta$ 
          for some substitution $\delta$ that differs from $\sigma'$ at most on the variables that are introduced by $\renterm$.
          W.l.o.g.\ we can assume that $\sigma'$ is equal to $\delta$ on all these fresh variables, and since the ADPs are variable-renamed, 
          we can also assume that $\sigma$ is equal to $\delta$ on all the fresh variables and all the variables from $\annoEps(\ell')$.
          Hence, $\annoEps(r_j|_{\tau}) \sigma = \renterm(\capterm_{\QQ}(\annoEps(\ell'))) \sigma$ shows there is an mgu $\gamma$ 
          of $\annoEps(r_j|_{\tau})$ and $\renterm(\capterm_{\QQ}(\annoEps(\ell')))$ with $\sigma = \gamma \zeta$ for some substitution $\zeta$.
          Hence, we can apply the new ADP $\ell \gamma \ruleArr{}{}{} \{ p_1:r_{1}\gamma, \ldots, p_k: r_k\gamma\}^{m} \in N$ 
          with the position $\pi$ and the substitution $\zeta$.
          This means that we have $t'_v \tored{}{}{\PP} \{\tfrac{p_{w_1}}{p_v}:t'_{w_1}, \ldots, \tfrac{p_{w_k}}{p_v}:t'_{w_j}\}$ 
          with $t_{w_j}' = \disannoPos{\pi}(t_v'[\sharp_{\posT(r_j) \cup   \Psi_j}(r_j\gamma \zeta)]_{\pi})$ 
          if $m = \ttrue$, or $t_{w_j}' = t_v'[\sharp_{\posT(r_j) \cup   \Psi_j}(r_j\gamma \zeta)]_{\pi}$, otherwise.
          Since, $\sigma = \gamma \zeta$ we directly get $t_{w_j} \flateq t_{w_j}'$ and $\posT(t_{w_j}) \subseteq \posT(t_{w_j}')$ so that \eqref{soundness-inst-induction-hypothesis} is satisfied again

    \item \underline{$\mathtt{iAST}$:} Let $\QQ' = \nonprob(\urules_{\PP}(\annoEps(r_j|_{\tau})))$ and let $\QQ = \QQ'^{-1}$.
          Again
          $\annoEps(\ell') \sigma' \to_{\QQ}^* \annoEps(r_j|_{\tau}) \sigma$, since $\annoEps(r_j|_{\tau}) \sigma \to_{\QQ'}^* \annoEps(\ell') \sigma'$.
          By \cref{lemma:cap-props} one directly obtains $\annoEps(r_j|_{\tau}) \sigma = \renterm(\capterm_{\QQ}(\annoEps(\ell'))) \delta$ 
          for some substitution $\delta$ that differs from $\sigma'$ at most on the variables that are introduced by $\renterm$ and $\capterm_{\QQ}$.
          W.l.o.g.\ we can assume that $\sigma'$ is equal to $\delta$ on all these fresh variables, and since the ADPs 
          are variable-renamed, we can also assume that $\sigma$ is equal to $\delta$ on all the fresh variables and all the variables from $\annoEps(\ell')$.
          Hence, $\annoEps(r_j|_{\tau}) \sigma = \renterm(\capterm_{\QQ}(\annoEps(\ell'))) \sigma$ shows there is 
          an mgu $\gamma$ of $\annoEps(r_j|_{\tau})$ and $\renterm(\capterm_{\QQ}(\annoEps(\ell')))$ with $\sigma = \gamma \zeta$ for some substitution $\zeta$.
          Moreover, the property $\{\ell \sigma, \ell' \sigma'\} \subseteq \ANF_{\PP}$
          must remain true
          when replacing $\sigma$ and $\sigma'$ by the more general substitution $\gamma$, i.e., $\{\ell \gamma, \ell' \gamma\} \subseteq \ANF_{\PP}$.
          Hence, we can apply the new ADP $\ell \gamma \ruleArr{}{}{} \{ p_1:r_{1}\gamma, \ldots, p_k: r_k\gamma\}^{m} \in N$ 
          with the position $\pi$ and the substitution $\zeta$.
          This means that we have $t'_v \tored{}{}{\PP} \{\tfrac{p_{w_1}}{p_v}:t'_{w_1}, \ldots, \tfrac{p_{w_k}}{p_v}:t'_{w_j}\}$ 
          with $t_{w_j}' = \disannoPos{\pi}(t_v'[\sharp_{\posT(r_j) \cup   \Psi_j}(r_j\gamma \zeta)]_{\pi})$ if $m = \ttrue$, 
          or $t_{w_j}' = t_v'[\sharp_{\posT(r_j) \cup   \Psi_j}(r_j\gamma \zeta)]_{\pi}$, otherwise.
          Since, $\sigma = \gamma \zeta$ we directly get $t_{w_j} \flateq t_{w_j}'$ and $\posT(t_{w_j}) \subseteq \posT(t_{w_j}')$ so that \eqref{soundness-inst-induction-hypothesis} is satisfied again.
  \end{itemize}

  \medskip

  \noindent
  \underline{\emph{Completeness: }} Completely the same as for \cref{theorem:inst-proc-ast}.
\end{myproof}

\RuleOverlapInstProc*

\begin{myproof}
  \smallskip

  \noindent
  \underline{\emph{Soundness:}} We use the same idea as in the proof of soundness for \cref{theorem:forward-inst-proc-ast}
  but at a node $v$ that uses the position $\pi \in \posT(t_v)$ we do not look at the next node that rewrites at an annotated position below or equal to $\pi$, but at all such nodes that can be either annotated or not.

  Let $\PP$ be not $\mathtt{iAST}$.
  Then by \Cref{lemma:starting} there exists an innermost $\PP$-CT $\F{T} = (V,E,L)$ that converges with probability $<1$ that starts with $(1:t)$ and $t = s \theta \in \ANF_{\PP}$ for a substitution $\theta$ and an ADP $s \to \ldots \in \PP$, and $\posT(t) = \{\varepsilon\}$.
  Let $\overline{\PP'} = \PP' \cup N$.
  We will now create an innermost $\overline{\PP'}$-CT $\F{T}' = (V,E,L')$, with the same underlying tree structure, and an adjusted labeling such that $p_v^{\F{T}} = p_v^{\F{T}'}$ for all $v \in V$.
  Since the tree structure and the probabilities are the same, we then get $|\F{T}'| = |\F{T}|$, and hence $\overline{\PP'}$ is not $\mathtt{iAST}$ either.

  The core idea of this construction is that every rewrite step with $\ell \ruleArr{}{}{} \{ p_1:r_{1}, \ldots, p_k: r_k\}^{m}$ can also be done with a rule from $N$.
  If we use $\ell \ruleArr{}{}{} \{ p_1:\anno_{\capt_1(\delta_1,\ldots,\delta_d)}(r_{1}), \ldots, p_k: \anno_{\capt_k(\delta_1,\ldots,\delta_d)}(r_k)\}^{m} \in N$, we may create fewer annotations than we did when using the old ADP $\ell \ruleArr{}{}{} \{ p_1:r_{1}, \ldots, p_k: r_k\}^{m}$.
  However, in the $\PP$-CT $\F{T}$
  we never rewrite at the position of those annotations that do not get created in the CT
  $\F{T}'$, hence we can ignore them.
  We construct the new labeling $L'$ for the $\overline{\PP'}$-CT $\F{T}'$ inductively such that for all nodes $v \in V \setminus \ctleaf$ with $vE = \{w_1, \ldots, w_m\}$ we have $t'_v \itored{}{}{\overline{\PP'}} \{\tfrac{p_{w_1}}{p_v}:t'_{w_1}, \ldots, \tfrac{p_{w_m}}{p_v}:t'_{w_m}\}$.
  Let $W \subseteq V$ be the set of nodes $v$ where we have already defined the labeling $L'(x)$.
  During our construction, we ensure that for every $v \in W$ we have
  \begin{equation} \label{soundness-roi-induction-hypothesis}
    t_v \flateq t'_v \text{ and } \posT(t_v) \setminus \Junk(t_v) \subseteq \posT(t'_v).
  \end{equation}
  In this case, we define $\pi \in \Junk(t_v)$:$\Leftrightarrow$ there is no ADP $\ell \to \{p_1: r_1, \ldots, p_k: r_k\}^{m} \in \PP$ 
  and no substitution $\sigma$ such that $\annoEps(t_v|_{\pi}) \to_{\nonprob(\PP)}^* \ell^\sharp \sigma$.

  For the construction, we start with the same term at the root.
  Here, \eqref{soundness-roi-induction-hypothesis} obviously holds.
  As long as there is still an inner node $v \in W$ such that its successors are not contained in $W$, we do the following.
  Let $vE = \{w_1, \ldots, w_m\}$ be the set of its successors.
  We need to define the corresponding sets $t_{w_1}', \ldots, t_{w_m}'$ for the nodes $w_1, \ldots, w_m$.
  Since $v$ is not a leaf and $\F{T}$ is a $\PP$-CT, we have $t_v \itored{}{}{\PP} \{\tfrac{p_{w_1}}{p_v}:t_{w_1}, \ldots, \tfrac{p_{w_m}}{p_v}:t_{w_m}\}$.
  We have the following three cases:
  \begin{enumerate}
    \item[(A)] If it is a step with $\itored{}{}{\PP}$ using an ADP that is different from $\ell \ruleArr{}{}{} \{ p_1:r_{1}, \ldots, p_k: r_k\}^{m}$ in $\F{T}$, then we perform a rewrite step with the same ADP, the same redex, and the same substitution in $\F{T}'$.
      Analogous to Case (A) of the soundness proof for \Cref{theorem:inst-proc-ast}, we can show \eqref{soundness-roi-induction-hypothesis} for the resulting terms.
    \item[(B)] If it is a step with $\itored{}{}{\PP}$ using the ADP $\ell \ruleArr{}{}{} \{ p_1:r_{1}, \ldots, p_k: r_k\}^{m}$ at a position $\pi \notin \posT(t_v)$ in $\F{T}$, then we perform a rewrite step with $\ell \ruleArr{}{}{} \{ p_1:\anno_{\capt_1(\delta_1,\ldots,\delta_d)}(r_{1}), \ldots, p_k: \anno_{\capt_k(\delta_1,\ldots,\delta_d)}(r_k)\}^{m}$, same redex, same substitution, and same position in $\F{T}'$.
      Analogous to Case (B) of the soundness proof for \Cref{theorem:inst-proc-ast}, we can show \eqref{soundness-roi-induction-hypothesis} for the resulting terms.
      Note that the rule that we use contains fewer annotations than the original rule, but since $\pi \notin \posT(t_v)$, we remove all annotations from the rule during the application of the rewrite step anyway.
    \item[(C)] If it is a step with $\itored{}{}{\PP}$ using the ADP $\ell \ruleArr{}{}{} \{ p_1:r_{1}, \ldots, p_k: r_k\}^{m}$ at a position $\pi \in \posT(t_v)$ in $\F{T}$, then we look at specific successors to find a substitution $\delta$ such that $\ell \delta \ruleArr{}{}{} \{ p_1:r_{1}\delta, \ldots, p_k: r_k\delta\}^{m} \in N$ or we detect that we can use the ADP $\ell \ruleArr{}{}{} \{ p_1:\anno_{\capt_1(\delta_1,\ldots,\delta_d)}(r_{1}), \ldots, p_k: \anno_{\capt_k(\delta_1,\ldots,\delta_d)}(r_k)\}^{m}$ and perform a rewrite step with this new ADP, at the same position in $\F{T}'$.
  \end{enumerate}

  \noindent
  So it remains to consider Case (C) in detail.
  Here, we have $t_v \itored{}{}{\PP} \{\tfrac{p_{w_1}}{p_v}:t_{w_1}, \ldots, \tfrac{p_{w_k}}{p_v}:t_{w_k}\}$ using the ADP $\ell \ruleArr{}{}{} \{ p_1:r_{1}, \ldots, p_k: r_k\}^{m}$, the position $\pi \in \posT(t_v)$, and a substitution $\sigma$ such that $\flat(t_v|_{\pi}) = \ell \sigma \in \ANF_{\PP}$.

  We first consider the case where there is no successor $z$ of $v$ where an ADP is applied at an annotated position below or at $\pi$, or an ADP is applied on a position above $\pi$ before reaching such a node $v$.
  Then, we can use $\ell \ruleArr{}{}{} \{ p_1:\anno_{\capt_1(\delta_1,\ldots,\delta_d)}(r_{1}), \ldots, p_k: \anno_{\capt_k(\delta_1,\ldots,\delta_d)}(r_{k})\}^{m}$ instead, because the annotations will never be used, so they do not matter.

  Otherwise, there exists a successor $z$ of $v$ where an ADP is applied at an annotated position below or at $\pi$, and no ADP is applied on a position above $\pi$ before.
  Let $v_1, \ldots, v_n$ be all (not necessarily direct) successors that rewrite below position $\pi$, or rewrite at position $\pi$, and on the path from $v$ to $z$ there is no other node with this property, and no node that performs a rewrite step above $\pi$.
  Furthermore, let $t_1, \ldots, t_n$ be the used redexes and $\rho_1, \ldots, \rho_n$ be the used substitutions.
  \begin{itemize}
    \item \textbf{(C1) If} none of the redexes $t_1, \ldots, t_n$ are covered by $t$, then we use the ADP $\ell \ruleArr{}{}{} \{ p_1:\anno_{\capt_1(\delta_1,\ldots,\delta_d)}(r_{1}), \ldots, p_k: \anno_{\capt_k(\delta_1,\ldots,\delta_d)}(r_k)\}^{m}$ with the position $\pi \in \posT(t_v) \setminus \Junk(t_v) \subseteq_{(IH)} \posT(t'_v)$ and the substitution $\sigma$.
          Once again, \eqref{soundness-roi-induction-hypothesis} is satisfied for our resulting terms.

    \item \textbf{(C2) If} $t = t_i$ for some $1 \leq i \leq n$, then we can find a narrowing substitution $\delta_e$ of $t$ that is more general than $\sigma$, i.e., we have $\delta_e \gamma = \sigma$.
          Now, we use the ADP $\ell \delta_e \ruleArr{}{}{} \{ p_1:r_{1}\delta_e, \ldots, p_k: r_k\delta_e\}^{m}$ with the position $\pi \in \posT(t_v) \setminus \Junk(t_v) \subseteq_{(IH)} \posT(t'_v)$ and the substitution $\gamma$ such that $\flat(t_v|_{\pi}) = \ell \delta_e \gamma = \ell \sigma \in \ANF_{\PP}$.
          Once again, \eqref{soundness-roi-induction-hypothesis} is satisfied for our resulting terms.

    \item \textbf{(C3) If} $t \neq t_i$ for all $1 \leq i \leq n$ but there is an $1 \leq i \leq n$ such that $t_i$ is covered, then, since $t_i$ is covered, there exists a narrowing substitution $\delta_e$ of $t$ that is more general than $\rho_i$, i.e., there exists a substitution $\kappa_1$ with $\delta_e \kappa_1 = \rho_i$, and since we use $\rho_i$ later on we additionally have that $\rho_i$ is more general than $\sigma$, i.e., there exists a substitution $\kappa_2$ with $\rho_i \kappa_2 = \sigma$.
          Now, we use the ADP $\ell \delta_e \ruleArr{}{}{} \{ p_1:r_{1}\delta_e, \ldots, p_k: r_k\delta_e\}^{m}$ with the position $\pi \in \posT(t_v) \setminus \Junk(t_v) \subseteq_{(IH)} \posT(t'_v)$ and the substitution $\kappa_1 \kappa_2$ such that $\flat(t_v|_{\pi}) = \ell \delta_e \kappa_1 \kappa_2 = \ell \sigma \in \ANF_{\PP}$.
          Once again, \eqref{soundness-roi-induction-hypothesis} is satisfied for our resulting terms.
  \end{itemize}

  \medskip

  \noindent
  \underline{\emph{Completeness:}} The proof is analogous to the completeness proof of \cref{theorem:inst-proc-ast}.
  We can replace each ADP $\ell \delta_e \ruleArr{}{}{} \{ p_1:r_{1}\delta_e, \ldots, p_k:
  r_k\delta_e\}^{m}$ with the more general one $\ell \ruleArr{}{}{} \{ p_1:r_{1}, \ldots,
  p_k: r_k\}^{m}$, and each ADP $\ell \ruleArr{}{}{} \{
  p_1:\anno_{\capt_1(\delta_1,\ldots,\delta_d)}(r_{1}), \ldots,\linebreak p_k:
  \anno_{\capt_k(\delta_1,\ldots,\delta_d)}(r_k)\}^{m}$ can be replaced by $\ell
  \ruleArr{}{}{} \{ p_1:r_{1}, \ldots, p_k: r_k\}^{m}$ as well, leading to more
  annotations than before. 
\end{myproof}

\end{document}